\documentclass[12pt,twoside,english]{article}
\usepackage{ifthen}                
\newboolean{boldeqnitalic}
\setboolean{boldeqnitalic}{true} 

\usepackage{caption}
\usepackage{umlaute}
\usepackage[intlimits] {amsmath}   
\usepackage{defjs2}                
\usepackage{epsfig}
\usepackage{graphicx}
\usepackage{psfrag}
\usepackage{ifthen}
\usepackage{fancyhdr}
\usepackage{rotating}
\usepackage{multirow}
\usepackage{booktabs}              
\usepackage{amssymb}
\usepackage{amsbsy}
\usepackage{bm}                    
\usepackage{bbm}
\usepackage{babel}
\usepackage{theorem}
\usepackage{upgreek}  
\usepackage{pifont}   
\usepackage{pdfpages}
\usepackage{float}

\definecolor{mygray}{RGB}{140,140,140}
\definecolor{myorange}{RGB}{255,206,162}
\usepackage[authoryear]{natbib}
\bibpunct{[}{]}{,}{n}{}{;}
\fancypagestyle{plain}{%
  \fancyhead{}%
  \fancyfoot[c]{\sffamily\thepage}%
}
\makeatletter                           
\def\cleardoublepage{\clearpage\if@twoside \ifodd\c@page\else
  \hbox{}
  \vspace*{\fill}
  \thispagestyle{empty}
  \newpage
  \if@twocolumn\hbox{}\newpage\fi\fi\fi}
\makeatother
\renewcommand{\cite}[1]{\mbox{\citet{#1}}} 
\let\OrigBibitem\bibitem
\renewcommand{\bibitem}[2][]{\OrigBibitem[{#1}]{#2}}

\begin{document}
\unitlength1.0cm
\frenchspacing
\thispagestyle{empty}
\ce{\bf \large
The exponentiated Hencky energy:
}
\ce{\bf \large
Anisotropic extension and case studies}

\vspace{4mm}
\ce{J\"org Schr\"oder$^1$ and Markus von Hoegen$^1$ and Patrizio Neff$^2$}

\vspace{4mm}
\ce{$^1$Institut f\"ur Mechanik, Fakult\"at f\"ur Ingenieurwissenschaften 
    / Abtl. Bauwissenschaften}
\ce{Universit\"at Duisburg-Essen, 45141 Essen, Universit\"atsstr. 15,
Germany}
\ce{\small e-mail: j.schroeder@uni-due.de,
    phone: +49 201 183 2708,
    fax: +49 201 183 2680}
    \ce{\small e-mail: markus.von-hoegen@uni-due.de,
    phone: +49 201 183 3091,
    fax: +49 201 183 2680}

\vspace{4mm}
\ce{$^2$Lehrstuhl f\"ur Nichtlineare Analysis und Modellierung, Fakult\"at f\"ur Mathematik}
\ce{Universit\"at Duisburg-Essen, 45127 Essen, Thea-Leymann-Stra{\ss}e 9, Germany}
\ce{\small e-mail: patrizio.neff@uni-due.de,
    phone: +49 201 183 4243,
    fax: +49 201 183 4394}

\vspace{4mm}
\begin{center}
{\bf \large Abstract}
\bigskip

{\footnotesize
\begin{minipage}{14.5cm}
\noindent
In this paper we propose an anisotropic extension of the isotropic exponentiated Hencky energy, based on logarithmic
strain invariants. Unlike other elastic formulations, the isotropic exponentiated Hencky elastic energy has been derived solely on differential geometric grounds,
involving the geodesic distance of the deformation gradient $\bF$ to the group of rotations. We formally extend this approach 
towards anisotropy by defining additional anisotropic logarithmic strain invariants with the help of suitable structural tensors and consider our findings for 
selected case studies.
\end{minipage}
}
\end{center}

{\bf Keywords:}
Hencky energy, logarithmic strain tensor, anisotropy, strain invariants,\\ \hspace*{2.2cm} biomechanics

\sect{\hspace{-5mm}. Introduction \label{sec: intro}}
\vspace{-4mm}

In this article we consider a novel Hencky-type hyperelasticity model, the exponential Hencky-logarithmic
strain energy proposed by \cite{NefGhiLan:2015:teh}, \cite{NefLanGhiMarSte:2015:teh} and \cite{NefGhi:2016:teh}. Here, we focus on an extension to anisotropy in a coordinate
invariant setting. Therefore, we apply the concept of structural tensors and introduce additional mixed invariants.
The flexibility of the proposed formulation is demonstrated by identifying the linearized fourth-order elasticity 
tensor with the well-known coordinate dependent representations. Furthermore, we propose an anisotropic exponential
Hencky model suitable for the description of soft biological tissues. The performance of this model is demonstrated
by the analysis of a patient specific artery. \\

The modeling of anisotropic soft tissues in the framework of nonlinear elasticity has made considerable progress in the last decades. 
From the mathematical side, the polyconvexity condition introduced by John Ball in his seminal paper \cite{Bal:1977:cca} is a strong mathematical
requirement which implies Legendre-Hadamard ellipticity (rank-one convexity) at all deformation gradients $\bF$. In the early time after its introduction, 
polyconvexity was exclusively used in the isotropic setting and it was unclear how to extend the framework to anisotropy. In \cite{HarNef:2003:etf} a large variety of isotropic 
strain energy functions have been discussed. Two of the present authors have been able to solve one of Ball's major open problems, see \cite{Bal:2002:sop}, namely the meaningful 
application of polyconvexity to anisotropic materials. In a series of papers, \cite{SchNef:2003:ifo}, \cite{SchNefBal:2005:ava}, \cite{BalNefSchHol:2006:apf}, \cite{SchNefEbb:2008:ape}, \cite{EbbSchNef:2009:aoa}, 
the concept has been elaborated in detail, with papers from other authors 
following in due course, see e.g. \cite{ItsEhrMav:2006:apa}, \cite{EhrIts:2007:aph}.
It is, however, clear that polyconvexity (or ellipticity) alone is not sufficient to characterize physically reasonable material response: there exist polyconvex strain energies with unacceptable
non-monotone pressure-compression relation. Moreover, the identification 
of material parameters in the above proposed anisotropic extensions heavily relies on nonlinear optimization strategies after which the physical meaning of the obtained parameters is doubtful: a completely different 
set of material parameters may equally well fit the available experimental data. This already occurs for the isotropic Ogden-model, see \cite{Ogd:1972:ldi}. 
The situation for the anisotropic response can only be worse in general. Therefore, the need is to construct strain-energy functions whose possibly few parameters have a clear
physical meaning and which are uniquely and easily identified from experiments. At the same time the proposed strain energy should be Legendre-Hadamard elliptic at least in that range
of deformation which is typically encountered in the applications. In this paper we numerically explore such a formulation based on the well known logarithmic Hencky strain. \\

In 1928 Heinrich \cite{Hen:udf:1928} proposed the after him named strain-energy 
function $W_{\rm H}$ for finite isotropic elasticity. He replaced the small strain tensor
$\Bvarepsilon$ in classical linear isotropic energy by the Hencky or logarithmic
strain measure $\log \bU$, with the right stretch tensor $\bU$. For moderate deformations, this simple function $W_{\rm H}$ with the two classical Lam\'e constants is
useful for a wide class of materials, see {\cite{Ana:1979:ohh,Ana:1986:mdi} and \cite{BruXiaMey:2000:hem}.
However, Hencky's energy function is not rank-one convex, i.e. it does not fulfill the well known
Legendre-Hadamard, or ellipticity, condition. \cite{BruXiaMey:2001:cif} derived necessary and sufficient conditions for
ellipticity in terms of principal stretches and computed the largest common ellipticity region. They showed, in the case 
of positive Lam\'e constants, that $W_{\rm H}$ is elliptic whenever every principal stretch is in the range $[0.21162...,~1.39561...]$.
Furthermore, Hencky's strain-energy automatically satisfies the Baker-Ericksen inequality (\cite{BakEri:1954:irt}) and Hill's inequality (\cite{Hil:1968:oci,Hil:1970:cif}), see \cite{BruXiaMey:2001:cif}, \cite{GhiNefMar:2015:aed}.

\subsection{The exponentiated Hencky energy}
The exponentiated Hencky-logarithmic model was recently introduced by \cite{NefGhiLan:2015:teh}. It is induced by the \emph{exponentiated Hencky strain energy}
\begin{align}
\nonumber {W}_\mathrm{eH}(\bm{F}) &=
             \frac{\mu}{k} \exp \left[ k \,\lVert\mbox{dev}_n \log\bU\rVert^2 \right]
           + \frac{\kappa}{2 \hat{k}} \exp \left\{ \hat{k} \,[\tr(\log\bU)]^2 \right\}\\
\nonumber &= \frac{\mu}{k}\exp \left[ k \,\biggl\lVert \log\frac{\bU}{(\det\bU)^{1/n}}\biggr\rVert^2 \right]  + \frac{\kappa}{2 \hat{k}} \exp \left[ \hat{k} \,(\log \det\bm{U})^2 \right] \,
\end{align}
where $\mu>0$ is the (infinitesimal) \emph{shear modulus}, $\kappa>0$ is the \emph{bulk modulus}, $k$ and $\hat{k}$ are additional dimensionless material parameters, $\bm{U}=\sqrt{\bm{F}^T\bm{F}}$ is the \emph{right stretch tensor} corresponding to the \emph{deformation gradient} $\bm{F}$, $\log$ denotes the \emph{principal matrix logarithm} on the set of positive definite symmetric matrices, $\mbox{dev}_n \bm{X} = \bm{X}-\frac{1}{n}\mbox{tr} \bm{X}\bm{1}$ and $\lVert \bm{X} \rVert = \sqrt{\mbox{tr}(\bm{X}^T\bm{X}})$ are the \emph{deviatoric part} and the \emph{Frobenius matrix norm} of an $n\times n$-matrix $\bm{X}$, respectively, and $\mbox{tr}$ denotes the \emph{trace operator}.

The exponentiated Hencky energy is based on the so-called \emph{volumetric and isochoric logarithmic strain measures}
$\omega_{\textrm{iso}} = \lVert\mbox{dev}_n \log\bm{U}\rVert$ and $\omega_{\textrm{vol}} = \lvert\mbox{tr}\log\bm{U}\rvert = \vert\log\det \bm{U}\vert\,,$ respectively
which have recently been given a purely geometric characterization not shared by any other set of isotropic invariants (\cite{NefEidMar:2016:gol}): consider the general linear group $\mathrm{GL}(n)$ endowed with the canonical left-invariant Riemannian metric $g$, which for $\bm{A}\in\mathrm{GL}(n)$ and $\bm{X},\bm{Y}\in \mathfrak{gl}(n) = T_{\bm{A}}\mathrm{GL}(n) \cong \mathbb{R}^{n\times n}$ is given by \cite{agn_martin2014minimal}
\[
	g_{\bm{A}}(\bm{X},\bm{Y}) = \langle \bm{A}^{-1}\bm{X},\, \bm{A}^{-1}\bm{Y} \rangle
\]
where $\langle \bm{X},\bm{Y} \rangle = \tr(\bm{X}\bm{Y}^T)$ is the canonical inner product on the space of real $n\times n$-matrices. Then the logarithmic strain measures are the geodesic distance of the \emph{isochoric} part $\frac{\bm{F}}{\det\bm{F}^{1\!/\!n}}$ and the \emph{volumetric} part $(\det\bm{F})^{1\!/\!n}\bm{1}$ to $\bm{1}$ of the deformation gradient to the special orthogonal group $\mathrm{SO}(n)$, respectively, see \cite{NefEidMar:2016:gol} (Theorem 3.7):
\begin{align}\label{geoprop}
\lVert\mbox{dev}_n \log \bm{U}\rVert &= {\rm dist}_{{\rm geod}}\left( \frac{\bm{F}}{(\det \bm{F})^{1\!/\!n}}, {\rm SO}(n)\right)\,,\notag\\ 
\lvert\log \det \bm{F}\rvert &= {\rm dist}_{{\rm geod}}\left((\det \bm{F})^{1\!/\!n}\cdot \bm{1}, {\rm SO}(n)\right)\,.
\end{align}
These two quantities are thereby identified as the ``natural'' measures of strain in any deformation, an observation which strongly suggests that an idealized elastic strain energy function may depend on these quantities alone.\footnote{Note that not every objective and isotropic energy function can be expressed in terms of the logarithmic strain measures alone, see \cite{NefEidMar:2016:gol}, whereas every such energy can be expressed in terms of the logarithmic strain tensor $\log\bm{U}$.} An important example of such an energy function is the classical \emph{quadratic Hencky energy}
\begin{align}
\nonumber  W_\mathrm{H}(\bF) &=
           \mu\,\lVert\mbox{dev}_n \log\bU\rVert^2
           + \frac{\kappa}{2}\, [\tr (\log \bU)]^2\\
           & = \mu \biggl\lVert\log \frac{\bU}{(\det \bU)^{1/n}} \biggr\rVert^2 + \frac{\kappa}{2}\, (\log \det\bm{U})^2\,,
\end{align}
which was introduced by Heinrich Hencky in 1929 (\cite{Hen:1929:wub,agn_neff2014axiomatic}). While the elasticity model induced by the Hencky energy is in very good agreement with experimental observations for up to moderate strains for a large number of materials \cite{NefGhiLan:2015:teh,Ana:1979:ohh}, there are some major shortcomings of this model. For example, the qualitative behavior of materials under very large deformations is not modeled accurately, and since the energy function is neither polyconvex nor quasiconvex or rank-one convex (\cite{agn_neff2000diss}, \cite{GhiNefMar:2015:aed}), no known methods are available to ensure the existence of energy minimizers for general boundary value problems.
Moreover, the pressure-compression relation is not monotone.

In order to alleviate some of these shortcomings, Neff et al.\ introduced the exponentiated Hencky energy ${W}_\mathrm{eH}$ in a series of articles \cite{NefGhiLan:2015:teh,NefLanGhiMarSte:2015:teh}, \cite{NefGhi:2016:teh}, \cite{agn_ghiba2015exponentiated}. This energy function closely approximates the classical quadratic Hencky energy for small deformations, but aims to provide a more accurate model for large deformations as well as an improvement in terms of common constitutive requirements; for example, ${W}_\mathrm{eH}$ is polyconvex in the two-dimensional case \cite{NefLanGhiMarSte:2015:teh}, and in the three-dimensional case the rank-one convexity domain contains the 
extremely large set  $\{\bF\in {\rm GL}^+(3)|~\lVert\dev_3 \log\bU \rVert\leq6\}$.
Moreover, the induced mapping $\bm{B}\mapsto\bm{\sigma}$ of the Finger tensor $\bm{B}=\bm{F}\bm{F}^T$ to the Cauchy stress tensor $\bm{\sigma}$ is invertible \cite{agn_neff2016injectivity,MihNef:2016:hbu,MihNef:2016:hbu2,jog2013}, as is the case for suitable variants of the Neo-Hooke and Mooney-Rivlin energies for slightly compressible materials like rubber.

The low number of additional material parameters in the exponentiated Hencky model also suggests that a good material fitting could be possible even without extensive experimental measurements. Additionally, the exponentiated Hencky energy allows for the modeling of a zero apparent Poisson's modulus $\nu=\frac{3\,\kappa-2\,\mu}{2(3\,\kappa+\mu)}$ in the finite strain regime: if the additional parameters $k,\hat{k}$ are chosen such that $k=\frac23\,\hat{k}$, then ${W}_\mathrm{eH}$ can be written as
\[
	\frac{1}{2\,k}\,\left( \frac{E}{1+\nu}\,\exp \biggl[ k \,\lVert\mbox{dev}_n \log\bm{U}\rVert^2 \biggr] + \frac{E}{2\,(1-2\,\nu)}\,\exp \biggl[ \frac23\,k \,(\log \det\bm{U})^2 \biggr] \right)\,,
\]
where $E = \frac{9\,\kappa\,\mu}{3\,\kappa + \mu}$ is \emph{Young's modulus}, and for $\nu=0$ we obtain a model with zero lateral contraction under finite strains \cite{NefGhiLan:2015:teh}.

A variant of the exponentiated Hencky energy has previously been applied to so-called \emph{tire derived materials} and was found to be in good agreement with experimental data, see \cite{MonGovNef:2016:teh}. In particular, the highly nonlinear equation of state (EOS) relating pressure to purely volumetric deformations has been captured extraordinarily well.
The extra appearing non-dimensional parameters $k$ and $\hat{k}$ have an intuitive meaning: larger $k$, $\hat{k}$ lead to monotonically increased strain hardening. In principal these parameters can be fitted independent of the shear and bulk modulus. 
Next, we extend the exponential Hencky energy to the anisotropic case.
\section{\hspace{-5mm}. Theoretical framework}
\vspace{-4mm}
\subsection{Kinematics}

\begin{table}[htb]
\centering
\caption{Kinematic and constitutive quantities.}
\begin{tabular}{p{4.0cm}p{7.0cm}}
\toprule
Symbol & Continuum mechanical description \\
\midrule
$\bu$ & displacement vector \\
$\boldsymbol{F}=\bone + {\rm Grad}\bu$ & deformation gradient\\
$\boldsymbol{C}=\bF^T\bF$ & right Cauchy-Green tensor\\
$\boldsymbol{B}=\bF\bF^T$ & left  Cauchy-Green tensor\\
$\boldsymbol{U}=\sqrt{\bF^T\bF}$ & right stretch tensor\\
$\boldsymbol{V}=\sqrt{\bF\bF^T}$ & left  stretch tensor\\
$\log\bU$ & right Hencky strain tensor\\
$\log\bV$ & left  Hencky strain tensor\\
\midrule
$\psi$ &  generic elastic energy\\
$W_{\rm H}$ & isotropic Hencky energy \\
${W}_{\rm eH}$ & exponentiated Hencky energy\\
\midrule
$\boldsymbol{S} = 2\partial_{\bC} \psi(\bC)$ & second Piola-Kirchoff stress tensor \\
$\boldsymbol{\tau} = \partial_{\log\bV}\psi(\log\bV)$ & Kirchoff stress tensor (see \cite{Val:1978:ldc}) \\
$\boldsymbol{\sigma} = (\det\bF)^{-1}\Btau$ & Cauchy stress tensor\\
\bottomrule
\end{tabular}
\label{tab:kinematical}
\end{table}

For a better overview, the 
continuum-mechanical kinematic and constitutive quantities are listed in Table 
\ref{tab:kinematical}.
Let ${\cal B} \subset \IR^3$ be the body of interest in the
reference placement, parametrized in  $\bX$, and
let ${\cal S}$ be the body in the current placement,
parametrized in $\bx$.
The boundary $\partial{\cal B}$ of ${\cal B}$ is decomposed
in $\partial{{\cal B}_u}$ and $\partial{{\cal B}_t}$ with
$\partial{{\cal B}_u} \cup \partial{{\cal B}_t} = \partial{\cal B}$ 
and $\partial{{\cal B}_u} \cap \partial{{\cal B}_t} = \emptyset $.
The nonlinear deformation map is given by
$ \bx= {\Bvarphi} (\bX)$.
As basic kinematical quantities we define the deformation
gradient and the right Cauchy-Green tensor
\eb
\bF =\Grad { {\Bvarphi} (\bX)}
\qquad\mbox{and}\qquad
\bC=\bF^T\bF = \bU^2\;,
\ee
respectively. Here, $\bone$ denotes the second-order
identity tensor. The Jacobian of the deformation
gradient has to satisfy $J:= \det\bF > 0$.
The deformation gradient may be split into 
\eb
\bF = \bR\,\bU = \bV\bR\,,
\ee
where $\bR \in {\rm SO(3)}$ denotes a pure rotation tensor and $\bU$ and $\bV$
are the right and left stretch tensors, respectively.
In order to fulfill the principle of material
frame indifference a priori, 
we formulate the generic free energy function~$\psi$ in terms
of the right
Cauchy-Green tensor, i.e. $\psi = \psi (\bC)$.
In spectral decomposition the right Cauchy-Green tensor $\bC$ and the left Cauchy-Green tensor $\bB$ may be written as
\eb
\bC = \sum_{k=1}^3 \widehat{\lambda}_k \bN^{k}\otimes\bN^{k}\,, \qquad
\bB = \sum_{k=1}^3 \widehat{\lambda}_k \bn^{k}\otimes\bn^{k}\,
\ee
where $\widehat{\lambda}_k$ denote the eigenvalues of $\bC$ and $\bB$. The
eigenvectors  are expressed through $\bN^k$ and $\bn^k$
associated to $\bC$ and $\bB$, respectively.
Therefore, we obtain the tensor functions
\begin{align}
\bU &=\sqrt{\bC} =\sum_{k=1}^3 \sqrt{\widehat{\lambda}_k}\, \bN^{k}\otimes\bN^{k} = \sum_{k=1}^3 {{\lambda}_k}\, \bN^{k}\otimes\bN^{k}\,,\\
\bV &=\sqrt{\bB} = \sum_{k=1}^3 \sqrt{\widehat{\lambda}_k}\, \bn^{k}\otimes\bn^{k}~~ = \sum_{k=1}^3 {{\lambda}_k}\, \bn^{k}\otimes\bn^{k}\,,\\
\log\bU &=\log(\sqrt{\bC}) = \frac{1}{2}\log\bC \stackrel{!}{=}\sum_{k=1}^3 \frac{1}{2}\log{\widehat{\lambda}_k}\, \bN^{k}\otimes\bN^{k}  \stackrel{!}{=} \sum_{k=1}^3 \log{{\lambda}_k}\, \bN^{k}\otimes\bN^{k}\,, \\
\log\bV &=\log(\sqrt{\bB}) = \frac{1}{2}\log\bB \stackrel{!}{=} \sum_{k=1}^3 \frac{1}{2}\log{\widehat{\lambda}_k}\, \bn^{k}\otimes\bn^{k} ~~\stackrel{!}{=} \sum_{k=1}^3 \log{{\lambda}_k}\, \bn^{k}\otimes\bn^{k}\,,
\end{align}
where ${\lambda}_k$ denote the eigenvalues of $\bU$ and $\bV$.
The tensor $\log{\bU}$ is called right Hencky strain tensor.

\subsection{Stress measures}
Let $\partial_{\log\bU}\widehat{\psi}(\log\bU)$ be the stress measure work conjugate to
$\log\bU=\frac{1}{2}\log\bC$, then the transformation rule for the second Piola-Kirchhoff stress tensor $\bS$ is given by
\begin{align}
\nonumber &\bS = 2\pp{\widehat{\psi}(\log\bU)}{\bC}=2\frac{\partial \widehat{\psi}(\log\bU)}{\partial \log\bU} : \pp{\log\bU}{\bC}= \frac{\partial \widehat{\psi}(\log\bU)}{\partial \log\bU} : \mathbb{P}_{\rm H} \\
&\mbox{with} \quad  \mathbb{P}_{\rm H} = 2\frac{\partial \log\bU}{\partial \bC }\,. \label{eq: 2nd PK}
\end{align}
The fourth-order tensor $\mathbb{P}_{\rm H}$ can only be derived with the help of the spectral decomposition and yields
\begin{equation}
\boxed{
\begin{aligned}
\mathbb{P}_{\rm H}  = &\sum_{k=1}^3 \sum_{j=1}^3 P_{kkjj} \bN^k\otimes\bN^k\otimes\bN^j\otimes\bN^j  \\
                      &+ 2\sum_{k=1}^3 \sum_{k\ne j}^3 P_{kjkj} (\bN^k\otimes\bN^j)\otimes(\bN^k\otimes\bN^j + \bN^j\otimes\bN^k) \\
     P_{kkjj} =       &\, \delta_{kj}\, \widehat{\lambda}_k^{-1} \\
     P_{kjkj} =       &\begin{cases} 
                      \frac{\frac{1}{2}{\rm log}\widehat{\lambda}_k - \frac{1}{2}{\rm log}\widehat{\lambda}_j}{\widehat{\lambda}_k-\widehat{\lambda}_j} \quad&\mbox{for}~\widehat{\lambda}_k\ne\widehat{\lambda}_j \\
                      \lim\limits_{\widehat{\lambda}_k \rightarrow \widehat{\lambda}_j}{\frac{\frac{1}{2}\log\widehat{\lambda}_k-\frac{1}{2}\log\widehat{\lambda}_j}{\widehat{\lambda}_k-\widehat{\lambda}_j}}=:\partial_{\widehat{\lambda}_k}(\frac{1}{2}\log\widehat{\lambda}_k)= \left(2\widehat{\lambda}_k\right)^{-1} \quad &\mbox{for}~\widehat{\lambda}_k = \widehat{\lambda}_j
                      \end{cases}     \label{eq: material tangent}
\end{aligned}
}
\end{equation}
see also \cite{Ogd:1997:nle} and \cite{Sim:1998:naa}.
The first part of $\mathbb{P}_{\rm H}$ is related to the derivative of the eigenvalues of $\log\bU$ with respect to $\bC$, while the second part is related to
$\partial_{\bC}(\bN^k\otimes\bN^k)$.
In the isotropic case the following relations regarding the Kirchoff stress
\begin{align}
\nonumber \Btau &= \pp{\widetilde{\psi}(\log \bV)}{\log \bV} = \bR\pp{\widehat{\psi}(\log \bU)}{\log\bU}\bR^T =  \bR\pp{\psi^\#(\bU)}{\bU}\bU\bR^T \\
                &= 2\bF\pp{\bar{\psi}(\bC)}{\bC}\bF^T =2\bB\pp{\psi^+(\bB)}{\bB} = \pp{\psi(\bF)}{\bF}\bF^T
\end{align}
hold true. But if anisotropic behavior is considered only the relations
\eb
\Btau = 2\bF\pp{\bar\psi(\bC)}{\bC}\bF^T = \pp{\psi(\bF)}{\bF}\bF^T
\ee
remain valid. For the derivation of the above mentioned relations the reader is referred to the appendix.
The expression $\bT_{\rm Biot}=\partial_{\bU}{\psi^\#(\bU)}$ is also known as Biot-stress.
For the linearization of the weak form, we need the tangent moduli 
\eb
\mathbb{C} =4\frac{\partial^2\widehat{\psi}(\log\bU)}{\partial\bC\partial\bC} =\mathbb{P}_{\rm H}:\mathbb{C}^{\rm H}:\mathbb{P}_{\rm H} + \,\pp{\widehat\psi(\log\bU)}{\log\bU}:\mathbb{K} \label{eq: tangent}
\ee
with
\eb
\mathbb{C}^{\rm H} = \frac{\partial^2  \hat{\psi}(\log\bU)}{\partial \log\bU \partial \log\bU} \quad \mbox{and} \quad
\mathbb{K}= 2 \frac{\partial \mathbb{P}_{\rm H}}{\partial\bC} = 4 \frac{\partial^2\log\bU}{\partial\bC\partial \bC}\;.
\ee
 The multiplicative volumetric isochoric decomposition of the deformation gradient
\eb
\widetilde{\bF} = J^{-1/3}\bF \quad \mbox{and} \quad \widetilde{\bC} = J^{-2/3} \bC 
\ee
was first proposed by Hans \cite{Ric:1948:die}, see also \cite{Flo:1961:trf}.
In doing so we can express the volumetric Hencky strain tensor with help of an additive split according to
\eb
{\log\bU} = \dev(\log\bU) + \frac{1}{3}\tr (\log\bU) \bone\,, \quad \mbox{with} \quad {\rm tr} (\dev{\log\bU}) = 0\,.  
\ee
For the numerical treatment of an energy function $\psi^*(\dev{\log\bU})$ we need the derivative
\eb
\pp{\dev{\log\bU}}{\log\bU}  = \bone\boxtimes\bone - \frac{1}{3}\bone\otimes\bone =\IP\,,
\ee
where $\boxtimes$ denotes the Kronecker product of second-order tensors. Let $\bG$ and $\bH$ denote two second-order tensors and $\bg$ and $\bh$ 
two first-order tensors, then the operator is defined by $(\bG\boxtimes\bH):(\bg\otimes\bh)=(\bG\bg)\otimes(\bH\bh)$.
Formulating a strain energy in $\dev{\log\bU}$, before projecting the stress tensor and tangent moduli on $\bC$ we first need to project them on
the Hencky strain $\log\bU$.
Therefore, we define 
\eb
\pp{\psi^*(\dev\log\bU)}{\log\bU} = \pp{\psi^*(\dev\log\bU)}{\dev\log\bU} : \pp{\dev\log\bU}{\log\bU} =  \pp{\psi^*(\dev\log\bU)}{\dev\log\bU} :\IP \,,
\ee
and for the linearization of the weak form it follows
\eb
\mathbb{C}^{\rm H} =\frac{\partial^2\psi^*(\dev\log\bU)}{\partial \log\bU \partial\log\bU} =\IP:\widetilde{\mathbb{C}}^{\rm H}:\IP\,,
\ee
with
\eb
\widetilde{\mathbb{C}}^H = \frac{\partial^2\psi^*(\dev\log\bU)}{\partial\dev\log\bU\partial\dev\log\bU}\,.
\ee
The corresponding tensors are to be inserted in Eq.~(\ref{eq: 2nd PK}) and Eq.~(\ref{eq: tangent}).

\subsection{Isotropic and anisotropic invariants \label{sec: iso aniso invarianten}}
The principal isotropic invariants of the right Cauchy-Green tensor $\bC$ are given by
\begin{align}
\nonumber I_1^{\rm C} &= {\rm tr} [\bC] = \lVert \bF\rVert^2 \:,
\quad     I_2^{\rm C} = {\rm tr}[{\rm Cof}\, \bC] = \lVert\Cof\bF \rVert^2
\quad\mbox{and} \\
          I_3^{\rm C}&={\rm det}\, \bC = (\det\bF)^2= J^2 \;.
\label{equ:I}
\end{align}
Further, we introduce the basic invariants of the Hencky strain tensor $\log\bU$
\begin{align}
\nonumber J_1^{\rm H}  &= \log(\det\bU)= {\rm tr} (\log\bU)\,, \quad J_2^{\rm H} = \lVert \log\bU \rVert^2 = {\rm tr} [(\log\bU)^2] \quad\mbox{and}\\
          J_3^{\rm H} &= {\rm tr} [(\log\bU)^3]\,,
\end{align}
already used by \cite{Ric:1948:die}.
Let $\bA$, with $\lVert \bA\rVert = 1$, be the preferred direction of the transversely isotropic material, then the material symmetry group
is defined by 
\eb
{\cal G}_{ti}\,:=\,\{\pm\bone;\,\bQ(\alpha,\bA)\,|\,0<\alpha<2\pi\}\,,
\ee
where $\bQ(\alpha,\bA)$ are all rotations along the $\bA$-axis. The structural tensor $\bM$
whose invariance group preserves the material symmetry group ${\cal G}_{ti}$ is given by the 
rank-one tensor 
\eb
\bM = \bA\otimes\bA\,,
\ee
see \cite{Boe:1978:ldc} and \cite{Boe:1979:asd} regarding the concept of structural tensors.
Based on the structural tensor we define the mixed invariants
\begin{equation}
\begin{alignedat}{2}
          I_4^{{\rm C}^i} &= \langle \bC^i,\bM\rangle \,,  \qquad &&J_5^{\rm C} = \langle{\rm Cof}\, \bC,\bM\rangle\,,  \\
          I_4^{{\rm H}^i} &= \langle (\log\bU)^i,\bM\rangle \,, \qquad &&J_5^{\rm H} = \langle\log(\Cof\bU),\bM\rangle\,,  \label{eq: I4Hi} 
\end{alignedat}
\end{equation}
where $i\in \mathbb{N}$, $i>0$, denotes an exponent.
Note that the cofactor  ${\rm Cof}(\log\bU)$ has no physical meaning and that for $\bU \in {\rm Sym}^+(3)$, $\Cof\bU$ is also positive definite. Because of that we instead consider
the logarithmic cofactor function
\begin{align}
\nonumber \log(\Cof\bU) =&\, {\rm log} \left[({\rm det}\bU)\bU^{-1}\right] = \sum_{k=1}^3{\rm log} \left(\frac{{\rm det}\sqrt{\bC}}{\widehat{\lambda}_k^{1/2}}\right) \bN^k\otimes\bN^k \\
\nonumber               =& \sum_{k=1}^3 \left[{\rm log} ({\rm det}\sqrt{\bC})- \frac{1}{2}{\rm log}\widehat{\lambda}_k\right]  \bN^k\otimes\bN^k\\
\nonumber               =&\, (\frac{1}{2}{\rm log}\widehat{\lambda}_2+\frac{1}{2}\log\widehat{\lambda}_3)\bN^1\otimes\bN^1 + (\frac{1}{2}\log\widehat{\lambda}_1+\frac{1}{2}\log\widehat{\lambda}_3)\bN^2\otimes\bN^2 \\
\nonumber                &+ (\frac{1}{2}\log\widehat{\lambda}_1+\frac{1}{2}\log\widehat{\lambda}_2)\bN^3\otimes\bN^3\\
                        =&\,  {\rm tr}(\log\bU)\bone-\log\bU 
\end{align}
and finally we observe the following
\eb
J_5^{\rm H} = \langle\log(\Cof\bU),\bM\rangle = \tr(\log\bU)\underbrace{\langle\bM,\bone\rangle}_1-\langle\log\bU,\bM\rangle = J_1^{\rm H}-I_4^{\rm H^1}\,.
\ee

\section{Isotropic strain energy functions}
\subsection{\hspace{-5mm}. Isotropic Hencky Energy}
The isotropic Hencky energy was  introduced in \cite{Hen:1929:wub}. It measures the geodesic distance
of the deformation gradient to the special orthogonal group ${\rm SO}(n)$, as it was discovered in \cite{Nef:EidOstMar:2013:ths}. The Hencky strain energy
\eb
 {W}_{\rm H} (\log\bU) = \mu \left\norm{\mbox{dev} \log\bU\right}^2 + \frac{\kappa}{2} \left[\tr (\log\bU)\right]^2 =  \mu\norm{\log\bU}^2 + \frac{\lambda}{2} [\tr(\log\bU)]^2    \label{eq: iso Hencky}         
\ee
can be reformulated in  principal logarithmic strains 
\begin{align}
\nonumber  {W}_{\rm H} (\log\bU) =\, &\mu \left[\left(\frac{1}{2}\log\widehat{\lambda}_1\right)^2 + \left(\frac{1}{2}\log\widehat{\lambda}_2\right)^2 + \left(\frac{1}{2}\log\widehat{\lambda}_3\right)^2\right] \\
                                        &+ \frac{\lambda}{2}  \left[\frac{1}{2}\log\widehat{\lambda}_1 + \frac{1}{2}\log\widehat{\lambda}_2 + \frac{1}{2}\log\widehat{\lambda}_3\right]^2
\label{eq:energy_eigen_isoH}
\end{align}
based on the eigenvalues $\hat{\lambda}$ of $\bC$,
where the Lam\'{e} parameters $\lambda$ and  $\mu$ as well as the bulk modulus  $\kappa$ are used. Note that $\kappa = \frac{3 \lambda + 2 \mu}{3}$ 
and
\begin{align}
\nonumber \left\norm{\dev \left(\log \bU\right)\right}^2 \,=\,& \frac{1}{3}\left[\left(\log\sqrt{\frac{{\widehat{\lambda}_1}}{{\widehat{\lambda}_2}}}\right)^2+\left(\log\sqrt{\frac{\widehat{\lambda}_1}{\widehat{\lambda}_3}}\right)^2+\left(\log\sqrt{\frac{\widehat{\lambda}_2}{\widehat{\lambda}_3}}\right)^2\right]\\
\nonumber                                                  =\,& \frac{2}{3}\left[\left(\frac{1}{2}\log\widehat{\lambda}_1\right)^2+\left(\frac{1}{2}\log\widehat{\lambda}_2\right)^2+\left(\frac{1}{2}\log\widehat{\lambda}_3\right)^2\right] \\
                                                &-\frac{2}{3}\left[\frac{1}{4}\log\widehat{\lambda}_1\log\widehat{\lambda}_2 + \frac{1}{4}\log\widehat{\lambda}_1\log\widehat{\lambda}_3+\frac{1}{4}\log\widehat{\lambda}_2\log\widehat{\lambda}_3\right]   \,.
\end{align}
The function ${W}_{\rm H}$ 
is not polyconvex, not quasiconvex, not coercive and not rank-one-elliptic, even for every admissible deformation state, see  \cite{NefGhiLan:2015:teh}. 
However, it holds that ${W}_{\rm H}(\bF) = W_{\rm H}(\bF^{-1})$.
The first and second derivative with respect to the Hencky strain yield 
\begin{align}
\frac{\partial {W}_{\rm H}}{\partial \log\bU} &= 2 \mu\,\mbox{dev} \left(\log\bU \right) + \kappa\,{\rm tr}(\log\bU) \bone \,, \label{eq:dpsi_isoHencky} \\
\mathbb{C}^{\rm H} &= \frac{\partial^2 {W}_{\rm H}}{\partial \log\bU \partial \log\bU} = 2\mu\,\IP + \kappa\,\bone\otimes\bone \,. \label{eq:d2psi_isoHencky} 
\end{align}

In the reference configuration with $\bC = \bone$ the final material tangent $\IC$, in Voigt-notation\footnote{In the contracted notation the tensorial indices are allocated to the matrix indexes as follows $\{11,22,33,12,23,13\}\rightarrow\{1,2,3,4,5,6\}$.} denoted as $\IC^V$, according to Eq.~(\ref{eq: tangent}) simplifies to 
\eb
\IC^{\rm V}|_{\bC=\bone}=4\frac{\partial^2{W}_{\rm H}}{\partial\bC\partial\bC}\Big|_{\bC=\bone} =  \begin{pmatrix}
\kappa+\frac{4}{3}\mu & \kappa-\frac{2}{3}\mu & \kappa-\frac{2}{3}\mu & 0 & 0& 0\\
\kappa-\frac{2}{3}\mu & \kappa+\frac{4}{3}\mu & \kappa-\frac{2}{3}\mu & 0 & 0& 0\\
\kappa-\frac{2}{3}\mu & \kappa-\frac{2}{3}\mu & \kappa+\frac{4}{3}\mu & 0 & 0& 0\\
0 & 0 & 0 & \mu &0 &0\\
0 & 0 & 0 &0 &\mu  &0\\
0 & 0 & 0 &0 &0 &\mu 
\end{pmatrix}\,. \label{eq: elasti_W_H}
\ee

\subsection{\hspace{-5mm}. Exponentiated Hencky energy}

The exponentiated Hencky energy 
\begin{align}
\nonumber {W}_{\rm eH} &= \frac{\mu}{k} {\rm exp}\left[k\left\norm{{\rm dev}\left(\log\bU\right)\right}^2\right]  + \frac{\kappa}{2 \hat{k}} {\rm exp}\left[\hat{k} \left({\rm tr} \log\bU\right)^2\right],  \label{eq: exponentiated Hencky}\\
                                &= \frac{\mu}{k} {\rm exp}\left[k\left\norm{{\rm dev}\left(\frac{1}{2}\log \bC\right)\right}^2\right]  + \frac{\kappa}{2 \hat{k}} {\rm exp}\left[\hat{k} \biggl\langle\bone,\frac{1}{2}\log \bC\biggr\rangle^2\right]\,,~~  k>\frac{1}{3}, ~\hat{k}>\frac{1}{8} 
\end{align}

was introduced and described in \cite{NefGhiLan:2015:teh}. It is still volumetric-isochoric decoupled
and polyconvex in 2D if $k>\frac{1}{3}$ and  $\hat{k}>\frac{1}{8}$, cf. \cite{NefLanGhiMarSte:2015:teh}. Rank-one convexity is not preserved in 3D, see \cite{NefGhiLan:2015:teh}. 
However, numerical calculations show that the ellipticity domain contains the extremely large set $\{\bF\in {\rm GL}^+(3)|~\lVert\dev_3 \log\bU \rVert\leq6\}$.
In the small strain regime for principal stretches $\lambda_i \in(0.7,1.4)$ it approximates the aforementioned isotropic Hencky energy quite well.

Reformulation in terms of the Lam\'{e} parameters $\mu$  and $\lambda$ yields
\begin{align}
 {W}_{\rm eH} &= \frac{\mu}{k}\,{\rm exp}\left[k\left\norm{{\rm dev}\left(\frac{1}{2}\log \bC\right)\right}^2\right]  + \frac{(3\lambda+2\mu)/3}{2 \hat{k}} \,{\rm exp}\left[\hat{k} \biggl\langle\bone,\frac{1}{2}\log \bC\biggr\rangle^2\right] \,.
\end{align}
The derivatives with respect to the Hencky strain yield
\begin{align}
\nonumber \frac{\partial {W}_{\rm eH}}{\partial \log\bU} =&\, 2\mu~\mbox{exp}\left[ k \left\norm{\mbox{dev} \left(\log\bU\right)\right}^2  \right]~\mbox{dev} \left(\log\bU\right) \\
                                                    &+ \kappa~\mbox{exp}\left[\hat{k} \left({\rm tr}\log\bU\right)^2\right] \left({\rm tr}\log\bU \right) \bone\,,
\label{eq:dpsi_expHenck}                                                     
\end{align}
\begin{align}
\nonumber \mathbb{C}^{\rm H} = \frac{\partial^2 {W}_{\rm eH}}{\partial \log\bU \partial \log\bU} =\, & 4\mu~\mbox{exp}\left[ k \left\norm{\mbox{dev} \left(\frac{1}{2}\log \bC\right)\right}^2  \right]~\mbox{dev} \left(\frac{1}{2}\log \bC\right)\otimes\mbox{dev} \left(\frac{1}{2}\log \bC\right) \\
\nonumber                                               &+ 2\mu~\mbox{exp}\left[ k \left\norm{\mbox{dev} \left(\frac{1}{2}\log \bC\right)\right}^2  \right] \IP \\
\nonumber                                               &+ \kappa~\mbox{exp}\left[\hat{k} \left\langle\bone,\frac{1}{2}\log \bC\right\rangle^2\right] \bone \otimes\bone \\
                                                        &+ 2\kappa\hat{k}~\mbox{exp}\left[\hat{k} \left\langle\bone,\frac{1}{2}\log \bC\right\rangle^2\right] \left\langle\bone,\frac{1}{2}\log \bC\right\rangle^2 \bone\otimes\bone\,.
\label{eq:d2psi_expHenck}                                                        
\end{align}
In the reference configuration with $\bC = \bone$ and $\log\bU =\bzero$ the above equation simplifies to 
\eb
\frac{\partial^2 {W}_{\rm eH}}{\partial^2 \log\bU}\Big|_{\bC=\bone} = 2\mu\,\IP + \kappa\, \bone \otimes \bone 
\ee
and the final tangent according to Eq.~(\ref{eq: tangent}) becomes
\eb
\IC^{\rm V}|_{\bC=\bone}=4\frac{\partial^2{W}_{\rm eH}}{\partial\bC\partial\bC}\Big|_{\bC=\bone} =  \begin{pmatrix}
\kappa+\frac{4}{3}\mu & \kappa-\frac{2}{3}\mu & \kappa-\frac{2}{3}\mu & 0 & 0& 0\\
\kappa-\frac{2}{3}\mu & \kappa+\frac{4}{3}\mu & \kappa-\frac{2}{3}\mu & 0 & 0& 0\\
\kappa-\frac{2}{3}\mu & \kappa-\frac{2}{3}\mu & \kappa+\frac{4}{3}\mu & 0 & 0& 0\\
0 & 0 & 0 & \mu &0 &0\\
0 & 0 & 0 &0 &\mu  &0\\
0 & 0 & 0 &0 &0 &\mu \\
\end{pmatrix}\,,
\ee
which is identical to the elasticity tensor of ${W}_H$, provided in Eq.~(\ref{eq: elasti_W_H}).

\section{\hspace{-5mm}. Anisotropic extension}
\vspace{-4mm}

\subsection{\hspace{-5mm}.  Transverse isotropic Hencky and exponentiated Hencky models}

In a first step we aim to investigate a strain energy function ${W}_1(J_1^{\rm H},\lVert\dev\log\bU\rVert,I_4^{\rm H^{1}},I_4^{\rm H^{2}})$
which basically extends the classical Hencky-strain energy by the basic mixed invariants, introduced in chapter~\ref{sec: iso aniso invarianten}:
\begin{align}
\nonumber {W}_1(\log\bU) =&\, \mu_{\rm T}\, \lVert\dev\log\bU\rVert^2 + \frac{\kappa}{2}\,[{\rm tr}(\log\bU)]^2  +\alpha \langle\log\bU,\bM\rangle({\rm tr}\log\bU) \\
               &+2(\mu_{\rm L}-\mu_{\rm T})\langle\bM,(\log\bU)^2\rangle +\frac{1}{2}\beta\langle\bM,\log\bU\rangle^2 \label{eq: psi_1}\,.
\end{align}
The parameters are chosen in analogy to \cite{Spe:1987:kcc}. Here, $\mu_T$ and $\mu_L$ are associated to the shear moduli in the transverse isotropy plane and perpendicular to that, $\kappa$ is associated to the bulk modulus, $\beta$
is associated to the stiffness in fiber direction.
In a small strain framework, i.e. replacing $\log\bU$ by $\Bvarepsilon = \frac{1}{2}(\Grad \bu + \Grad^T\bu)$, the above given energy function would refer to transversely isotropic linear elasticity. 
In an exponential framework the exponentiated transversely isotropic strain-energy function 
\begin{align}
\nonumber  {W}_2(\log\bU) =&\, \frac{\mu_{\rm T}}{k_1}\,{\rm exp}[k_1 \lVert{\rm dev}\log\bU\rVert^2] + \frac{\kappa}{2k_2}\,{\rm exp}[k_2({\rm tr}\log\bU)^2]\\
\nonumber                  &+ \frac{\alpha}{k_3}\,{\rm exp}[k_3\langle\log\bU,\bM\rangle({\rm tr}\log\bU)] + \frac{2(\mu_{\rm L}-\mu_{\rm T})}{k_4}\,{\rm exp}[k_4\langle\bM,(\log\bU)^2\rangle]\\
                           &+ \frac{\beta}{2 k_5}\,{\rm exp}[k_5\langle\bM,\log\bU\rangle^2] \label{eq: psi_2}
\end{align}
will result in the same elasticity tensor $\IC|_{\bC=\bone}$ at the identity, whereby $k_i>0$ are further non-dimensional parameters. 
We also note that while the isotropic invariants have the proposed differential geometric meaning, the novel exponential terms are formulated on an ad hoc basis.

Due to the non-linearity of the above given equations we aim to identify the general material parameters in the reference configuration
with $\bC=\bone$ and $\log\bU =\bzero$. In case of linearized transversely isotropic materials the elasticity tensor $\IC$ may be formulated in terms of 
five material parameters, see Eq.~\eqref{eq: classical transverse isotropy}. If we choose the isotropic-plane to be spanned by the $X_1$ and $X_2$ axis and the preferred direction to coincide with the coordinate axis $X_3$, perpendicular to the isotropic plane,
and $\bM = {\rm diag}(0,0,1)$ we may write
\eb
 \IC^{\rm V} = \begin{pmatrix}
       \IC^{\rm V}_{11} & \IC^{\rm V}_{12} & \IC^{\rm V}_{13} &0&0&0\\ 
       \IC^{\rm V}_{12} & \IC^{\rm V}_{11} & \IC^{\rm V}_{13} &0&0&0\\ 
       \IC^{\rm V}_{13} & \IC^{\rm V}_{13} & \IC^{\rm V}_{33} &0&0&0\\   
       0&0&0&\frac{1}{2}(\IC^{\rm V}_{11}-\IC^{\rm V}_{12}) &0&0\\
       0&0&0& 0&\IC^{\rm V}_{44}&0\\
       0&0&0& 0&0&\IC^{\rm V}_{44}\\       
      \end{pmatrix}  \label{eq: classical transverse isotropy}
\ee
in Voigt-notation.
In the reference configuration, the formulated transversely isotropic strain energy functions ${W}_1$ and ${W}_2$,
both yield
\eb
\nonumber \IC^{\rm V}|_{\bC=\bone} = \begin{pmatrix}
       \lambda+2\mu_{\rm T}    & \lambda         & \lambda+\alpha &0&0&0\\ 
       \lambda           & \lambda+2\mu_{\rm T}  & \lambda+\alpha &0&0&0\\ 
       \lambda + \alpha  & \lambda+\alpha  & \lambda-2\mu_{\rm T}+2\alpha +4\mu_{\rm L} +\beta &0&0&0\\   
       0&0&0&\mu_{\rm T}&0 \\
       0&0&0& 0&\mu_{\rm L}&0\\
       0&0&0& 0&0&\mu_{\rm L}\\       
      \end{pmatrix}\,,
\ee
where the conversion $\lambda = (3\kappa-2\mu_T)/3$ was used.
The components of the  above presented scheme are related to the five classical components in Eq.~(\ref{eq: classical transverse isotropy})
through
\begin{equation}
\boxed{
\begin{aligned} 
 \mu_{\rm L} &= \IC^{\rm V}_{44}\\
 \mu_{\rm T} &= \frac{1}{2}(\IC^{\rm V}_{11}-\IC^{\rm V}_{12})\\
 \lambda &= \IC^{\rm V}_{12}\\
 \alpha &= \IC^{\rm V}_{13}-\IC^{\rm V}_{12}\\
 \beta &= \IC^{\rm V}_{11}+\IC^{\rm V}_{33}-2 \IC^{\rm V}_{13}-4 \IC^{\rm V}_{44}
\end{aligned}}\,,
\end{equation}
in analogy to \cite{SchGro:2004:ifo}.
The anisotropic characteristic of the strain energy function ${W}_2$ in Eq.~(\ref{eq: psi_2}) will be outlined on a number of numerical examples, where different 
material parameters according to Table~\ref{tab:Parameter 1} are used.

\begin{Table}[h!]
\centering
\begin{tabular}{|c|c|c|c|c|c|c|c|c|c|c|}
\hline
Set&$\lambda$ & $\mu_{\rm T}$ & $\alpha$ & $\beta$ & $\mu_{\rm L}$ & $k_1$ & $k_2$ & $k_3$ & $k_4$ & $k_5$\\
\hline
Set 1&1000&175 &10   &10   &375 &1&1&1&1&1\\      
Set 2&5.64&2.64&1.27 &0.29 &5.66 &1&1&1&1&1  \\
Set 3&5.5 &2.5 &0.00 &0.00 &2.5 &1&1&75&25&45\\
Set 4&5.5 &14  &40.75&0.00 &14 &1&1&75&25&45    \\
Set 5&5.5 &2.5 &0.00 &104.5&2.5&1&1&75&25&45    \\
Set 6&5.5 &2.5 &0.00 &0.00 &28.625&1&1&75&25&45 \\
\hline
\end{tabular}
\caption{Different parameter sets for the numerical examples.\label{tab:Parameter 1}}
\end{Table}

Further, the sets 4, 5  and 6 are chosen such that only one term involving a structural tensor
in Eq.~(\ref{eq: psi_2}) is active. Therefore, Set 4 is directly associated with the term $\langle\log\bU,\bM\rangle({\rm tr}\log\bU)$,
Set 5 with $\langle\log\bU,\bM\rangle$ and Set 6 with $\langle(\log\bU)^2,\bM\rangle = \lVert \dev (\log\bU)\, \bM\rVert^2$.
In contrast, Set 3 will serve as the isotropic reference case. 
In order to obtain comparable results a similar level of distinct anisotropy is chosen for each set. To achieve this, $\IC^{\rm V}_{33}$ is the same for all three
sets regarding the reference configuration.  All parameter sets have been checked to be positive definite for $\IC^{\rm V}|_{\bC=\bone}$.
Note that in case of Set 4 the parameter $\mu_{\rm T}$ needed to be increased in order to guarantee the positive definiteness
of $\IC^{\rm V}|_{\bC=\bone}$. The corresponding matrices are listed below:

\eb
\nonumber {\rm Set~4:}~~\IC^{\rm V}|_{\bC=\bone} = \begin{pmatrix}
   33.5&   5.5&   46.25&         0&         0&         0\\
   5.5&   33.5&   46.25&         0&         0&         0\\
   46.25&   46.25&  115&         0&         0&         0\\
         0&         0&         0&   14&         0&         0\\
         0&         0&         0&         0&   14&         0\\
         0&         0&         0&         0&         0&   14\\       
      \end{pmatrix} \qquad
\ee
\eb
\nonumber {\rm Set~5:}~~\IC^{\rm V}|_{\bC=\bone} = \begin{pmatrix}
    10.5&     5.5&           5.5&         0&         0&         0\\
     5.5&    10.5&           5.5&         0&         0&         0\\
     5.5&     5.5&           115&         0&         0&         0\\
         0&         0&         0&    2.5&         0&         0\\
         0&         0&         0&         0&    2.5&         0\\
         0&         0&         0&         0&         0&    2.5\\       
      \end{pmatrix} \qquad   
\ee
\eb
\nonumber {\rm Set~6:}~~\IC^{\rm V}|_{\bC=\bone} = \begin{pmatrix}
    10.5&     5.5&           5.5&         0&         0&         0\\
     5.5&    10.5&           5.5&         0&         0&         0\\
     5.5&     5.5&           115&         0&         0&         0\\
         0&         0&         0&    2.5&         0&         0\\
         0&         0&         0&         0&    28.625&         0\\
         0&         0&         0&         0&         0&    28.625\\       
      \end{pmatrix}      
\ee
The implementation in a finite element framework in this work was done according to the 
formulation in \cite{SchGruLob:2002:aso} and \cite{LobSchGru:2003:aog}. 
The weak form of balance of momentum 
\eb
G(\bu,\,\delta\bu):= \int_{\cal B} \langle \Div(\partial_{\bF}\psi(\bF)) +\rho_0 (\bb-\ddot{\bx}),\delta\bu\rangle\,{\rm d}V
\ee
required for the finite element code
may be reformulated such that we obtain
\eb
G(\bu,\,\delta\bu) = \underbrace{\int_{\cal B}\langle\Btau,\nabla(\delta\bu)\rangle\,{\rm d}V}_{G^{\rm int}} - \underbrace{\left(\,\,\int_{\partial {\cal B}_t} \langle \mathbf{f},\delta\bu \rangle\,{\rm d}A + \int_{\cal B} \langle \rho_0 (\bb-\ddot{\bx}),\delta\bu\rangle \,{\rm d}V\right)}_{G^{\rm ext}} \,=\, 0\,.
\ee
Here, the body force in the reference configuration is denoted by $\mathbf{f}$, $\delta\bu$ is the variation of the displacement field and $\ddot\bx$ the acceleration.
For the solution scheme in a finite element framework a Newton iteration is required. Therefore, the linearization 
\eb
LinG(\bar{\bu},\delta\bu,\Delta\bu) := G(\bar{\bu},\delta\bu) + \Delta G(\bar{\bu},\delta\bu,\Delta\bu)
\ee
at $\bu=\bar{\bu}$ is required where the increment $\Delta G$ is defined through
\eb
\Delta G = \int_{\cal B} \langle\nabla_{\sym}(\delta\bu),\mathbbm{c}:\nabla_{\sym}(\Delta\bu)\rangle \,{\rm d}V + \int_{\cal B} \langle \nabla(\Delta\bu)\,\Btau,\nabla(\delta\bu) \rangle\,{\rm dV}\,,
\ee
where the Eulerian tangent moduli $\mathbbm{c}$ is obtained by the push-forward of the Lagrangian tangent moduli $\IC$, i.e. 
\eb
\mathbbm{c} = (\bF\boxtimes\bF):\IC:(\bF^T\boxtimes\bF^T)
\ee
and $\nabla_{\sym}(\bullet)=1/2\,[\grad (\bullet)+\grad^T(\bullet)]$.
For the numerical treatment the weak formulations of the aforementioned balance equation has been implemented in the finite element analysis program FEAP of R.L. Taylor, University of California.
For all of the following examples quadratic triangular elements with six nodes per element were used.

\textbf{Tension test:}
In a first example the transversely isotropic material behavior is 
to be explained on the basis of a tensile test under plane strain conditions,
see Fig.~\ref{fig: Tensile vary fiber}a).
Material parameter Set 1 was chosen and the preferred direction $\bA$, defined with help
of the fiber angle $\beta_{\rm f}$, was varied. The computed displacements of the nodes 1, 2 and 3 
over the fiber angle are plotted in Fig.~\ref{fig: Tensile vary fiber}b).The displacements 
$\delta_{\rm h_1}$, $\delta_{\rm h_2}$ and $\delta_{\rm v_3}$
are symmetric concerning $\beta_{\rm f} = 90^{\circ}$, while the vertical displacements 
$\delta_{\rm v_1}$ and $\delta_{\rm v_2}$ are antisymmetric. Surprisingly, the horizontal displacements
don't reach their maximum value for $\beta_{\rm f} = 90^{\circ}$.
The load $p_0$ has been chosen such that
large deformations are present.

\begin{Figure}[!htb]
\unitlength 1 cm
\begin{picture}(14,6.1)
\put(1.0,0.25){\includegraphics[width = 6.5cm]{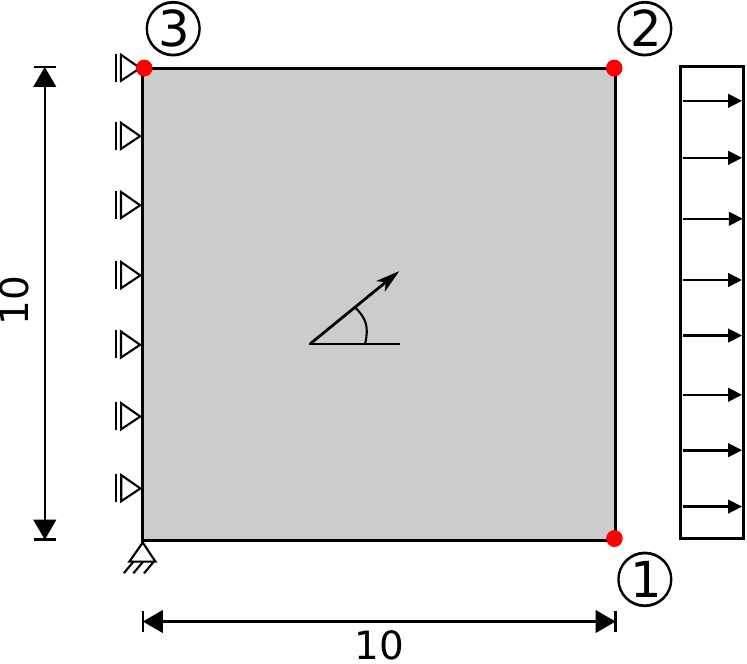}}
\put(8.5,0.0){
\begingroup
  \makeatletter
  \providecommand\color[2][]{%
    \GenericError{(gnuplot) \space\space\space\@spaces}{%
      Package color not loaded in conjunction with
      terminal option `colourtext'%
    }{See the gnuplot documentation for explanation.%
    }{Either use 'blacktext' in gnuplot or load the package
      color.sty in LaTeX.}%
    \renewcommand\color[2][]{}%
  }%
  \providecommand\includegraphics[2][]{%
    \GenericError{(gnuplot) \space\space\space\@spaces}{%
      Package graphicx or graphics not loaded%
    }{See the gnuplot documentation for explanation.%
    }{The gnuplot epslatex terminal needs graphicx.sty or graphics.sty.}%
    \renewcommand\includegraphics[2][]{}%
  }%
  \providecommand\rotatebox[2]{#2}%
  \@ifundefined{ifGPcolor}{%
    \newif\ifGPcolor
    \GPcolortrue
  }{}%
  \@ifundefined{ifGPblacktext}{%
    \newif\ifGPblacktext
    \GPblacktexttrue
  }{}%
  \let\gplgaddtomacro\g@addto@macro
  \gdef\gplbacktext{}%
  \gdef\gplfronttext{}%
  \makeatother
  \ifGPblacktext
    \def\colorrgb#1{}%
    \def\colorgray#1{}%
  \else
    \ifGPcolor
      \def\colorrgb#1{\color[rgb]{#1}}%
      \def\colorgray#1{\color[gray]{#1}}%
      \expandafter\def\csname LTw\endcsname{\color{white}}%
      \expandafter\def\csname LTb\endcsname{\color{black}}%
      \expandafter\def\csname LTa\endcsname{\color{black}}%
      \expandafter\def\csname LT0\endcsname{\color[rgb]{1,0,0}}%
      \expandafter\def\csname LT1\endcsname{\color[rgb]{0,1,0}}%
      \expandafter\def\csname LT2\endcsname{\color[rgb]{0,0,1}}%
      \expandafter\def\csname LT3\endcsname{\color[rgb]{1,0,1}}%
      \expandafter\def\csname LT4\endcsname{\color[rgb]{0,1,1}}%
      \expandafter\def\csname LT5\endcsname{\color[rgb]{1,1,0}}%
      \expandafter\def\csname LT6\endcsname{\color[rgb]{0,0,0}}%
      \expandafter\def\csname LT7\endcsname{\color[rgb]{1,0.3,0}}%
      \expandafter\def\csname LT8\endcsname{\color[rgb]{0.5,0.5,0.5}}%
    \else
      \def\colorrgb#1{\color{black}}%
      \def\colorgray#1{\color[gray]{#1}}%
      \expandafter\def\csname LTw\endcsname{\color{white}}%
      \expandafter\def\csname LTb\endcsname{\color{black}}%
      \expandafter\def\csname LTa\endcsname{\color{black}}%
      \expandafter\def\csname LT0\endcsname{\color{black}}%
      \expandafter\def\csname LT1\endcsname{\color{black}}%
      \expandafter\def\csname LT2\endcsname{\color{black}}%
      \expandafter\def\csname LT3\endcsname{\color{black}}%
      \expandafter\def\csname LT4\endcsname{\color{black}}%
      \expandafter\def\csname LT5\endcsname{\color{black}}%
      \expandafter\def\csname LT6\endcsname{\color{black}}%
      \expandafter\def\csname LT7\endcsname{\color{black}}%
      \expandafter\def\csname LT8\endcsname{\color{black}}%
    \fi
  \fi
  \setlength{\unitlength}{0.0500bp}%
  \begin{picture}(4534.40,3310.40)%
    \gplgaddtomacro\gplbacktext{%
      \csname LTb\endcsname%
      \put(682,704){\makebox(0,0)[r]{\strut{}-2}}%
      \csname LTb\endcsname%
      \put(682,1094){\makebox(0,0)[r]{\strut{}-1}}%
      \csname LTb\endcsname%
      \put(682,1485){\makebox(0,0)[r]{\strut{} 0}}%
      \csname LTb\endcsname%
      \put(682,1875){\makebox(0,0)[r]{\strut{} 1}}%
      \csname LTb\endcsname%
      \put(682,2265){\makebox(0,0)[r]{\strut{} 2}}%
      \csname LTb\endcsname%
      \put(682,2656){\makebox(0,0)[r]{\strut{} 3}}%
      \csname LTb\endcsname%
      \put(682,3046){\makebox(0,0)[r]{\strut{} 4}}%
      \csname LTb\endcsname%
      \put(814,484){\makebox(0,0){\strut{} 0}}%
      \csname LTb\endcsname%
      \put(1368,484){\makebox(0,0){\strut{} 30}}%
      \csname LTb\endcsname%
      \put(1922,484){\makebox(0,0){\strut{} 60}}%
      \csname LTb\endcsname%
      \put(2476,484){\makebox(0,0){\strut{} 90}}%
      \csname LTb\endcsname%
      \put(3029,484){\makebox(0,0){\strut{} 120}}%
      \csname LTb\endcsname%
      \put(3583,484){\makebox(0,0){\strut{} 150}}%
      \csname LTb\endcsname%
      \put(4137,484){\makebox(0,0){\strut{} 180}}%
      \put(176,1875){\rotatebox{-270}{\makebox(0,0){\strut{}$\delta_{\rm h_{i}}|_{\rm i =1,2},~\delta_{\rm v_{i}}|_{\rm i = 1,2,3}$}}}%
      \put(2475,154){\makebox(0,0){\strut{}$\beta_{\rm f}$}}%
    }%
    \gplgaddtomacro\gplfronttext{%
    }%
    \gplbacktext
    \put(0,0){\includegraphics{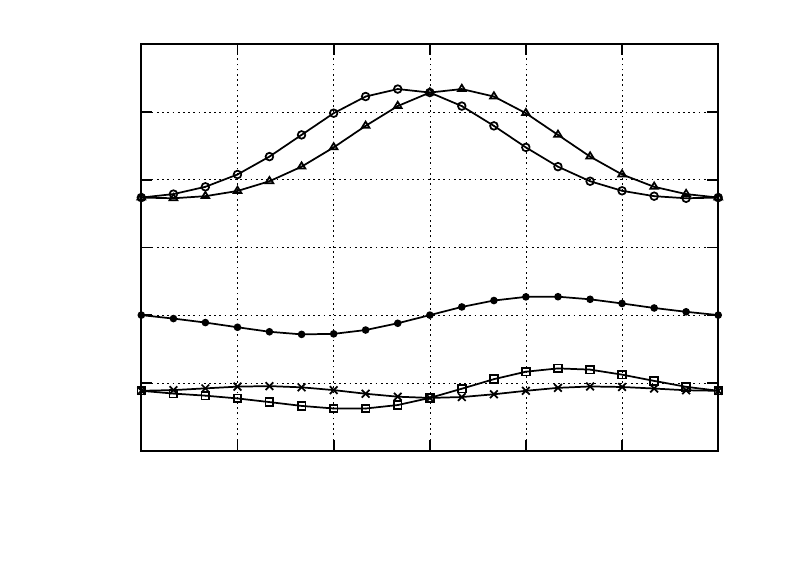}}%
    \gplfronttext
  \end{picture}%
\endgroup
}
\put(4.1,3.7){$\bA$}
\put(4.3,3.12){${\beta_{\rm f}}$}
\put(14.1,2.95){$\delta_{{\rm v}_{1}}$}
\put(14,2.2){$\delta_{{\rm v}_{2}}$}
\put(14,1.45){$\delta_{{\rm v}_{3}}$}
\put(13.8,4.8){$\delta_{{\rm h}_{2}}$}
\put(11.3,4.8){$\delta_{{\rm h}_{1}}$}
\put(7.1,0.9){$p_0 = 200$}
\put(1.0,0.0){a)}
\put(9.0,0.0){b)}
\end{picture}
\setlength{\baselineskip}{11pt}
\caption{Example 1: a) Tension test with one preferred direction and varying fiber orientation. b) Nodal displacements depending on the fiber orientations are plotted for parameter Set 1.\label{fig: Tensile vary fiber}%
}
\end{Figure}

\textbf{Cooks Membrane:}
In a second example we consider the Cooks Membrane problem, as depicted in Fig.~\ref{fig: Cooke vary fiber}a) which
is dominated by non-homogenous stress distributions.
Again the fiber direction is to be varied and the body will undergo large deformations during loading.
In Fig.~\ref{fig: Cooke vary fiber}b), the vertical displacements $\delta_{\rm V}$ of the node at the top right are plotted 
for different fiber angles. Parameter Set 2 was considered. The anisotropic effect due to the different fiber orientation
clearly becomes apparent.

\begin{Figure}[!htb]
\unitlength 1 cm
\begin{picture}(14,6.8)
\put(1.0,0.3){\includegraphics[width = 5.5cm]{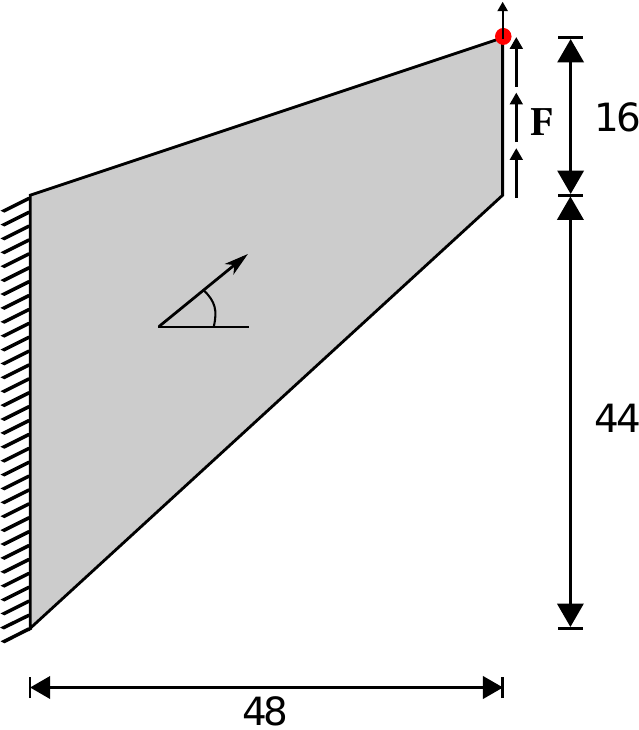}}
\put(2.8,4.4){$\bA$}
\put(2.95,3.85){${\beta_{\rm f}}$}
\put(5.2,6.6){$\delta_V$}
\put(8.5,0.15){
\begingroup
  \makeatletter
  \providecommand\color[2][]{%
    \GenericError{(gnuplot) \space\space\space\@spaces}{%
      Package color not loaded in conjunction with
      terminal option `colourtext'%
    }{See the gnuplot documentation for explanation.%
    }{Either use 'blacktext' in gnuplot or load the package
      color.sty in LaTeX.}%
    \renewcommand\color[2][]{}%
  }%
  \providecommand\includegraphics[2][]{%
    \GenericError{(gnuplot) \space\space\space\@spaces}{%
      Package graphicx or graphics not loaded%
    }{See the gnuplot documentation for explanation.%
    }{The gnuplot epslatex terminal needs graphicx.sty or graphics.sty.}%
    \renewcommand\includegraphics[2][]{}%
  }%
  \providecommand\rotatebox[2]{#2}%
  \@ifundefined{ifGPcolor}{%
    \newif\ifGPcolor
    \GPcolortrue
  }{}%
  \@ifundefined{ifGPblacktext}{%
    \newif\ifGPblacktext
    \GPblacktexttrue
  }{}%
  \let\gplgaddtomacro\g@addto@macro
  \gdef\gplbacktext{}%
  \gdef\gplfronttext{}%
  \makeatother
  \ifGPblacktext
    \def\colorrgb#1{}%
    \def\colorgray#1{}%
  \else
    \ifGPcolor
      \def\colorrgb#1{\color[rgb]{#1}}%
      \def\colorgray#1{\color[gray]{#1}}%
      \expandafter\def\csname LTw\endcsname{\color{white}}%
      \expandafter\def\csname LTb\endcsname{\color{black}}%
      \expandafter\def\csname LTa\endcsname{\color{black}}%
      \expandafter\def\csname LT0\endcsname{\color[rgb]{1,0,0}}%
      \expandafter\def\csname LT1\endcsname{\color[rgb]{0,1,0}}%
      \expandafter\def\csname LT2\endcsname{\color[rgb]{0,0,1}}%
      \expandafter\def\csname LT3\endcsname{\color[rgb]{1,0,1}}%
      \expandafter\def\csname LT4\endcsname{\color[rgb]{0,1,1}}%
      \expandafter\def\csname LT5\endcsname{\color[rgb]{1,1,0}}%
      \expandafter\def\csname LT6\endcsname{\color[rgb]{0,0,0}}%
      \expandafter\def\csname LT7\endcsname{\color[rgb]{1,0.3,0}}%
      \expandafter\def\csname LT8\endcsname{\color[rgb]{0.5,0.5,0.5}}%
    \else
      \def\colorrgb#1{\color{black}}%
      \def\colorgray#1{\color[gray]{#1}}%
      \expandafter\def\csname LTw\endcsname{\color{white}}%
      \expandafter\def\csname LTb\endcsname{\color{black}}%
      \expandafter\def\csname LTa\endcsname{\color{black}}%
      \expandafter\def\csname LT0\endcsname{\color{black}}%
      \expandafter\def\csname LT1\endcsname{\color{black}}%
      \expandafter\def\csname LT2\endcsname{\color{black}}%
      \expandafter\def\csname LT3\endcsname{\color{black}}%
      \expandafter\def\csname LT4\endcsname{\color{black}}%
      \expandafter\def\csname LT5\endcsname{\color{black}}%
      \expandafter\def\csname LT6\endcsname{\color{black}}%
      \expandafter\def\csname LT7\endcsname{\color{black}}%
      \expandafter\def\csname LT8\endcsname{\color{black}}%
    \fi
  \fi
  \setlength{\unitlength}{0.0500bp}%
  \begin{picture}(4534.40,3310.40)%
    \gplgaddtomacro\gplbacktext{%
      \csname LTb\endcsname%
      \put(946,704){\makebox(0,0)[r]{\strut{} 0}}%
      \csname LTb\endcsname%
      \put(946,1172){\makebox(0,0)[r]{\strut{} 0.2}}%
      \csname LTb\endcsname%
      \put(946,1641){\makebox(0,0)[r]{\strut{} 0.4}}%
      \csname LTb\endcsname%
      \put(946,2109){\makebox(0,0)[r]{\strut{} 0.6}}%
      \csname LTb\endcsname%
      \put(946,2578){\makebox(0,0)[r]{\strut{} 0.8}}%
      \csname LTb\endcsname%
      \put(946,3046){\makebox(0,0)[r]{\strut{} 1}}%
      \csname LTb\endcsname%
      \put(1078,484){\makebox(0,0){\strut{} 0}}%
      \csname LTb\endcsname%
      \put(1418,484){\makebox(0,0){\strut{} 2}}%
      \csname LTb\endcsname%
      \put(1758,484){\makebox(0,0){\strut{} 4}}%
      \csname LTb\endcsname%
      \put(2098,484){\makebox(0,0){\strut{} 6}}%
      \csname LTb\endcsname%
      \put(2438,484){\makebox(0,0){\strut{} 8}}%
      \csname LTb\endcsname%
      \put(2777,484){\makebox(0,0){\strut{} 10}}%
      \csname LTb\endcsname%
      \put(3117,484){\makebox(0,0){\strut{} 12}}%
      \csname LTb\endcsname%
      \put(3457,484){\makebox(0,0){\strut{} 14}}%
      \csname LTb\endcsname%
      \put(3797,484){\makebox(0,0){\strut{} 16}}%
      \csname LTb\endcsname%
      \put(4137,484){\makebox(0,0){\strut{} 18}}%
      \put(176,1875){\rotatebox{-270}{\makebox(0,0){\strut{}$||F||$}}}%
      \put(2607,154){\makebox(0,0){\strut{}$\delta_{\rm v}$}}%
    }%
    \gplgaddtomacro\gplfronttext{%
      \csname LTb\endcsname%
      \put(2134,2873){\makebox(0,0)[r]{\strut{}$\beta_{\rm f} = {0^\circ}$}}%
      \csname LTb\endcsname%
      \put(2134,2653){\makebox(0,0)[r]{\strut{}$\beta_{\rm f} = {30^\circ}$}}%
      \csname LTb\endcsname%
      \put(2134,2433){\makebox(0,0)[r]{\strut{}$\beta_{\rm f} = {60^\circ}$}}%
      \csname LTb\endcsname%
      \put(2134,2213){\makebox(0,0)[r]{\strut{}$\beta_{\rm f} = {90^\circ}$}}%
    }%
    \gplbacktext
    \put(0,0){\includegraphics{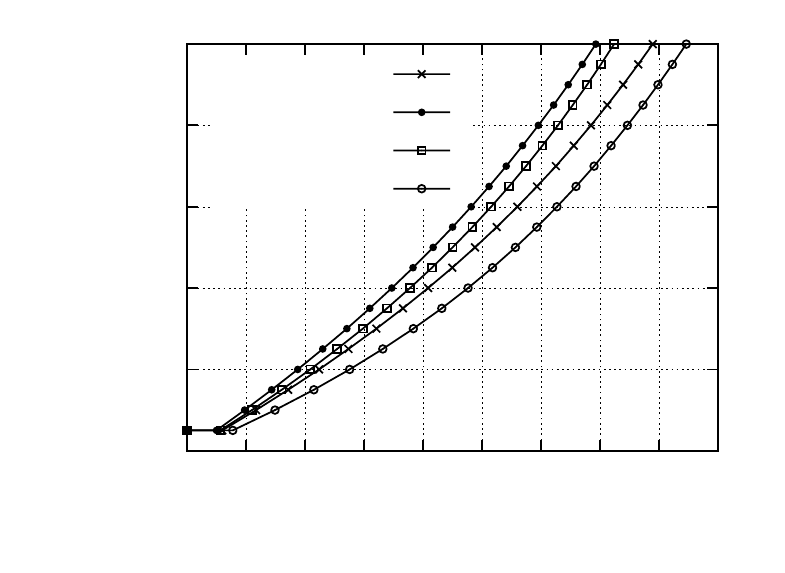}}%
    \gplfronttext
  \end{picture}%
\endgroup
}
\put(1.0,0.0){a)}
\put(9.0,0.0){b)}
\end{picture}
\setlength{\baselineskip}{11pt}
\caption{Example 2: a) Cooks Membrane with one preferred direction and varying fiber orientation. b) Plot of the vertical displacement for different fiber orientations and parameter Set 2.\label{fig: Cooke vary fiber}%
}
\end{Figure}

\newpage
\textbf{Perforated Plate:}
Lastly, we want to consider three perforated plates, again under consideration of plane strain conditions. 
The first two plates are squares and share the same geometry. The first one, referred to as Example 3a, is subject to uniaxial tensile loads, see Fig.~\ref{fig: Example perforated plate uniax}a),
while the second one, referred to as example 3b, is subject to biaxial tensile loads, see Fig.~\ref{fig: Example perforated plate biax}a). In example 3c we consider a circular disk, depicted in Fig.~\ref{fig: Example perforated plate circle}a), 
which is expanded on the inner ring, i.e. we apply a radial displacement of 3.5.
All simulations are displacement driven, i.e. only boundary conditions of Dirichlet type are present. Further, all three bodies have  one preferred direction $\bA$ with an angle of $45^{\circ}$
to the horizontal axis. 
The deformed bodies for the parameter Sets 3, 4, 5 and 6 are plotted next to the boundary value problems in Fig.~\ref{fig: Example perforated plate uniax},
Fig.~\ref{fig: Example perforated plate biax} and Fig.~\ref{fig: Example perforated plate circle}. 
The contour plots of the squares show the horizontal displacements $u_1$. In order to highlight the anisotropic characteristic 
of the circular disk, the circumferential stretch $\lambda_{\phi} =\sqrt{\langle\bC,\bN_{\varphi}\otimes\bN_{\varphi}\rangle}$,
where $\bN_{\varphi}$ denotes the circumferential direction, is plotted for the third plate. The black lines in each plot indicate 
the shape of the body in the undeformed configuration.

The comparison of the different parameter sets are intended to demonstrate different anisotropic characteristics of different strain measures. 
Therefore, the parameter sets were chosen such that only one anisotropic part of the strain energy in Eq.~\eqref{eq: psi_2} is active.
First of all from Fig.~\ref{fig: Example perforated plate uniax} and Fig.~\ref{fig: Example perforated plate biax} it becomes apparent that for the anisotropic Sets 4, 5 and 6
the displacements $u_1$ are not symmetric with respect to the horizontal axis, which is different for the isotropic Set 3. 
The different shapes of the holes very well emphasize that the usage of different anisotropic invariants may lead to very different deformations, even if the preferred direction is identical. 
In example 3c the inner ring of the disk is exposed to predefined deformations, which is why the shapes of the holes are the same for each parameter set. Nevertheless, the distribution of the circumferential 
stretch $\lambda_\varphi$ is considerably different. Considering a polar coordinate system the circumferential stretch is only depending on the radius, but independent of the polar angle for the isotropic Set 3.
For Set 4 and Set 6 it can be seen that $\lambda_\varphi$ is smallest in the regions were the preferred direction $\bA$ and the circumferential direction $\bN_\varphi$ coincide and largest in the regions were $\bA$ and $\bN_\varphi$
are perpendicular. For Set 5 only the first of these two observations holds true. We conclude that a considerable stiffening effect in the preferred direction is visible for Set 4, 5 and 6.

\begin{Figure}[!htb]
\unitlength 1 cm
\begin{picture}(14,7.6)
\put(0,-1){
\put(0.3,4.92){\includegraphics[height = 3.732cm]{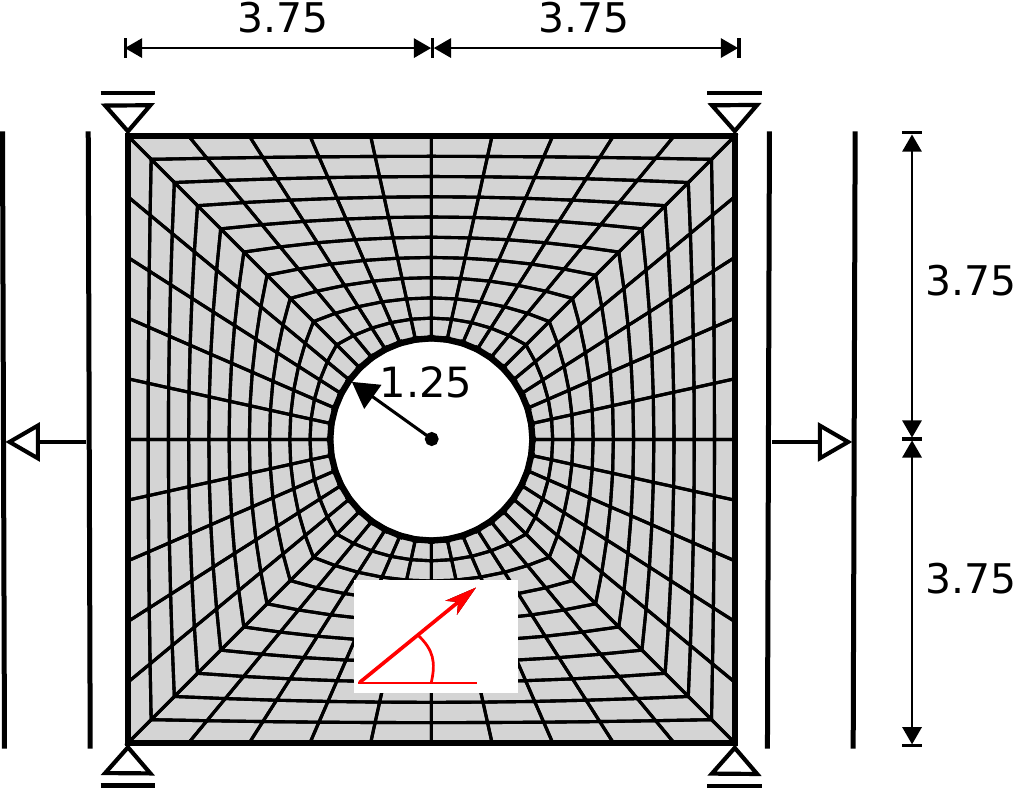}}
\put(5.44,1.03){\includegraphics[height = 3.4611cm]{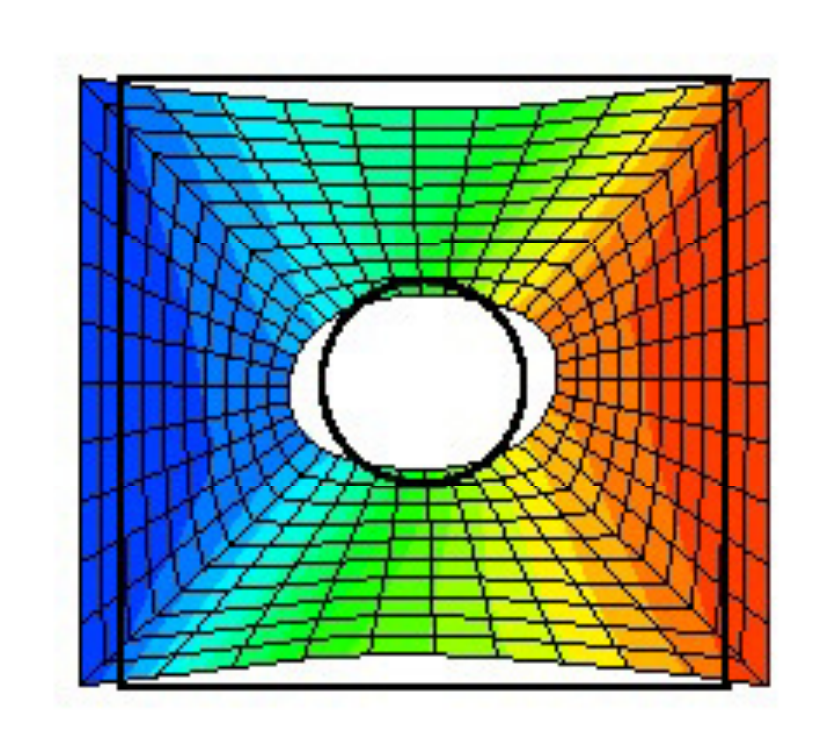}}
\put(10.52,1.03){\includegraphics[height = 3.4611cm]{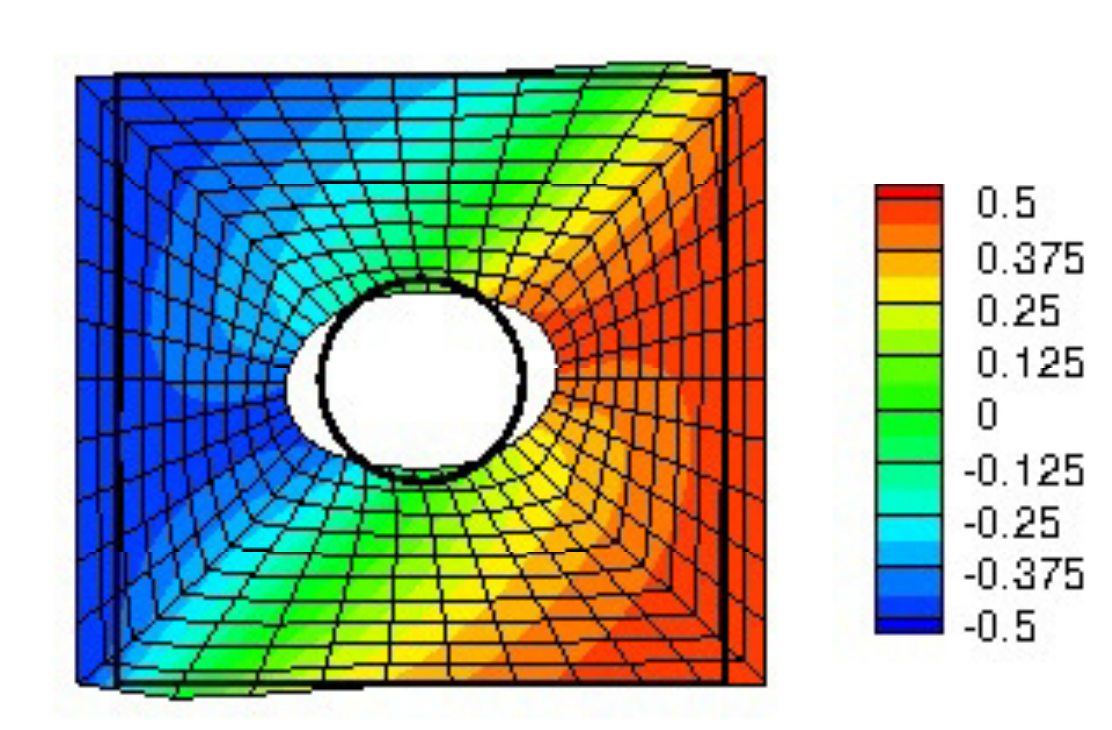}}
\put(5.44,4.9){\includegraphics[height = 3.4611cm]{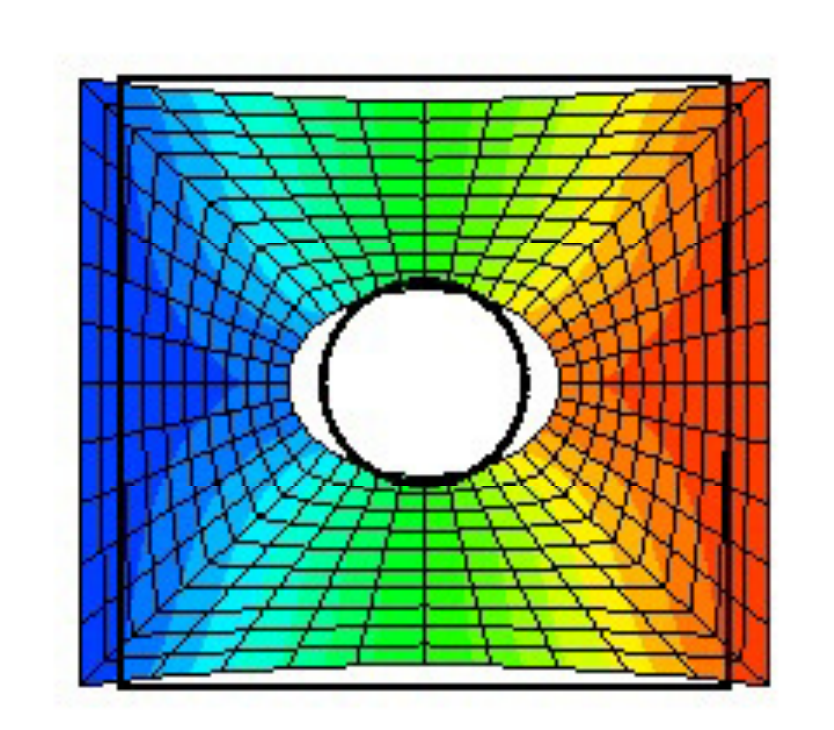}}
\put(10.52,4.9){\includegraphics[height = 3.4611cm]{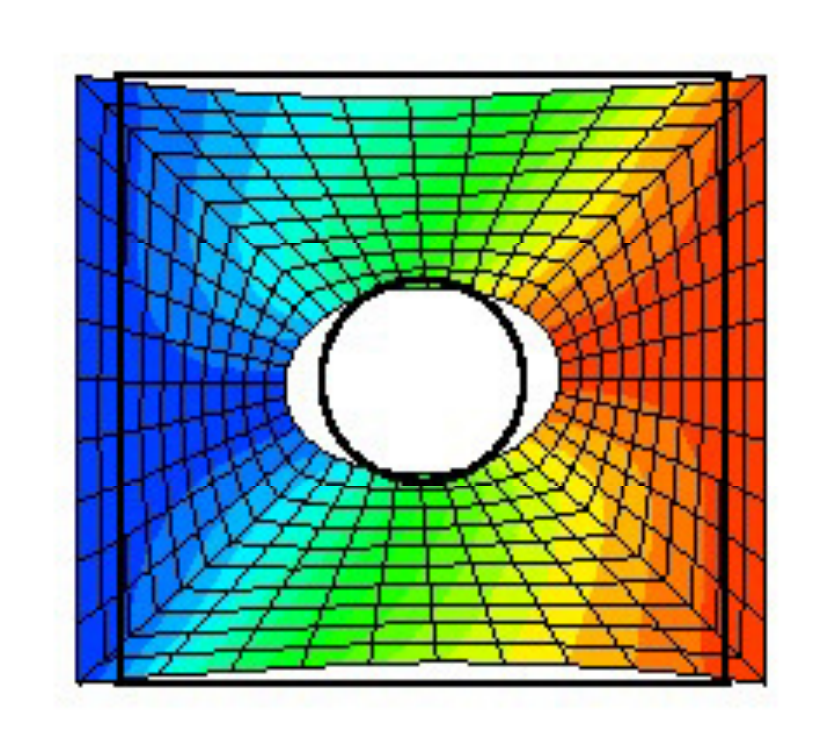}}
\put(7.02,6.42){\small Set 3}
\put(12.11,6.42){\small Set 4}
\put(7.02,2.6){\small Set 5}
\put(12.11,2.6){\small Set 6}
\put(0.3,4.6){a)}
\put(5.2,5.2){b)}
\put(10.3,5.2){c)}
\put(5.2,1.4){d)}
\put(10.3,1.4){e)}
\put(14.7,3.85){$u_1$}
\put(0.34,6.75){$u_1$}
\put(3.95,6.75){$u_1$}
\put(2.0,5.705){\textcolor{red}{\tiny $\bA$}}
\put(2.4,5.525){\textcolor{red}{\tiny $45^{\circ}$}}
}
\end{picture}
\setlength{\baselineskip}{11pt}
\caption{Example 3a: a) Boundary conditions and b)-e) deformed bodies of a perforated plate under uniaxial tension for parameter Set 3, 4, 5 and 6.\label{fig: Example perforated plate uniax}%
}
\end{Figure}

\begin{Figure}[!htb]
\unitlength 1 cm
\begin{picture}(14,10)
\put(5.1,0){\includegraphics[height = 4.45cm]{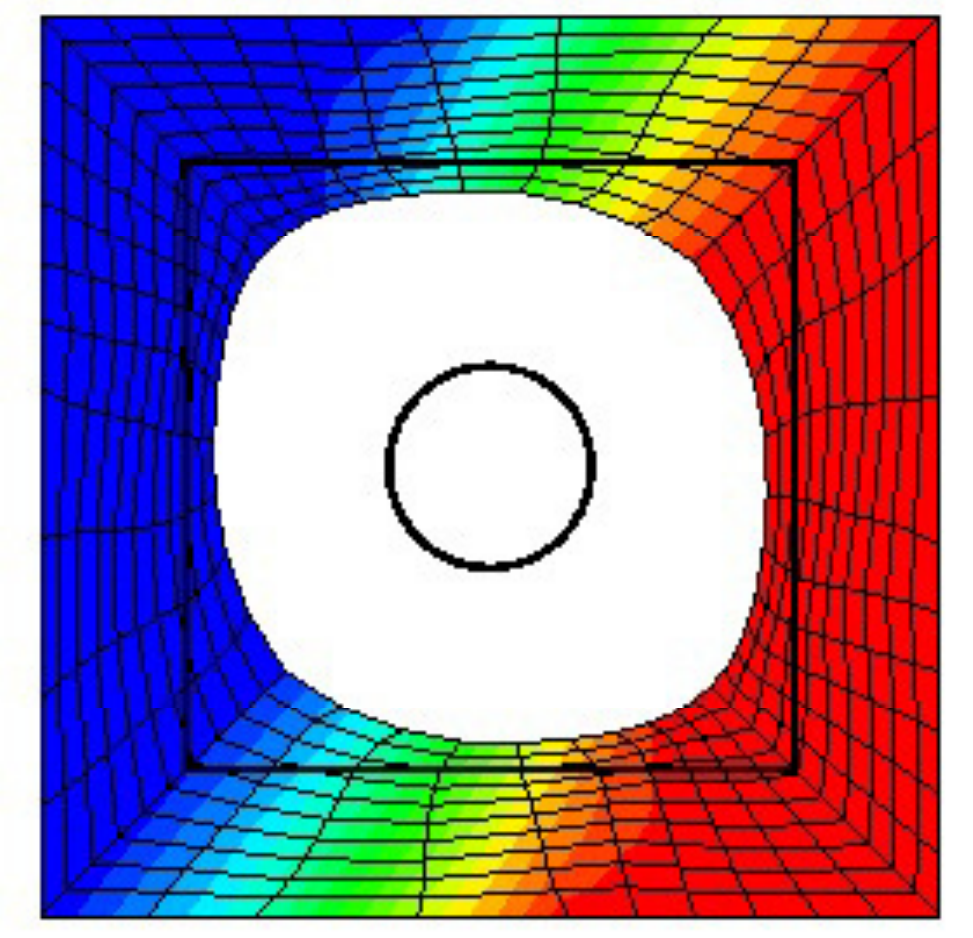}}
\put(10.2,5.0){\includegraphics[height = 4.45cm]{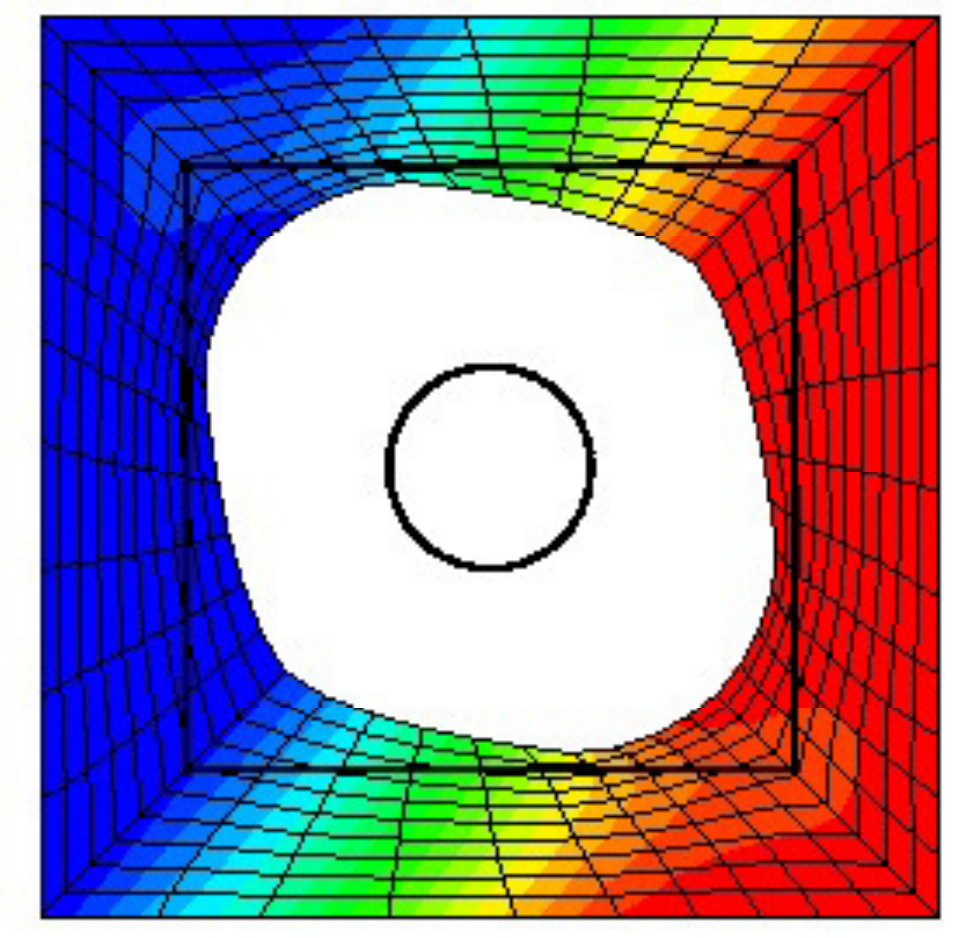}}
\put(5.125,5.03){\includegraphics[height = 4.45cm]{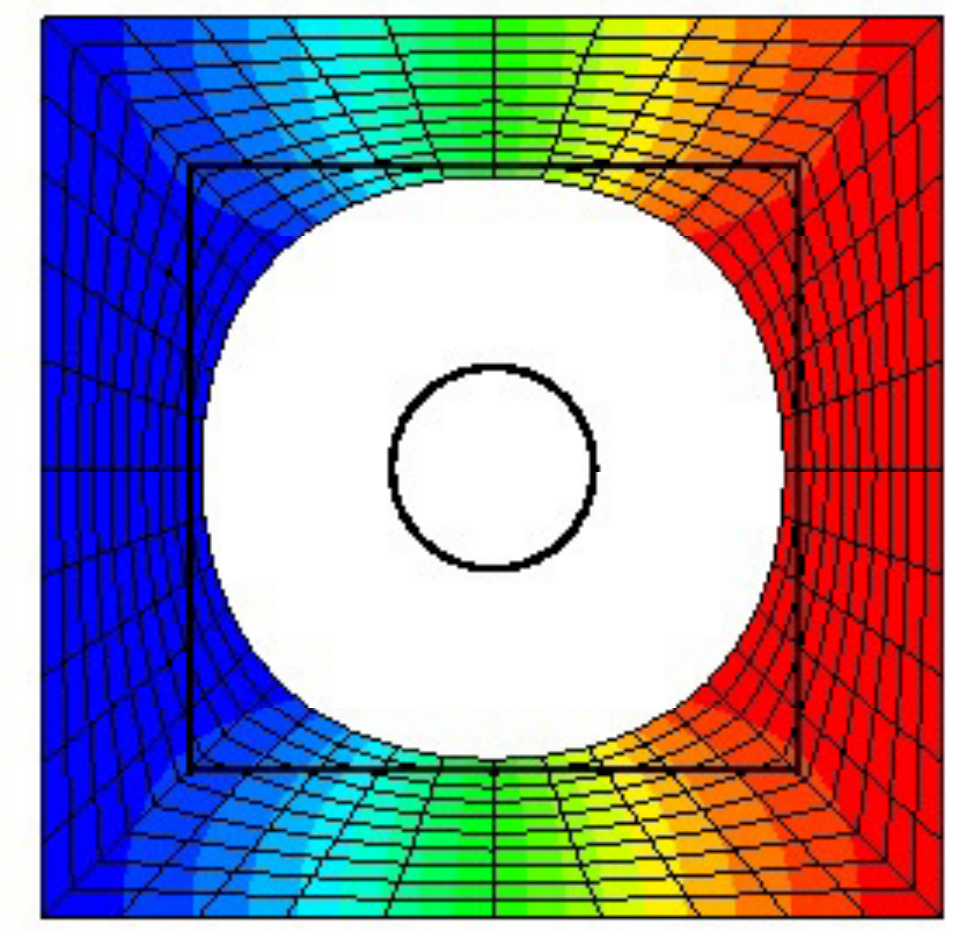}}
\put(10.2,0){\includegraphics[height = 4.45cm]{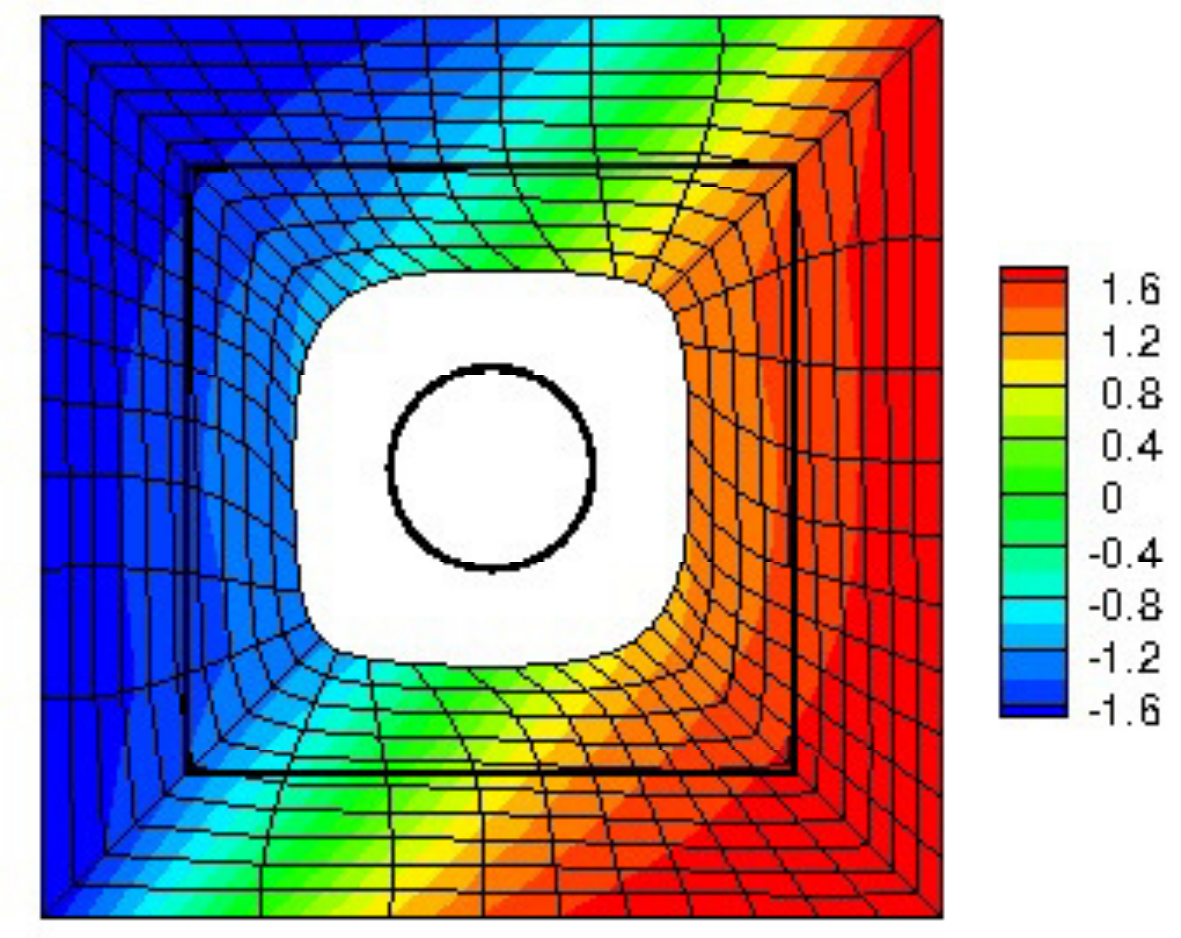}}
\put(0.3,5.27){\includegraphics[height = 4.355cm]{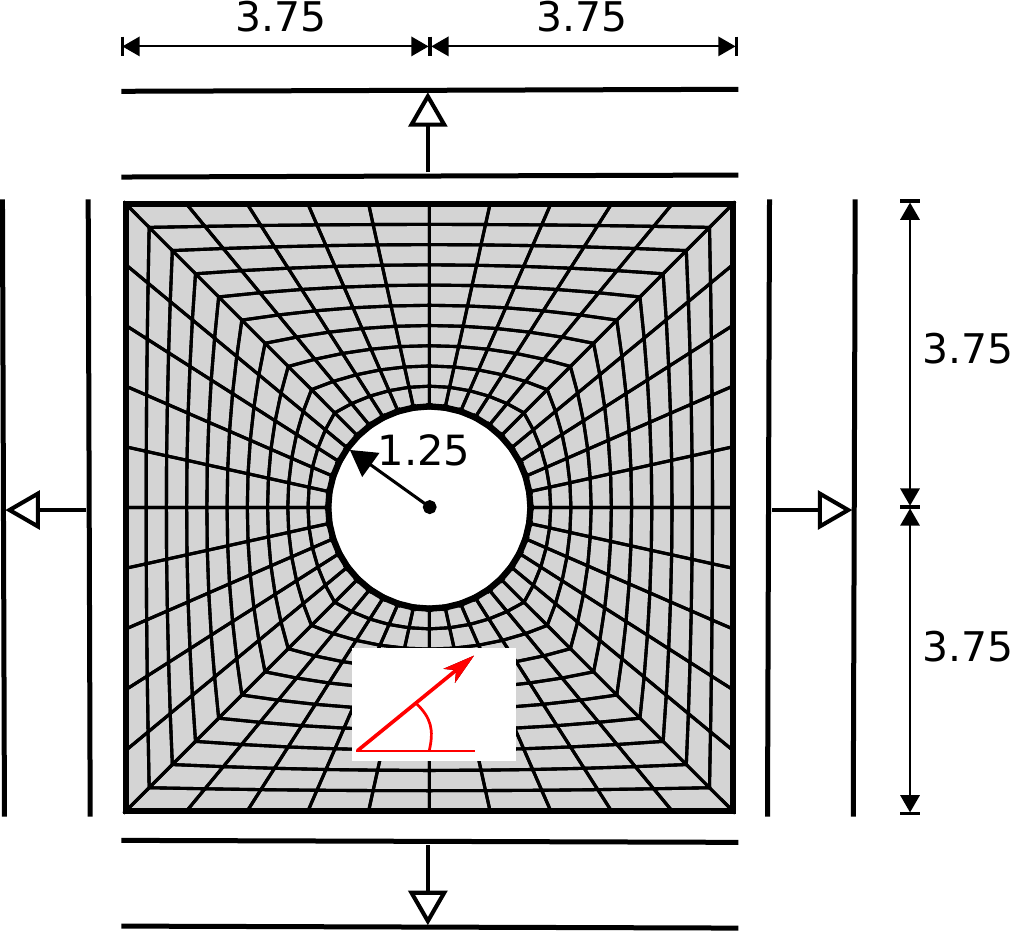}}
\put(7.04,7.09){\small Set 3}
\put(12.12,7.09){\small Set 4}
\put(7.03,2.08){\small Set 5}
\put(12.13,2.08){\small Set 6}
\put(0.25,5.2){a)}
\put(4.8,5.2){b)}
\put(10.0,5.2){c)}
\put(4.8,0.1){d)}
\put(10.0,0.1){e)}
\put(2,6.38){\textcolor{red}{\tiny $\bA$}}
\put(2.34,6.2){\textcolor{red}{\tiny $45^{\circ}$}}
\put(15.00,3.35){$u_1$}
\put(0.34,7.45){$u_1$}
\put(3.92,7.45){$u_1$}
\put(2.55,5.377){$u_2=u_1$}
\put(2.55,8.9){$u_2=u_1$}
\end{picture}
\setlength{\baselineskip}{11pt}
\caption{Example 3b: a) Boundary conditions and b)-e) deformed bodies of a perforated plate under biaxial tension for parameter Set 3, 4, 5 and 6.\label{fig: Example perforated plate biax}%
}
\end{Figure}

\clearpage

\begin{Figure}[t]
\unitlength 1 cm
\begin{picture}(14,9.5)
\put(10.25,5.0){\includegraphics[height = 4.35cm]{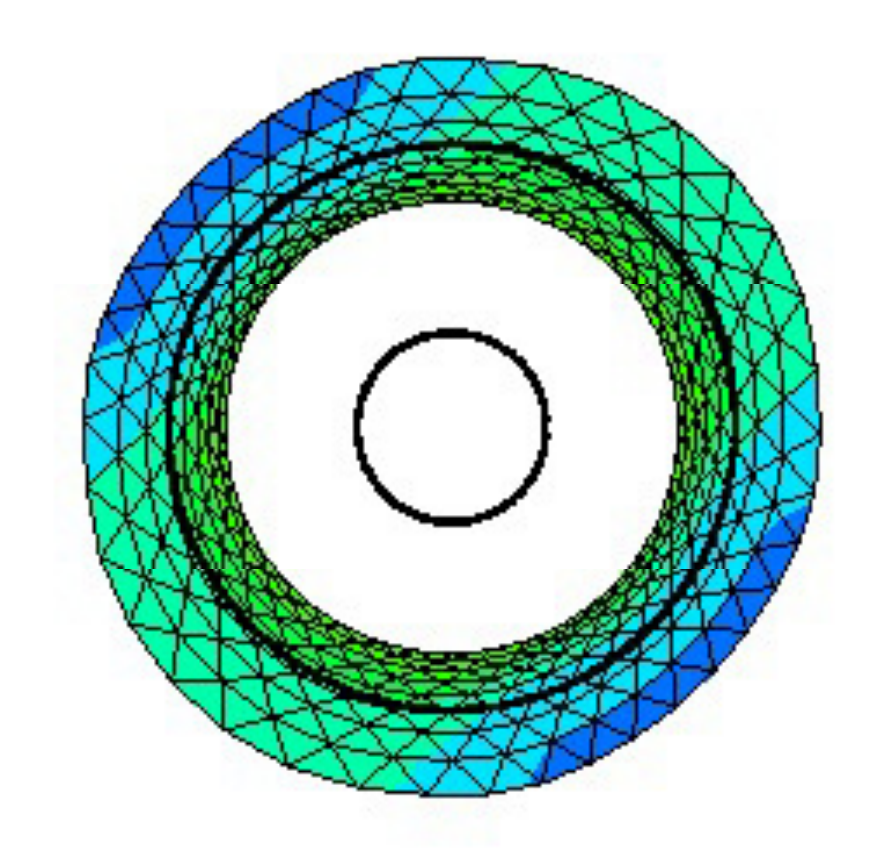}}
\put(5.2,5.0){\includegraphics[height = 4.35cm]{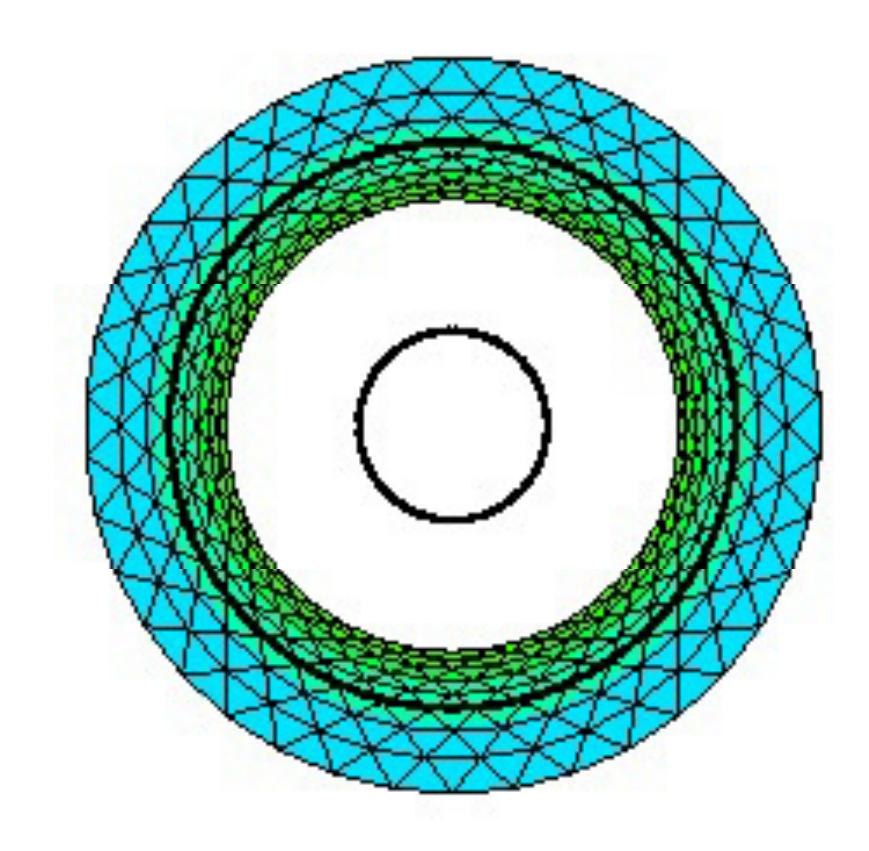}}
\put(5.2,0.25){\includegraphics[height = 4.35cm]{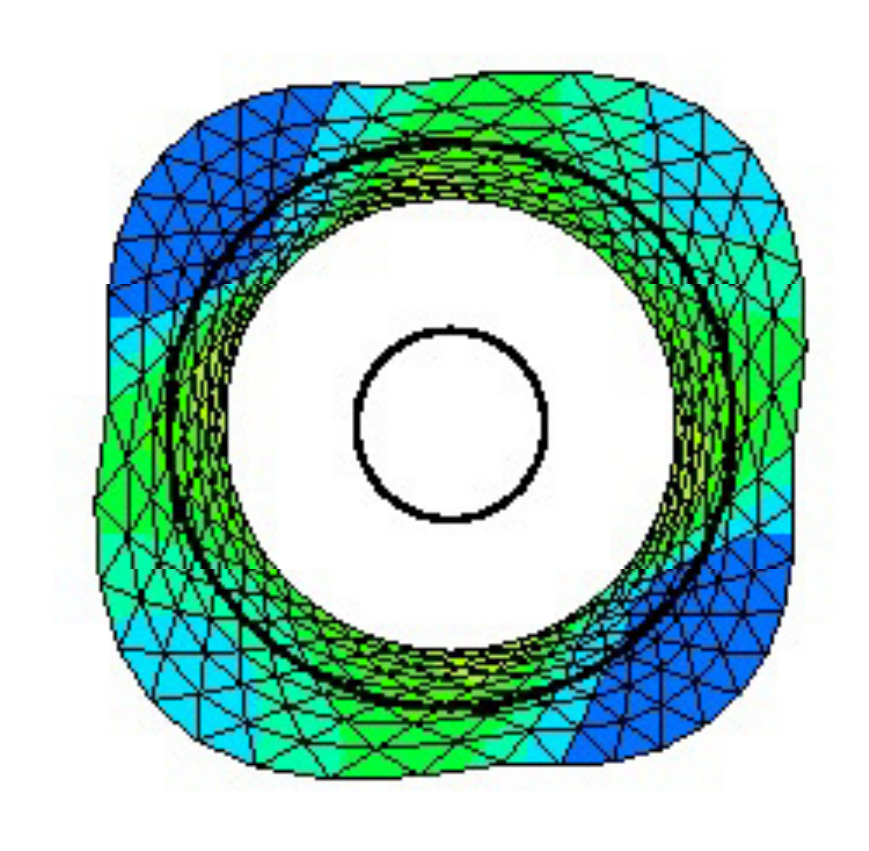}}
\put(10.25,0.25){\includegraphics[height = 4.35cm]{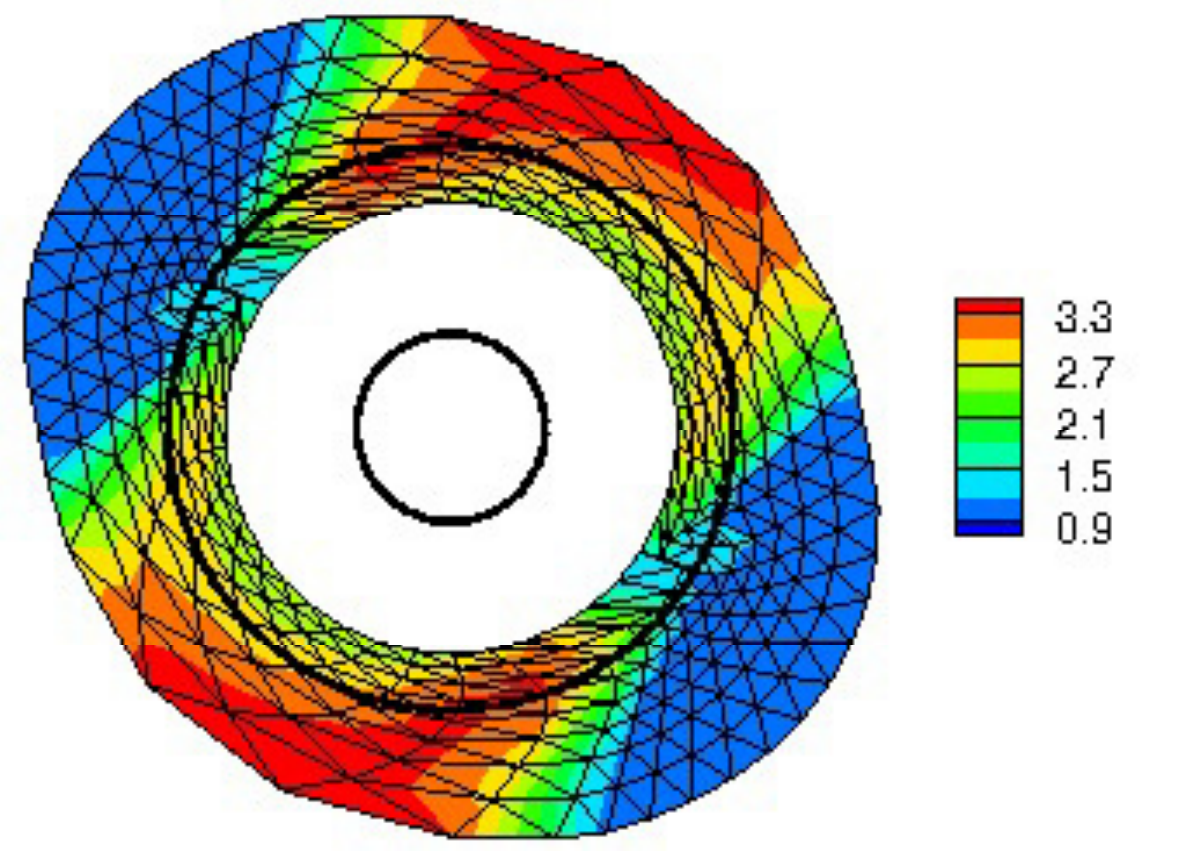}}
\put(0.3,5.76){\includegraphics[height = 3.4cm]{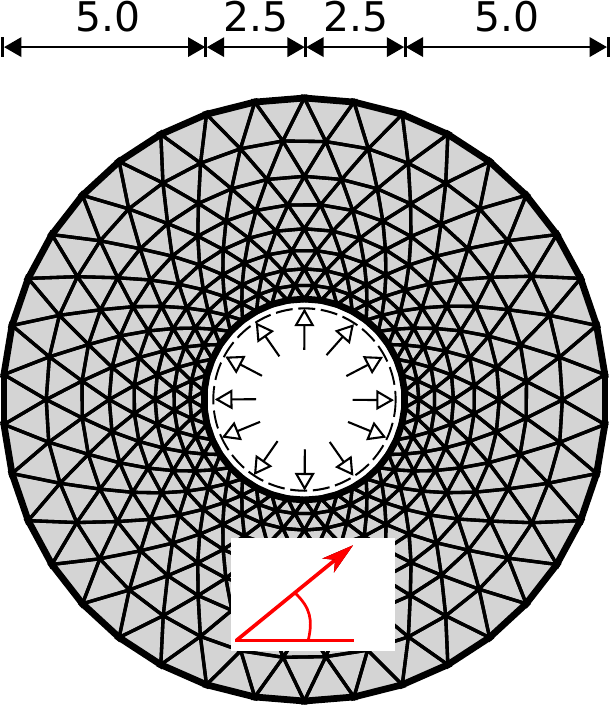}}
\put(7.09,7.05){\small Set 3}
\put(12.14,7.05){\small Set 4}
\put(7.08,2.27){\small Set 5}
\put(12.15,2.27){\small Set 6}
\put(0.2,5.7){a)}
\put(5.2,5.7){b)}
\put(10.4 ,5.7){c)}
\put(5.2,0.6){d)}
\put(10.4,0.6){e)}
\put(1.43,6.31){\textcolor{red}{\tiny $\bA$}}
\put(1.8,6.13){\textcolor{red}{\tiny $45^{\circ}$}}
\put(15.2,3.25){$\lambda_{\varphi}$}
\put(1.6,7.16){\small $u_{\rm r}$}
\end{picture}
\setlength{\baselineskip}{11pt}
\caption{Example 3c: a) Boundary conditions and b)-e) deformed bodies of a circular perforated plate for parameter Set 3, 4, 5 and 6.\label{fig: Example perforated plate circle}%
}
\end{Figure}

\subsection{Orthotropic Hencky and exponentiated Hencky models \label{sec: ortho}}

Orthotropic material behavior is symmetric regarding three orthogonal planes. These three planes are spanned by the three 
preferred directions $\bA_1$, $\bA_2$ and $\bA_3$, which are unit vectors. The material symmetry group is then defined by
\eb
{\cal G}_{o}\,:=\,\{\pm\bone;\,\bR_1,\,\bR_2,\,\bR_3\}\,,
\ee
where $\bR_1$, $\bR_2$ and $\bR_3$ are the reflections with respect to the planes spanned by $(\bA_2,\,\bA_3)$, $(\bA_1,\,\bA_3)$ and $(\bA_1,\,\bA_2)$, respectively.
The three preferred directions $\bA_i|i=1,2,3$ are orthogonal and form the three structural tensors $\bM_i=\bA_i\otimes\bA_i |i=1,2,3$, which satisfy the orthogonality condition $\langle\bM_i,\bM_j\rangle=\delta_{ij}$. 
However,  it is sufficient to formulate orthotropic strain energies
with two structural tensors $\bM_1$ and $\bM_2$ and additional isotropic principal or main invariants.
In the same  manner as in the previous section we may formulate the orthotropic energy function
\begin{align}
\nonumber {W}_3(\log\bU) =&\, \mu \,\lVert\dev \log\bU\rVert^2 + \frac{\kappa}{2}\,[{\rm tr}(\log\bU)]^2 \\
\nonumber                 & +\alpha_1 \langle\log\bU,\bM_1\rangle({\rm tr}\log\bU) +\alpha_2 \langle\log\bU,\bM_2\rangle({\rm tr}\log\bU) \\
\nonumber                 & +2\mu_{1}\langle\bM_1,(\log\bU)^2\rangle +2\mu_{2}\langle\bM_2,(\log\bU)^2\rangle \\
                 & + \frac{1}{2}\beta_1\langle\bM_1,\log\bU\rangle^2 + \frac{1}{2}\beta_2\langle\bM_2,\log\bU\rangle^2 + \frac{1}{2}\beta_3\langle\bM_1,\log\bU\rangle\langle\bM_2,\log\bU\rangle\label{eq: psi_3}
\end{align}
and the corresponding exponentiated version
\begin{align}
 \nonumber {W}_4(\log\bU) =&\, \frac{\mu}{k_1}\,{\rm exp}[k_1 \lVert{\rm dev}\log\bU\rVert^2] + \frac{\kappa}{2k_2}\,{\rm exp}[k_2({\rm tr}\log\bU)^2]\\
\nonumber                  &+ \frac{\alpha_1}{k_3}\,{\rm exp}[k_3\langle\log\bU,\bM_1\rangle({\rm tr}\log\bU)] +  \frac{\alpha_1}{k_4}\,{\rm exp}[k_4\langle\log\bU,\bM_2\rangle({\rm tr}\log\bU)]\\
\nonumber                  &+ \frac{2\mu_1}{k_5}\,{\rm exp}[k_5\langle\bM_1,(\log\bU)^2\rangle]+\frac{2\mu_2}{k_6}\,{\rm exp}[k_6\langle\bM_2,(\log\bU)^2\rangle]\\
\nonumber                  &+ \frac{\beta_1}{2 k_7}\,{\rm exp}[k_7\langle\bM_1,\log\bU\rangle^2] + \frac{\beta_2}{2 k_8}\,{\rm exp}[k_8\langle\bM_2,\log\bU\rangle^2]\\
                           &+ \frac{\beta_3}{2 k_9}\,{\rm exp}[k_9\langle\bM_1,\log\bU\rangle\langle\bM_2,\log\bU\rangle]\,.
\end{align}
The orthotropic elasticity tensor $\IC^{\rm V}$ has nine independent variables. 
Choosing the structural tensors
$\bM_1= {\rm diag}(1,0,0)$, $\bM_2 ={\rm diag}(0,1,0)$ we obtain the general form
\eb
 \IC^{\rm V} = \begin{pmatrix}
       \IC^{\rm V}_{11} & \IC^{\rm V}_{12} & \IC^{\rm V}_{13} &0&0&0\\ 
       \IC^{\rm V}_{12} & \IC^{\rm V}_{22} & \IC^{\rm V}_{23} &0&0&0\\ 
       \IC^{\rm V}_{13} & \IC^{\rm V}_{23} & \IC^{\rm V}_{33} &0&0&0\\   
       0&0&0&\IC^{\rm V}_{44} &0&0\\
       0&0&0& 0&\IC^{\rm V}_{55}&0\\
       0&0&0& 0&0&\IC^{\rm V}_{66}\\       
      \end{pmatrix}\,.  \label{eq: classical orthotropic isotropy}
\ee
The material tangent of both functions, ${W}_3$ and ${W}_4$ both yield
\eb
\nonumber
\IC^{\rm V}|_{\bC =\bone} = \begin{pmatrix}
                    \begin{matrix}2\mu+\lambda+\\2\alpha_1+4\mu_1+\beta_1\end{matrix} &\lambda+\alpha_1+\alpha_2+\beta_3 &\lambda+\alpha_1 &0 &0 &0  \\
                     \lambda+\alpha_1+\alpha_2+\beta_3 & \begin{matrix}2\mu+\lambda+\\ 2\alpha_2+4\mu_2+\beta_2\end{matrix} &\lambda+\alpha_2 &0 &0 &0  \\
                     \lambda +\alpha_1&\lambda+\alpha_2 &2\mu+\lambda &0 &0 &0  \\
                     0 &0 &0 &\mu+\mu_2 &0 &0  \\
                     0 &0 &0 &0 &\mu+\mu_1 &0  \\
                     0 &0 &0 &0 &0 &\mu+\mu_1+\mu_2
                    \end{pmatrix}
\ee
in the reference configuration, with $\lambda = (3\kappa-2\mu)/3$ . Following the same scheme as in the transversely isotropic case the parameter identification
gives the following relations
\begin{equation}
\boxed{
\begin{aligned} 
 \mu &= \IC^{\rm V}_{44}+\IC^{\rm V}_{55}-\IC^{\rm V}_{66}\\
 \mu_1 &= \IC^{\rm V}_{66}-\IC^{\rm V}_{44}\\
 \mu_2 &= \IC^{\rm V}_{66}-\IC^{\rm V}_{55}\\
 \lambda &= \IC^{\rm V}_{33} + 2(\IC^{\rm V}_{66}-\IC^{\rm V}_{44}-\IC^{\rm V}_{55})\\
 \alpha_1 &= \IC^{\rm V}_{13}-\IC^{\rm V}_{33}-2(\IC^{\rm V}_{66}-\IC^{\rm V}_{44}-\IC^{\rm V}_{55})\\
 \alpha_2 &= \IC^{\rm V}_{23}-\IC^{\rm V}_{33}-2(\IC^{\rm V}_{66}-\IC^{\rm V}_{44}-\IC^{\rm V}_{55})\\
 \beta_1  &= \IC^{\rm V}_{11}+\IC^{\rm V}_{33}-2 \IC^{\rm V}_{13}-4 \IC^{\rm V}_{55}\\
 \beta_2  &= \IC^{\rm V}_{22}+\IC^{\rm V}_{33}-2 \IC^{\rm V}_{23}-4 \IC^{\rm V}_{44}\\
 \beta_3  &= \IC^{\rm V}_{12}-\IC^{\rm V}_{13}-\IC^{\rm V}_{23}+\IC^{\rm V}_{33}+2(\IC^{\rm V}_{66}-\IC^{\rm V}_{44}-\IC^{\rm V}_{55})
\end{aligned}}\,.
\end{equation}
A viscoelastic, orthotropic material model based on finite logarithmic strains has been recently proposed by \cite{LatMon:2015:afs}.

\clearpage
\subsection{Case study of the transversely isotropic model in logarithmic strain space\label{sec: academic}}

In order to study the anisotropic properties under compression we further introduce the strain energy functions
\begin{align}
\psi^{\rm ti}_{\rm \mbox{\small \ding{182}}_C} &=  \dfrac{\mu_1}{2 k_1}\left\{\exp \left[k_1 (\underbrace{\langle\bC^i,\bM \rangle}_{I_4^{{\rm C}^i}}-1)^2\right]-1\right\} \label{eq: psi_ti_C_no}\,,\\
\psi^{\rm ti}_{\rm \mbox{\small \ding{183}}_H} &=  \dfrac{\mu_1}{2 k_1}\left\{\exp \left[k_1 {\underbrace{\langle (\log\bU)^i,\bM \rangle}_{I_4^{{\rm H}^i}}}^2\right]-1\right\}\;. \label{eq: psi_ti_H_no} 
\end{align}
In the following we aim to investigate the performance of the anisotropic invariants $I_4^{{\rm C}^i}$,
$I_4^{{\rm H}^i}$, see Eq.~(\ref{eq: I4Hi}), respectively. The evolution of the invariants, as well as 
the stress response of the transversely isotropic strain energy functions $\psi^{\rm ti}(I_4^{{\rm C}^i})$ and $\psi^{\rm ti}(I_4^{{\rm H}^i})$, 
are plotted for different loading scenarios. The examples are evaluated such that the results are independent of any chosen isotropic strain energy function, since only the
anisotropic stress response will be plotted.
The case distinction for compression and tension included in the energy functions will be neglected, i.e. the fibers are allowed to induce stresses under compression
and the energy functions according to Eq.~(\ref{eq: psi_ti_C_no}) and Eq.~(\ref{eq: psi_ti_H_no}) will be used. The parameters $\mu_1$ and $k_1$ 
are set to one and the plotted evolution of the Cauchy stress 
\eb
\Bsigma^{\rm aniso} = \frac{1}{J}2\bF\pp{\psi^{\rm ti}}{\bC}\bF^T 
\ee
will be normalized by the occurring maximum stress at the final deformation state, to allow for a 
better comparison.

This study is restricted to classical homogenous deformation states, i.e. uniaxial tension and compression, simple shear and biaxial loading conditions. 
During uniaxial tension, uniaxial compression and biaxial loading the angles $\theta_k$ will remain constant and the body is free
of rotations, i.e. $\bF = \bU$. Only during 
the shear test they will change with a change in the deformation.

\subsubsection{Uniaxial tension and compression}

The considered problem is depicted in Fig.~\ref{fig: Uniax comp ten}. 
In this case the fiber direction is aligned with the loading direction. The reference configuration as well as the 
deformed configurations under tension and compression are shown on the right.
The component $F_{11}$ of the deformation gradient refers to the stretch in fiber direction.
Since the body is considered to be incompressible we find that $F_{22} = F_{33} = 1/\sqrt{F_{11}}$.

The results for the transversely anisotropic Hencky function $\psi^{\rm ti}_{\rm \mbox{\small \ding{183}}_H}$ are displayed in Fig.~\ref{fig: uniax_tension_aniso_invarianten}.
As already discussed in the previous section, in Fig.~\ref{fig: uniax_tension_aniso_invarianten}a) it becomes apparent that for even 
exponents of $i$ in $I_4^{\rm H^i}$ the values of the invariant are also positive under compression, i.e. the sign of the 
invariant is not the right choice to distinguish between tensile and compressive stretches.
Nevertheless, the stress response 
seems to be adequate from a physical point of view for each of the considered invariants. For $I_4^{\rm H^1}=\langle\log\bU,\bM \rangle$ we obtain a perfectly 
linear material behavior. Due to the logarithmic 
framework the stress function is generally more sensitive to compression than to tension, see Fig.~\ref{fig: uniax_tension_aniso_invarianten}b).  
Note that the stress function was normalized with the corresponding highest stress ${\rm max}(\lVert\sigma_{11}^{\rm aniso})\rVert$ to allow for a better comparison.  
All stress-strain responses show the potential to exhibit significant strain stiffening for both tension and compression. This is especially remarkable 
for the compressive case. 
\cite{HyuNak:2003:acp} for example found that porous copper fabricated by unidirectional solidification behaves strongly anisotropic under compression and exhibits considerable
stiffening under large strains up to $80\%$  due to the alignment of the pores. 
Classical anisotropic material laws based on the  invariant $\langle \bC^i,\bM\rangle$ struggle to reproduce this effect. 
The normalized stresses under compression according to Eq.~(\ref{eq: psi_ti_C_no}) are plotted in Fig.~\ref{fig: uniax_compression_sig11_aniso}a).
The plot reveals a strain softening behavior. Moreover, after a certain point the stresses will begin to increase although the body is further compressed
which is strictly unphysical. Also classical polynomial laws of the form
\eb
\psi^{\rm ti}_{\rm \mbox{\small \ding{184}}_C} = \frac{\mu_1}{2 k_1}(\langle\bC^i,\bM\rangle-1)^{k_1}
\ee
suffer from this effect, see Fig.~\ref{fig: uniax_compression_sig11_aniso}b). Here, $\mu_1$ was set to one and $k_1$ to two. 
In addition the domain of definition of the above energy is restricted to even values of $k_1$ in compression which significantly limits the 
parameter fitting properties.

\begin{Figure}[!htb]
\unitlength 1 cm
\begin{picture}(14,5.5)
\put(2.0,0.0){\includegraphics[height = 5.5cm]{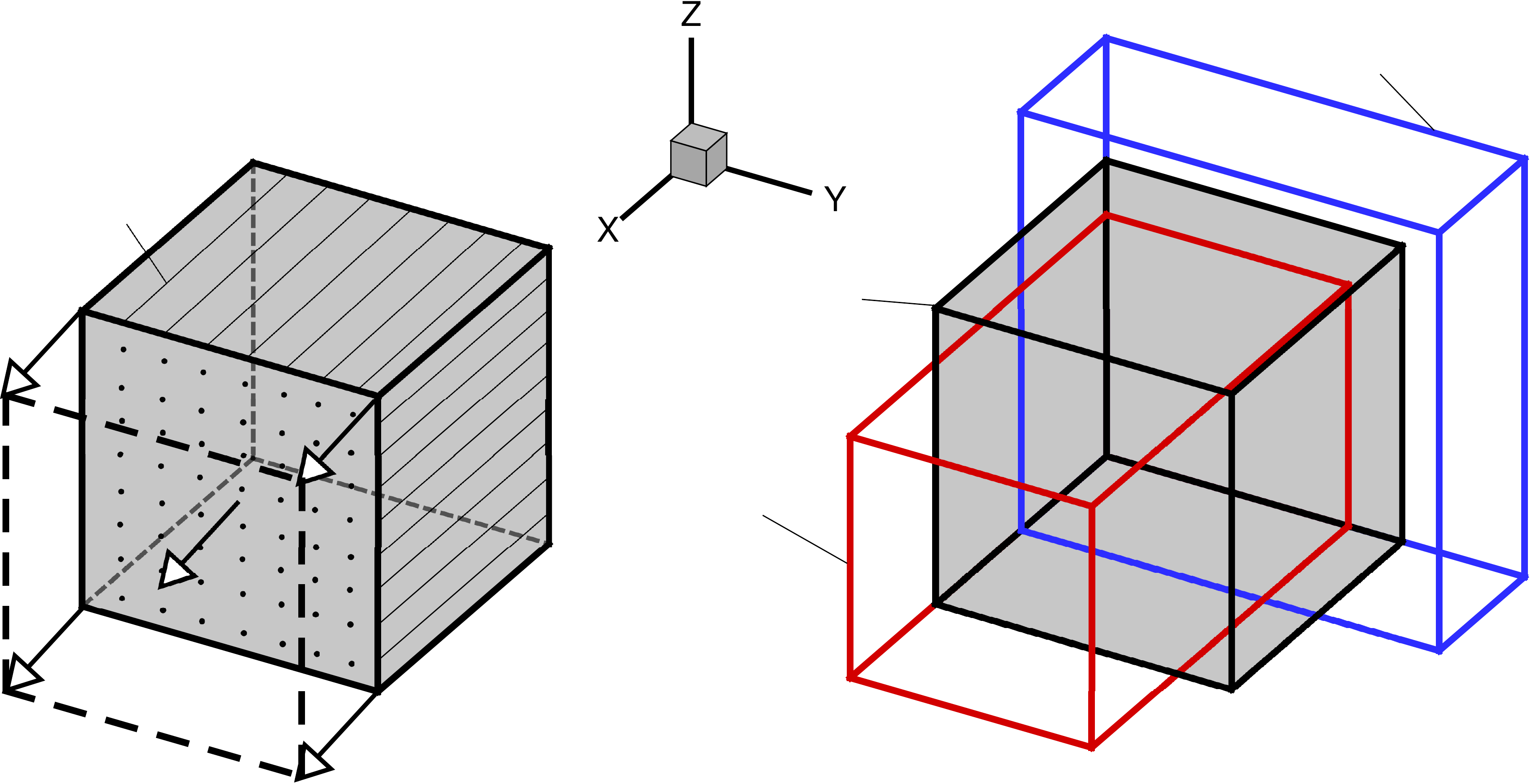}}
\put(2.7,4.0){$\bA$}
\put(11.2,5.05){comp.}
\put(7.42,3.27){ref.}
\put(6.5,1.97){tension}
\end{picture}
\setlength{\baselineskip}{11pt}
\caption{Uniaxial compression and tension test, where the preferred direction and the loading are aligned. The body is assumed to be incompressible 
with $V_{\rm ref}\,=\,V_{\rm tension}\,=\,V_{\rm comp}$. \label{fig: Uniax comp ten}%
}
\end{Figure}

\begin{Figure}[!htb]
\unitlength 1 cm
\begin{picture}(14,5.8)
\put(0.0,0.0){
\begingroup
  \makeatletter
  \providecommand\color[2][]{%
    \GenericError{(gnuplot) \space\space\space\@spaces}{%
      Package color not loaded in conjunction with
      terminal option `colourtext'%
    }{See the gnuplot documentation for explanation.%
    }{Either use 'blacktext' in gnuplot or load the package
      color.sty in LaTeX.}%
    \renewcommand\color[2][]{}%
  }%
  \providecommand\includegraphics[2][]{%
    \GenericError{(gnuplot) \space\space\space\@spaces}{%
      Package graphicx or graphics not loaded%
    }{See the gnuplot documentation for explanation.%
    }{The gnuplot epslatex terminal needs graphicx.sty or graphics.sty.}%
    \renewcommand\includegraphics[2][]{}%
  }%
  \providecommand\rotatebox[2]{#2}%
  \@ifundefined{ifGPcolor}{%
    \newif\ifGPcolor
    \GPcolortrue
  }{}%
  \@ifundefined{ifGPblacktext}{%
    \newif\ifGPblacktext
    \GPblacktexttrue
  }{}%
  \let\gplgaddtomacro\g@addto@macro
  \gdef\gplbacktext{}%
  \gdef\gplfronttext{}%
  \makeatother
  \ifGPblacktext
    \def\colorrgb#1{}%
    \def\colorgray#1{}%
  \else
    \ifGPcolor
      \def\colorrgb#1{\color[rgb]{#1}}%
      \def\colorgray#1{\color[gray]{#1}}%
      \expandafter\def\csname LTw\endcsname{\color{white}}%
      \expandafter\def\csname LTb\endcsname{\color{black}}%
      \expandafter\def\csname LTa\endcsname{\color{black}}%
      \expandafter\def\csname LT0\endcsname{\color[rgb]{1,0,0}}%
      \expandafter\def\csname LT1\endcsname{\color[rgb]{0,1,0}}%
      \expandafter\def\csname LT2\endcsname{\color[rgb]{0,0,1}}%
      \expandafter\def\csname LT3\endcsname{\color[rgb]{1,0,1}}%
      \expandafter\def\csname LT4\endcsname{\color[rgb]{0,1,1}}%
      \expandafter\def\csname LT5\endcsname{\color[rgb]{1,1,0}}%
      \expandafter\def\csname LT6\endcsname{\color[rgb]{0,0,0}}%
      \expandafter\def\csname LT7\endcsname{\color[rgb]{1,0.3,0}}%
      \expandafter\def\csname LT8\endcsname{\color[rgb]{0.5,0.5,0.5}}%
    \else
      \def\colorrgb#1{\color{black}}%
      \def\colorgray#1{\color[gray]{#1}}%
      \expandafter\def\csname LTw\endcsname{\color{white}}%
      \expandafter\def\csname LTb\endcsname{\color{black}}%
      \expandafter\def\csname LTa\endcsname{\color{black}}%
      \expandafter\def\csname LT0\endcsname{\color{black}}%
      \expandafter\def\csname LT1\endcsname{\color{black}}%
      \expandafter\def\csname LT2\endcsname{\color{black}}%
      \expandafter\def\csname LT3\endcsname{\color{black}}%
      \expandafter\def\csname LT4\endcsname{\color{black}}%
      \expandafter\def\csname LT5\endcsname{\color{black}}%
      \expandafter\def\csname LT6\endcsname{\color{black}}%
      \expandafter\def\csname LT7\endcsname{\color{black}}%
      \expandafter\def\csname LT8\endcsname{\color{black}}%
    \fi
  \fi
  \setlength{\unitlength}{0.0500bp}%
  \begin{picture}(4534.40,3310.40)%
    \gplgaddtomacro\gplbacktext{%
      \csname LTb\endcsname%
      \put(946,704){\makebox(0,0)[r]{\strut{}-0.8}}%
      \csname LTb\endcsname%
      \put(946,1039){\makebox(0,0)[r]{\strut{}-0.6}}%
      \csname LTb\endcsname%
      \put(946,1373){\makebox(0,0)[r]{\strut{}-0.4}}%
      \csname LTb\endcsname%
      \put(946,1708){\makebox(0,0)[r]{\strut{}-0.2}}%
      \csname LTb\endcsname%
      \put(946,2042){\makebox(0,0)[r]{\strut{} 0}}%
      \csname LTb\endcsname%
      \put(946,2377){\makebox(0,0)[r]{\strut{} 0.2}}%
      \csname LTb\endcsname%
      \put(946,2711){\makebox(0,0)[r]{\strut{} 0.4}}%
      \csname LTb\endcsname%
      \put(946,3046){\makebox(0,0)[r]{\strut{} 0.6}}%
      \csname LTb\endcsname%
      \put(1384,484){\makebox(0,0){\strut{} 0.6}}%
      \csname LTb\endcsname%
      \put(1996,484){\makebox(0,0){\strut{} 0.8}}%
      \csname LTb\endcsname%
      \put(2608,484){\makebox(0,0){\strut{} 1}}%
      \csname LTb\endcsname%
      \put(3219,484){\makebox(0,0){\strut{} 1.2}}%
      \csname LTb\endcsname%
      \put(3831,484){\makebox(0,0){\strut{} 1.4}}%
      \put(176,1875){\rotatebox{-270}{\makebox(0,0){\strut{}$I_4^{H^i}$}}}%
      \put(2607,154){\makebox(0,0){\strut{}$F_{11}$}}%
    }%
    \gplgaddtomacro\gplfronttext{%
      \csname LTb\endcsname%
      \put(3150,1537){\makebox(0,0)[r]{\strut{}$i=1$}}%
      \csname LTb\endcsname%
      \put(3150,1317){\makebox(0,0)[r]{\strut{}$i=2$}}%
      \csname LTb\endcsname%
      \put(3150,1097){\makebox(0,0)[r]{\strut{}$i=3$}}%
      \csname LTb\endcsname%
      \put(3150,877){\makebox(0,0)[r]{\strut{}$i=4$}}%
    }%
    \gplbacktext
    \put(0,0){\includegraphics{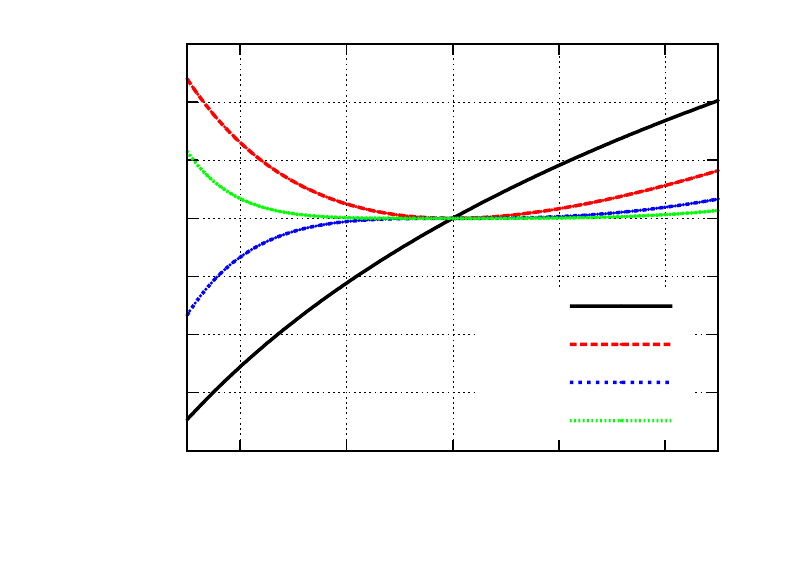}}%
    \gplfronttext
  \end{picture}%
\endgroup
}
\put(8.5,0.0){
\begingroup
  \makeatletter
  \providecommand\color[2][]{%
    \GenericError{(gnuplot) \space\space\space\@spaces}{%
      Package color not loaded in conjunction with
      terminal option `colourtext'%
    }{See the gnuplot documentation for explanation.%
    }{Either use 'blacktext' in gnuplot or load the package
      color.sty in LaTeX.}%
    \renewcommand\color[2][]{}%
  }%
  \providecommand\includegraphics[2][]{%
    \GenericError{(gnuplot) \space\space\space\@spaces}{%
      Package graphicx or graphics not loaded%
    }{See the gnuplot documentation for explanation.%
    }{The gnuplot epslatex terminal needs graphicx.sty or graphics.sty.}%
    \renewcommand\includegraphics[2][]{}%
  }%
  \providecommand\rotatebox[2]{#2}%
  \@ifundefined{ifGPcolor}{%
    \newif\ifGPcolor
    \GPcolortrue
  }{}%
  \@ifundefined{ifGPblacktext}{%
    \newif\ifGPblacktext
    \GPblacktexttrue
  }{}%
  \let\gplgaddtomacro\g@addto@macro
  \gdef\gplbacktext{}%
  \gdef\gplfronttext{}%
  \makeatother
  \ifGPblacktext
    \def\colorrgb#1{}%
    \def\colorgray#1{}%
  \else
    \ifGPcolor
      \def\colorrgb#1{\color[rgb]{#1}}%
      \def\colorgray#1{\color[gray]{#1}}%
      \expandafter\def\csname LTw\endcsname{\color{white}}%
      \expandafter\def\csname LTb\endcsname{\color{black}}%
      \expandafter\def\csname LTa\endcsname{\color{black}}%
      \expandafter\def\csname LT0\endcsname{\color[rgb]{1,0,0}}%
      \expandafter\def\csname LT1\endcsname{\color[rgb]{0,1,0}}%
      \expandafter\def\csname LT2\endcsname{\color[rgb]{0,0,1}}%
      \expandafter\def\csname LT3\endcsname{\color[rgb]{1,0,1}}%
      \expandafter\def\csname LT4\endcsname{\color[rgb]{0,1,1}}%
      \expandafter\def\csname LT5\endcsname{\color[rgb]{1,1,0}}%
      \expandafter\def\csname LT6\endcsname{\color[rgb]{0,0,0}}%
      \expandafter\def\csname LT7\endcsname{\color[rgb]{1,0.3,0}}%
      \expandafter\def\csname LT8\endcsname{\color[rgb]{0.5,0.5,0.5}}%
    \else
      \def\colorrgb#1{\color{black}}%
      \def\colorgray#1{\color[gray]{#1}}%
      \expandafter\def\csname LTw\endcsname{\color{white}}%
      \expandafter\def\csname LTb\endcsname{\color{black}}%
      \expandafter\def\csname LTa\endcsname{\color{black}}%
      \expandafter\def\csname LT0\endcsname{\color{black}}%
      \expandafter\def\csname LT1\endcsname{\color{black}}%
      \expandafter\def\csname LT2\endcsname{\color{black}}%
      \expandafter\def\csname LT3\endcsname{\color{black}}%
      \expandafter\def\csname LT4\endcsname{\color{black}}%
      \expandafter\def\csname LT5\endcsname{\color{black}}%
      \expandafter\def\csname LT6\endcsname{\color{black}}%
      \expandafter\def\csname LT7\endcsname{\color{black}}%
      \expandafter\def\csname LT8\endcsname{\color{black}}%
    \fi
  \fi
  \setlength{\unitlength}{0.0500bp}%
  \begin{picture}(4534.40,3310.40)%
    \gplgaddtomacro\gplbacktext{%
      \csname LTb\endcsname%
      \put(946,704){\makebox(0,0)[r]{\strut{}-1}}%
      \csname LTb\endcsname%
      \put(946,997){\makebox(0,0)[r]{\strut{}-0.8}}%
      \csname LTb\endcsname%
      \put(946,1290){\makebox(0,0)[r]{\strut{}-0.6}}%
      \csname LTb\endcsname%
      \put(946,1582){\makebox(0,0)[r]{\strut{}-0.4}}%
      \csname LTb\endcsname%
      \put(946,1875){\makebox(0,0)[r]{\strut{}-0.2}}%
      \csname LTb\endcsname%
      \put(946,2168){\makebox(0,0)[r]{\strut{} 0}}%
      \csname LTb\endcsname%
      \put(946,2461){\makebox(0,0)[r]{\strut{} 0.2}}%
      \csname LTb\endcsname%
      \put(946,2753){\makebox(0,0)[r]{\strut{} 0.4}}%
      \csname LTb\endcsname%
      \put(946,3046){\makebox(0,0)[r]{\strut{} 0.6}}%
      \csname LTb\endcsname%
      \put(1384,484){\makebox(0,0){\strut{} 0.6}}%
      \csname LTb\endcsname%
      \put(1996,484){\makebox(0,0){\strut{} 0.8}}%
      \csname LTb\endcsname%
      \put(2608,484){\makebox(0,0){\strut{} 1}}%
      \csname LTb\endcsname%
      \put(3219,484){\makebox(0,0){\strut{} 1.2}}%
      \csname LTb\endcsname%
      \put(3831,484){\makebox(0,0){\strut{} 1.4}}%
      \put(176,1875){\rotatebox{-270}{\makebox(0,0){\strut{}$\sigma_{11}^{\rm aniso}/{\rm max(||\sigma_{11}^{\rm aniso}||)}$}}}%
      \put(2607,154){\makebox(0,0){\strut{}$F_{11}$}}%
    }%
    \gplgaddtomacro\gplfronttext{%
      \csname LTb\endcsname%
      \put(3150,1537){\makebox(0,0)[r]{\strut{}$i=1$}}%
      \csname LTb\endcsname%
      \put(3150,1317){\makebox(0,0)[r]{\strut{}$i=2$}}%
      \csname LTb\endcsname%
      \put(3150,1097){\makebox(0,0)[r]{\strut{}$i=3$}}%
      \csname LTb\endcsname%
      \put(3150,877){\makebox(0,0)[r]{\strut{}$i=4$}}%
    }%
    \gplbacktext
    \put(0,0){\includegraphics{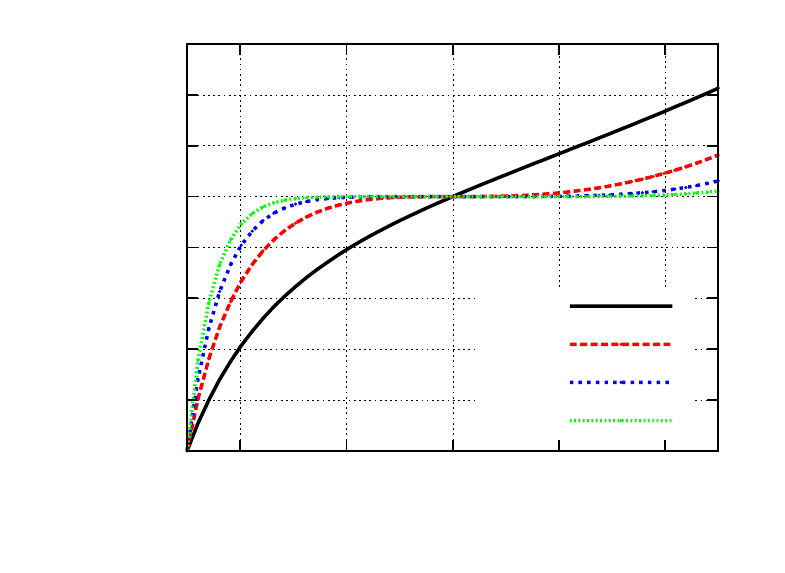}}%
    \gplfronttext
  \end{picture}%
\endgroup
}
\put(3.8,4.3){\tiny  \setlength{\fboxsep}{2.0mm}\fcolorbox{mygray}{mygray}{$I_4^{{\rm H}^i}$}}
\put(11.3,4.3){\tiny  \setlength{\fboxsep}{2.0mm}\fcolorbox{mygray}{mygray}{$I_4^{{\rm H}^i}$}}
\put(0.3,0.15){a)}
\put(8.5,0.15){b)}
\end{picture}
\setlength{\baselineskip}{11pt}
\caption{Monotonicity of the Cauchy stress $\sigma_{11}$ as function of uniaxial stretch.
Evolution of a) the anisotropic Hencky-type invariants $I_4^{{\rm H}^i}$ for $i=1,2,3,4$ and b) corresponding anisotropic stress response under uniaxial tension
and compression for $\psi^{\rm ti}_{\rm \mbox{\small \ding{183}}_H}$. The stresses are normalized by the absolute value of the maximum arising compressive stress at $F_{11} = 0.5$. \label{fig: uniax_tension_aniso_invarianten}}
\end{Figure}

\begin{Figure}[!htb]
\unitlength 1 cm
\begin{picture}(14,5.8)
\put(0.0,0.0){
\begingroup
  \makeatletter
  \providecommand\color[2][]{%
    \GenericError{(gnuplot) \space\space\space\@spaces}{%
      Package color not loaded in conjunction with
      terminal option `colourtext'%
    }{See the gnuplot documentation for explanation.%
    }{Either use 'blacktext' in gnuplot or load the package
      color.sty in LaTeX.}%
    \renewcommand\color[2][]{}%
  }%
  \providecommand\includegraphics[2][]{%
    \GenericError{(gnuplot) \space\space\space\@spaces}{%
      Package graphicx or graphics not loaded%
    }{See the gnuplot documentation for explanation.%
    }{The gnuplot epslatex terminal needs graphicx.sty or graphics.sty.}%
    \renewcommand\includegraphics[2][]{}%
  }%
  \providecommand\rotatebox[2]{#2}%
  \@ifundefined{ifGPcolor}{%
    \newif\ifGPcolor
    \GPcolortrue
  }{}%
  \@ifundefined{ifGPblacktext}{%
    \newif\ifGPblacktext
    \GPblacktexttrue
  }{}%
  \let\gplgaddtomacro\g@addto@macro
  \gdef\gplbacktext{}%
  \gdef\gplfronttext{}%
  \makeatother
  \ifGPblacktext
    \def\colorrgb#1{}%
    \def\colorgray#1{}%
  \else
    \ifGPcolor
      \def\colorrgb#1{\color[rgb]{#1}}%
      \def\colorgray#1{\color[gray]{#1}}%
      \expandafter\def\csname LTw\endcsname{\color{white}}%
      \expandafter\def\csname LTb\endcsname{\color{black}}%
      \expandafter\def\csname LTa\endcsname{\color{black}}%
      \expandafter\def\csname LT0\endcsname{\color[rgb]{1,0,0}}%
      \expandafter\def\csname LT1\endcsname{\color[rgb]{0,1,0}}%
      \expandafter\def\csname LT2\endcsname{\color[rgb]{0,0,1}}%
      \expandafter\def\csname LT3\endcsname{\color[rgb]{1,0,1}}%
      \expandafter\def\csname LT4\endcsname{\color[rgb]{0,1,1}}%
      \expandafter\def\csname LT5\endcsname{\color[rgb]{1,1,0}}%
      \expandafter\def\csname LT6\endcsname{\color[rgb]{0,0,0}}%
      \expandafter\def\csname LT7\endcsname{\color[rgb]{1,0.3,0}}%
      \expandafter\def\csname LT8\endcsname{\color[rgb]{0.5,0.5,0.5}}%
    \else
      \def\colorrgb#1{\color{black}}%
      \def\colorgray#1{\color[gray]{#1}}%
      \expandafter\def\csname LTw\endcsname{\color{white}}%
      \expandafter\def\csname LTb\endcsname{\color{black}}%
      \expandafter\def\csname LTa\endcsname{\color{black}}%
      \expandafter\def\csname LT0\endcsname{\color{black}}%
      \expandafter\def\csname LT1\endcsname{\color{black}}%
      \expandafter\def\csname LT2\endcsname{\color{black}}%
      \expandafter\def\csname LT3\endcsname{\color{black}}%
      \expandafter\def\csname LT4\endcsname{\color{black}}%
      \expandafter\def\csname LT5\endcsname{\color{black}}%
      \expandafter\def\csname LT6\endcsname{\color{black}}%
      \expandafter\def\csname LT7\endcsname{\color{black}}%
      \expandafter\def\csname LT8\endcsname{\color{black}}%
    \fi
  \fi
  \setlength{\unitlength}{0.0500bp}%
  \begin{picture}(4534.40,3310.40)%
    \gplgaddtomacro\gplbacktext{%
      \csname LTb\endcsname%
      \put(946,704){\makebox(0,0)[r]{\strut{}-1}}%
      \csname LTb\endcsname%
      \put(946,1172){\makebox(0,0)[r]{\strut{}-0.8}}%
      \csname LTb\endcsname%
      \put(946,1641){\makebox(0,0)[r]{\strut{}-0.6}}%
      \csname LTb\endcsname%
      \put(946,2109){\makebox(0,0)[r]{\strut{}-0.4}}%
      \csname LTb\endcsname%
      \put(946,2578){\makebox(0,0)[r]{\strut{}-0.2}}%
      \csname LTb\endcsname%
      \put(946,3046){\makebox(0,0)[r]{\strut{} 0}}%
      \csname LTb\endcsname%
      \put(1078,484){\makebox(0,0){\strut{} 0.5}}%
      \csname LTb\endcsname%
      \put(1690,484){\makebox(0,0){\strut{} 0.6}}%
      \csname LTb\endcsname%
      \put(2302,484){\makebox(0,0){\strut{} 0.7}}%
      \csname LTb\endcsname%
      \put(2913,484){\makebox(0,0){\strut{} 0.8}}%
      \csname LTb\endcsname%
      \put(3525,484){\makebox(0,0){\strut{} 0.9}}%
      \csname LTb\endcsname%
      \put(4137,484){\makebox(0,0){\strut{} 1}}%
      \put(176,1875){\rotatebox{-270}{\makebox(0,0){\strut{}$\sigma_{11}^{\rm aniso}/{\rm max(||\sigma_{11}^{\rm aniso}||)}$}}}%
      \put(2607,154){\makebox(0,0){\strut{}$F_{11}$}}%
    }%
    \gplgaddtomacro\gplfronttext{%
      \csname LTb\endcsname%
      \put(2976,2936){\makebox(0,0)[r]{\strut{}$i=1$}}%
      \csname LTb\endcsname%
      \put(2976,2716){\makebox(0,0)[r]{\strut{}$i=2$}}%
      \csname LTb\endcsname%
      \put(2976,2496){\makebox(0,0)[r]{\strut{}$i=3$}}%
      \csname LTb\endcsname%
      \put(2976,2276){\makebox(0,0)[r]{\strut{}$i=4$}}%
    }%
    \gplbacktext
    \put(0,0){\includegraphics{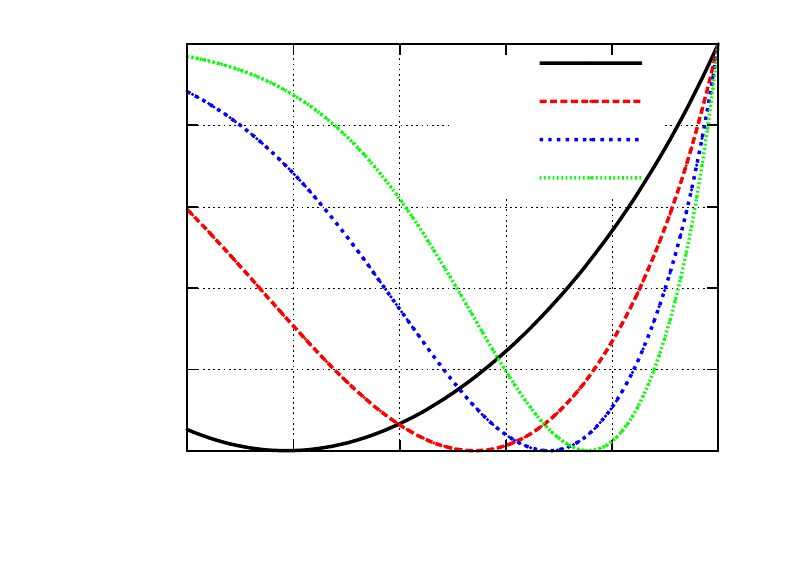}}%
    \gplfronttext
  \end{picture}%
\endgroup
}
\put(2.1,1.83){\tiny  \setlength{\fboxsep}{2.0mm}\fcolorbox{mygray}{mygray}{$I_4^{C^i}$}}
\put(8.5,0.0){
\begingroup
  \makeatletter
  \providecommand\color[2][]{%
    \GenericError{(gnuplot) \space\space\space\@spaces}{%
      Package color not loaded in conjunction with
      terminal option `colourtext'%
    }{See the gnuplot documentation for explanation.%
    }{Either use 'blacktext' in gnuplot or load the package
      color.sty in LaTeX.}%
    \renewcommand\color[2][]{}%
  }%
  \providecommand\includegraphics[2][]{%
    \GenericError{(gnuplot) \space\space\space\@spaces}{%
      Package graphicx or graphics not loaded%
    }{See the gnuplot documentation for explanation.%
    }{The gnuplot epslatex terminal needs graphicx.sty or graphics.sty.}%
    \renewcommand\includegraphics[2][]{}%
  }%
  \providecommand\rotatebox[2]{#2}%
  \@ifundefined{ifGPcolor}{%
    \newif\ifGPcolor
    \GPcolortrue
  }{}%
  \@ifundefined{ifGPblacktext}{%
    \newif\ifGPblacktext
    \GPblacktexttrue
  }{}%
  \let\gplgaddtomacro\g@addto@macro
  \gdef\gplbacktext{}%
  \gdef\gplfronttext{}%
  \makeatother
  \ifGPblacktext
    \def\colorrgb#1{}%
    \def\colorgray#1{}%
  \else
    \ifGPcolor
      \def\colorrgb#1{\color[rgb]{#1}}%
      \def\colorgray#1{\color[gray]{#1}}%
      \expandafter\def\csname LTw\endcsname{\color{white}}%
      \expandafter\def\csname LTb\endcsname{\color{black}}%
      \expandafter\def\csname LTa\endcsname{\color{black}}%
      \expandafter\def\csname LT0\endcsname{\color[rgb]{1,0,0}}%
      \expandafter\def\csname LT1\endcsname{\color[rgb]{0,1,0}}%
      \expandafter\def\csname LT2\endcsname{\color[rgb]{0,0,1}}%
      \expandafter\def\csname LT3\endcsname{\color[rgb]{1,0,1}}%
      \expandafter\def\csname LT4\endcsname{\color[rgb]{0,1,1}}%
      \expandafter\def\csname LT5\endcsname{\color[rgb]{1,1,0}}%
      \expandafter\def\csname LT6\endcsname{\color[rgb]{0,0,0}}%
      \expandafter\def\csname LT7\endcsname{\color[rgb]{1,0.3,0}}%
      \expandafter\def\csname LT8\endcsname{\color[rgb]{0.5,0.5,0.5}}%
    \else
      \def\colorrgb#1{\color{black}}%
      \def\colorgray#1{\color[gray]{#1}}%
      \expandafter\def\csname LTw\endcsname{\color{white}}%
      \expandafter\def\csname LTb\endcsname{\color{black}}%
      \expandafter\def\csname LTa\endcsname{\color{black}}%
      \expandafter\def\csname LT0\endcsname{\color{black}}%
      \expandafter\def\csname LT1\endcsname{\color{black}}%
      \expandafter\def\csname LT2\endcsname{\color{black}}%
      \expandafter\def\csname LT3\endcsname{\color{black}}%
      \expandafter\def\csname LT4\endcsname{\color{black}}%
      \expandafter\def\csname LT5\endcsname{\color{black}}%
      \expandafter\def\csname LT6\endcsname{\color{black}}%
      \expandafter\def\csname LT7\endcsname{\color{black}}%
      \expandafter\def\csname LT8\endcsname{\color{black}}%
    \fi
  \fi
  \setlength{\unitlength}{0.0500bp}%
  \begin{picture}(4534.40,3310.40)%
    \gplgaddtomacro\gplbacktext{%
      \csname LTb\endcsname%
      \put(946,704){\makebox(0,0)[r]{\strut{}-1}}%
      \csname LTb\endcsname%
      \put(946,1172){\makebox(0,0)[r]{\strut{}-0.8}}%
      \csname LTb\endcsname%
      \put(946,1641){\makebox(0,0)[r]{\strut{}-0.6}}%
      \csname LTb\endcsname%
      \put(946,2109){\makebox(0,0)[r]{\strut{}-0.4}}%
      \csname LTb\endcsname%
      \put(946,2578){\makebox(0,0)[r]{\strut{}-0.2}}%
      \csname LTb\endcsname%
      \put(946,3046){\makebox(0,0)[r]{\strut{} 0}}%
      \csname LTb\endcsname%
      \put(1078,484){\makebox(0,0){\strut{} 0.5}}%
      \csname LTb\endcsname%
      \put(1690,484){\makebox(0,0){\strut{} 0.6}}%
      \csname LTb\endcsname%
      \put(2302,484){\makebox(0,0){\strut{} 0.7}}%
      \csname LTb\endcsname%
      \put(2913,484){\makebox(0,0){\strut{} 0.8}}%
      \csname LTb\endcsname%
      \put(3525,484){\makebox(0,0){\strut{} 0.9}}%
      \csname LTb\endcsname%
      \put(4137,484){\makebox(0,0){\strut{} 1}}%
      \put(176,1875){\rotatebox{-270}{\makebox(0,0){\strut{}$\sigma_{11}^{\rm aniso}/{\rm max(||\sigma_{11}^{\rm aniso}||)}$}}}%
      \put(2607,154){\makebox(0,0){\strut{}$F_{11}$}}%
    }%
    \gplgaddtomacro\gplfronttext{%
      \csname LTb\endcsname%
      \put(2976,2936){\makebox(0,0)[r]{\strut{}$i=1$}}%
      \csname LTb\endcsname%
      \put(2976,2716){\makebox(0,0)[r]{\strut{}$i=2$}}%
      \csname LTb\endcsname%
      \put(2976,2496){\makebox(0,0)[r]{\strut{}$i=3$}}%
      \csname LTb\endcsname%
      \put(2976,2276){\makebox(0,0)[r]{\strut{}$i=4$}}%
    }%
    \gplbacktext
    \put(0,0){\includegraphics{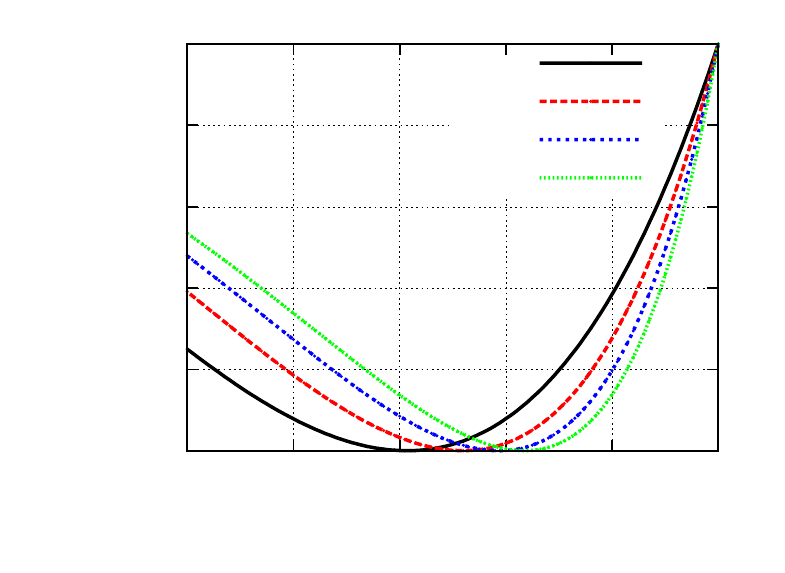}}%
    \gplfronttext
  \end{picture}%
\endgroup
}
\put(12.5,3.0){\tiny  \setlength{\fboxsep}{2.0mm}\fcolorbox{mygray}{mygray}{$I_4^{C^i}$}}
\put(0.3,0.15){a)}
\put(8.5,0.15){b)}
\end{picture}
\setlength{\baselineskip}{11pt}
\caption{Pathological non-monotonicity of the Cauchy stress $\sigma_{11}$ as function of uniaxial stretch.
Normalized anisotropic stress response under uniaxial compression. In a) the exponential strain energy function $\psi^{\rm ti}_{\rm \mbox{\small \ding{182}}_C}$ 
and in b) the polynomial function $\psi^{\rm ti}_{\rm \mbox{\small \ding{184}}_C}$ is plotted. \label{fig: uniax_compression_sig11_aniso}}
\end{Figure}

\clearpage
\subsubsection{Simple shear}
Next we investigate the behavior for simple shear, according to Fig.~\ref{fig: shear}. The shear direction will be aligned
with the fiber direction and the amount of shear 
\eb
\gamma = \frac{u}{L}
\ee
is defined as the quotient of the displacements by the length. Note that in this example the fibers are not elongated at all, i.e.
$I_4^{\rm C^1} = 1$.
The results for the transversely anisotropic Hencky function are displayed in Fig.~\ref{fig: shear_aniso_H_invarianten_sig13}
and Fig.~\ref{fig: shear_aniso_H_sig11_sig33}. Again the invariants of even and odd powers take a different sign. The stress quantities 
which are not plotted in Fig.~\ref{fig: shear_aniso_H_sig11_sig33}a) are equal to zero. That  means for even powers $i$, 
$\sigma^{\rm aniso}_{11}$ will be equal to zero.

\begin{Figure}[!htb]
\unitlength 1 cm
\begin{picture}(14,5.5)
\put(2.0,0.0){\includegraphics[height = 5.5cm]{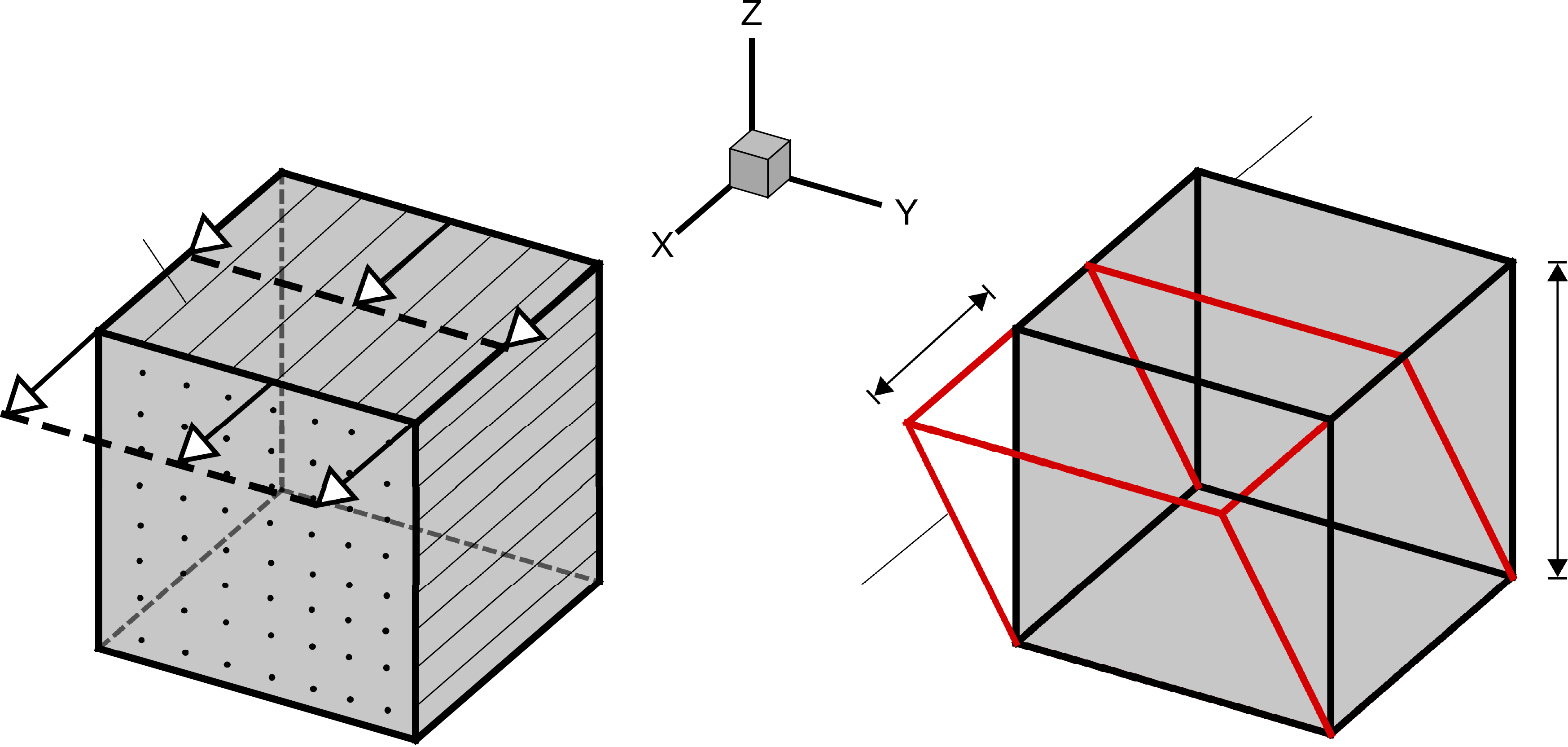}}
\put(2.7,4.0){$\bA$}
\put(11.35,4.7){ref.}
\put(7.7,0.9){sheared}
\begin{turn}{45}
\put(8.3,-4.05){$u$}
\end{turn}
\put(10.78,2.3){$L$}
\end{picture}
\setlength{\baselineskip}{11pt}
\caption{Shear test, where the preferred direction and the shear direction are aligned. Simple shear deformation is incompressible 
with $V_{\rm ref}\,=\,V_{\rm sheared}$. \label{fig: shear}%
}
\end{Figure}

\begin{Figure}[!htb]
\unitlength 1 cm
\begin{picture}(14,5.8)
\put(0.0,0.0){
\begingroup
  \makeatletter
  \providecommand\color[2][]{%
    \GenericError{(gnuplot) \space\space\space\@spaces}{%
      Package color not loaded in conjunction with
      terminal option `colourtext'%
    }{See the gnuplot documentation for explanation.%
    }{Either use 'blacktext' in gnuplot or load the package
      color.sty in LaTeX.}%
    \renewcommand\color[2][]{}%
  }%
  \providecommand\includegraphics[2][]{%
    \GenericError{(gnuplot) \space\space\space\@spaces}{%
      Package graphicx or graphics not loaded%
    }{See the gnuplot documentation for explanation.%
    }{The gnuplot epslatex terminal needs graphicx.sty or graphics.sty.}%
    \renewcommand\includegraphics[2][]{}%
  }%
  \providecommand\rotatebox[2]{#2}%
  \@ifundefined{ifGPcolor}{%
    \newif\ifGPcolor
    \GPcolortrue
  }{}%
  \@ifundefined{ifGPblacktext}{%
    \newif\ifGPblacktext
    \GPblacktexttrue
  }{}%
  \let\gplgaddtomacro\g@addto@macro
  \gdef\gplbacktext{}%
  \gdef\gplfronttext{}%
  \makeatother
  \ifGPblacktext
    \def\colorrgb#1{}%
    \def\colorgray#1{}%
  \else
    \ifGPcolor
      \def\colorrgb#1{\color[rgb]{#1}}%
      \def\colorgray#1{\color[gray]{#1}}%
      \expandafter\def\csname LTw\endcsname{\color{white}}%
      \expandafter\def\csname LTb\endcsname{\color{black}}%
      \expandafter\def\csname LTa\endcsname{\color{black}}%
      \expandafter\def\csname LT0\endcsname{\color[rgb]{1,0,0}}%
      \expandafter\def\csname LT1\endcsname{\color[rgb]{0,1,0}}%
      \expandafter\def\csname LT2\endcsname{\color[rgb]{0,0,1}}%
      \expandafter\def\csname LT3\endcsname{\color[rgb]{1,0,1}}%
      \expandafter\def\csname LT4\endcsname{\color[rgb]{0,1,1}}%
      \expandafter\def\csname LT5\endcsname{\color[rgb]{1,1,0}}%
      \expandafter\def\csname LT6\endcsname{\color[rgb]{0,0,0}}%
      \expandafter\def\csname LT7\endcsname{\color[rgb]{1,0.3,0}}%
      \expandafter\def\csname LT8\endcsname{\color[rgb]{0.5,0.5,0.5}}%
    \else
      \def\colorrgb#1{\color{black}}%
      \def\colorgray#1{\color[gray]{#1}}%
      \expandafter\def\csname LTw\endcsname{\color{white}}%
      \expandafter\def\csname LTb\endcsname{\color{black}}%
      \expandafter\def\csname LTa\endcsname{\color{black}}%
      \expandafter\def\csname LT0\endcsname{\color{black}}%
      \expandafter\def\csname LT1\endcsname{\color{black}}%
      \expandafter\def\csname LT2\endcsname{\color{black}}%
      \expandafter\def\csname LT3\endcsname{\color{black}}%
      \expandafter\def\csname LT4\endcsname{\color{black}}%
      \expandafter\def\csname LT5\endcsname{\color{black}}%
      \expandafter\def\csname LT6\endcsname{\color{black}}%
      \expandafter\def\csname LT7\endcsname{\color{black}}%
      \expandafter\def\csname LT8\endcsname{\color{black}}%
    \fi
  \fi
  \setlength{\unitlength}{0.0500bp}%
  \begin{picture}(4534.40,3310.40)%
    \gplgaddtomacro\gplbacktext{%
      \csname LTb\endcsname%
      \put(946,704){\makebox(0,0)[r]{\strut{}-0.5}}%
      \csname LTb\endcsname%
      \put(946,938){\makebox(0,0)[r]{\strut{}-0.4}}%
      \csname LTb\endcsname%
      \put(946,1172){\makebox(0,0)[r]{\strut{}-0.3}}%
      \csname LTb\endcsname%
      \put(946,1407){\makebox(0,0)[r]{\strut{}-0.2}}%
      \csname LTb\endcsname%
      \put(946,1641){\makebox(0,0)[r]{\strut{}-0.1}}%
      \csname LTb\endcsname%
      \put(946,1875){\makebox(0,0)[r]{\strut{} 0}}%
      \csname LTb\endcsname%
      \put(946,2109){\makebox(0,0)[r]{\strut{} 0.1}}%
      \csname LTb\endcsname%
      \put(946,2343){\makebox(0,0)[r]{\strut{} 0.2}}%
      \csname LTb\endcsname%
      \put(946,2578){\makebox(0,0)[r]{\strut{} 0.3}}%
      \csname LTb\endcsname%
      \put(946,2812){\makebox(0,0)[r]{\strut{} 0.4}}%
      \csname LTb\endcsname%
      \put(946,3046){\makebox(0,0)[r]{\strut{} 0.5}}%
      \csname LTb\endcsname%
      \put(1078,484){\makebox(0,0){\strut{} 0}}%
      \csname LTb\endcsname%
      \put(1486,484){\makebox(0,0){\strut{} 0.2}}%
      \csname LTb\endcsname%
      \put(1894,484){\makebox(0,0){\strut{} 0.4}}%
      \csname LTb\endcsname%
      \put(2302,484){\makebox(0,0){\strut{} 0.6}}%
      \csname LTb\endcsname%
      \put(2709,484){\makebox(0,0){\strut{} 0.8}}%
      \csname LTb\endcsname%
      \put(3117,484){\makebox(0,0){\strut{} 1}}%
      \csname LTb\endcsname%
      \put(3525,484){\makebox(0,0){\strut{} 1.2}}%
      \csname LTb\endcsname%
      \put(3933,484){\makebox(0,0){\strut{} 1.4}}%
      \put(176,1875){\rotatebox{-270}{\makebox(0,0){\strut{}$I_4^{H^i}$}}}%
      \put(2607,154){\makebox(0,0){\strut{}amount of shear $\gamma$}}%
    }%
    \gplgaddtomacro\gplfronttext{%
      \csname LTb\endcsname%
      \put(1606,2873){\makebox(0,0)[r]{\strut{}$i=1$}}%
      \csname LTb\endcsname%
      \put(1606,2653){\makebox(0,0)[r]{\strut{}$i=2$}}%
      \csname LTb\endcsname%
      \put(1606,2433){\makebox(0,0)[r]{\strut{}$i=3$}}%
      \csname LTb\endcsname%
      \put(1606,2213){\makebox(0,0)[r]{\strut{}$i=4$}}%
    }%
    \gplbacktext
    \put(0,0){\includegraphics{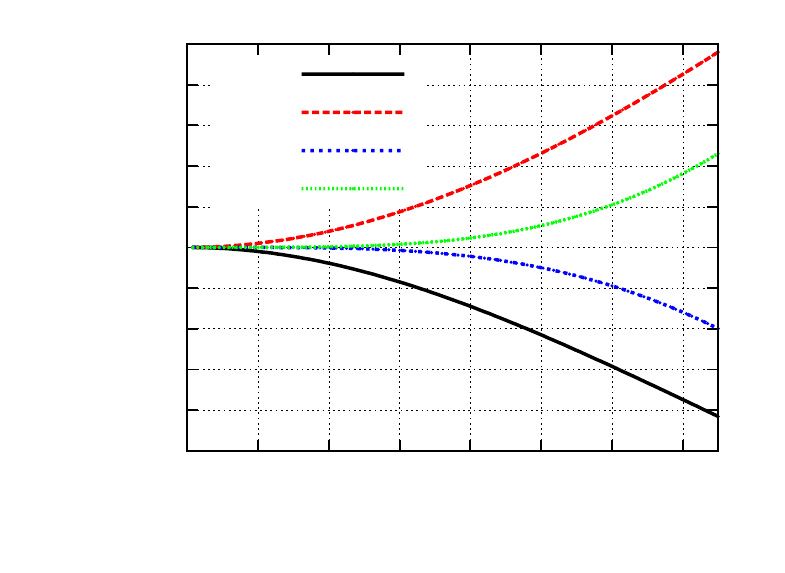}}%
    \gplfronttext
  \end{picture}%
\endgroup
}
\put(8.5,0.0){
\begingroup
  \makeatletter
  \providecommand\color[2][]{%
    \GenericError{(gnuplot) \space\space\space\@spaces}{%
      Package color not loaded in conjunction with
      terminal option `colourtext'%
    }{See the gnuplot documentation for explanation.%
    }{Either use 'blacktext' in gnuplot or load the package
      color.sty in LaTeX.}%
    \renewcommand\color[2][]{}%
  }%
  \providecommand\includegraphics[2][]{%
    \GenericError{(gnuplot) \space\space\space\@spaces}{%
      Package graphicx or graphics not loaded%
    }{See the gnuplot documentation for explanation.%
    }{The gnuplot epslatex terminal needs graphicx.sty or graphics.sty.}%
    \renewcommand\includegraphics[2][]{}%
  }%
  \providecommand\rotatebox[2]{#2}%
  \@ifundefined{ifGPcolor}{%
    \newif\ifGPcolor
    \GPcolortrue
  }{}%
  \@ifundefined{ifGPblacktext}{%
    \newif\ifGPblacktext
    \GPblacktexttrue
  }{}%
  \let\gplgaddtomacro\g@addto@macro
  \gdef\gplbacktext{}%
  \gdef\gplfronttext{}%
  \makeatother
  \ifGPblacktext
    \def\colorrgb#1{}%
    \def\colorgray#1{}%
  \else
    \ifGPcolor
      \def\colorrgb#1{\color[rgb]{#1}}%
      \def\colorgray#1{\color[gray]{#1}}%
      \expandafter\def\csname LTw\endcsname{\color{white}}%
      \expandafter\def\csname LTb\endcsname{\color{black}}%
      \expandafter\def\csname LTa\endcsname{\color{black}}%
      \expandafter\def\csname LT0\endcsname{\color[rgb]{1,0,0}}%
      \expandafter\def\csname LT1\endcsname{\color[rgb]{0,1,0}}%
      \expandafter\def\csname LT2\endcsname{\color[rgb]{0,0,1}}%
      \expandafter\def\csname LT3\endcsname{\color[rgb]{1,0,1}}%
      \expandafter\def\csname LT4\endcsname{\color[rgb]{0,1,1}}%
      \expandafter\def\csname LT5\endcsname{\color[rgb]{1,1,0}}%
      \expandafter\def\csname LT6\endcsname{\color[rgb]{0,0,0}}%
      \expandafter\def\csname LT7\endcsname{\color[rgb]{1,0.3,0}}%
      \expandafter\def\csname LT8\endcsname{\color[rgb]{0.5,0.5,0.5}}%
    \else
      \def\colorrgb#1{\color{black}}%
      \def\colorgray#1{\color[gray]{#1}}%
      \expandafter\def\csname LTw\endcsname{\color{white}}%
      \expandafter\def\csname LTb\endcsname{\color{black}}%
      \expandafter\def\csname LTa\endcsname{\color{black}}%
      \expandafter\def\csname LT0\endcsname{\color{black}}%
      \expandafter\def\csname LT1\endcsname{\color{black}}%
      \expandafter\def\csname LT2\endcsname{\color{black}}%
      \expandafter\def\csname LT3\endcsname{\color{black}}%
      \expandafter\def\csname LT4\endcsname{\color{black}}%
      \expandafter\def\csname LT5\endcsname{\color{black}}%
      \expandafter\def\csname LT6\endcsname{\color{black}}%
      \expandafter\def\csname LT7\endcsname{\color{black}}%
      \expandafter\def\csname LT8\endcsname{\color{black}}%
    \fi
  \fi
  \setlength{\unitlength}{0.0500bp}%
  \begin{picture}(4534.40,3310.40)%
    \gplgaddtomacro\gplbacktext{%
      \csname LTb\endcsname%
      \put(946,704){\makebox(0,0)[r]{\strut{} 0}}%
      \csname LTb\endcsname%
      \put(946,1172){\makebox(0,0)[r]{\strut{} 0.2}}%
      \csname LTb\endcsname%
      \put(946,1641){\makebox(0,0)[r]{\strut{} 0.4}}%
      \csname LTb\endcsname%
      \put(946,2109){\makebox(0,0)[r]{\strut{} 0.6}}%
      \csname LTb\endcsname%
      \put(946,2578){\makebox(0,0)[r]{\strut{} 0.8}}%
      \csname LTb\endcsname%
      \put(946,3046){\makebox(0,0)[r]{\strut{} 1}}%
      \csname LTb\endcsname%
      \put(1078,484){\makebox(0,0){\strut{} 0}}%
      \csname LTb\endcsname%
      \put(1486,484){\makebox(0,0){\strut{} 0.2}}%
      \csname LTb\endcsname%
      \put(1894,484){\makebox(0,0){\strut{} 0.4}}%
      \csname LTb\endcsname%
      \put(2302,484){\makebox(0,0){\strut{} 0.6}}%
      \csname LTb\endcsname%
      \put(2709,484){\makebox(0,0){\strut{} 0.8}}%
      \csname LTb\endcsname%
      \put(3117,484){\makebox(0,0){\strut{} 1}}%
      \csname LTb\endcsname%
      \put(3525,484){\makebox(0,0){\strut{} 1.2}}%
      \csname LTb\endcsname%
      \put(3933,484){\makebox(0,0){\strut{} 1.4}}%
      \put(176,1875){\rotatebox{-270}{\makebox(0,0){\strut{}$\sigma_{13}^{\rm aniso}/{\rm max(||\sigma_{13}^{\rm aniso}||)}$}}}%
      \put(2607,154){\makebox(0,0){\strut{}amount of shear $\gamma$}}%
    }%
    \gplgaddtomacro\gplfronttext{%
      \csname LTb\endcsname%
      \put(1606,2873){\makebox(0,0)[r]{\strut{}$i=1$}}%
      \csname LTb\endcsname%
      \put(1606,2653){\makebox(0,0)[r]{\strut{}$i=2$}}%
      \csname LTb\endcsname%
      \put(1606,2433){\makebox(0,0)[r]{\strut{}$i=3$}}%
      \csname LTb\endcsname%
      \put(1606,2213){\makebox(0,0)[r]{\strut{}$i=4$}}%
    }%
    \gplbacktext
    \put(0,0){\includegraphics{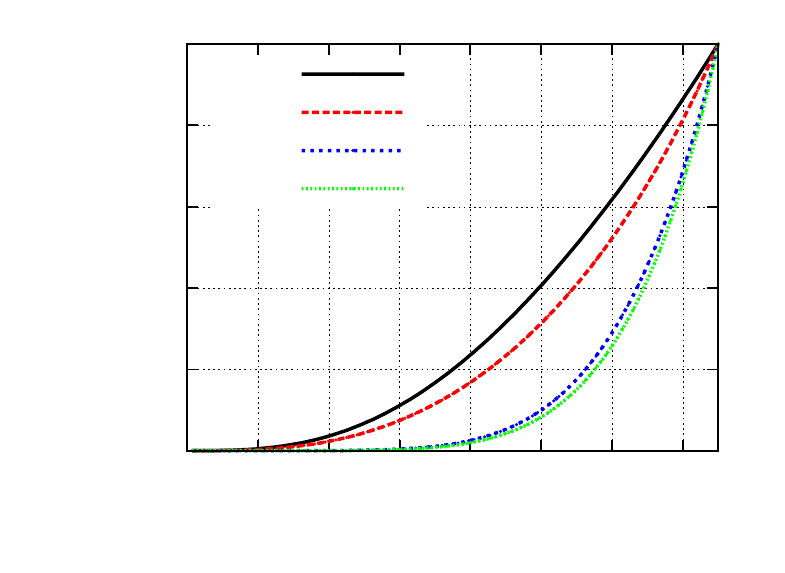}}%
    \gplfronttext
  \end{picture}%
\endgroup
}
\put(3.2,2.0){\tiny  \setlength{\fboxsep}{2.0mm}\fcolorbox{mygray}{mygray}{$I_4^{{\rm H}^i}$}}
\put(11.2,2.9){\tiny  \setlength{\fboxsep}{2.0mm}\fcolorbox{mygray}{mygray}{$I_4^{{\rm H}^i}$}}
\put(0.3,0.15){a)}
\put(8.5,0.15){b)}
\end{picture}
\setlength{\baselineskip}{11pt}
\caption{Evolution of the anisotropic Hencky-type invariants $I_4^{{\rm H}^i}$ for $i=1,2,3,4$ and b) corresponding Cauchy stresses $\sigma_{13}$ under simple shear. 
The stresses are normalized by the absolute value of the maximum arising shear stress at $\gamma = 1.5$.
\label{fig: shear_aniso_H_invarianten_sig13}}
\end{Figure}
\clearpage
\begin{Figure}[!htb]
\unitlength 1 cm
\begin{picture}(14,5.8)
\put(0.0,0.0){
\begingroup
  \makeatletter
  \providecommand\color[2][]{%
    \GenericError{(gnuplot) \space\space\space\@spaces}{%
      Package color not loaded in conjunction with
      terminal option `colourtext'%
    }{See the gnuplot documentation for explanation.%
    }{Either use 'blacktext' in gnuplot or load the package
      color.sty in LaTeX.}%
    \renewcommand\color[2][]{}%
  }%
  \providecommand\includegraphics[2][]{%
    \GenericError{(gnuplot) \space\space\space\@spaces}{%
      Package graphicx or graphics not loaded%
    }{See the gnuplot documentation for explanation.%
    }{The gnuplot epslatex terminal needs graphicx.sty or graphics.sty.}%
    \renewcommand\includegraphics[2][]{}%
  }%
  \providecommand\rotatebox[2]{#2}%
  \@ifundefined{ifGPcolor}{%
    \newif\ifGPcolor
    \GPcolortrue
  }{}%
  \@ifundefined{ifGPblacktext}{%
    \newif\ifGPblacktext
    \GPblacktexttrue
  }{}%
  \let\gplgaddtomacro\g@addto@macro
  \gdef\gplbacktext{}%
  \gdef\gplfronttext{}%
  \makeatother
  \ifGPblacktext
    \def\colorrgb#1{}%
    \def\colorgray#1{}%
  \else
    \ifGPcolor
      \def\colorrgb#1{\color[rgb]{#1}}%
      \def\colorgray#1{\color[gray]{#1}}%
      \expandafter\def\csname LTw\endcsname{\color{white}}%
      \expandafter\def\csname LTb\endcsname{\color{black}}%
      \expandafter\def\csname LTa\endcsname{\color{black}}%
      \expandafter\def\csname LT0\endcsname{\color[rgb]{1,0,0}}%
      \expandafter\def\csname LT1\endcsname{\color[rgb]{0,1,0}}%
      \expandafter\def\csname LT2\endcsname{\color[rgb]{0,0,1}}%
      \expandafter\def\csname LT3\endcsname{\color[rgb]{1,0,1}}%
      \expandafter\def\csname LT4\endcsname{\color[rgb]{0,1,1}}%
      \expandafter\def\csname LT5\endcsname{\color[rgb]{1,1,0}}%
      \expandafter\def\csname LT6\endcsname{\color[rgb]{0,0,0}}%
      \expandafter\def\csname LT7\endcsname{\color[rgb]{1,0.3,0}}%
      \expandafter\def\csname LT8\endcsname{\color[rgb]{0.5,0.5,0.5}}%
    \else
      \def\colorrgb#1{\color{black}}%
      \def\colorgray#1{\color[gray]{#1}}%
      \expandafter\def\csname LTw\endcsname{\color{white}}%
      \expandafter\def\csname LTb\endcsname{\color{black}}%
      \expandafter\def\csname LTa\endcsname{\color{black}}%
      \expandafter\def\csname LT0\endcsname{\color{black}}%
      \expandafter\def\csname LT1\endcsname{\color{black}}%
      \expandafter\def\csname LT2\endcsname{\color{black}}%
      \expandafter\def\csname LT3\endcsname{\color{black}}%
      \expandafter\def\csname LT4\endcsname{\color{black}}%
      \expandafter\def\csname LT5\endcsname{\color{black}}%
      \expandafter\def\csname LT6\endcsname{\color{black}}%
      \expandafter\def\csname LT7\endcsname{\color{black}}%
      \expandafter\def\csname LT8\endcsname{\color{black}}%
    \fi
  \fi
  \setlength{\unitlength}{0.0500bp}%
  \begin{picture}(4534.40,3310.40)%
    \gplgaddtomacro\gplbacktext{%
      \csname LTb\endcsname%
      \put(946,704){\makebox(0,0)[r]{\strut{}-1}}%
      \csname LTb\endcsname%
      \put(946,1172){\makebox(0,0)[r]{\strut{}-0.8}}%
      \csname LTb\endcsname%
      \put(946,1641){\makebox(0,0)[r]{\strut{}-0.6}}%
      \csname LTb\endcsname%
      \put(946,2109){\makebox(0,0)[r]{\strut{}-0.4}}%
      \csname LTb\endcsname%
      \put(946,2578){\makebox(0,0)[r]{\strut{}-0.2}}%
      \csname LTb\endcsname%
      \put(946,3046){\makebox(0,0)[r]{\strut{} 0}}%
      \csname LTb\endcsname%
      \put(1078,484){\makebox(0,0){\strut{} 0}}%
      \csname LTb\endcsname%
      \put(1486,484){\makebox(0,0){\strut{} 0.2}}%
      \csname LTb\endcsname%
      \put(1894,484){\makebox(0,0){\strut{} 0.4}}%
      \csname LTb\endcsname%
      \put(2302,484){\makebox(0,0){\strut{} 0.6}}%
      \csname LTb\endcsname%
      \put(2709,484){\makebox(0,0){\strut{} 0.8}}%
      \csname LTb\endcsname%
      \put(3117,484){\makebox(0,0){\strut{} 1}}%
      \csname LTb\endcsname%
      \put(3525,484){\makebox(0,0){\strut{} 1.2}}%
      \csname LTb\endcsname%
      \put(3933,484){\makebox(0,0){\strut{} 1.4}}%
      \put(176,1875){\rotatebox{-270}{\makebox(0,0){\strut{}$\sigma_{11}^{\rm aniso}/{\rm max(||\sigma_{11}^{\rm aniso}||)}$}}}%
      \put(2607,154){\makebox(0,0){\strut{}amount of shear $\gamma$}}%
    }%
    \gplgaddtomacro\gplfronttext{%
      \csname LTb\endcsname%
      \put(1606,1097){\makebox(0,0)[r]{\strut{}$i=1$}}%
      \csname LTb\endcsname%
      \put(1606,877){\makebox(0,0)[r]{\strut{}$i=3$}}%
    }%
    \gplbacktext
    \put(0,0){\includegraphics{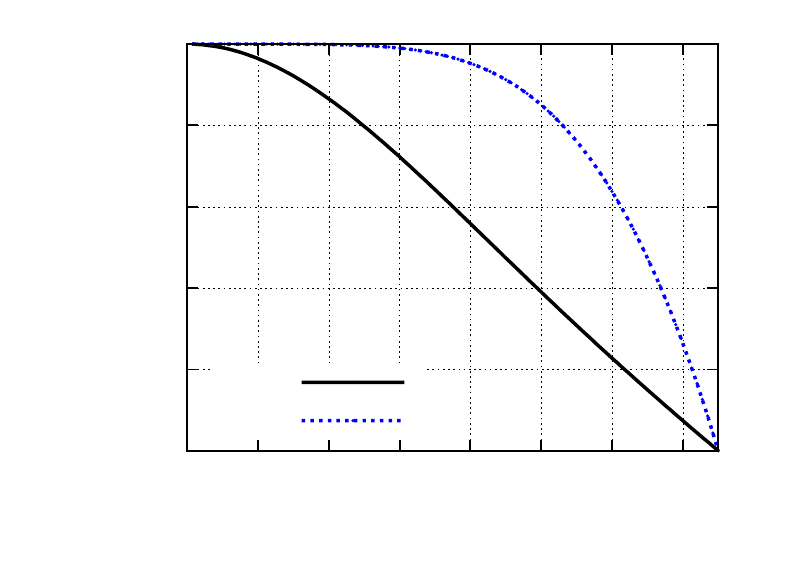}}%
    \gplfronttext
  \end{picture}%
\endgroup
}
\put(2.7,3.4){\tiny  \setlength{\fboxsep}{2.0mm}\fcolorbox{mygray}{mygray}{$I_4^{{\rm H}^i}$}}
\put(8.5,0.0){
\begingroup
  \makeatletter
  \providecommand\color[2][]{%
    \GenericError{(gnuplot) \space\space\space\@spaces}{%
      Package color not loaded in conjunction with
      terminal option `colourtext'%
    }{See the gnuplot documentation for explanation.%
    }{Either use 'blacktext' in gnuplot or load the package
      color.sty in LaTeX.}%
    \renewcommand\color[2][]{}%
  }%
  \providecommand\includegraphics[2][]{%
    \GenericError{(gnuplot) \space\space\space\@spaces}{%
      Package graphicx or graphics not loaded%
    }{See the gnuplot documentation for explanation.%
    }{The gnuplot epslatex terminal needs graphicx.sty or graphics.sty.}%
    \renewcommand\includegraphics[2][]{}%
  }%
  \providecommand\rotatebox[2]{#2}%
  \@ifundefined{ifGPcolor}{%
    \newif\ifGPcolor
    \GPcolortrue
  }{}%
  \@ifundefined{ifGPblacktext}{%
    \newif\ifGPblacktext
    \GPblacktexttrue
  }{}%
  \let\gplgaddtomacro\g@addto@macro
  \gdef\gplbacktext{}%
  \gdef\gplfronttext{}%
  \makeatother
  \ifGPblacktext
    \def\colorrgb#1{}%
    \def\colorgray#1{}%
  \else
    \ifGPcolor
      \def\colorrgb#1{\color[rgb]{#1}}%
      \def\colorgray#1{\color[gray]{#1}}%
      \expandafter\def\csname LTw\endcsname{\color{white}}%
      \expandafter\def\csname LTb\endcsname{\color{black}}%
      \expandafter\def\csname LTa\endcsname{\color{black}}%
      \expandafter\def\csname LT0\endcsname{\color[rgb]{1,0,0}}%
      \expandafter\def\csname LT1\endcsname{\color[rgb]{0,1,0}}%
      \expandafter\def\csname LT2\endcsname{\color[rgb]{0,0,1}}%
      \expandafter\def\csname LT3\endcsname{\color[rgb]{1,0,1}}%
      \expandafter\def\csname LT4\endcsname{\color[rgb]{0,1,1}}%
      \expandafter\def\csname LT5\endcsname{\color[rgb]{1,1,0}}%
      \expandafter\def\csname LT6\endcsname{\color[rgb]{0,0,0}}%
      \expandafter\def\csname LT7\endcsname{\color[rgb]{1,0.3,0}}%
      \expandafter\def\csname LT8\endcsname{\color[rgb]{0.5,0.5,0.5}}%
    \else
      \def\colorrgb#1{\color{black}}%
      \def\colorgray#1{\color[gray]{#1}}%
      \expandafter\def\csname LTw\endcsname{\color{white}}%
      \expandafter\def\csname LTb\endcsname{\color{black}}%
      \expandafter\def\csname LTa\endcsname{\color{black}}%
      \expandafter\def\csname LT0\endcsname{\color{black}}%
      \expandafter\def\csname LT1\endcsname{\color{black}}%
      \expandafter\def\csname LT2\endcsname{\color{black}}%
      \expandafter\def\csname LT3\endcsname{\color{black}}%
      \expandafter\def\csname LT4\endcsname{\color{black}}%
      \expandafter\def\csname LT5\endcsname{\color{black}}%
      \expandafter\def\csname LT6\endcsname{\color{black}}%
      \expandafter\def\csname LT7\endcsname{\color{black}}%
      \expandafter\def\csname LT8\endcsname{\color{black}}%
    \fi
  \fi
  \setlength{\unitlength}{0.0500bp}%
  \begin{picture}(4534.40,3310.40)%
    \gplgaddtomacro\gplbacktext{%
      \csname LTb\endcsname%
      \put(946,704){\makebox(0,0)[r]{\strut{}-1}}%
      \csname LTb\endcsname%
      \put(946,1172){\makebox(0,0)[r]{\strut{}-0.8}}%
      \csname LTb\endcsname%
      \put(946,1641){\makebox(0,0)[r]{\strut{}-0.6}}%
      \csname LTb\endcsname%
      \put(946,2109){\makebox(0,0)[r]{\strut{}-0.4}}%
      \csname LTb\endcsname%
      \put(946,2578){\makebox(0,0)[r]{\strut{}-0.2}}%
      \csname LTb\endcsname%
      \put(946,3046){\makebox(0,0)[r]{\strut{} 0}}%
      \csname LTb\endcsname%
      \put(1078,484){\makebox(0,0){\strut{} 0}}%
      \csname LTb\endcsname%
      \put(1486,484){\makebox(0,0){\strut{} 0.2}}%
      \csname LTb\endcsname%
      \put(1894,484){\makebox(0,0){\strut{} 0.4}}%
      \csname LTb\endcsname%
      \put(2302,484){\makebox(0,0){\strut{} 0.6}}%
      \csname LTb\endcsname%
      \put(2709,484){\makebox(0,0){\strut{} 0.8}}%
      \csname LTb\endcsname%
      \put(3117,484){\makebox(0,0){\strut{} 1}}%
      \csname LTb\endcsname%
      \put(3525,484){\makebox(0,0){\strut{} 1.2}}%
      \csname LTb\endcsname%
      \put(3933,484){\makebox(0,0){\strut{} 1.4}}%
      \put(176,1875){\rotatebox{-270}{\makebox(0,0){\strut{}$\sigma_{33}^{\rm aniso}/{\rm max(||\sigma_{33}^{\rm aniso}||)}$}}}%
      \put(2607,154){\makebox(0,0){\strut{}amount of shear $\gamma$}}%
    }%
    \gplgaddtomacro\gplfronttext{%
      \csname LTb\endcsname%
      \put(1606,1537){\makebox(0,0)[r]{\strut{}$i=1$}}%
      \csname LTb\endcsname%
      \put(1606,1317){\makebox(0,0)[r]{\strut{}$i=2$}}%
      \csname LTb\endcsname%
      \put(1606,1097){\makebox(0,0)[r]{\strut{}$i=3$}}%
      \csname LTb\endcsname%
      \put(1606,877){\makebox(0,0)[r]{\strut{}$i=4$}}%
    }%
    \gplbacktext
    \put(0,0){\includegraphics{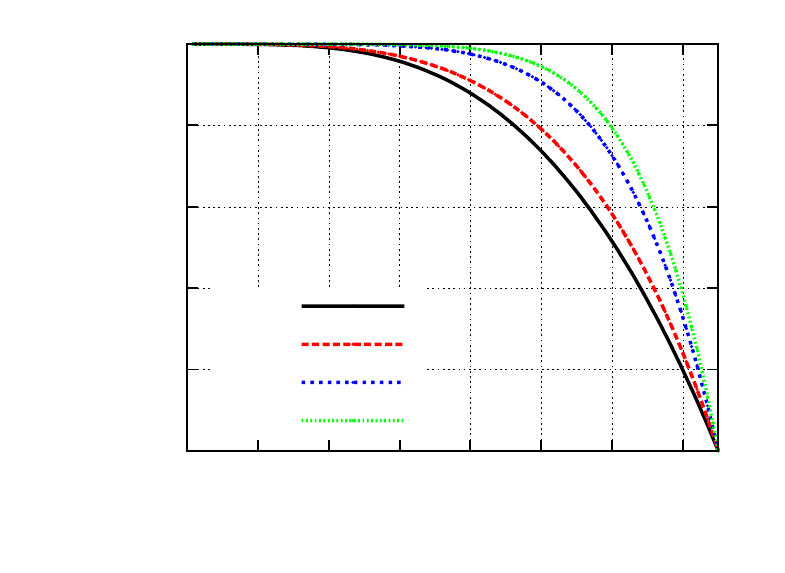}}%
    \gplfronttext
  \end{picture}%
\endgroup
}
\put(11.3,3.4){\tiny  \setlength{\fboxsep}{2.0mm}\fcolorbox{mygray}{mygray}{$I_4^{{\rm H}^i}$}}
\put(0.3,0.15){a)}
\put(8.5,0.15){b)}
\end{picture}
\setlength{\baselineskip}{11pt}
\caption{Evolution of the Cauchy stresses a) $\sigma_{11}$ and b) $\sigma_{33}$ corresponding to the anisotropic strain invariants plotted in Fig.~\ref{fig: shear_aniso_H_invarianten_sig13}a), 
under simple shear. Note that for $i=2$ and $i=4$, $\sigma_{11}$ is equal to zero. \label{fig: shear_aniso_H_sig11_sig33}}
\end{Figure}

\subsubsection{Biaxial tension \label{sec: Biax tension}}

In this case the body is exposed to biaxial tensile displacements, see Fig.~\ref{fig: biax}. 
We consider two fiber families which are orientated symmetric regarding the X-axis.
Further, as indicated in the figure three different stretch ratios will be compared.
The stretch ratio is here defined as the quotient $F_{11}/F_{22}$.
For the three different stretch ratios the evolution of the quotient $\sigma^{\rm aniso}_{22}/\sigma^{\rm aniso}_{11}$
is plotted in Fig.~\ref{fig: biax_17_17}, \ref{fig: biax_17_135} and \ref{fig: biax_135_17}. In each of the figures we find 
the results for the anisotropic invariants of $\bC^i$ on the left hand side and the results for $(\log\bU)^i$ on the right hand side.
For the equi-biaxial test in Fig.~\ref{fig: biax_17_17} the stress ratio is the same for each model, at each time. The stress ratio 
directly follows from the fiber angle with respect to the x-axis to be $\sigma^{\rm aniso}_{22}/\sigma^{\rm aniso}_{11}=\tan^2 30^\circ =1/3$.
This ratio remains exactly the same for the other stretch ratios only if $I_4^{\rm H^1}$ is used, which seems to be unreasonable.
Regarding the invariants of $\bC$ we see that the starting point at nearly zero deformation is always defined by $\sigma^{\rm aniso}_{22}/\sigma^{\rm aniso}_{11}=1/3$,
which is different for the Hencky-type strain measures. Further, the change in the stress ratio with increasing deformation is less pronounced in the case that 
Hencky strains are used. This behavior can be explained, when taking a look on Eq.~(\ref{eq: invariant physical}). Since the angles $\theta_l|\,l=1,2,3$ are constant 
only the logarithmic stretches are of interest. Due to the logarithmic function the slope is decreasing when the stretch is increasing, i.e. for higher strains the slope
is smaller than for lower strains which is also reflected by the shown stress ratios.
Generally the slope of the stress ratios seem to have the opposite sign, regarding the basic strain measure. 
But all stresses appear to have the same sign, independent of the stress measure.

The stress ratios of the $I_4^{\rm C^1}$ model can be exactly reproduced by the computation of the fiber angle
\eb
\beta_{\rm act} = \arccos\left(\frac{\langle\bF\bA,\be_{\rm x}\rangle}{\lVert\bF\bA\rVert~\lVert\be_{\rm x}\rVert}\right)\,,
\ee
where $\be_{\rm x}$ denotes the direction of the X-axis. Then  $\sigma^{\rm aniso}_{22}/\sigma^{\rm aniso}_{11}$ is equal to $\tan^2 (\beta_{\rm act})$.

\begin{Figure}[!htb]
\unitlength 1 cm
\begin{picture}(14,5.5)
\put(2.0,0.0){\includegraphics[height = 5.5cm]{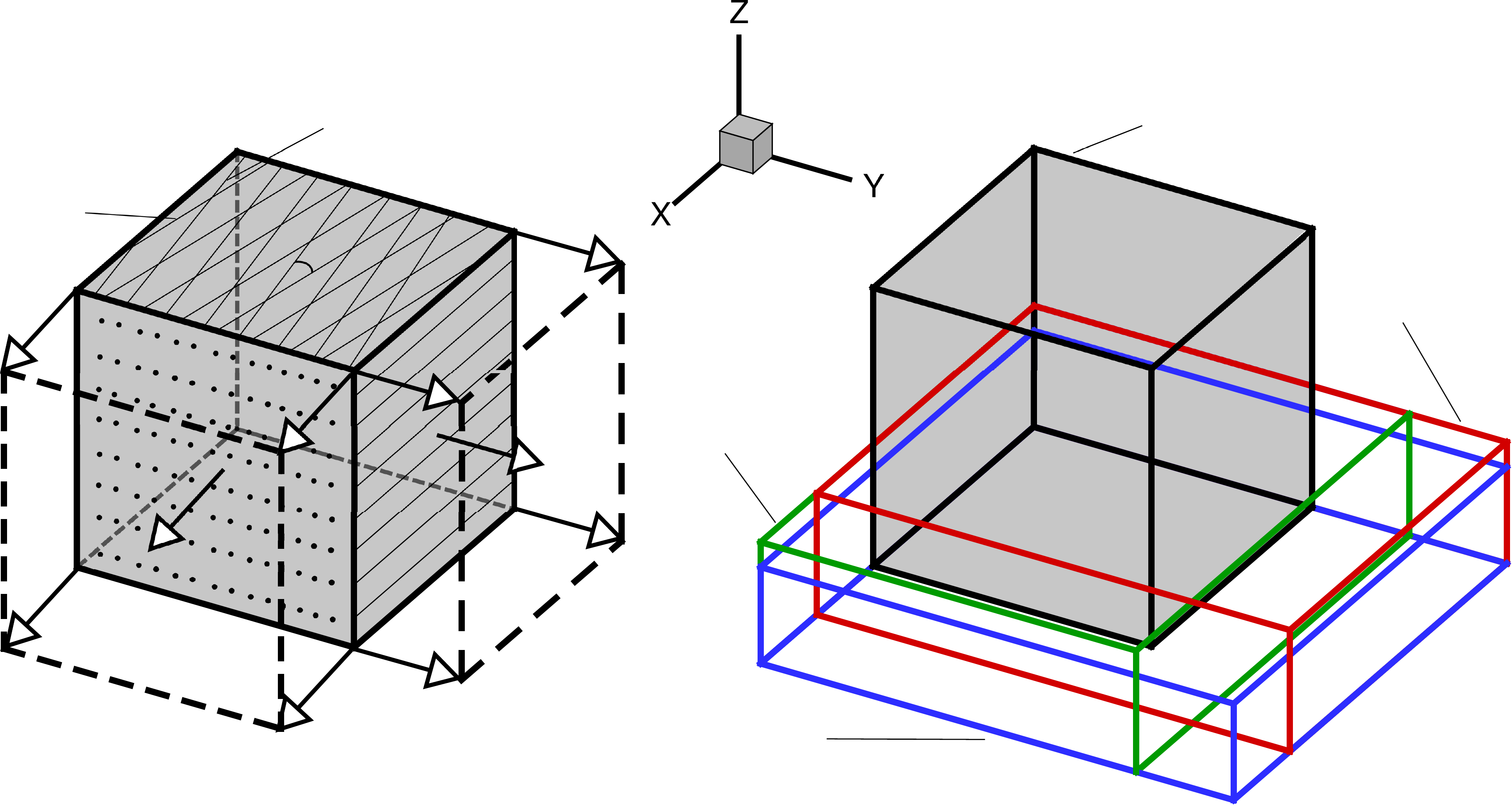}}
\put(2.2,4.0){$\bA^{(1)}$}
\put(4.05,4.65){$\bA^{(2)}$}
\put(4.1,3.72){$60^{\circ}$}
\put(9.85,4.55){ref.}
\put(6.4,2.5){1.7:1.35}
\put(6.4,0.35){1.7:1.7}
\put(11,3.4){1.35:1.7}
\end{picture}
\setlength{\baselineskip}{11pt}
\caption{Biaxial tension test with two fiber families. The body is assumed to be incompressible 
with $F_{33} = 1/(F_{11} F_{22})$ \label{fig: biax}%
}
\end{Figure}

\begin{Figure}[!htb]
\unitlength 1 cm
\begin{picture}(14.0,5.8)
\put(0.0,0.0){
\begingroup
  \makeatletter
  \providecommand\color[2][]{%
    \GenericError{(gnuplot) \space\space\space\@spaces}{%
      Package color not loaded in conjunction with
      terminal option `colourtext'%
    }{See the gnuplot documentation for explanation.%
    }{Either use 'blacktext' in gnuplot or load the package
      color.sty in LaTeX.}%
    \renewcommand\color[2][]{}%
  }%
  \providecommand\includegraphics[2][]{%
    \GenericError{(gnuplot) \space\space\space\@spaces}{%
      Package graphicx or graphics not loaded%
    }{See the gnuplot documentation for explanation.%
    }{The gnuplot epslatex terminal needs graphicx.sty or graphics.sty.}%
    \renewcommand\includegraphics[2][]{}%
  }%
  \providecommand\rotatebox[2]{#2}%
  \@ifundefined{ifGPcolor}{%
    \newif\ifGPcolor
    \GPcolortrue
  }{}%
  \@ifundefined{ifGPblacktext}{%
    \newif\ifGPblacktext
    \GPblacktexttrue
  }{}%
  \let\gplgaddtomacro\g@addto@macro
  \gdef\gplbacktext{}%
  \gdef\gplfronttext{}%
  \makeatother
  \ifGPblacktext
    \def\colorrgb#1{}%
    \def\colorgray#1{}%
  \else
    \ifGPcolor
      \def\colorrgb#1{\color[rgb]{#1}}%
      \def\colorgray#1{\color[gray]{#1}}%
      \expandafter\def\csname LTw\endcsname{\color{white}}%
      \expandafter\def\csname LTb\endcsname{\color{black}}%
      \expandafter\def\csname LTa\endcsname{\color{black}}%
      \expandafter\def\csname LT0\endcsname{\color[rgb]{1,0,0}}%
      \expandafter\def\csname LT1\endcsname{\color[rgb]{0,1,0}}%
      \expandafter\def\csname LT2\endcsname{\color[rgb]{0,0,1}}%
      \expandafter\def\csname LT3\endcsname{\color[rgb]{1,0,1}}%
      \expandafter\def\csname LT4\endcsname{\color[rgb]{0,1,1}}%
      \expandafter\def\csname LT5\endcsname{\color[rgb]{1,1,0}}%
      \expandafter\def\csname LT6\endcsname{\color[rgb]{0,0,0}}%
      \expandafter\def\csname LT7\endcsname{\color[rgb]{1,0.3,0}}%
      \expandafter\def\csname LT8\endcsname{\color[rgb]{0.5,0.5,0.5}}%
    \else
      \def\colorrgb#1{\color{black}}%
      \def\colorgray#1{\color[gray]{#1}}%
      \expandafter\def\csname LTw\endcsname{\color{white}}%
      \expandafter\def\csname LTb\endcsname{\color{black}}%
      \expandafter\def\csname LTa\endcsname{\color{black}}%
      \expandafter\def\csname LT0\endcsname{\color{black}}%
      \expandafter\def\csname LT1\endcsname{\color{black}}%
      \expandafter\def\csname LT2\endcsname{\color{black}}%
      \expandafter\def\csname LT3\endcsname{\color{black}}%
      \expandafter\def\csname LT4\endcsname{\color{black}}%
      \expandafter\def\csname LT5\endcsname{\color{black}}%
      \expandafter\def\csname LT6\endcsname{\color{black}}%
      \expandafter\def\csname LT7\endcsname{\color{black}}%
      \expandafter\def\csname LT8\endcsname{\color{black}}%
    \fi
  \fi
  \setlength{\unitlength}{0.0500bp}%
  \begin{picture}(4534.40,3310.40)%
    \gplgaddtomacro\gplbacktext{%
      \csname LTb\endcsname%
      \put(1078,704){\makebox(0,0)[r]{\strut{} 0}}%
      \csname LTb\endcsname%
      \put(1078,997){\makebox(0,0)[r]{\strut{} 0.05}}%
      \csname LTb\endcsname%
      \put(1078,1290){\makebox(0,0)[r]{\strut{} 0.1}}%
      \csname LTb\endcsname%
      \put(1078,1582){\makebox(0,0)[r]{\strut{} 0.15}}%
      \csname LTb\endcsname%
      \put(1078,1875){\makebox(0,0)[r]{\strut{} 0.2}}%
      \csname LTb\endcsname%
      \put(1078,2168){\makebox(0,0)[r]{\strut{} 0.25}}%
      \csname LTb\endcsname%
      \put(1078,2461){\makebox(0,0)[r]{\strut{} 0.3}}%
      \csname LTb\endcsname%
      \put(1078,2753){\makebox(0,0)[r]{\strut{} 0.35}}%
      \csname LTb\endcsname%
      \put(1078,3046){\makebox(0,0)[r]{\strut{} 0.4}}%
      \csname LTb\endcsname%
      \put(1210,484){\makebox(0,0){\strut{} 1}}%
      \csname LTb\endcsname%
      \put(1628,484){\makebox(0,0){\strut{} 1.1}}%
      \csname LTb\endcsname%
      \put(2046,484){\makebox(0,0){\strut{} 1.2}}%
      \csname LTb\endcsname%
      \put(2464,484){\makebox(0,0){\strut{} 1.3}}%
      \csname LTb\endcsname%
      \put(2883,484){\makebox(0,0){\strut{} 1.4}}%
      \csname LTb\endcsname%
      \put(3301,484){\makebox(0,0){\strut{} 1.5}}%
      \csname LTb\endcsname%
      \put(3719,484){\makebox(0,0){\strut{} 1.6}}%
      \csname LTb\endcsname%
      \put(4137,484){\makebox(0,0){\strut{} 1.7}}%
      \put(176,1875){\rotatebox{-270}{\makebox(0,0){\strut{}$\sigma_{22}^{\rm aniso}/{\sigma_{11}^{\rm aniso}}$}}}%
      \put(2673,154){\makebox(0,0){\strut{}$F_{11}$}}%
    }%
    \gplgaddtomacro\gplfronttext{%
      \csname LTb\endcsname%
      \put(3150,1537){\makebox(0,0)[r]{\strut{}$i=1$}}%
      \csname LTb\endcsname%
      \put(3150,1317){\makebox(0,0)[r]{\strut{}$i=2$}}%
      \csname LTb\endcsname%
      \put(3150,1097){\makebox(0,0)[r]{\strut{}$i=3$}}%
      \csname LTb\endcsname%
      \put(3150,877){\makebox(0,0)[r]{\strut{}$i=4$}}%
    }%
    \gplbacktext
    \put(0,0){\includegraphics{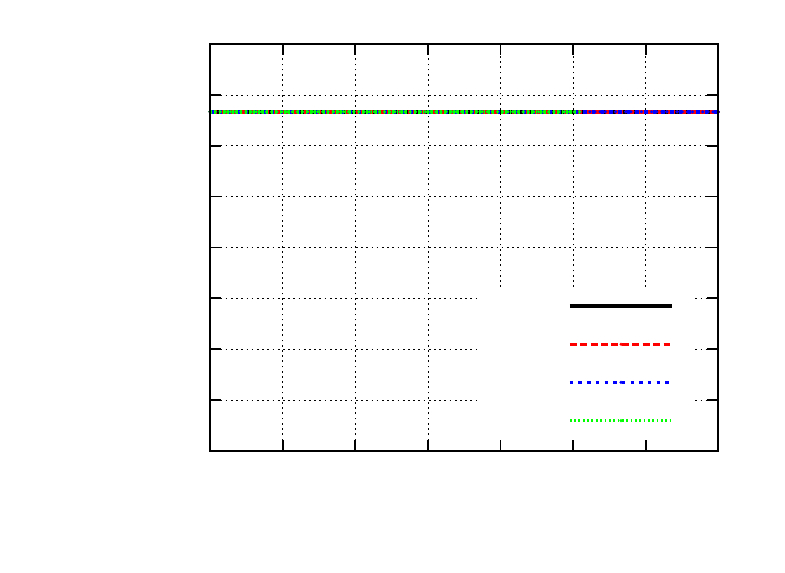}}%
    \gplfronttext
  \end{picture}%
\endgroup
}
\put(8.5,0.0){
\begingroup
  \makeatletter
  \providecommand\color[2][]{%
    \GenericError{(gnuplot) \space\space\space\@spaces}{%
      Package color not loaded in conjunction with
      terminal option `colourtext'%
    }{See the gnuplot documentation for explanation.%
    }{Either use 'blacktext' in gnuplot or load the package
      color.sty in LaTeX.}%
    \renewcommand\color[2][]{}%
  }%
  \providecommand\includegraphics[2][]{%
    \GenericError{(gnuplot) \space\space\space\@spaces}{%
      Package graphicx or graphics not loaded%
    }{See the gnuplot documentation for explanation.%
    }{The gnuplot epslatex terminal needs graphicx.sty or graphics.sty.}%
    \renewcommand\includegraphics[2][]{}%
  }%
  \providecommand\rotatebox[2]{#2}%
  \@ifundefined{ifGPcolor}{%
    \newif\ifGPcolor
    \GPcolortrue
  }{}%
  \@ifundefined{ifGPblacktext}{%
    \newif\ifGPblacktext
    \GPblacktexttrue
  }{}%
  \let\gplgaddtomacro\g@addto@macro
  \gdef\gplbacktext{}%
  \gdef\gplfronttext{}%
  \makeatother
  \ifGPblacktext
    \def\colorrgb#1{}%
    \def\colorgray#1{}%
  \else
    \ifGPcolor
      \def\colorrgb#1{\color[rgb]{#1}}%
      \def\colorgray#1{\color[gray]{#1}}%
      \expandafter\def\csname LTw\endcsname{\color{white}}%
      \expandafter\def\csname LTb\endcsname{\color{black}}%
      \expandafter\def\csname LTa\endcsname{\color{black}}%
      \expandafter\def\csname LT0\endcsname{\color[rgb]{1,0,0}}%
      \expandafter\def\csname LT1\endcsname{\color[rgb]{0,1,0}}%
      \expandafter\def\csname LT2\endcsname{\color[rgb]{0,0,1}}%
      \expandafter\def\csname LT3\endcsname{\color[rgb]{1,0,1}}%
      \expandafter\def\csname LT4\endcsname{\color[rgb]{0,1,1}}%
      \expandafter\def\csname LT5\endcsname{\color[rgb]{1,1,0}}%
      \expandafter\def\csname LT6\endcsname{\color[rgb]{0,0,0}}%
      \expandafter\def\csname LT7\endcsname{\color[rgb]{1,0.3,0}}%
      \expandafter\def\csname LT8\endcsname{\color[rgb]{0.5,0.5,0.5}}%
    \else
      \def\colorrgb#1{\color{black}}%
      \def\colorgray#1{\color[gray]{#1}}%
      \expandafter\def\csname LTw\endcsname{\color{white}}%
      \expandafter\def\csname LTb\endcsname{\color{black}}%
      \expandafter\def\csname LTa\endcsname{\color{black}}%
      \expandafter\def\csname LT0\endcsname{\color{black}}%
      \expandafter\def\csname LT1\endcsname{\color{black}}%
      \expandafter\def\csname LT2\endcsname{\color{black}}%
      \expandafter\def\csname LT3\endcsname{\color{black}}%
      \expandafter\def\csname LT4\endcsname{\color{black}}%
      \expandafter\def\csname LT5\endcsname{\color{black}}%
      \expandafter\def\csname LT6\endcsname{\color{black}}%
      \expandafter\def\csname LT7\endcsname{\color{black}}%
      \expandafter\def\csname LT8\endcsname{\color{black}}%
    \fi
  \fi
  \setlength{\unitlength}{0.0500bp}%
  \begin{picture}(4534.40,3310.40)%
    \gplgaddtomacro\gplbacktext{%
      \csname LTb\endcsname%
      \put(1078,704){\makebox(0,0)[r]{\strut{} 0}}%
      \csname LTb\endcsname%
      \put(1078,997){\makebox(0,0)[r]{\strut{} 0.05}}%
      \csname LTb\endcsname%
      \put(1078,1290){\makebox(0,0)[r]{\strut{} 0.1}}%
      \csname LTb\endcsname%
      \put(1078,1582){\makebox(0,0)[r]{\strut{} 0.15}}%
      \csname LTb\endcsname%
      \put(1078,1875){\makebox(0,0)[r]{\strut{} 0.2}}%
      \csname LTb\endcsname%
      \put(1078,2168){\makebox(0,0)[r]{\strut{} 0.25}}%
      \csname LTb\endcsname%
      \put(1078,2461){\makebox(0,0)[r]{\strut{} 0.3}}%
      \csname LTb\endcsname%
      \put(1078,2753){\makebox(0,0)[r]{\strut{} 0.35}}%
      \csname LTb\endcsname%
      \put(1078,3046){\makebox(0,0)[r]{\strut{} 0.4}}%
      \csname LTb\endcsname%
      \put(1210,484){\makebox(0,0){\strut{} 1}}%
      \csname LTb\endcsname%
      \put(1628,484){\makebox(0,0){\strut{} 1.1}}%
      \csname LTb\endcsname%
      \put(2046,484){\makebox(0,0){\strut{} 1.2}}%
      \csname LTb\endcsname%
      \put(2464,484){\makebox(0,0){\strut{} 1.3}}%
      \csname LTb\endcsname%
      \put(2883,484){\makebox(0,0){\strut{} 1.4}}%
      \csname LTb\endcsname%
      \put(3301,484){\makebox(0,0){\strut{} 1.5}}%
      \csname LTb\endcsname%
      \put(3719,484){\makebox(0,0){\strut{} 1.6}}%
      \csname LTb\endcsname%
      \put(4137,484){\makebox(0,0){\strut{} 1.7}}%
      \put(176,1875){\rotatebox{-270}{\makebox(0,0){\strut{}$\sigma_{22}^{\rm aniso}/{\sigma_{11}^{\rm aniso}}$}}}%
      \put(2673,154){\makebox(0,0){\strut{}$F_{11}$}}%
    }%
    \gplgaddtomacro\gplfronttext{%
      \csname LTb\endcsname%
      \put(3150,1537){\makebox(0,0)[r]{\strut{}$i=1$}}%
      \csname LTb\endcsname%
      \put(3150,1317){\makebox(0,0)[r]{\strut{}$i=2$}}%
      \csname LTb\endcsname%
      \put(3150,1097){\makebox(0,0)[r]{\strut{}$i=3$}}%
      \csname LTb\endcsname%
      \put(3150,877){\makebox(0,0)[r]{\strut{}$i=4$}}%
    }%
    \gplbacktext
    \put(0,0){\includegraphics{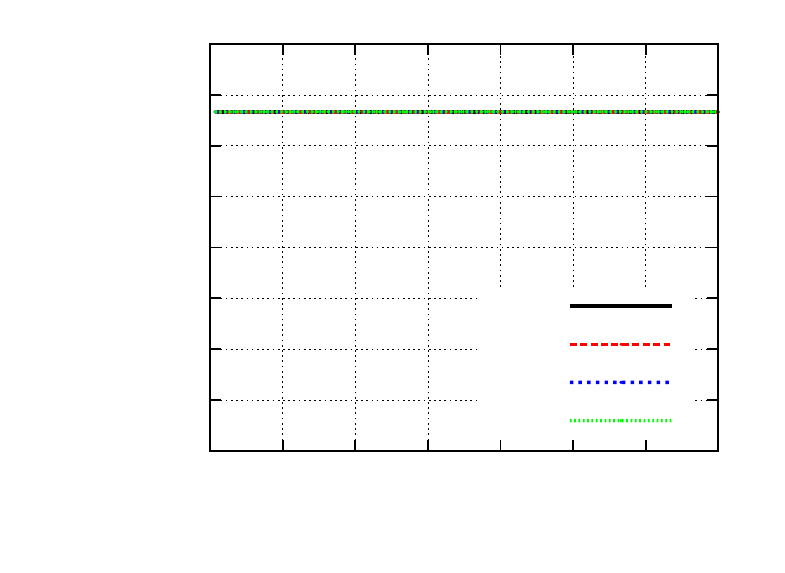}}%
    \gplfronttext
  \end{picture}%
\endgroup
}
\put(2.7,3.4){\tiny  \setlength{\fboxsep}{2.0mm}\fcolorbox{mygray}{mygray}{$I_4^{{\rm C}^i}$}}
\put(11.3,3.4){\tiny  \setlength{\fboxsep}{2.0mm}\fcolorbox{mygray}{mygray}{$I_4^{{\rm H}^i}$}}
\put(0.3,0.15){a)}
\put(8.5,0.15){b)}
\end{picture}
\setlength{\baselineskip}{11pt}
\caption{Evolution of the stress ratio for the equi-biaxial test with the stretch ratio $1.7:1.7$ for a) the invariants $I_4^{{\rm C}^i}$ and
b) the Hencky-type invariants $I_4^{{\rm H}^i}$ with the exponent $i=1,2,3,4$. \label{fig: biax_17_17}}
\end{Figure}

\begin{Figure}[!htb]
\unitlength 1 cm
\begin{picture}(14,5.8)
\put(0.0,0.0){
\begingroup
  \makeatletter
  \providecommand\color[2][]{%
    \GenericError{(gnuplot) \space\space\space\@spaces}{%
      Package color not loaded in conjunction with
      terminal option `colourtext'%
    }{See the gnuplot documentation for explanation.%
    }{Either use 'blacktext' in gnuplot or load the package
      color.sty in LaTeX.}%
    \renewcommand\color[2][]{}%
  }%
  \providecommand\includegraphics[2][]{%
    \GenericError{(gnuplot) \space\space\space\@spaces}{%
      Package graphicx or graphics not loaded%
    }{See the gnuplot documentation for explanation.%
    }{The gnuplot epslatex terminal needs graphicx.sty or graphics.sty.}%
    \renewcommand\includegraphics[2][]{}%
  }%
  \providecommand\rotatebox[2]{#2}%
  \@ifundefined{ifGPcolor}{%
    \newif\ifGPcolor
    \GPcolortrue
  }{}%
  \@ifundefined{ifGPblacktext}{%
    \newif\ifGPblacktext
    \GPblacktexttrue
  }{}%
  \let\gplgaddtomacro\g@addto@macro
  \gdef\gplbacktext{}%
  \gdef\gplfronttext{}%
  \makeatother
  \ifGPblacktext
    \def\colorrgb#1{}%
    \def\colorgray#1{}%
  \else
    \ifGPcolor
      \def\colorrgb#1{\color[rgb]{#1}}%
      \def\colorgray#1{\color[gray]{#1}}%
      \expandafter\def\csname LTw\endcsname{\color{white}}%
      \expandafter\def\csname LTb\endcsname{\color{black}}%
      \expandafter\def\csname LTa\endcsname{\color{black}}%
      \expandafter\def\csname LT0\endcsname{\color[rgb]{1,0,0}}%
      \expandafter\def\csname LT1\endcsname{\color[rgb]{0,1,0}}%
      \expandafter\def\csname LT2\endcsname{\color[rgb]{0,0,1}}%
      \expandafter\def\csname LT3\endcsname{\color[rgb]{1,0,1}}%
      \expandafter\def\csname LT4\endcsname{\color[rgb]{0,1,1}}%
      \expandafter\def\csname LT5\endcsname{\color[rgb]{1,1,0}}%
      \expandafter\def\csname LT6\endcsname{\color[rgb]{0,0,0}}%
      \expandafter\def\csname LT7\endcsname{\color[rgb]{1,0.3,0}}%
      \expandafter\def\csname LT8\endcsname{\color[rgb]{0.5,0.5,0.5}}%
    \else
      \def\colorrgb#1{\color{black}}%
      \def\colorgray#1{\color[gray]{#1}}%
      \expandafter\def\csname LTw\endcsname{\color{white}}%
      \expandafter\def\csname LTb\endcsname{\color{black}}%
      \expandafter\def\csname LTa\endcsname{\color{black}}%
      \expandafter\def\csname LT0\endcsname{\color{black}}%
      \expandafter\def\csname LT1\endcsname{\color{black}}%
      \expandafter\def\csname LT2\endcsname{\color{black}}%
      \expandafter\def\csname LT3\endcsname{\color{black}}%
      \expandafter\def\csname LT4\endcsname{\color{black}}%
      \expandafter\def\csname LT5\endcsname{\color{black}}%
      \expandafter\def\csname LT6\endcsname{\color{black}}%
      \expandafter\def\csname LT7\endcsname{\color{black}}%
      \expandafter\def\csname LT8\endcsname{\color{black}}%
    \fi
  \fi
  \setlength{\unitlength}{0.0500bp}%
  \begin{picture}(4534.40,3310.40)%
    \gplgaddtomacro\gplbacktext{%
      \csname LTb\endcsname%
      \put(1078,704){\makebox(0,0)[r]{\strut{} 0}}%
      \csname LTb\endcsname%
      \put(1078,1039){\makebox(0,0)[r]{\strut{} 0.05}}%
      \csname LTb\endcsname%
      \put(1078,1373){\makebox(0,0)[r]{\strut{} 0.1}}%
      \csname LTb\endcsname%
      \put(1078,1708){\makebox(0,0)[r]{\strut{} 0.15}}%
      \csname LTb\endcsname%
      \put(1078,2042){\makebox(0,0)[r]{\strut{} 0.2}}%
      \csname LTb\endcsname%
      \put(1078,2377){\makebox(0,0)[r]{\strut{} 0.25}}%
      \csname LTb\endcsname%
      \put(1078,2711){\makebox(0,0)[r]{\strut{} 0.3}}%
      \csname LTb\endcsname%
      \put(1078,3046){\makebox(0,0)[r]{\strut{} 0.35}}%
      \csname LTb\endcsname%
      \put(1210,484){\makebox(0,0){\strut{} 1}}%
      \csname LTb\endcsname%
      \put(1628,484){\makebox(0,0){\strut{} 1.1}}%
      \csname LTb\endcsname%
      \put(2046,484){\makebox(0,0){\strut{} 1.2}}%
      \csname LTb\endcsname%
      \put(2464,484){\makebox(0,0){\strut{} 1.3}}%
      \csname LTb\endcsname%
      \put(2883,484){\makebox(0,0){\strut{} 1.4}}%
      \csname LTb\endcsname%
      \put(3301,484){\makebox(0,0){\strut{} 1.5}}%
      \csname LTb\endcsname%
      \put(3719,484){\makebox(0,0){\strut{} 1.6}}%
      \csname LTb\endcsname%
      \put(4137,484){\makebox(0,0){\strut{} 1.7}}%
      \put(176,1875){\rotatebox{-270}{\makebox(0,0){\strut{}$\sigma_{22}^{\rm aniso}/{\sigma_{11}^{\rm aniso}}$}}}%
      \put(2673,154){\makebox(0,0){\strut{}$F_{11}$}}%
    }%
    \gplgaddtomacro\gplfronttext{%
      \csname LTb\endcsname%
      \put(1738,1537){\makebox(0,0)[r]{\strut{}$i=1$}}%
      \csname LTb\endcsname%
      \put(1738,1317){\makebox(0,0)[r]{\strut{}$i=2$}}%
      \csname LTb\endcsname%
      \put(1738,1097){\makebox(0,0)[r]{\strut{}$i=3$}}%
      \csname LTb\endcsname%
      \put(1738,877){\makebox(0,0)[r]{\strut{}$i=4$}}%
    }%
    \gplbacktext
    \put(0,0){\includegraphics{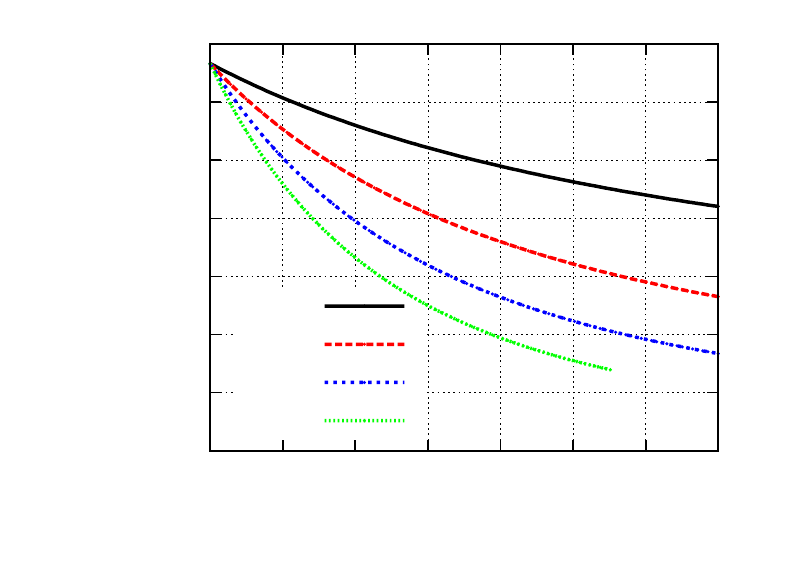}}%
    \gplfronttext
  \end{picture}%
\endgroup
}
\put(8.5,0.0){
\begingroup
  \makeatletter
  \providecommand\color[2][]{%
    \GenericError{(gnuplot) \space\space\space\@spaces}{%
      Package color not loaded in conjunction with
      terminal option `colourtext'%
    }{See the gnuplot documentation for explanation.%
    }{Either use 'blacktext' in gnuplot or load the package
      color.sty in LaTeX.}%
    \renewcommand\color[2][]{}%
  }%
  \providecommand\includegraphics[2][]{%
    \GenericError{(gnuplot) \space\space\space\@spaces}{%
      Package graphicx or graphics not loaded%
    }{See the gnuplot documentation for explanation.%
    }{The gnuplot epslatex terminal needs graphicx.sty or graphics.sty.}%
    \renewcommand\includegraphics[2][]{}%
  }%
  \providecommand\rotatebox[2]{#2}%
  \@ifundefined{ifGPcolor}{%
    \newif\ifGPcolor
    \GPcolortrue
  }{}%
  \@ifundefined{ifGPblacktext}{%
    \newif\ifGPblacktext
    \GPblacktexttrue
  }{}%
  \let\gplgaddtomacro\g@addto@macro
  \gdef\gplbacktext{}%
  \gdef\gplfronttext{}%
  \makeatother
  \ifGPblacktext
    \def\colorrgb#1{}%
    \def\colorgray#1{}%
  \else
    \ifGPcolor
      \def\colorrgb#1{\color[rgb]{#1}}%
      \def\colorgray#1{\color[gray]{#1}}%
      \expandafter\def\csname LTw\endcsname{\color{white}}%
      \expandafter\def\csname LTb\endcsname{\color{black}}%
      \expandafter\def\csname LTa\endcsname{\color{black}}%
      \expandafter\def\csname LT0\endcsname{\color[rgb]{1,0,0}}%
      \expandafter\def\csname LT1\endcsname{\color[rgb]{0,1,0}}%
      \expandafter\def\csname LT2\endcsname{\color[rgb]{0,0,1}}%
      \expandafter\def\csname LT3\endcsname{\color[rgb]{1,0,1}}%
      \expandafter\def\csname LT4\endcsname{\color[rgb]{0,1,1}}%
      \expandafter\def\csname LT5\endcsname{\color[rgb]{1,1,0}}%
      \expandafter\def\csname LT6\endcsname{\color[rgb]{0,0,0}}%
      \expandafter\def\csname LT7\endcsname{\color[rgb]{1,0.3,0}}%
      \expandafter\def\csname LT8\endcsname{\color[rgb]{0.5,0.5,0.5}}%
    \else
      \def\colorrgb#1{\color{black}}%
      \def\colorgray#1{\color[gray]{#1}}%
      \expandafter\def\csname LTw\endcsname{\color{white}}%
      \expandafter\def\csname LTb\endcsname{\color{black}}%
      \expandafter\def\csname LTa\endcsname{\color{black}}%
      \expandafter\def\csname LT0\endcsname{\color{black}}%
      \expandafter\def\csname LT1\endcsname{\color{black}}%
      \expandafter\def\csname LT2\endcsname{\color{black}}%
      \expandafter\def\csname LT3\endcsname{\color{black}}%
      \expandafter\def\csname LT4\endcsname{\color{black}}%
      \expandafter\def\csname LT5\endcsname{\color{black}}%
      \expandafter\def\csname LT6\endcsname{\color{black}}%
      \expandafter\def\csname LT7\endcsname{\color{black}}%
      \expandafter\def\csname LT8\endcsname{\color{black}}%
    \fi
  \fi
  \setlength{\unitlength}{0.0500bp}%
  \begin{picture}(4534.40,3310.40)%
    \gplgaddtomacro\gplbacktext{%
      \csname LTb\endcsname%
      \put(1078,704){\makebox(0,0)[r]{\strut{} 0}}%
      \csname LTb\endcsname%
      \put(1078,1039){\makebox(0,0)[r]{\strut{} 0.05}}%
      \csname LTb\endcsname%
      \put(1078,1373){\makebox(0,0)[r]{\strut{} 0.1}}%
      \csname LTb\endcsname%
      \put(1078,1708){\makebox(0,0)[r]{\strut{} 0.15}}%
      \csname LTb\endcsname%
      \put(1078,2042){\makebox(0,0)[r]{\strut{} 0.2}}%
      \csname LTb\endcsname%
      \put(1078,2377){\makebox(0,0)[r]{\strut{} 0.25}}%
      \csname LTb\endcsname%
      \put(1078,2711){\makebox(0,0)[r]{\strut{} 0.3}}%
      \csname LTb\endcsname%
      \put(1078,3046){\makebox(0,0)[r]{\strut{} 0.35}}%
      \csname LTb\endcsname%
      \put(1210,484){\makebox(0,0){\strut{} 1}}%
      \csname LTb\endcsname%
      \put(1628,484){\makebox(0,0){\strut{} 1.1}}%
      \csname LTb\endcsname%
      \put(2046,484){\makebox(0,0){\strut{} 1.2}}%
      \csname LTb\endcsname%
      \put(2464,484){\makebox(0,0){\strut{} 1.3}}%
      \csname LTb\endcsname%
      \put(2883,484){\makebox(0,0){\strut{} 1.4}}%
      \csname LTb\endcsname%
      \put(3301,484){\makebox(0,0){\strut{} 1.5}}%
      \csname LTb\endcsname%
      \put(3719,484){\makebox(0,0){\strut{} 1.6}}%
      \csname LTb\endcsname%
      \put(4137,484){\makebox(0,0){\strut{} 1.7}}%
      \put(176,1875){\rotatebox{-270}{\makebox(0,0){\strut{}$\sigma_{22}^{\rm aniso}/{\sigma_{11}^{\rm aniso}}$}}}%
      \put(2673,154){\makebox(0,0){\strut{}$F_{11}$}}%
    }%
    \gplgaddtomacro\gplfronttext{%
      \csname LTb\endcsname%
      \put(3289,2769){\makebox(0,0)[r]{\strut{}$i=1$}}%
      \csname LTb\endcsname%
      \put(3289,2549){\makebox(0,0)[r]{\strut{}$i=2$}}%
      \csname LTb\endcsname%
      \put(3289,2329){\makebox(0,0)[r]{\strut{}$i=3$}}%
      \csname LTb\endcsname%
      \put(3289,2109){\makebox(0,0)[r]{\strut{}$i=4$}}%
    }%
    \gplbacktext
    \put(0,0){\includegraphics{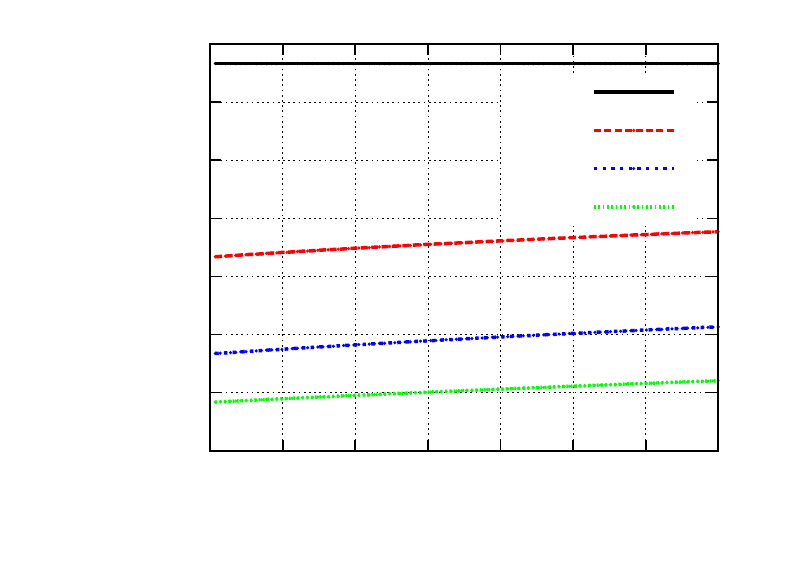}}%
    \gplfronttext
  \end{picture}%
\endgroup
}
\put(5.9,4.6){\tiny  \setlength{\fboxsep}{2.0mm}\fcolorbox{mygray}{mygray}{$I_4^{{\rm C}^i}$}}
\put(11.3,4.0){\tiny  \setlength{\fboxsep}{2.0mm}\fcolorbox{mygray}{mygray}{$I_4^{{\rm H}^i}$}}
\put(0.3,0.15){a)}
\put(8.5,0.15){b)}
\end{picture}
\setlength{\baselineskip}{11pt}
\caption{Evolution of the stress ratio for the biaxial test with the stretch ratio $1.7:1.35$ for a) the invariants $I_4^{{\rm C}^i}$ and
b) the Hencky-type invariants $I_4^{{\rm H}^i}$ with the exponent $i=1,2,3,4$.  \label{fig: biax_17_135}}
\end{Figure}

\clearpage

\begin{Figure}[!htb]
\unitlength 1 cm
\begin{picture}(14,5.8)
\put(0.0,0.0){
\begingroup
  \makeatletter
  \providecommand\color[2][]{%
    \GenericError{(gnuplot) \space\space\space\@spaces}{%
      Package color not loaded in conjunction with
      terminal option `colourtext'%
    }{See the gnuplot documentation for explanation.%
    }{Either use 'blacktext' in gnuplot or load the package
      color.sty in LaTeX.}%
    \renewcommand\color[2][]{}%
  }%
  \providecommand\includegraphics[2][]{%
    \GenericError{(gnuplot) \space\space\space\@spaces}{%
      Package graphicx or graphics not loaded%
    }{See the gnuplot documentation for explanation.%
    }{The gnuplot epslatex terminal needs graphicx.sty or graphics.sty.}%
    \renewcommand\includegraphics[2][]{}%
  }%
  \providecommand\rotatebox[2]{#2}%
  \@ifundefined{ifGPcolor}{%
    \newif\ifGPcolor
    \GPcolortrue
  }{}%
  \@ifundefined{ifGPblacktext}{%
    \newif\ifGPblacktext
    \GPblacktexttrue
  }{}%
  \let\gplgaddtomacro\g@addto@macro
  \gdef\gplbacktext{}%
  \gdef\gplfronttext{}%
  \makeatother
  \ifGPblacktext
    \def\colorrgb#1{}%
    \def\colorgray#1{}%
  \else
    \ifGPcolor
      \def\colorrgb#1{\color[rgb]{#1}}%
      \def\colorgray#1{\color[gray]{#1}}%
      \expandafter\def\csname LTw\endcsname{\color{white}}%
      \expandafter\def\csname LTb\endcsname{\color{black}}%
      \expandafter\def\csname LTa\endcsname{\color{black}}%
      \expandafter\def\csname LT0\endcsname{\color[rgb]{1,0,0}}%
      \expandafter\def\csname LT1\endcsname{\color[rgb]{0,1,0}}%
      \expandafter\def\csname LT2\endcsname{\color[rgb]{0,0,1}}%
      \expandafter\def\csname LT3\endcsname{\color[rgb]{1,0,1}}%
      \expandafter\def\csname LT4\endcsname{\color[rgb]{0,1,1}}%
      \expandafter\def\csname LT5\endcsname{\color[rgb]{1,1,0}}%
      \expandafter\def\csname LT6\endcsname{\color[rgb]{0,0,0}}%
      \expandafter\def\csname LT7\endcsname{\color[rgb]{1,0.3,0}}%
      \expandafter\def\csname LT8\endcsname{\color[rgb]{0.5,0.5,0.5}}%
    \else
      \def\colorrgb#1{\color{black}}%
      \def\colorgray#1{\color[gray]{#1}}%
      \expandafter\def\csname LTw\endcsname{\color{white}}%
      \expandafter\def\csname LTb\endcsname{\color{black}}%
      \expandafter\def\csname LTa\endcsname{\color{black}}%
      \expandafter\def\csname LT0\endcsname{\color{black}}%
      \expandafter\def\csname LT1\endcsname{\color{black}}%
      \expandafter\def\csname LT2\endcsname{\color{black}}%
      \expandafter\def\csname LT3\endcsname{\color{black}}%
      \expandafter\def\csname LT4\endcsname{\color{black}}%
      \expandafter\def\csname LT5\endcsname{\color{black}}%
      \expandafter\def\csname LT6\endcsname{\color{black}}%
      \expandafter\def\csname LT7\endcsname{\color{black}}%
      \expandafter\def\csname LT8\endcsname{\color{black}}%
    \fi
  \fi
  \setlength{\unitlength}{0.0500bp}%
  \begin{picture}(4534.40,3310.40)%
    \gplgaddtomacro\gplbacktext{%
      \csname LTb\endcsname%
      \put(946,704){\makebox(0,0)[r]{\strut{} 0}}%
      \csname LTb\endcsname%
      \put(946,1039){\makebox(0,0)[r]{\strut{} 0.5}}%
      \csname LTb\endcsname%
      \put(946,1373){\makebox(0,0)[r]{\strut{} 1}}%
      \csname LTb\endcsname%
      \put(946,1708){\makebox(0,0)[r]{\strut{} 1.5}}%
      \csname LTb\endcsname%
      \put(946,2042){\makebox(0,0)[r]{\strut{} 2}}%
      \csname LTb\endcsname%
      \put(946,2377){\makebox(0,0)[r]{\strut{} 2.5}}%
      \csname LTb\endcsname%
      \put(946,2711){\makebox(0,0)[r]{\strut{} 3}}%
      \csname LTb\endcsname%
      \put(946,3046){\makebox(0,0)[r]{\strut{} 3.5}}%
      \csname LTb\endcsname%
      \put(1078,484){\makebox(0,0){\strut{} 1}}%
      \csname LTb\endcsname%
      \put(1515,484){\makebox(0,0){\strut{} 1.05}}%
      \csname LTb\endcsname%
      \put(1952,484){\makebox(0,0){\strut{} 1.1}}%
      \csname LTb\endcsname%
      \put(2389,484){\makebox(0,0){\strut{} 1.15}}%
      \csname LTb\endcsname%
      \put(2826,484){\makebox(0,0){\strut{} 1.2}}%
      \csname LTb\endcsname%
      \put(3263,484){\makebox(0,0){\strut{} 1.25}}%
      \csname LTb\endcsname%
      \put(3700,484){\makebox(0,0){\strut{} 1.3}}%
      \csname LTb\endcsname%
      \put(4137,484){\makebox(0,0){\strut{} 1.35}}%
      \put(176,1875){\rotatebox{-270}{\makebox(0,0){\strut{}$\sigma_{22}^{\rm aniso}/{\sigma_{11}^{\rm aniso}}$}}}%
      \put(2607,154){\makebox(0,0){\strut{}$F_{11}$}}%
    }%
    \gplgaddtomacro\gplfronttext{%
      \csname LTb\endcsname%
      \put(1606,2873){\makebox(0,0)[r]{\strut{}$i=1$}}%
      \csname LTb\endcsname%
      \put(1606,2653){\makebox(0,0)[r]{\strut{}$i=2$}}%
      \csname LTb\endcsname%
      \put(1606,2433){\makebox(0,0)[r]{\strut{}$i=3$}}%
      \csname LTb\endcsname%
      \put(1606,2213){\makebox(0,0)[r]{\strut{}$i=4$}}%
    }%
    \gplbacktext
    \put(0,0){\includegraphics{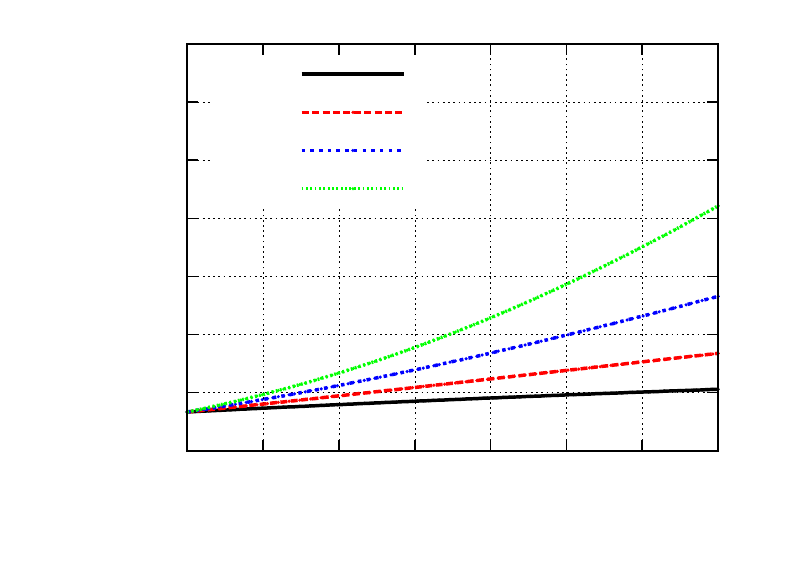}}%
    \gplfronttext
  \end{picture}%
\endgroup
}
\put(8.5,0.0){
\begingroup
  \makeatletter
  \providecommand\color[2][]{%
    \GenericError{(gnuplot) \space\space\space\@spaces}{%
      Package color not loaded in conjunction with
      terminal option `colourtext'%
    }{See the gnuplot documentation for explanation.%
    }{Either use 'blacktext' in gnuplot or load the package
      color.sty in LaTeX.}%
    \renewcommand\color[2][]{}%
  }%
  \providecommand\includegraphics[2][]{%
    \GenericError{(gnuplot) \space\space\space\@spaces}{%
      Package graphicx or graphics not loaded%
    }{See the gnuplot documentation for explanation.%
    }{The gnuplot epslatex terminal needs graphicx.sty or graphics.sty.}%
    \renewcommand\includegraphics[2][]{}%
  }%
  \providecommand\rotatebox[2]{#2}%
  \@ifundefined{ifGPcolor}{%
    \newif\ifGPcolor
    \GPcolortrue
  }{}%
  \@ifundefined{ifGPblacktext}{%
    \newif\ifGPblacktext
    \GPblacktexttrue
  }{}%
  \let\gplgaddtomacro\g@addto@macro
  \gdef\gplbacktext{}%
  \gdef\gplfronttext{}%
  \makeatother
  \ifGPblacktext
    \def\colorrgb#1{}%
    \def\colorgray#1{}%
  \else
    \ifGPcolor
      \def\colorrgb#1{\color[rgb]{#1}}%
      \def\colorgray#1{\color[gray]{#1}}%
      \expandafter\def\csname LTw\endcsname{\color{white}}%
      \expandafter\def\csname LTb\endcsname{\color{black}}%
      \expandafter\def\csname LTa\endcsname{\color{black}}%
      \expandafter\def\csname LT0\endcsname{\color[rgb]{1,0,0}}%
      \expandafter\def\csname LT1\endcsname{\color[rgb]{0,1,0}}%
      \expandafter\def\csname LT2\endcsname{\color[rgb]{0,0,1}}%
      \expandafter\def\csname LT3\endcsname{\color[rgb]{1,0,1}}%
      \expandafter\def\csname LT4\endcsname{\color[rgb]{0,1,1}}%
      \expandafter\def\csname LT5\endcsname{\color[rgb]{1,1,0}}%
      \expandafter\def\csname LT6\endcsname{\color[rgb]{0,0,0}}%
      \expandafter\def\csname LT7\endcsname{\color[rgb]{1,0.3,0}}%
      \expandafter\def\csname LT8\endcsname{\color[rgb]{0.5,0.5,0.5}}%
    \else
      \def\colorrgb#1{\color{black}}%
      \def\colorgray#1{\color[gray]{#1}}%
      \expandafter\def\csname LTw\endcsname{\color{white}}%
      \expandafter\def\csname LTb\endcsname{\color{black}}%
      \expandafter\def\csname LTa\endcsname{\color{black}}%
      \expandafter\def\csname LT0\endcsname{\color{black}}%
      \expandafter\def\csname LT1\endcsname{\color{black}}%
      \expandafter\def\csname LT2\endcsname{\color{black}}%
      \expandafter\def\csname LT3\endcsname{\color{black}}%
      \expandafter\def\csname LT4\endcsname{\color{black}}%
      \expandafter\def\csname LT5\endcsname{\color{black}}%
      \expandafter\def\csname LT6\endcsname{\color{black}}%
      \expandafter\def\csname LT7\endcsname{\color{black}}%
      \expandafter\def\csname LT8\endcsname{\color{black}}%
    \fi
  \fi
  \setlength{\unitlength}{0.0500bp}%
  \begin{picture}(4534.40,3310.40)%
    \gplgaddtomacro\gplbacktext{%
      \csname LTb\endcsname%
      \put(946,704){\makebox(0,0)[r]{\strut{} 0}}%
      \csname LTb\endcsname%
      \put(946,1039){\makebox(0,0)[r]{\strut{} 0.5}}%
      \csname LTb\endcsname%
      \put(946,1373){\makebox(0,0)[r]{\strut{} 1}}%
      \csname LTb\endcsname%
      \put(946,1708){\makebox(0,0)[r]{\strut{} 1.5}}%
      \csname LTb\endcsname%
      \put(946,2042){\makebox(0,0)[r]{\strut{} 2}}%
      \csname LTb\endcsname%
      \put(946,2377){\makebox(0,0)[r]{\strut{} 2.5}}%
      \csname LTb\endcsname%
      \put(946,2711){\makebox(0,0)[r]{\strut{} 3}}%
      \csname LTb\endcsname%
      \put(946,3046){\makebox(0,0)[r]{\strut{} 3.5}}%
      \csname LTb\endcsname%
      \put(1078,484){\makebox(0,0){\strut{} 1}}%
      \csname LTb\endcsname%
      \put(1515,484){\makebox(0,0){\strut{} 1.05}}%
      \csname LTb\endcsname%
      \put(1952,484){\makebox(0,0){\strut{} 1.1}}%
      \csname LTb\endcsname%
      \put(2389,484){\makebox(0,0){\strut{} 1.15}}%
      \csname LTb\endcsname%
      \put(2826,484){\makebox(0,0){\strut{} 1.2}}%
      \csname LTb\endcsname%
      \put(3263,484){\makebox(0,0){\strut{} 1.25}}%
      \csname LTb\endcsname%
      \put(3700,484){\makebox(0,0){\strut{} 1.3}}%
      \csname LTb\endcsname%
      \put(4137,484){\makebox(0,0){\strut{} 1.35}}%
      \put(176,1875){\rotatebox{-270}{\makebox(0,0){\strut{}$\sigma_{22}^{\rm aniso}/{\sigma_{11}^{\rm aniso}}$}}}%
      \put(2607,154){\makebox(0,0){\strut{}$F_{11}$}}%
    }%
    \gplgaddtomacro\gplfronttext{%
      \csname LTb\endcsname%
      \put(3150,2873){\makebox(0,0)[r]{\strut{}$i=1$}}%
      \csname LTb\endcsname%
      \put(3150,2653){\makebox(0,0)[r]{\strut{}$i=2$}}%
      \csname LTb\endcsname%
      \put(3150,2433){\makebox(0,0)[r]{\strut{}$i=3$}}%
      \csname LTb\endcsname%
      \put(3150,2213){\makebox(0,0)[r]{\strut{}$i=4$}}%
    }%
    \gplbacktext
    \put(0,0){\includegraphics{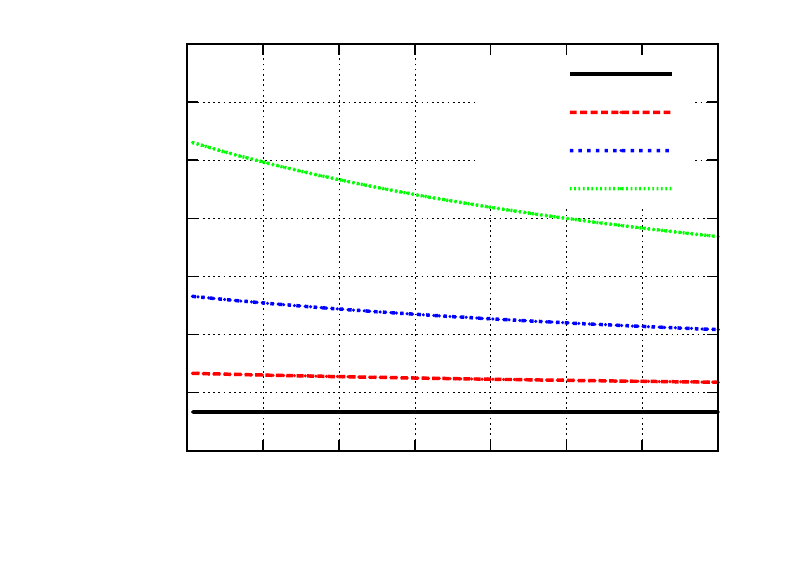}}%
    \gplfronttext
  \end{picture}%
\endgroup
}
\put(2.7,3.0){\tiny  \setlength{\fboxsep}{2.0mm}\fcolorbox{mygray}{mygray}{$I_4^{{\rm C}^i}$}}
\put(11.3,3.4){\tiny  \setlength{\fboxsep}{2.0mm}\fcolorbox{mygray}{mygray}{$I_4^{{\rm H}^i}$}}
\put(0.3,0.15){a)}
\put(8.5,0.15){b)}
\end{picture}
\setlength{\baselineskip}{11pt}
\caption{Evolution of the stress ratio for the biaxial test with the stretch ratio $1.35:1.7$ for a) the invariants $I_4^{{\rm C}^i}$ and
b) the Hencky-type invariants $I_4^{{\rm H}^i}$ with the exponent $i=1,2,3,4$.  \label{fig: biax_135_17}}
\end{Figure}

\subsubsection{Biaxial-tension-compression}

In order to complete the study of classical mechanical loading scenarios we consider a biaxial 
combined tension and compression test, assuming incompressibility. The problem is depicted in Fig.~\ref{fig: biax comp ten}
and the fiber arrangement is identical to the previous example in section~\ref{sec: Biax tension}.
While the stretches $\lambda_x$ and $\lambda_y$ are displacement driven, the principal stretch results form the incompressibility
condition $\lambda_x\lambda_y\lambda_z=1$. 

The fiber invariants, see Fig.~\ref{fig: biax_ten_comp_30_aniso_invarianten}, are identical for both fiber families and the evolution
is comparable to the uniaxial case. In Fig.~\ref{fig: biax_ten_comp_30_sig22_11_aniso}, again the evolution of the stress ratio is plotted, for both 
invariant sets. They appear to be quite different. For $I_4^{\rm C^i}$ the stress ratio is approaching zero for infinite strains. 
For $I_4^{\rm H^1}$ we obtain a constant line as it was already the case in the biaxial tension test. If $i$ in $I_4^{\rm H^i}$ is an even number 
$\sigma_{22}$ becomes negative, thus the stress ratio becomes negative.

\begin{Figure}[!htb]
\unitlength 1 cm
\begin{picture}(14,5.5)
\put(2.0,0.0){\includegraphics[height = 5.5cm]{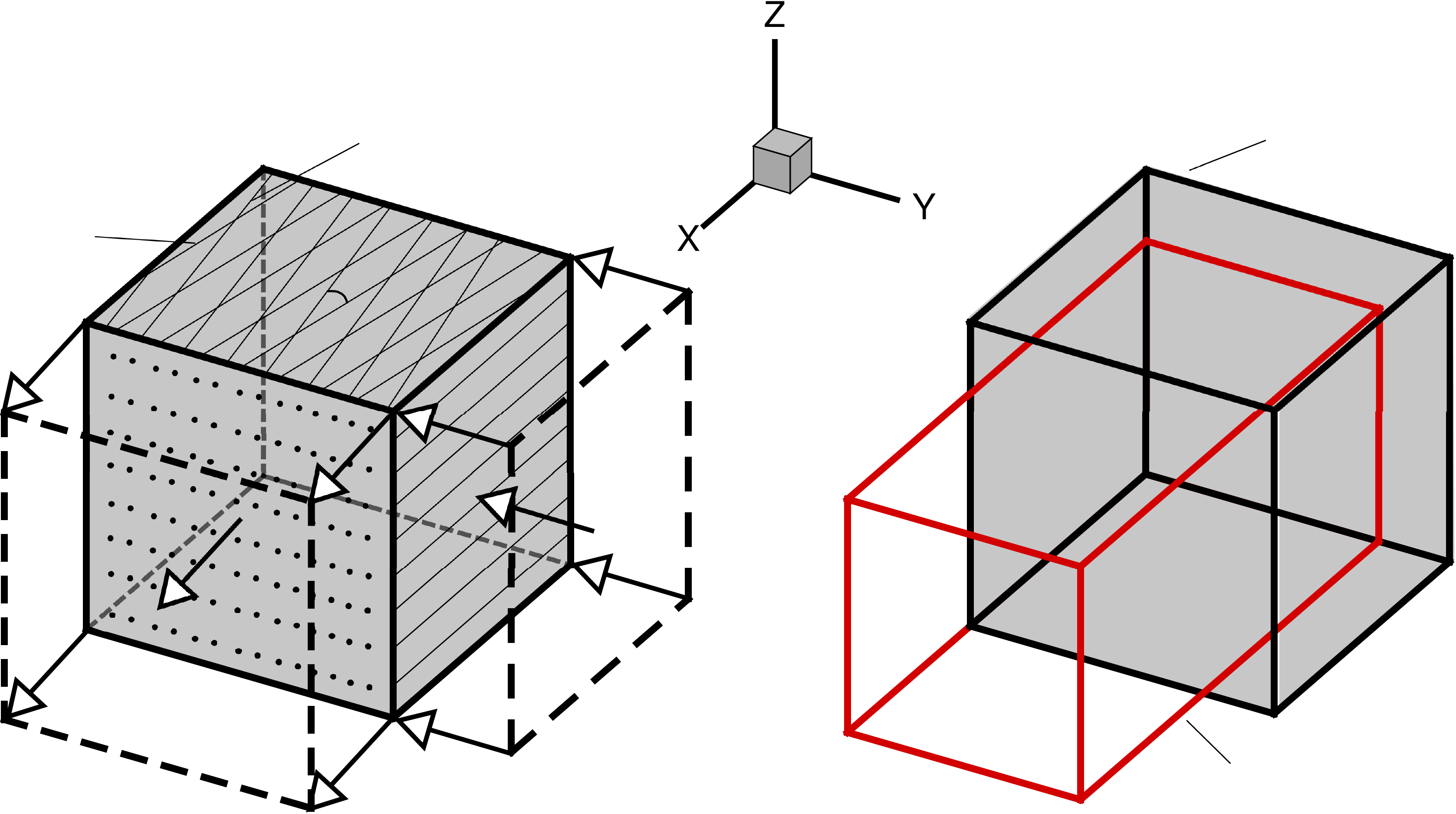}}
\put(2.2,3.85){$\bA^{(1)}$}
\put(4.27,4.6){$\bA^{(2)}$}
\put(4.2,3.6){$60^{\circ}$}
\put(10.3,4.62){ref.}
\put(10.4,0.0){$1.7:0.765$}
\end{picture}
\setlength{\baselineskip}{11pt}
\caption{Biaxial combined compression and tension test, where the preferred direction and the loading are aligned. The body is assumed to be incompressible 
with $F_{33} = 1/(F_{11} F_{22})$ and $F_{33} = F_{22}$. \label{fig: biax comp ten}%
}
\end{Figure}

\begin{Figure}[!htb]
\unitlength 1 cm
\begin{picture}(14,5.8)
\put(0.0,0.0){
\begingroup
  \makeatletter
  \providecommand\color[2][]{%
    \GenericError{(gnuplot) \space\space\space\@spaces}{%
      Package color not loaded in conjunction with
      terminal option `colourtext'%
    }{See the gnuplot documentation for explanation.%
    }{Either use 'blacktext' in gnuplot or load the package
      color.sty in LaTeX.}%
    \renewcommand\color[2][]{}%
  }%
  \providecommand\includegraphics[2][]{%
    \GenericError{(gnuplot) \space\space\space\@spaces}{%
      Package graphicx or graphics not loaded%
    }{See the gnuplot documentation for explanation.%
    }{The gnuplot epslatex terminal needs graphicx.sty or graphics.sty.}%
    \renewcommand\includegraphics[2][]{}%
  }%
  \providecommand\rotatebox[2]{#2}%
  \@ifundefined{ifGPcolor}{%
    \newif\ifGPcolor
    \GPcolortrue
  }{}%
  \@ifundefined{ifGPblacktext}{%
    \newif\ifGPblacktext
    \GPblacktexttrue
  }{}%
  \let\gplgaddtomacro\g@addto@macro
  \gdef\gplbacktext{}%
  \gdef\gplfronttext{}%
  \makeatother
  \ifGPblacktext
    \def\colorrgb#1{}%
    \def\colorgray#1{}%
  \else
    \ifGPcolor
      \def\colorrgb#1{\color[rgb]{#1}}%
      \def\colorgray#1{\color[gray]{#1}}%
      \expandafter\def\csname LTw\endcsname{\color{white}}%
      \expandafter\def\csname LTb\endcsname{\color{black}}%
      \expandafter\def\csname LTa\endcsname{\color{black}}%
      \expandafter\def\csname LT0\endcsname{\color[rgb]{1,0,0}}%
      \expandafter\def\csname LT1\endcsname{\color[rgb]{0,1,0}}%
      \expandafter\def\csname LT2\endcsname{\color[rgb]{0,0,1}}%
      \expandafter\def\csname LT3\endcsname{\color[rgb]{1,0,1}}%
      \expandafter\def\csname LT4\endcsname{\color[rgb]{0,1,1}}%
      \expandafter\def\csname LT5\endcsname{\color[rgb]{1,1,0}}%
      \expandafter\def\csname LT6\endcsname{\color[rgb]{0,0,0}}%
      \expandafter\def\csname LT7\endcsname{\color[rgb]{1,0.3,0}}%
      \expandafter\def\csname LT8\endcsname{\color[rgb]{0.5,0.5,0.5}}%
    \else
      \def\colorrgb#1{\color{black}}%
      \def\colorgray#1{\color[gray]{#1}}%
      \expandafter\def\csname LTw\endcsname{\color{white}}%
      \expandafter\def\csname LTb\endcsname{\color{black}}%
      \expandafter\def\csname LTa\endcsname{\color{black}}%
      \expandafter\def\csname LT0\endcsname{\color{black}}%
      \expandafter\def\csname LT1\endcsname{\color{black}}%
      \expandafter\def\csname LT2\endcsname{\color{black}}%
      \expandafter\def\csname LT3\endcsname{\color{black}}%
      \expandafter\def\csname LT4\endcsname{\color{black}}%
      \expandafter\def\csname LT5\endcsname{\color{black}}%
      \expandafter\def\csname LT6\endcsname{\color{black}}%
      \expandafter\def\csname LT7\endcsname{\color{black}}%
      \expandafter\def\csname LT8\endcsname{\color{black}}%
    \fi
  \fi
  \setlength{\unitlength}{0.0500bp}%
  \begin{picture}(4534.40,3310.40)%
    \gplgaddtomacro\gplbacktext{%
      \csname LTb\endcsname%
      \put(814,704){\makebox(0,0)[r]{\strut{} 0}}%
      \csname LTb\endcsname%
      \put(814,1039){\makebox(0,0)[r]{\strut{} 2}}%
      \csname LTb\endcsname%
      \put(814,1373){\makebox(0,0)[r]{\strut{} 4}}%
      \csname LTb\endcsname%
      \put(814,1708){\makebox(0,0)[r]{\strut{} 6}}%
      \csname LTb\endcsname%
      \put(814,2042){\makebox(0,0)[r]{\strut{} 8}}%
      \csname LTb\endcsname%
      \put(814,2377){\makebox(0,0)[r]{\strut{} 10}}%
      \csname LTb\endcsname%
      \put(814,2711){\makebox(0,0)[r]{\strut{} 12}}%
      \csname LTb\endcsname%
      \put(814,3046){\makebox(0,0)[r]{\strut{} 14}}%
      \csname LTb\endcsname%
      \put(946,484){\makebox(0,0){\strut{} 1}}%
      \csname LTb\endcsname%
      \put(1402,484){\makebox(0,0){\strut{} 1.1}}%
      \csname LTb\endcsname%
      \put(1858,484){\makebox(0,0){\strut{} 1.2}}%
      \csname LTb\endcsname%
      \put(2314,484){\makebox(0,0){\strut{} 1.3}}%
      \csname LTb\endcsname%
      \put(2769,484){\makebox(0,0){\strut{} 1.4}}%
      \csname LTb\endcsname%
      \put(3225,484){\makebox(0,0){\strut{} 1.5}}%
      \csname LTb\endcsname%
      \put(3681,484){\makebox(0,0){\strut{} 1.6}}%
      \csname LTb\endcsname%
      \put(4137,484){\makebox(0,0){\strut{} 1.7}}%
      \put(176,1875){\rotatebox{-270}{\makebox(0,0){\strut{}$I_4^{C^i}$}}}%
      \put(2541,154){\makebox(0,0){\strut{}$F_{11}$}}%
    }%
    \gplgaddtomacro\gplfronttext{%
      \csname LTb\endcsname%
      \put(1474,2873){\makebox(0,0)[r]{\strut{}$i=1$}}%
      \csname LTb\endcsname%
      \put(1474,2653){\makebox(0,0)[r]{\strut{}$i=2$}}%
      \csname LTb\endcsname%
      \put(1474,2433){\makebox(0,0)[r]{\strut{}$i=3$}}%
      \csname LTb\endcsname%
      \put(1474,2213){\makebox(0,0)[r]{\strut{}$i=4$}}%
    }%
    \gplbacktext
    \put(0,0){\includegraphics{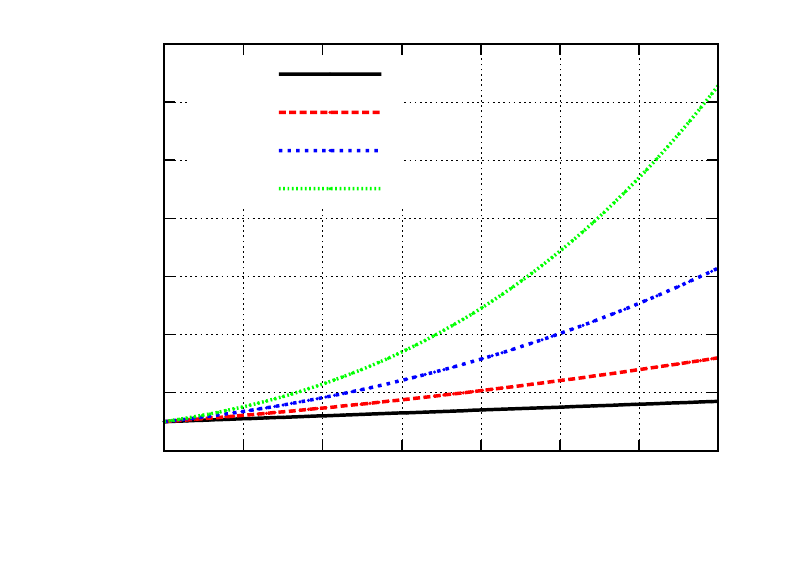}}%
    \gplfronttext
  \end{picture}%
\endgroup
}
\put(8.5,0.0){
\begingroup
  \makeatletter
  \providecommand\color[2][]{%
    \GenericError{(gnuplot) \space\space\space\@spaces}{%
      Package color not loaded in conjunction with
      terminal option `colourtext'%
    }{See the gnuplot documentation for explanation.%
    }{Either use 'blacktext' in gnuplot or load the package
      color.sty in LaTeX.}%
    \renewcommand\color[2][]{}%
  }%
  \providecommand\includegraphics[2][]{%
    \GenericError{(gnuplot) \space\space\space\@spaces}{%
      Package graphicx or graphics not loaded%
    }{See the gnuplot documentation for explanation.%
    }{The gnuplot epslatex terminal needs graphicx.sty or graphics.sty.}%
    \renewcommand\includegraphics[2][]{}%
  }%
  \providecommand\rotatebox[2]{#2}%
  \@ifundefined{ifGPcolor}{%
    \newif\ifGPcolor
    \GPcolortrue
  }{}%
  \@ifundefined{ifGPblacktext}{%
    \newif\ifGPblacktext
    \GPblacktexttrue
  }{}%
  \let\gplgaddtomacro\g@addto@macro
  \gdef\gplbacktext{}%
  \gdef\gplfronttext{}%
  \makeatother
  \ifGPblacktext
    \def\colorrgb#1{}%
    \def\colorgray#1{}%
  \else
    \ifGPcolor
      \def\colorrgb#1{\color[rgb]{#1}}%
      \def\colorgray#1{\color[gray]{#1}}%
      \expandafter\def\csname LTw\endcsname{\color{white}}%
      \expandafter\def\csname LTb\endcsname{\color{black}}%
      \expandafter\def\csname LTa\endcsname{\color{black}}%
      \expandafter\def\csname LT0\endcsname{\color[rgb]{1,0,0}}%
      \expandafter\def\csname LT1\endcsname{\color[rgb]{0,1,0}}%
      \expandafter\def\csname LT2\endcsname{\color[rgb]{0,0,1}}%
      \expandafter\def\csname LT3\endcsname{\color[rgb]{1,0,1}}%
      \expandafter\def\csname LT4\endcsname{\color[rgb]{0,1,1}}%
      \expandafter\def\csname LT5\endcsname{\color[rgb]{1,1,0}}%
      \expandafter\def\csname LT6\endcsname{\color[rgb]{0,0,0}}%
      \expandafter\def\csname LT7\endcsname{\color[rgb]{1,0.3,0}}%
      \expandafter\def\csname LT8\endcsname{\color[rgb]{0.5,0.5,0.5}}%
    \else
      \def\colorrgb#1{\color{black}}%
      \def\colorgray#1{\color[gray]{#1}}%
      \expandafter\def\csname LTw\endcsname{\color{white}}%
      \expandafter\def\csname LTb\endcsname{\color{black}}%
      \expandafter\def\csname LTa\endcsname{\color{black}}%
      \expandafter\def\csname LT0\endcsname{\color{black}}%
      \expandafter\def\csname LT1\endcsname{\color{black}}%
      \expandafter\def\csname LT2\endcsname{\color{black}}%
      \expandafter\def\csname LT3\endcsname{\color{black}}%
      \expandafter\def\csname LT4\endcsname{\color{black}}%
      \expandafter\def\csname LT5\endcsname{\color{black}}%
      \expandafter\def\csname LT6\endcsname{\color{black}}%
      \expandafter\def\csname LT7\endcsname{\color{black}}%
      \expandafter\def\csname LT8\endcsname{\color{black}}%
    \fi
  \fi
  \setlength{\unitlength}{0.0500bp}%
  \begin{picture}(4534.40,3310.40)%
    \gplgaddtomacro\gplbacktext{%
      \csname LTb\endcsname%
      \put(1078,704){\makebox(0,0)[r]{\strut{}-0.05}}%
      \csname LTb\endcsname%
      \put(1078,1094){\makebox(0,0)[r]{\strut{} 0}}%
      \csname LTb\endcsname%
      \put(1078,1485){\makebox(0,0)[r]{\strut{} 0.05}}%
      \csname LTb\endcsname%
      \put(1078,1875){\makebox(0,0)[r]{\strut{} 0.1}}%
      \csname LTb\endcsname%
      \put(1078,2265){\makebox(0,0)[r]{\strut{} 0.15}}%
      \csname LTb\endcsname%
      \put(1078,2656){\makebox(0,0)[r]{\strut{} 0.2}}%
      \csname LTb\endcsname%
      \put(1078,3046){\makebox(0,0)[r]{\strut{} 0.25}}%
      \csname LTb\endcsname%
      \put(1210,484){\makebox(0,0){\strut{} 1}}%
      \csname LTb\endcsname%
      \put(1628,484){\makebox(0,0){\strut{} 1.1}}%
      \csname LTb\endcsname%
      \put(2046,484){\makebox(0,0){\strut{} 1.2}}%
      \csname LTb\endcsname%
      \put(2464,484){\makebox(0,0){\strut{} 1.3}}%
      \csname LTb\endcsname%
      \put(2883,484){\makebox(0,0){\strut{} 1.4}}%
      \csname LTb\endcsname%
      \put(3301,484){\makebox(0,0){\strut{} 1.5}}%
      \csname LTb\endcsname%
      \put(3719,484){\makebox(0,0){\strut{} 1.6}}%
      \csname LTb\endcsname%
      \put(4137,484){\makebox(0,0){\strut{} 1.7}}%
      \put(176,1875){\rotatebox{-270}{\makebox(0,0){\strut{}$I_4^{H^i}$}}}%
      \put(2673,154){\makebox(0,0){\strut{}$F_{11}$}}%
    }%
    \gplgaddtomacro\gplfronttext{%
      \csname LTb\endcsname%
      \put(1738,2873){\makebox(0,0)[r]{\strut{}$i=1$}}%
      \csname LTb\endcsname%
      \put(1738,2653){\makebox(0,0)[r]{\strut{}$i=2$}}%
      \csname LTb\endcsname%
      \put(1738,2433){\makebox(0,0)[r]{\strut{}$i=3$}}%
      \csname LTb\endcsname%
      \put(1738,2213){\makebox(0,0)[r]{\strut{}$i=4$}}%
    }%
    \gplbacktext
    \put(0,0){\includegraphics{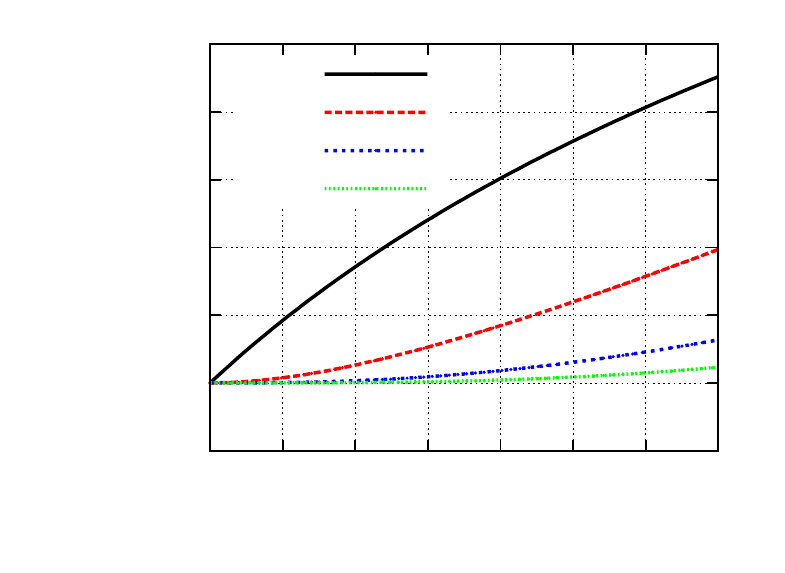}}%
    \gplfronttext
  \end{picture}%
\endgroup
}
\put(2.4,2.7){\tiny  \setlength{\fboxsep}{2.0mm}\fcolorbox{mygray}{mygray}{$I_4^{{\rm C}^i}$}}
\put(14.6,3.7){\tiny  \setlength{\fboxsep}{2.0mm}\fcolorbox{mygray}{mygray}{$I_4^{{\rm H}^i}$}}
\put(0.3,0.15){a)}
\put(8.5,0.15){b)}
\end{picture}
\setlength{\baselineskip}{11pt}
\caption{Evolution of the anisotropic invariants a) $I_4^{{\rm C}^i}$ and b) $I_4^{{\rm H}^i}$ for the exponent $i =1,2,3,4$ under biaxial-tension-compression. \label{fig: biax_ten_comp_30_aniso_invarianten}}
\end{Figure}

\begin{Figure}[!htb]
\unitlength 1 cm
\begin{picture}(14,5.8)
\put(0.0,0.0){
\begingroup
  \makeatletter
  \providecommand\color[2][]{%
    \GenericError{(gnuplot) \space\space\space\@spaces}{%
      Package color not loaded in conjunction with
      terminal option `colourtext'%
    }{See the gnuplot documentation for explanation.%
    }{Either use 'blacktext' in gnuplot or load the package
      color.sty in LaTeX.}%
    \renewcommand\color[2][]{}%
  }%
  \providecommand\includegraphics[2][]{%
    \GenericError{(gnuplot) \space\space\space\@spaces}{%
      Package graphicx or graphics not loaded%
    }{See the gnuplot documentation for explanation.%
    }{The gnuplot epslatex terminal needs graphicx.sty or graphics.sty.}%
    \renewcommand\includegraphics[2][]{}%
  }%
  \providecommand\rotatebox[2]{#2}%
  \@ifundefined{ifGPcolor}{%
    \newif\ifGPcolor
    \GPcolortrue
  }{}%
  \@ifundefined{ifGPblacktext}{%
    \newif\ifGPblacktext
    \GPblacktexttrue
  }{}%
  \let\gplgaddtomacro\g@addto@macro
  \gdef\gplbacktext{}%
  \gdef\gplfronttext{}%
  \makeatother
  \ifGPblacktext
    \def\colorrgb#1{}%
    \def\colorgray#1{}%
  \else
    \ifGPcolor
      \def\colorrgb#1{\color[rgb]{#1}}%
      \def\colorgray#1{\color[gray]{#1}}%
      \expandafter\def\csname LTw\endcsname{\color{white}}%
      \expandafter\def\csname LTb\endcsname{\color{black}}%
      \expandafter\def\csname LTa\endcsname{\color{black}}%
      \expandafter\def\csname LT0\endcsname{\color[rgb]{1,0,0}}%
      \expandafter\def\csname LT1\endcsname{\color[rgb]{0,1,0}}%
      \expandafter\def\csname LT2\endcsname{\color[rgb]{0,0,1}}%
      \expandafter\def\csname LT3\endcsname{\color[rgb]{1,0,1}}%
      \expandafter\def\csname LT4\endcsname{\color[rgb]{0,1,1}}%
      \expandafter\def\csname LT5\endcsname{\color[rgb]{1,1,0}}%
      \expandafter\def\csname LT6\endcsname{\color[rgb]{0,0,0}}%
      \expandafter\def\csname LT7\endcsname{\color[rgb]{1,0.3,0}}%
      \expandafter\def\csname LT8\endcsname{\color[rgb]{0.5,0.5,0.5}}%
    \else
      \def\colorrgb#1{\color{black}}%
      \def\colorgray#1{\color[gray]{#1}}%
      \expandafter\def\csname LTw\endcsname{\color{white}}%
      \expandafter\def\csname LTb\endcsname{\color{black}}%
      \expandafter\def\csname LTa\endcsname{\color{black}}%
      \expandafter\def\csname LT0\endcsname{\color{black}}%
      \expandafter\def\csname LT1\endcsname{\color{black}}%
      \expandafter\def\csname LT2\endcsname{\color{black}}%
      \expandafter\def\csname LT3\endcsname{\color{black}}%
      \expandafter\def\csname LT4\endcsname{\color{black}}%
      \expandafter\def\csname LT5\endcsname{\color{black}}%
      \expandafter\def\csname LT6\endcsname{\color{black}}%
      \expandafter\def\csname LT7\endcsname{\color{black}}%
      \expandafter\def\csname LT8\endcsname{\color{black}}%
    \fi
  \fi
  \setlength{\unitlength}{0.0500bp}%
  \begin{picture}(4534.40,3310.40)%
    \gplgaddtomacro\gplbacktext{%
      \csname LTb\endcsname%
      \put(946,938){\makebox(0,0)[r]{\strut{}-0.1}}%
      \csname LTb\endcsname%
      \put(946,1407){\makebox(0,0)[r]{\strut{} 0}}%
      \csname LTb\endcsname%
      \put(946,1875){\makebox(0,0)[r]{\strut{} 0.1}}%
      \csname LTb\endcsname%
      \put(946,2343){\makebox(0,0)[r]{\strut{} 0.2}}%
      \csname LTb\endcsname%
      \put(946,2812){\makebox(0,0)[r]{\strut{} 0.3}}%
      \csname LTb\endcsname%
      \put(1078,484){\makebox(0,0){\strut{} 1}}%
      \csname LTb\endcsname%
      \put(1515,484){\makebox(0,0){\strut{} 1.1}}%
      \csname LTb\endcsname%
      \put(1952,484){\makebox(0,0){\strut{} 1.2}}%
      \csname LTb\endcsname%
      \put(2389,484){\makebox(0,0){\strut{} 1.3}}%
      \csname LTb\endcsname%
      \put(2826,484){\makebox(0,0){\strut{} 1.4}}%
      \csname LTb\endcsname%
      \put(3263,484){\makebox(0,0){\strut{} 1.5}}%
      \csname LTb\endcsname%
      \put(3700,484){\makebox(0,0){\strut{} 1.6}}%
      \csname LTb\endcsname%
      \put(4137,484){\makebox(0,0){\strut{} 1.7}}%
      \put(176,1875){\rotatebox{-270}{\makebox(0,0){\strut{}$\sigma_{22}^{\rm aniso}/{\sigma_{11}^{\rm aniso}}$}}}%
      \put(2607,154){\makebox(0,0){\strut{}$F_{11}$}}%
    }%
    \gplgaddtomacro\gplfronttext{%
      \csname LTb\endcsname%
      \put(3150,2873){\makebox(0,0)[r]{\strut{}$i=1$}}%
      \csname LTb\endcsname%
      \put(3150,2653){\makebox(0,0)[r]{\strut{}$i=2$}}%
      \csname LTb\endcsname%
      \put(3150,2433){\makebox(0,0)[r]{\strut{}$i=3$}}%
      \csname LTb\endcsname%
      \put(3150,2213){\makebox(0,0)[r]{\strut{}$i=4$}}%
    }%
    \gplbacktext
    \put(0,0){\includegraphics{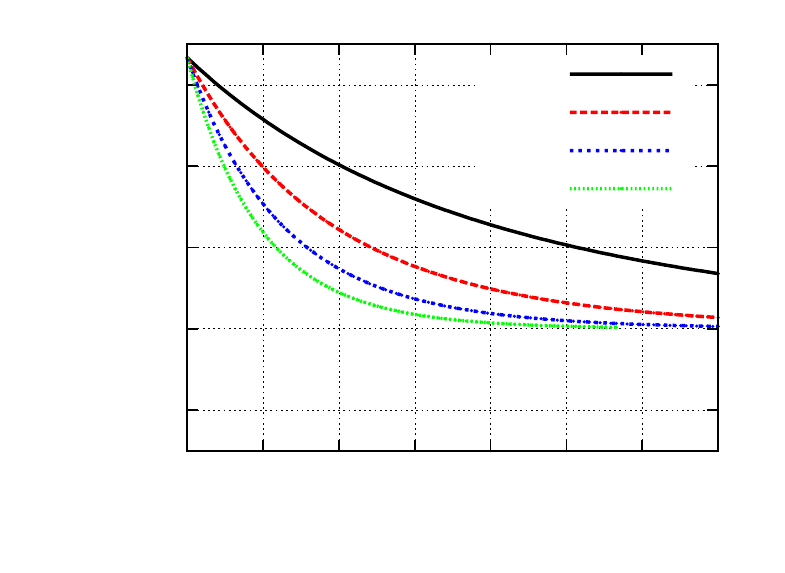}}%
    \gplfronttext
  \end{picture}%
\endgroup
}
\put(8.5,0.0){
\begingroup
  \makeatletter
  \providecommand\color[2][]{%
    \GenericError{(gnuplot) \space\space\space\@spaces}{%
      Package color not loaded in conjunction with
      terminal option `colourtext'%
    }{See the gnuplot documentation for explanation.%
    }{Either use 'blacktext' in gnuplot or load the package
      color.sty in LaTeX.}%
    \renewcommand\color[2][]{}%
  }%
  \providecommand\includegraphics[2][]{%
    \GenericError{(gnuplot) \space\space\space\@spaces}{%
      Package graphicx or graphics not loaded%
    }{See the gnuplot documentation for explanation.%
    }{The gnuplot epslatex terminal needs graphicx.sty or graphics.sty.}%
    \renewcommand\includegraphics[2][]{}%
  }%
  \providecommand\rotatebox[2]{#2}%
  \@ifundefined{ifGPcolor}{%
    \newif\ifGPcolor
    \GPcolortrue
  }{}%
  \@ifundefined{ifGPblacktext}{%
    \newif\ifGPblacktext
    \GPblacktexttrue
  }{}%
  \let\gplgaddtomacro\g@addto@macro
  \gdef\gplbacktext{}%
  \gdef\gplfronttext{}%
  \makeatother
  \ifGPblacktext
    \def\colorrgb#1{}%
    \def\colorgray#1{}%
  \else
    \ifGPcolor
      \def\colorrgb#1{\color[rgb]{#1}}%
      \def\colorgray#1{\color[gray]{#1}}%
      \expandafter\def\csname LTw\endcsname{\color{white}}%
      \expandafter\def\csname LTb\endcsname{\color{black}}%
      \expandafter\def\csname LTa\endcsname{\color{black}}%
      \expandafter\def\csname LT0\endcsname{\color[rgb]{1,0,0}}%
      \expandafter\def\csname LT1\endcsname{\color[rgb]{0,1,0}}%
      \expandafter\def\csname LT2\endcsname{\color[rgb]{0,0,1}}%
      \expandafter\def\csname LT3\endcsname{\color[rgb]{1,0,1}}%
      \expandafter\def\csname LT4\endcsname{\color[rgb]{0,1,1}}%
      \expandafter\def\csname LT5\endcsname{\color[rgb]{1,1,0}}%
      \expandafter\def\csname LT6\endcsname{\color[rgb]{0,0,0}}%
      \expandafter\def\csname LT7\endcsname{\color[rgb]{1,0.3,0}}%
      \expandafter\def\csname LT8\endcsname{\color[rgb]{0.5,0.5,0.5}}%
    \else
      \def\colorrgb#1{\color{black}}%
      \def\colorgray#1{\color[gray]{#1}}%
      \expandafter\def\csname LTw\endcsname{\color{white}}%
      \expandafter\def\csname LTb\endcsname{\color{black}}%
      \expandafter\def\csname LTa\endcsname{\color{black}}%
      \expandafter\def\csname LT0\endcsname{\color{black}}%
      \expandafter\def\csname LT1\endcsname{\color{black}}%
      \expandafter\def\csname LT2\endcsname{\color{black}}%
      \expandafter\def\csname LT3\endcsname{\color{black}}%
      \expandafter\def\csname LT4\endcsname{\color{black}}%
      \expandafter\def\csname LT5\endcsname{\color{black}}%
      \expandafter\def\csname LT6\endcsname{\color{black}}%
      \expandafter\def\csname LT7\endcsname{\color{black}}%
      \expandafter\def\csname LT8\endcsname{\color{black}}%
    \fi
  \fi
  \setlength{\unitlength}{0.0500bp}%
  \begin{picture}(4534.40,3310.40)%
    \gplgaddtomacro\gplbacktext{%
      \csname LTb\endcsname%
      \put(946,704){\makebox(0,0)[r]{\strut{}-0.2}}%
      \csname LTb\endcsname%
      \put(946,1130){\makebox(0,0)[r]{\strut{}-0.1}}%
      \csname LTb\endcsname%
      \put(946,1556){\makebox(0,0)[r]{\strut{} 0}}%
      \csname LTb\endcsname%
      \put(946,1981){\makebox(0,0)[r]{\strut{} 0.1}}%
      \csname LTb\endcsname%
      \put(946,2407){\makebox(0,0)[r]{\strut{} 0.2}}%
      \csname LTb\endcsname%
      \put(946,2833){\makebox(0,0)[r]{\strut{} 0.3}}%
      \csname LTb\endcsname%
      \put(1078,484){\makebox(0,0){\strut{} 1}}%
      \csname LTb\endcsname%
      \put(1515,484){\makebox(0,0){\strut{} 1.1}}%
      \csname LTb\endcsname%
      \put(1952,484){\makebox(0,0){\strut{} 1.2}}%
      \csname LTb\endcsname%
      \put(2389,484){\makebox(0,0){\strut{} 1.3}}%
      \csname LTb\endcsname%
      \put(2826,484){\makebox(0,0){\strut{} 1.4}}%
      \csname LTb\endcsname%
      \put(3263,484){\makebox(0,0){\strut{} 1.5}}%
      \csname LTb\endcsname%
      \put(3700,484){\makebox(0,0){\strut{} 1.6}}%
      \csname LTb\endcsname%
      \put(4137,484){\makebox(0,0){\strut{} 1.7}}%
      \put(176,1875){\rotatebox{-270}{\makebox(0,0){\strut{}$\sigma_{22}^{\rm aniso}/{\sigma_{11}^{\rm aniso}}$}}}%
      \put(2607,154){\makebox(0,0){\strut{}$F_{11}$}}%
    }%
    \gplgaddtomacro\gplfronttext{%
      \csname LTb\endcsname%
      \put(1971,2744){\makebox(0,0)[r]{\strut{}$i=1$}}%
      \csname LTb\endcsname%
      \put(1971,2524){\makebox(0,0)[r]{\strut{}$i=2$}}%
      \csname LTb\endcsname%
      \put(1971,2304){\makebox(0,0)[r]{\strut{}$i=3$}}%
      \csname LTb\endcsname%
      \put(1971,2084){\makebox(0,0)[r]{\strut{}$i=4$}}%
    }%
    \gplbacktext
    \put(0,0){\includegraphics{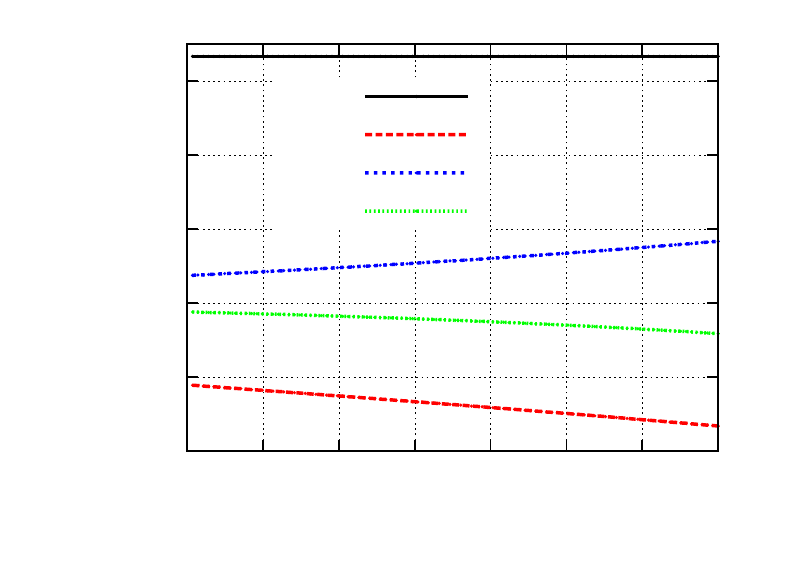}}%
    \gplfronttext
  \end{picture}%
\endgroup
}
\put(2.4,1.95){\tiny  \setlength{\fboxsep}{2.0mm}\fcolorbox{mygray}{mygray}{$I_4^{{\rm C}^i}$}}
\put(14.6,4.2){\tiny  \setlength{\fboxsep}{2.0mm}\fcolorbox{mygray}{mygray}{$I_4^{{\rm H}^i}$}}
\put(0.3,0.15){a)}
\put(8.5,0.15){b)}
\end{picture}
\setlength{\baselineskip}{11pt}
\caption{Evolution of the stress ratio for the biaxial-tension-compression test  for a) the invariants $I_4^{{\rm C}^i}$ and
b) the Hencky-type invariants $I_4^{{\rm H}^i}$ with the exponent $i=1,2,3,4$. \label{fig: biax_ten_comp_30_sig22_11_aniso}}
\end{Figure}

\subsection{Adaption to biological soft tissues\label{sec: bio}}

Soft biological tissues as they occur in arterial
walls have an anisotropic material behavior.
In biomechanical applications we often assume that the
material behaves nearly-incompressible.

The widely used anisotropic strain energy introduced in
\cite{HolGasOgd:2000:anc} is given by
\eb
\psi^{\rm ti}_{\rm HGO} = 
\begin{cases}
 \dfrac{\mu_1}{2 k_1}\left\{\exp \left[k_1 \left(\underbrace{\langle\bC,\bM\rangle}_{I_4^{{\rm C}^1}} -1\right)^2\right]-1\right\} &\quad\mbox{if}\quad \langle\bC,\bM\rangle = \lVert\bF\bA\rVert^2\geq 1\\
0 &\quad\mbox{if}\quad \langle\bC,\bM\rangle = \lVert\bF\bA\rVert^2< 1\; ,
\end{cases}
\label{eq: calssical HGO}
\ee
with the material parameters $\mu_1$ and $k_1$. Recall from Eq.~(\ref{eq: I4Hi}), that $I_4^{\rm C^1} = \langle \bC,\bM \rangle$.
The exponential function incorporated in 
the latter equation captures the material stiffening in the high strain domain, caused
by the fiber elongation. The case distinction for the quadratic fiber elongation $\langle\bC,\bM\rangle=\lVert\bF\bA\rVert^2$
in Eq.~(\ref{eq: calssical HGO}) aims to prevent the fibers from inducing stiffness under compression.
For further use below it is possible to rewrite the switching criterion for which the anisotropic fiber contribution is neglected,
as 
\eb
\langle \bC -\bone,\bM\rangle = \langle\bC,\bM\rangle-1<0\,.
\ee
When using anisotropic logarithmic invariants it seems natural to use criteria to switch off the compression regime which are 
themselves defined in terms of logarithmic invariants. 
However, as will be shown in the following, the case distinction for different anisotropic invariants 
(introduced  in  Eq.~(\ref{eq: I4Hi})) will lead to considerable differences. \\

Let us consider the right Cauchy-Green tensor
\begin{align}
\nonumber  \bC & = \sum_{k=1}^3 \hat{\lambda}_k \bN_k\otimes\bN_k\,, \qquad\mbox{with} \\
\nonumber   \hat{\lambda}_1 &= 0.9\,, \qquad \hat{\lambda}_2 = 1.65\,, \qquad \hat{\lambda}_3 =  \frac{1}{\hat{\lambda}_1\hat{\lambda}_2} \qquad\mbox{and} \\
\nonumber  \bN_1 &= \begin{pmatrix}
    1\\
    0\\
    0
          \end{pmatrix}\,, \qquad
 \bN_2 = \begin{pmatrix}
    0\\
    1\\
    0
          \end{pmatrix}\,, \qquad  
 \bN_3 = \begin{pmatrix}
    0 \\
    0\\
    1
          \end{pmatrix}\,, \qquad          
\end{align}
represented in the spectral decomposition. The considered deformation is incompressible, i.e. $\det\bC=1$.
The set of all possible preferred directions $\bA$ may be expressed with help of the spherical coordinates $(r,\phi,\theta)$ and
\begin{align}
\nonumber x &= r \sin\theta \cos\phi\,, \\
\nonumber y &= r \sin\theta \sin\phi\,, \\
          z &= r \cos \phi\,. 
\end{align}
Here, $r$ denotes the radius, $\phi\in[-\pi,\pi]$ the polar angle and $\theta\in[0,\pi]$ the azimuthal angle. Since $\lVert \bA \rVert=1$, we choose 
$r = 1$ and consequently 
\eb
\nonumber \bA = \begin{pmatrix}
                 \sin\theta \cos\phi\\
                 \sin\theta \sin\phi\\
                 \cos \phi
                \end{pmatrix}\,.
\ee
Then the anisotropic invariants can be computed from 
\begin{align}
\nonumber \langle \bC^i,\bM\rangle &= \widehat{\lambda}_1^i \cos^2\theta_1 + \widehat{\lambda}_2^i \cos^2\theta_2 +\widehat{\lambda}_3^i \cos^2\theta_3 \qquad\mbox{and}\\
           \langle (\log\bU)^i,\bM\rangle &= \left(\frac{1}{2}\log\widehat{\lambda}_1\right)^i \cos^2\theta_1 + \left(\frac{1}{2}\log\widehat{\lambda}_2\right)^i \cos^2\theta_2 +\left(\frac{1}{2}\log\widehat{\lambda}_3\right)^i \cos^2\theta_3\,, 
\label{eq: invariant physical}
\end{align}
where  
\eb
\nonumber \cos \theta_1 = \langle \bA,\bN_1\rangle \qquad \cos \theta_2 = \langle \bA,\bN_2\rangle \qquad \cos \theta_3 = \langle \bA,\bN_3\rangle \,.
\ee
The invariants, depending on the fiber orientation $\bA(x,y,z)$, are plotted in Fig. \ref{fig: fiber_invariants_sphere}. 
While the principal distributions are similar for different exponents of $\bC$, it appears that the distributions are very different for even and odd exponents $i$
for the logarithmic invariants, depending on $\log\bU$. 
Please note that a change in the eigenvectors $\bN_k|~k=1,2,3$ would merely lead to a rotation of the plotted sphere around 
the eigenvector base. Therefore, in this scheme the eigenvalues remain as the only predefined variables. 
As our main goal is to exclude the compression state from the anisotropic material response to induce any stiffening 
we consider the sign of the invariants $\langle (\log\bU)^i,\bM\rangle$ and $\langle\bC^i,\bM \rangle-1$ as the  determining criterion. 
The corresponding distributions are plotted in Fig.~\ref{fig: fiber_invariants_signum} over the azimuthal and polar angle, which are sufficient in order to uniquely define the 
fiber orientation. In that sense the black area representing negative values labels the fiber directions for which the fiber response will be switched off.
On the other hand the red areas of positive values cover the fiber directions for which the anisotropic strain energy function is switched on.   
For the invariants depending on $\bC^i$ the area of positive values will increase with the exponent $i$ and become more elliptic.
The plots for sign($\langle\bC,\bM \rangle-1$) and sign($\langle \log\bU,\bM\rangle$) are generally similar.
However, with help of Eq.~(\ref{eq: invariant physical}) it appears that 
\begin{align}
\nonumber  \langle\bC,\bM \rangle-1 - \langle\log\bC,\bM \rangle &= \langle\bC-\bone,\bM \rangle  - \langle\log\bC,\bM \rangle \\
\nonumber  &= \sum_k^3 (\underbrace{\widehat{\lambda}_k-1-\log\widehat{\lambda}_k}_{\geq0})\underbrace{\cos^2\theta_k}_{\geq0} \geq0
\end{align}
and therefore 
\eb
\langle\bC,\bM \rangle -1 \geq \langle\log\bC,\bM \rangle\,.
\ee
Note that $\cos^2\theta_j$ is the same for each invariant, because the eigenvectors $\bN_j$ for each considered strain measure are the same.
It immediately follows that 
\eb
\left[\langle\bC,\bM \rangle -1\right] \geq \frac{1}{2}\langle\log\bC,\bM \rangle = \langle\log\bU,\bM \rangle\quad \mbox{for}\quad \langle\bC,\bM \rangle \geq1\,.
\ee

That means there exists a transition zone, where $\langle \bC,\bM\rangle-1>0$, but $\langle (\log\bU),\bM\rangle<0$. In other words, one may conclude 
that it is possible that the fiber direction may be stretched and still the criterion  $\langle (\log\bU),\bM\rangle$ will switch off the anisotropic response.
This effect may be favorably used when it is assumed that initially crimped fibers don't exhibit significant stiffness until they are straightened out.
However, whenever the fiber direction is compressed, anisotropic material response is precluded for the discussed logarithmic transversely isotropic strain measure
based on the criterion $\langle\log\bU,\bM\rangle<0$. Thus, there is no anisotropic stiffening under compression.

For $\langle (\log\bU)^2,\bM\rangle$ and $\langle (\log\bU)^4,\bM\rangle$ (and also any other even exponent) the invariants will always have 
a positive sign, since both, the structural tensor $\bM$ as well as $(\log\bU)^2$ and $(\log\bU)^4$ are positive semidefinite. 
Consequently, these invariants may not serve as a switching criterion. 

\begin{Figure}[!htb]
\unitlength 1 cm
\begin{picture}(14,22)
\put(0.0,16.5){\includegraphics[height = 5.5cm]{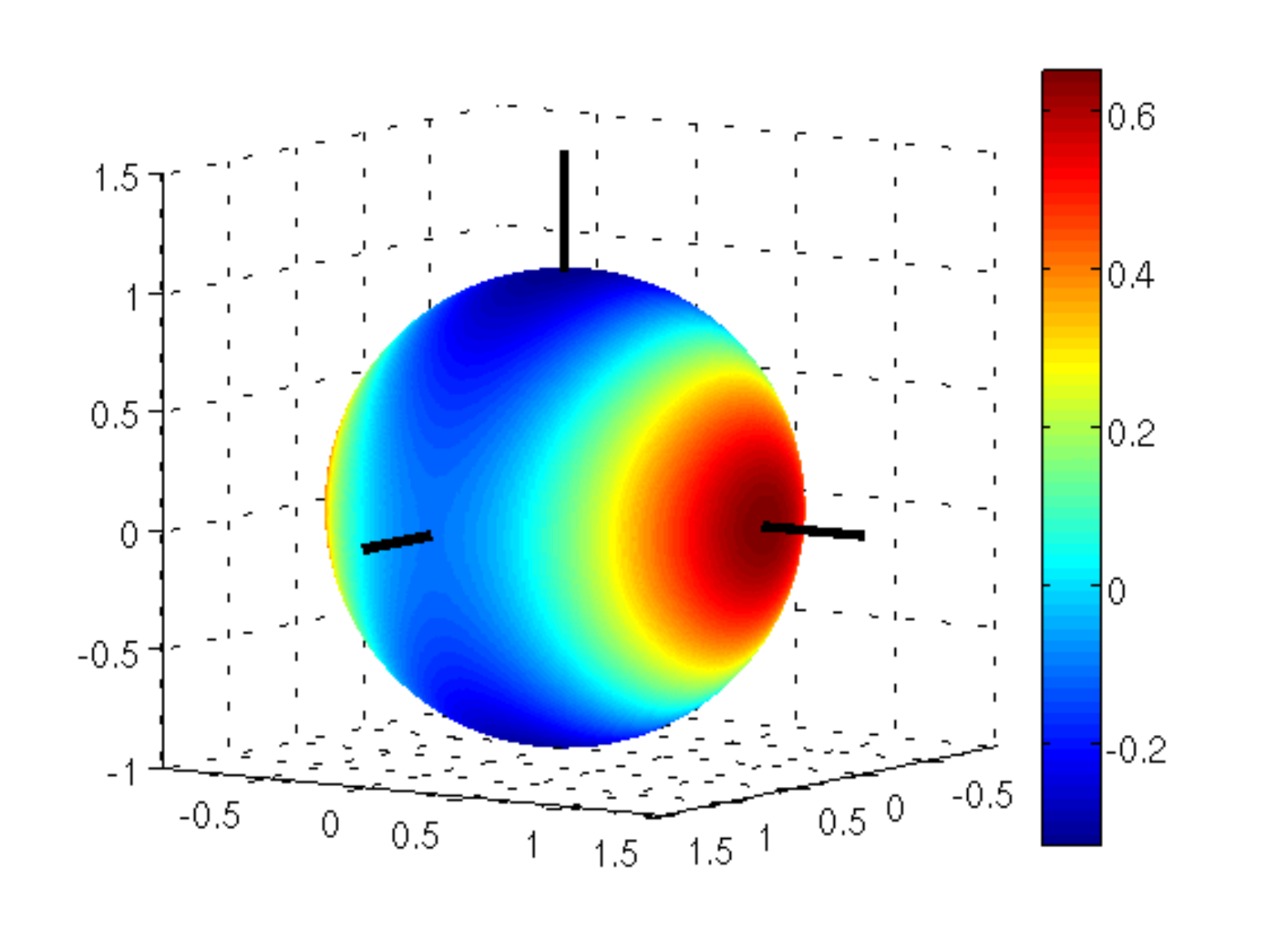}}
\put(8.0,16.5){\includegraphics[height = 5.5 cm]{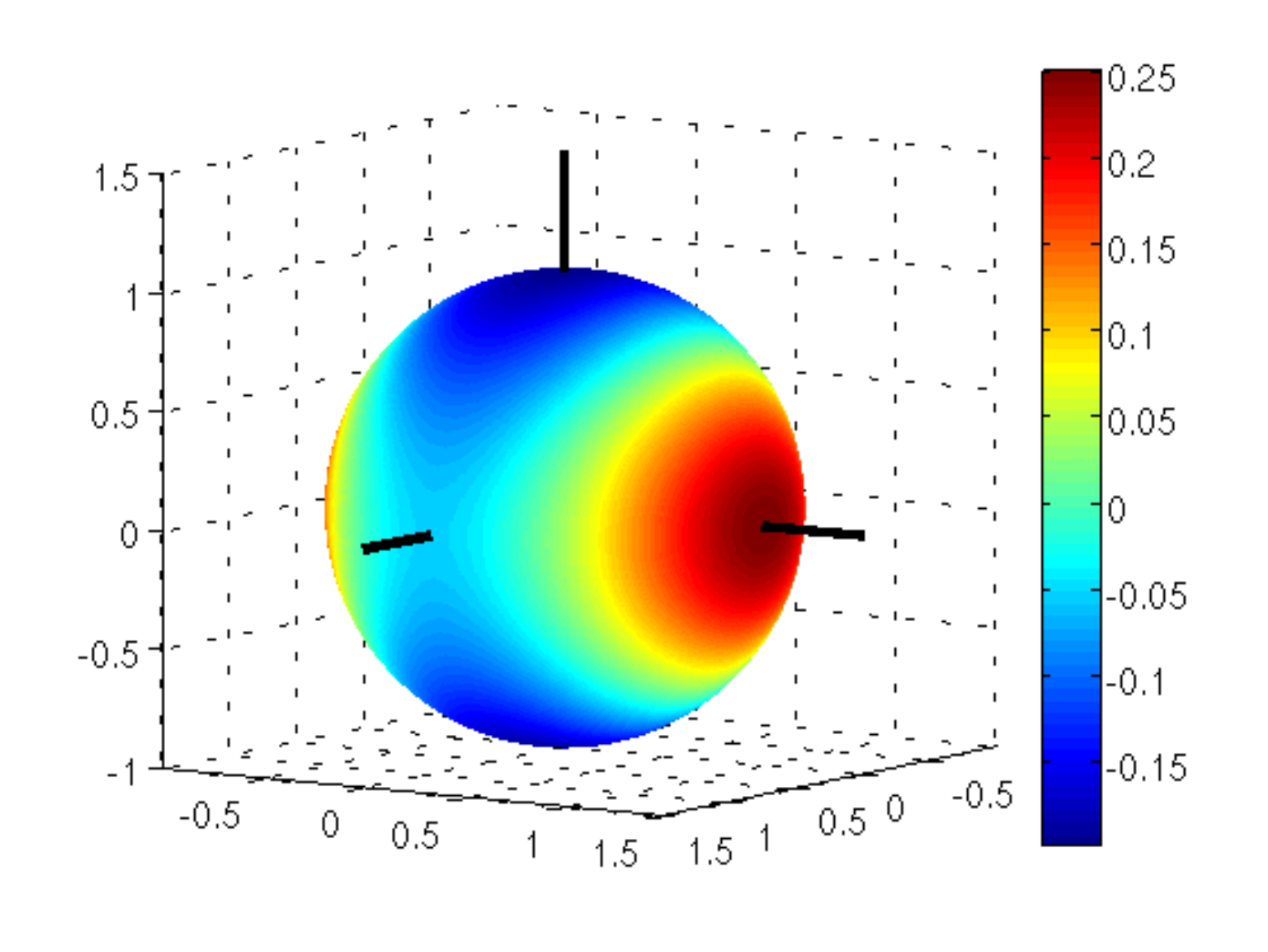}}
\put(0.0,11.0){\includegraphics[height = 5.5cm]{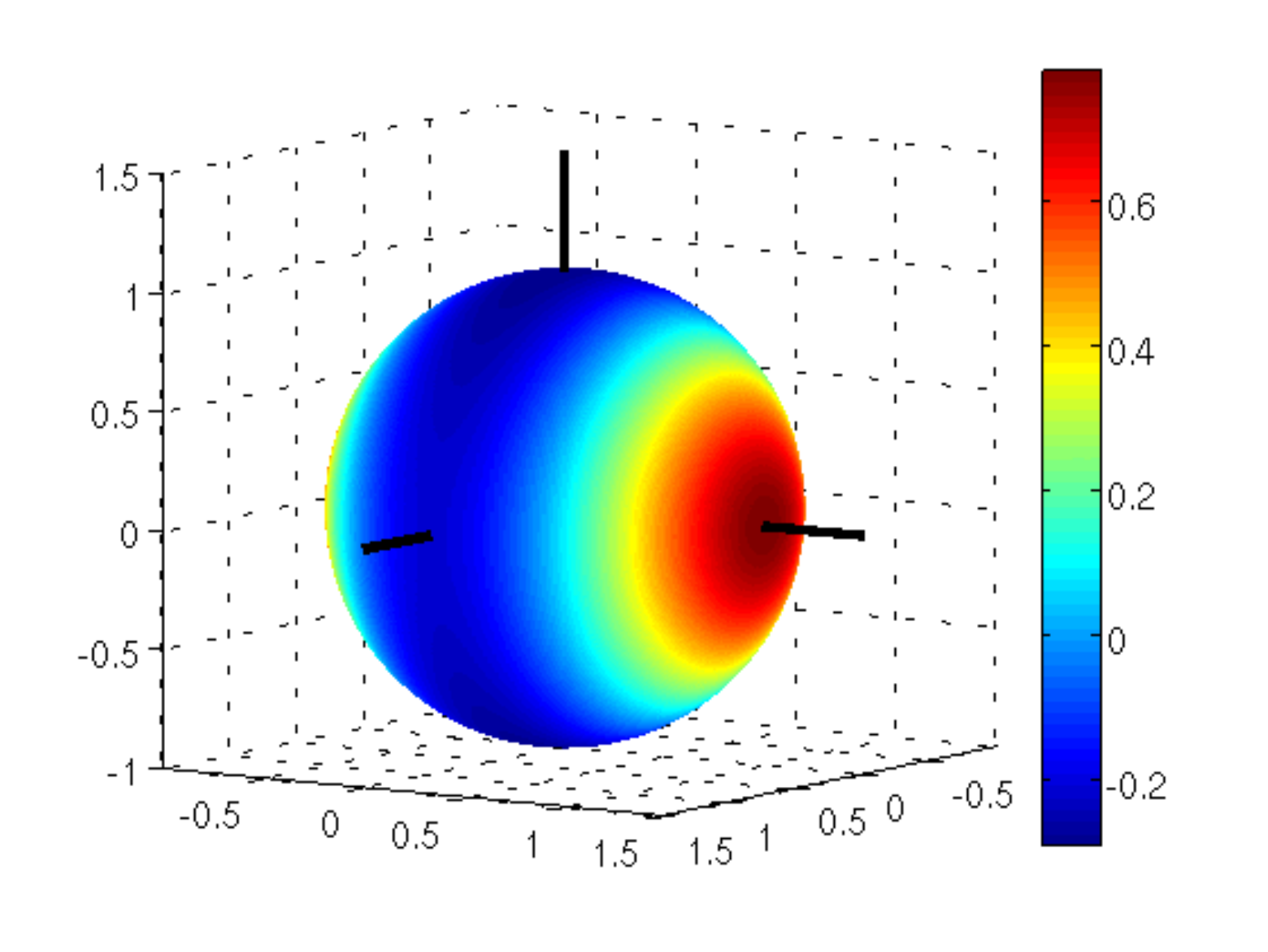}}
\put(8.0,11.0){\includegraphics[height = 5.5cm]{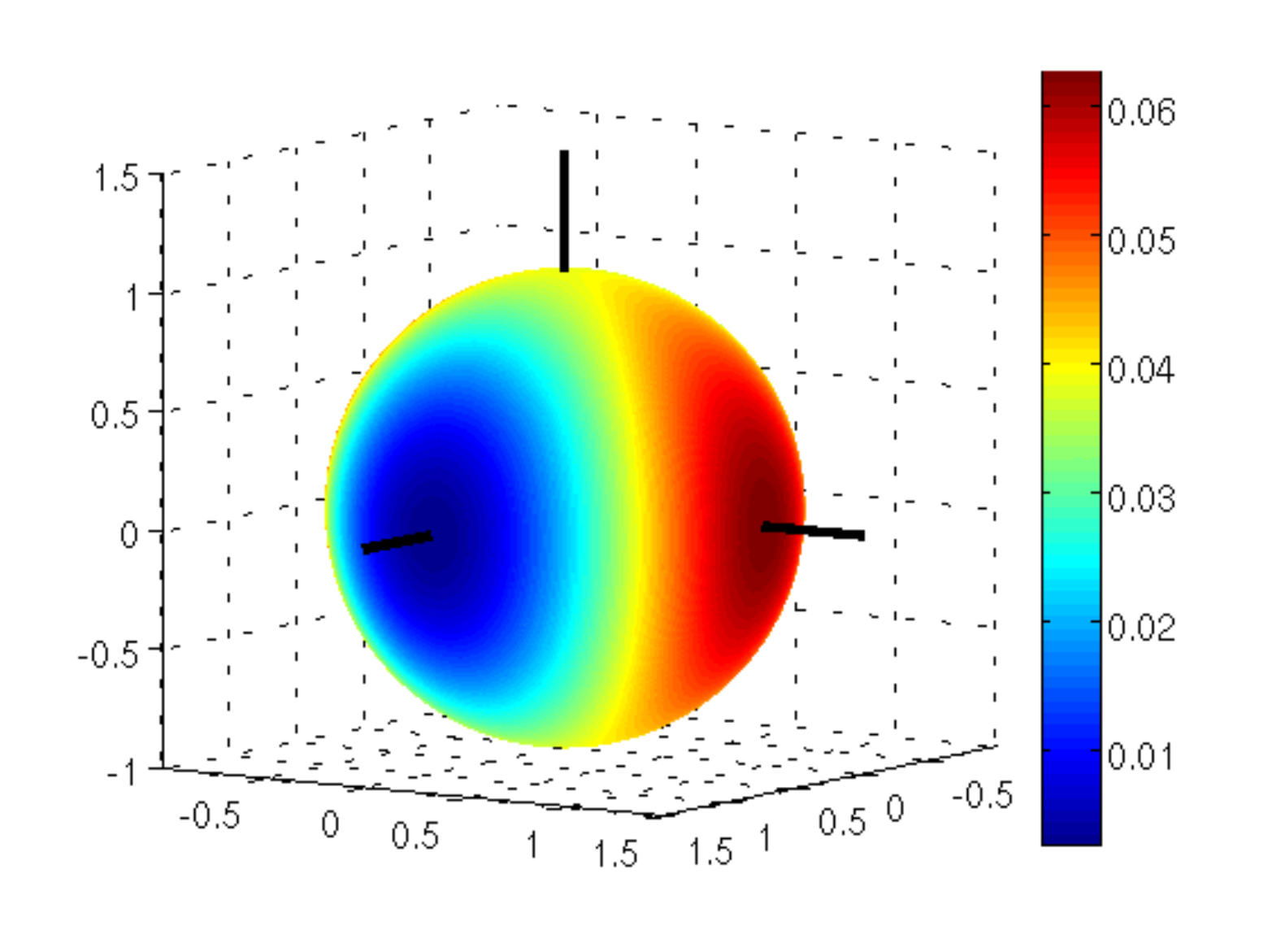}}
\put(0.0,5.5){\includegraphics[height = 5.5cm]{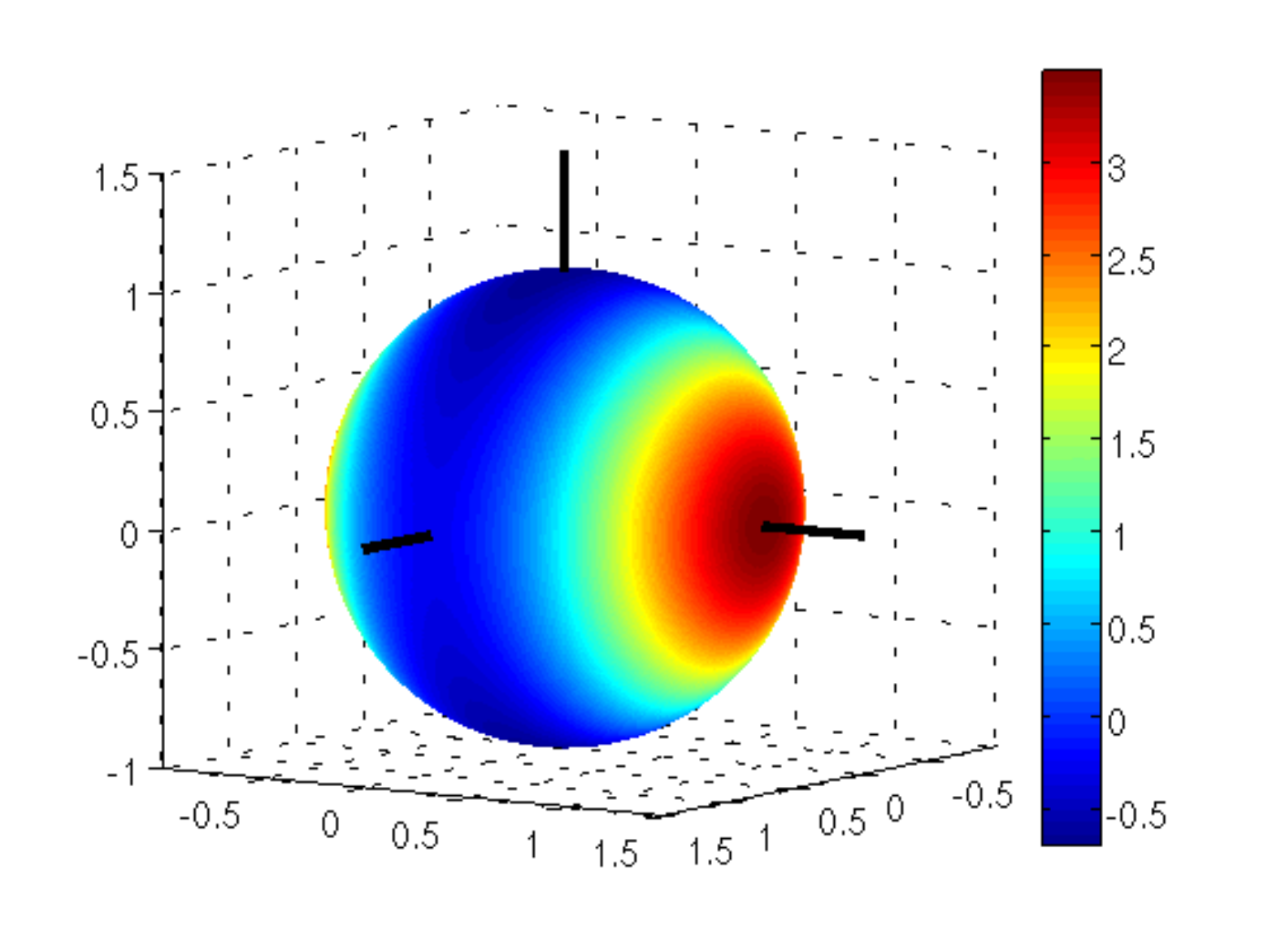}}
\put(8.0,5.5){\includegraphics[height = 5.5cm]{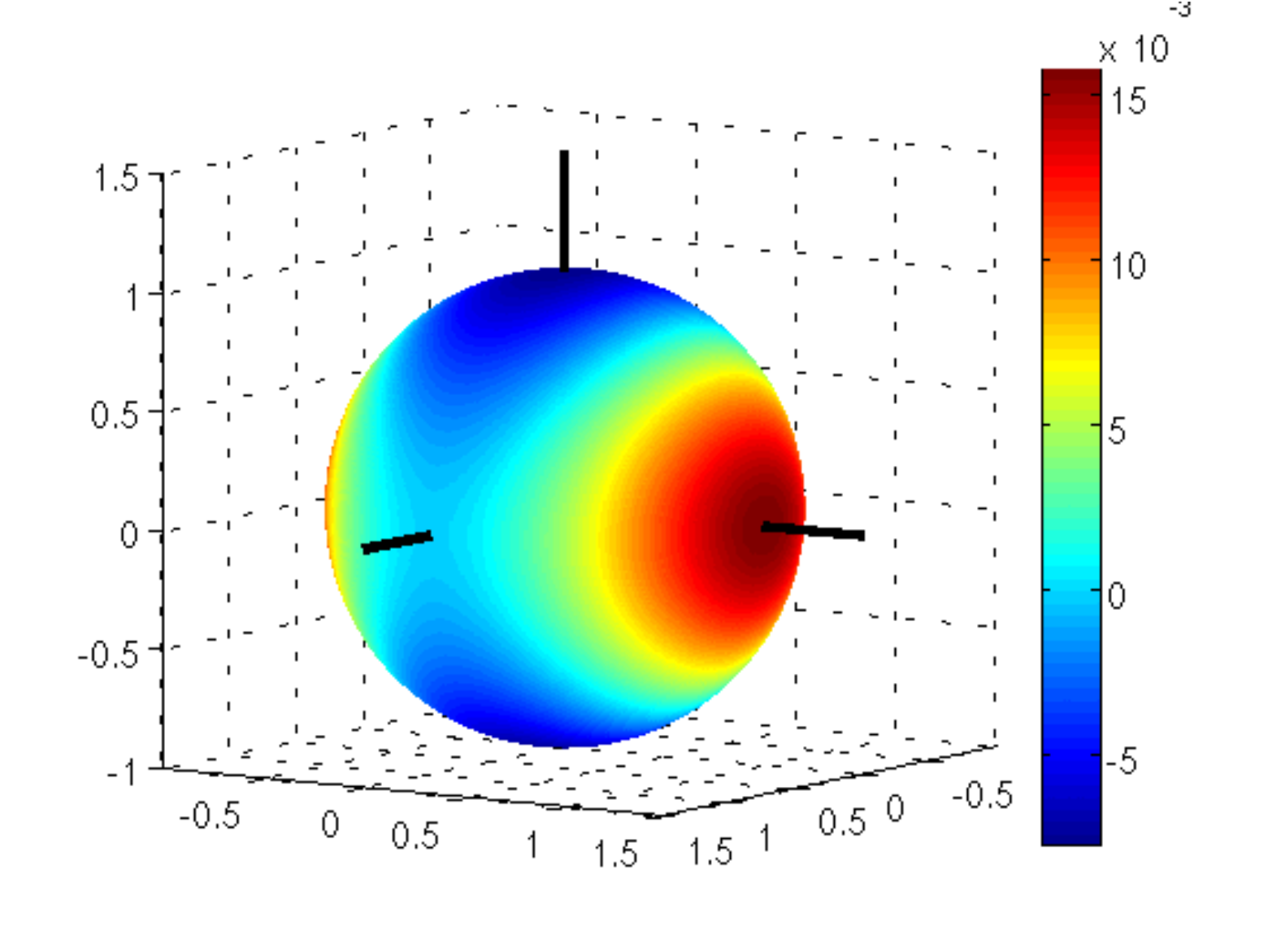}}
\put(0.0,0.0){\includegraphics[height = 5.5cm]{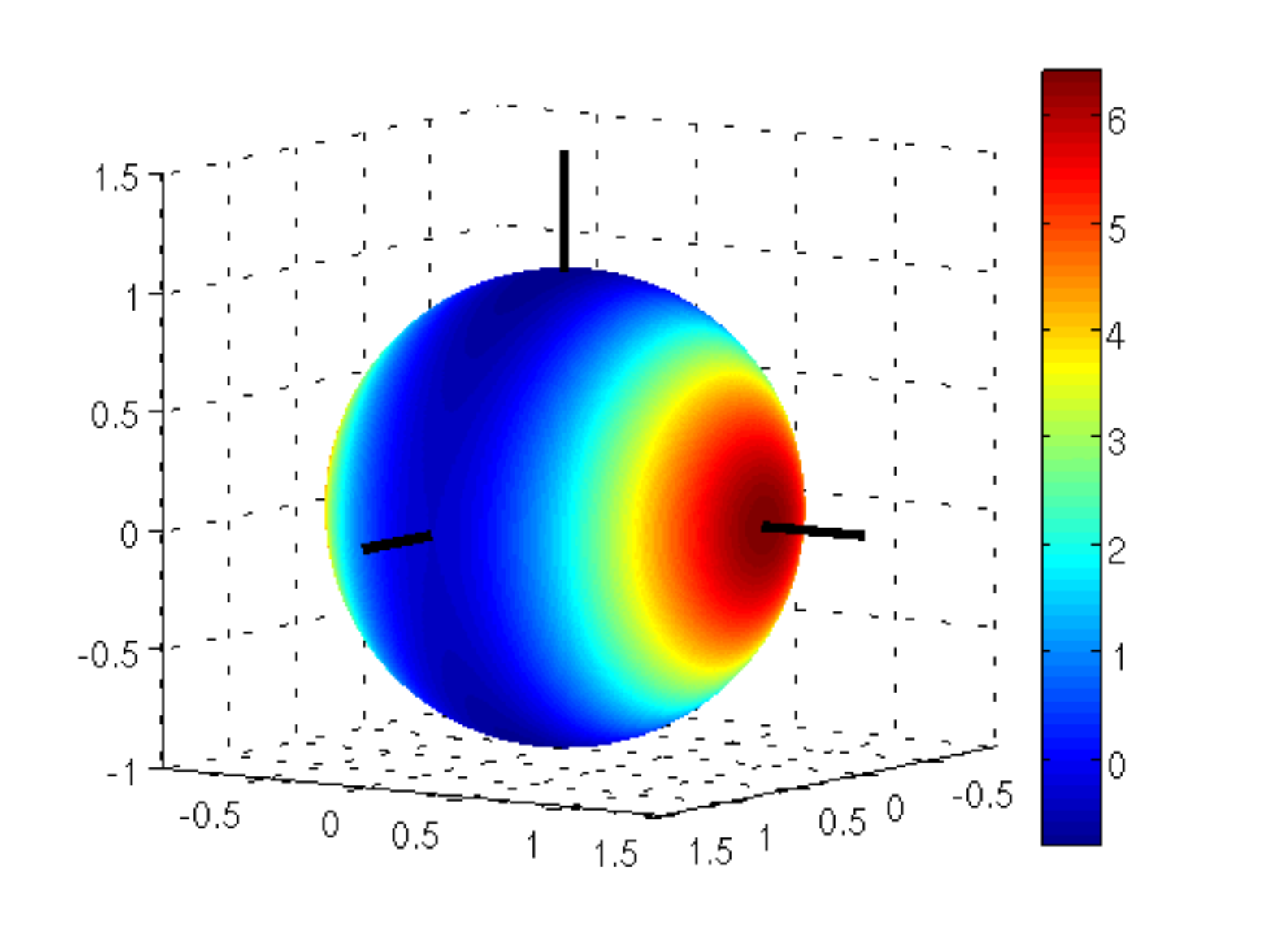}}
\put(8.0,0.0){\includegraphics[height = 5.5cm]{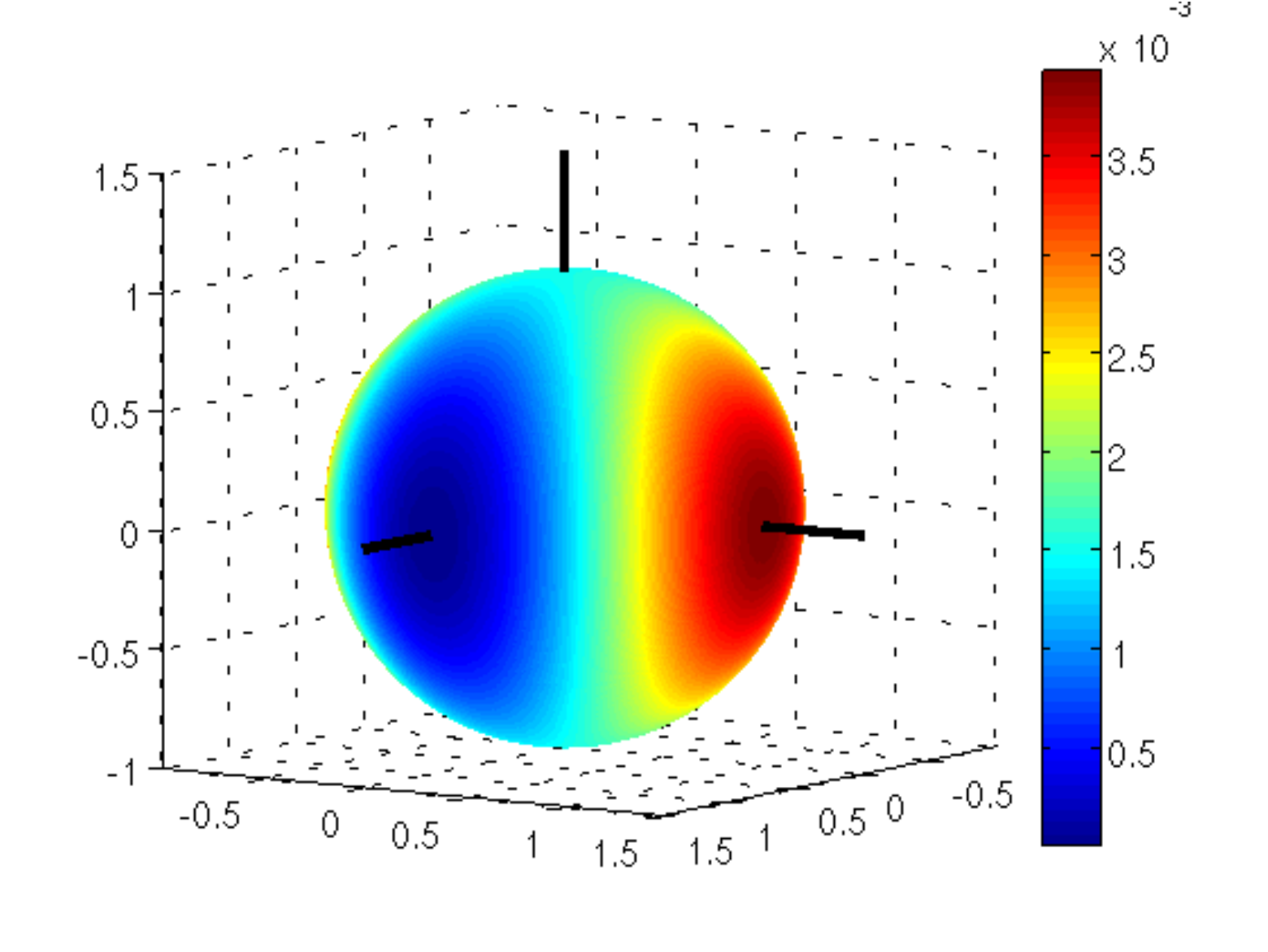}}
\put(2.64,4.5){\small $\bN_3$}
\put(1.45,2.03){\small $\bN_1$}
\put(5.1,2.25){\small $\bN_2$}
\put(0.9,4.7){\small $z$}
\put(2,0.22){\small $y$}
\put(4.8,0.3){\small $x$}
\put(2.64,10.0){\small $\bN_3$}
\put(1.45,7.53){\small $\bN_1$}
\put(5.1,7.75){\small $\bN_2$}
\put(0.9,10.2){\small $z$}
\put(2,5.72){\small $y$}
\put(4.8,5.8){\small $x$}
\put(2.64,15.5){\small $\bN_3$}
\put(1.45,13.03){\small $\bN_1$}
\put(5.1,13.25){\small $\bN_2$}
\put(0.9,15.7){\small $z$}
\put(2,11.32){\small $y$}
\put(4.8,11.3){\small $x$}
\put(2.64,21.0){\small $\bN_3$}
\put(1.45,18.53){\small $\bN_1$}
\put(5.1,18.75){\small $\bN_2$}
\put(0.9,21.2){\small $z$}
\put(2,16.82){\small $y$}
\put(4.8,16.8){\small $x$}
\put(10.64,4.5){\small $\bN_3$}
\put(9.45,2.03){\small $\bN_1$}
\put(13.1,2.25){\small $\bN_2$}
\put(8.9,4.7){\small $z$}
\put(10,0.22){\small $y$}
\put(12.8,0.3){\small $x$}
\put(10.64,10.0){\small $\bN_3$}
\put(9.45,7.53){\small $\bN_1$}
\put(13.1,7.75){\small $\bN_2$}
\put(8.9,10.2){\small $z$}
\put(10,5.72){\small $y$}
\put(12.8,5.8){\small $x$}
\put(10.64,15.5){\small $\bN_3$}
\put(9.45,13.03){\small $\bN_1$}
\put(13.1,13.25){\small $\bN_2$}
\put(8.9,15.7){\small $z$}
\put(10,11.32){\small $y$}
\put(12.8,11.3){\small $x$}
\put(10.64,21.0){\small $\bN_3$}
\put(9.55,18.6){\small $\bN_1$}
\put(13.1,18.75){\small $\bN_2$}
\put(8.9,21.2){\small $z$}
\put(10,16.82){\small $y$}
\put(12.8,16.8){\small $x$}
\begin{rotate}{90}
\put(2.0,-7.3){\small $\langle \bC^4,\bM\rangle-1$}
\put(7.5,-7.3){\small $\langle \bC^3,\bM\rangle-1$}
\put(13.0,-7.3){\small $\langle \bC^2,\bM\rangle-1$}
\put(18.5,-7.3){\small $\langle \bC,\bM\rangle-1$}
\put(1.8,-15.4){\small $\langle (\log\bU)^4,\bM\rangle$}
\put(7.3,-15.4){\small $\langle (\log\bU)^3,\bM\rangle$}
\put(12.8,-15.4){\small $\langle (\log\bU)^2,\bM\rangle$}
\put(18.3,-15.4){\small $\langle (\log\bU,\bM\rangle$}
\end{rotate}
\end{picture}
\setlength{\baselineskip}{11pt}
\caption{\label{fig: fiber_invariants_sphere}
Contour plot of different transversely isotropic invariants for a specific choice of eigenvalues and eigenvectors. 
The coordinates $(x,y,z)$ define the preferred direction $\bA$.}
\end{Figure}

\begin{Figure}[!htb]
\unitlength 1 cm
\begin{picture}(14,22)
\put(0.0,16.9){\includegraphics[height = 5.3cm]{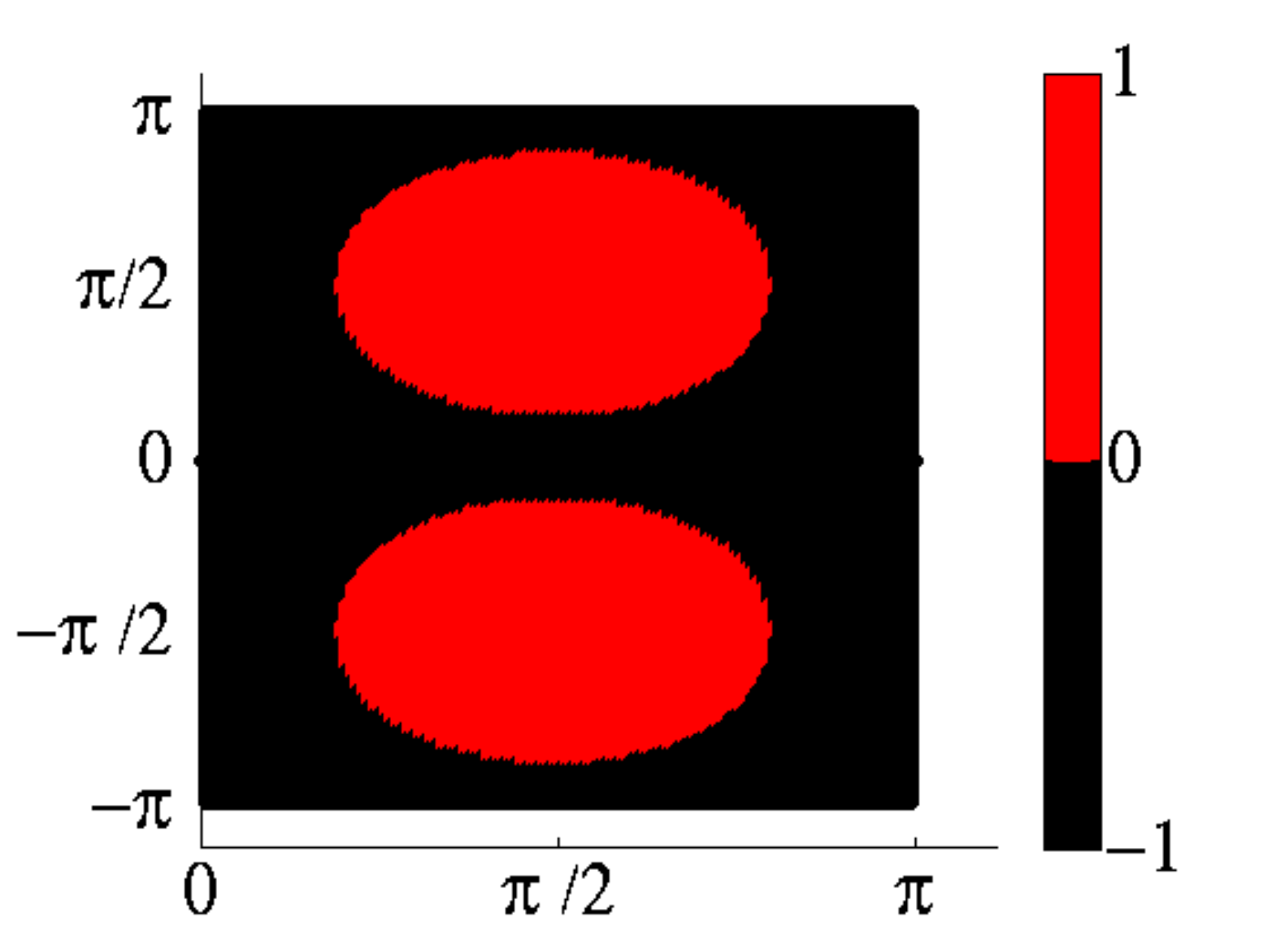}}
\put(8.0,16.9){\includegraphics[height = 5.3cm]{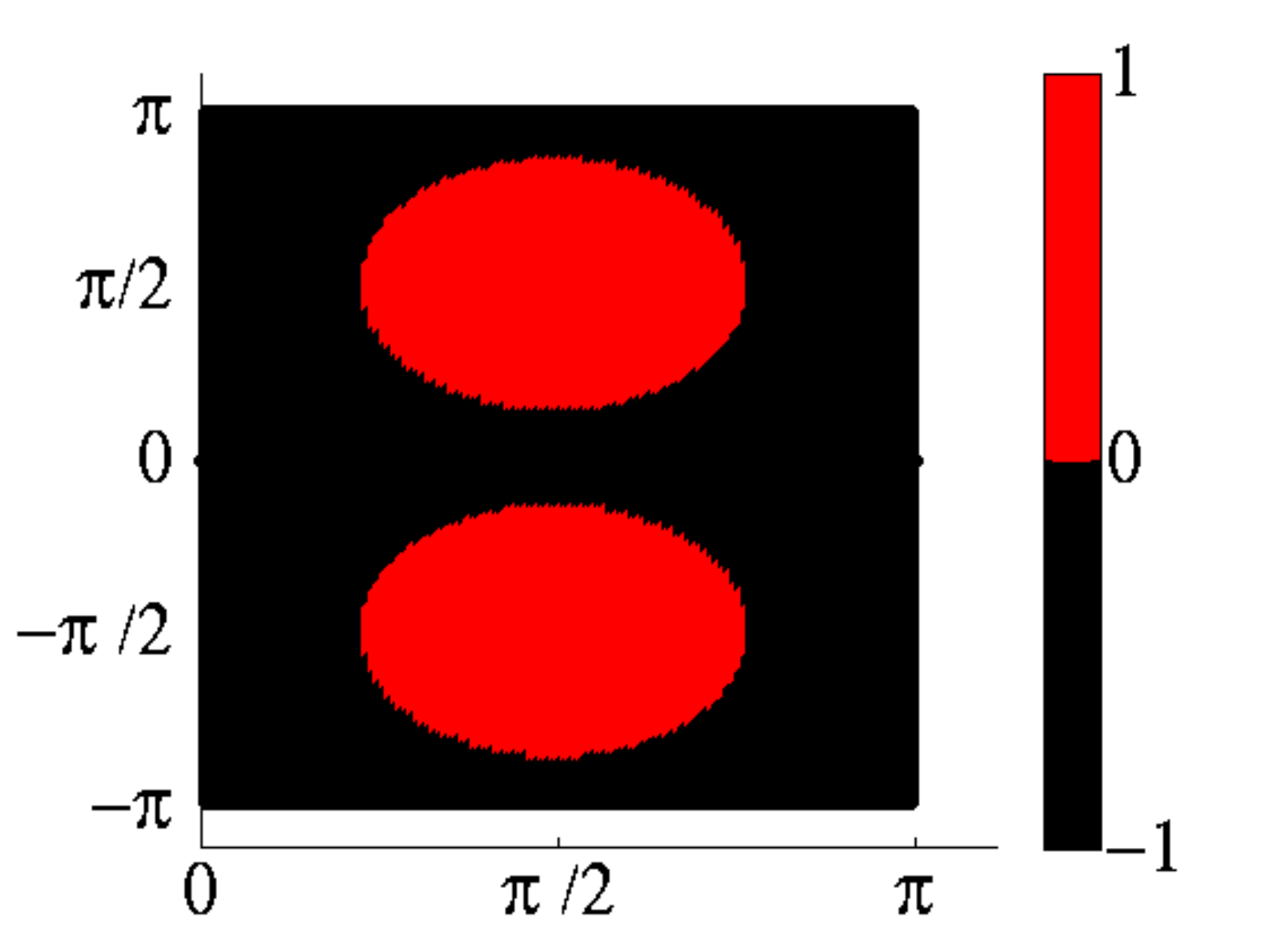}}
\put(0.0,11.3){\includegraphics[height = 5.3cm]{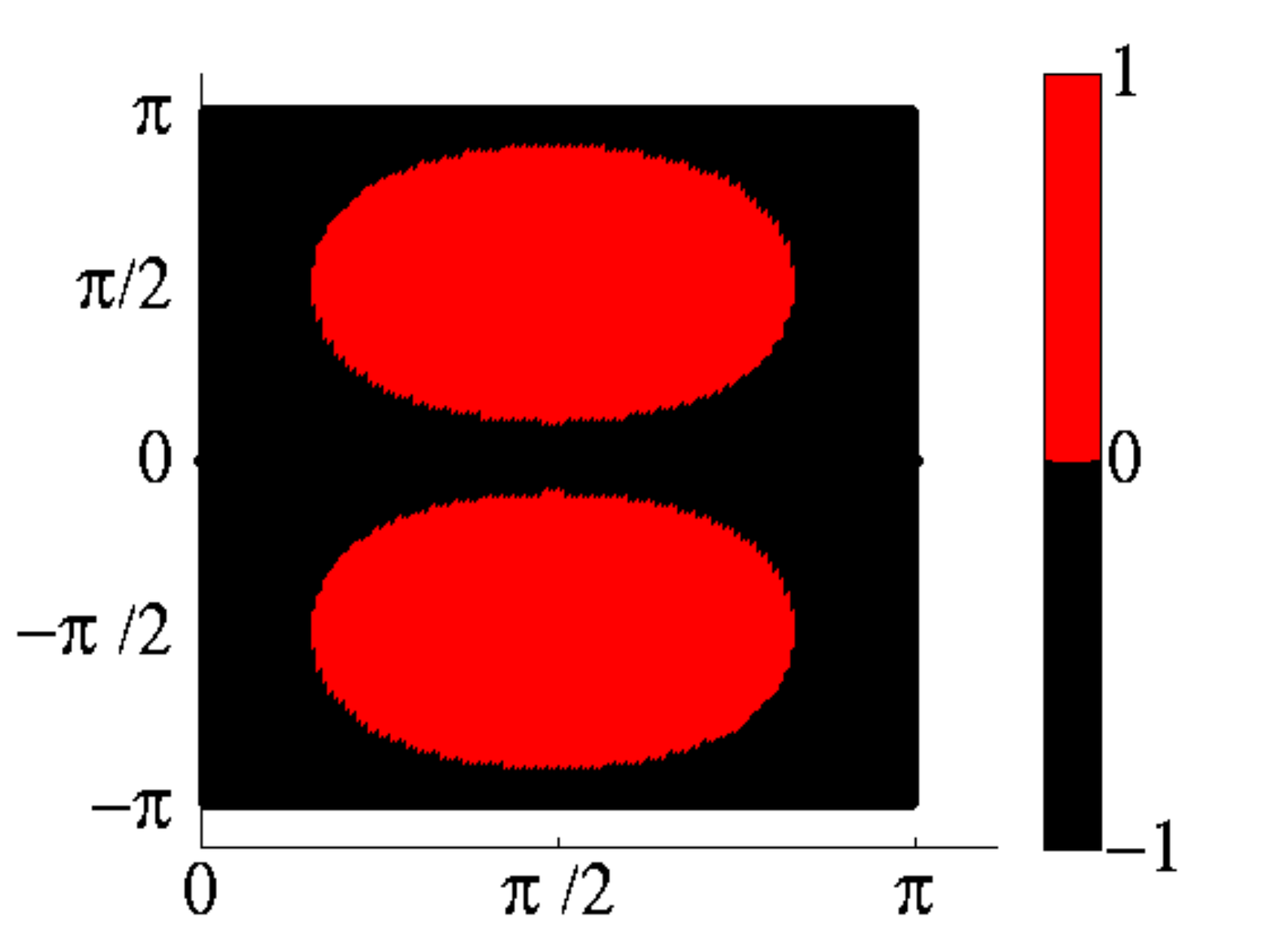}}
\put(8.0,11.3){\includegraphics[height = 5.3cm]{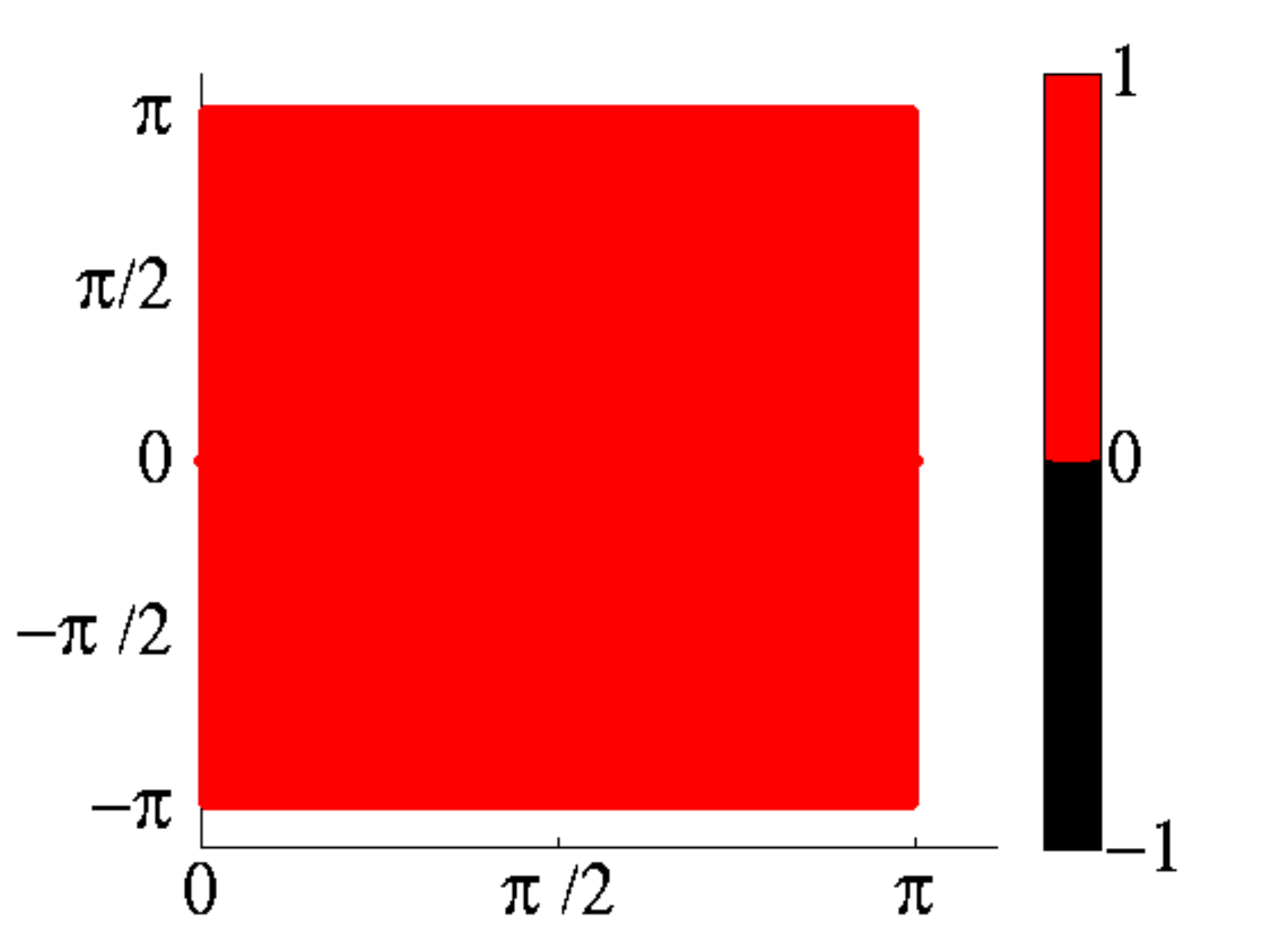}}
\put(0.0,5.8){\includegraphics[height = 5.3cm]{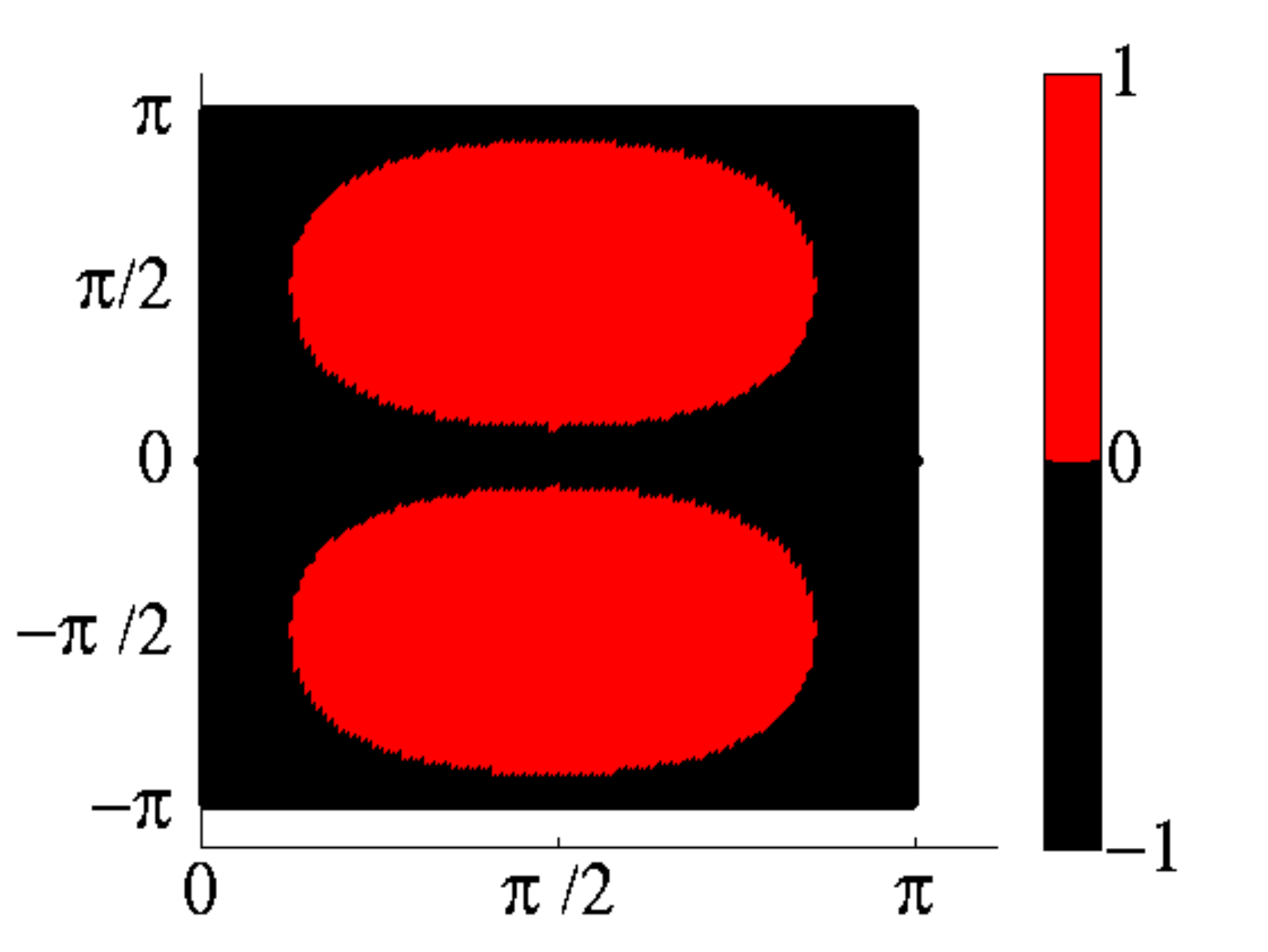}}
\put(8.0,5.8){\includegraphics[height = 5.3cm]{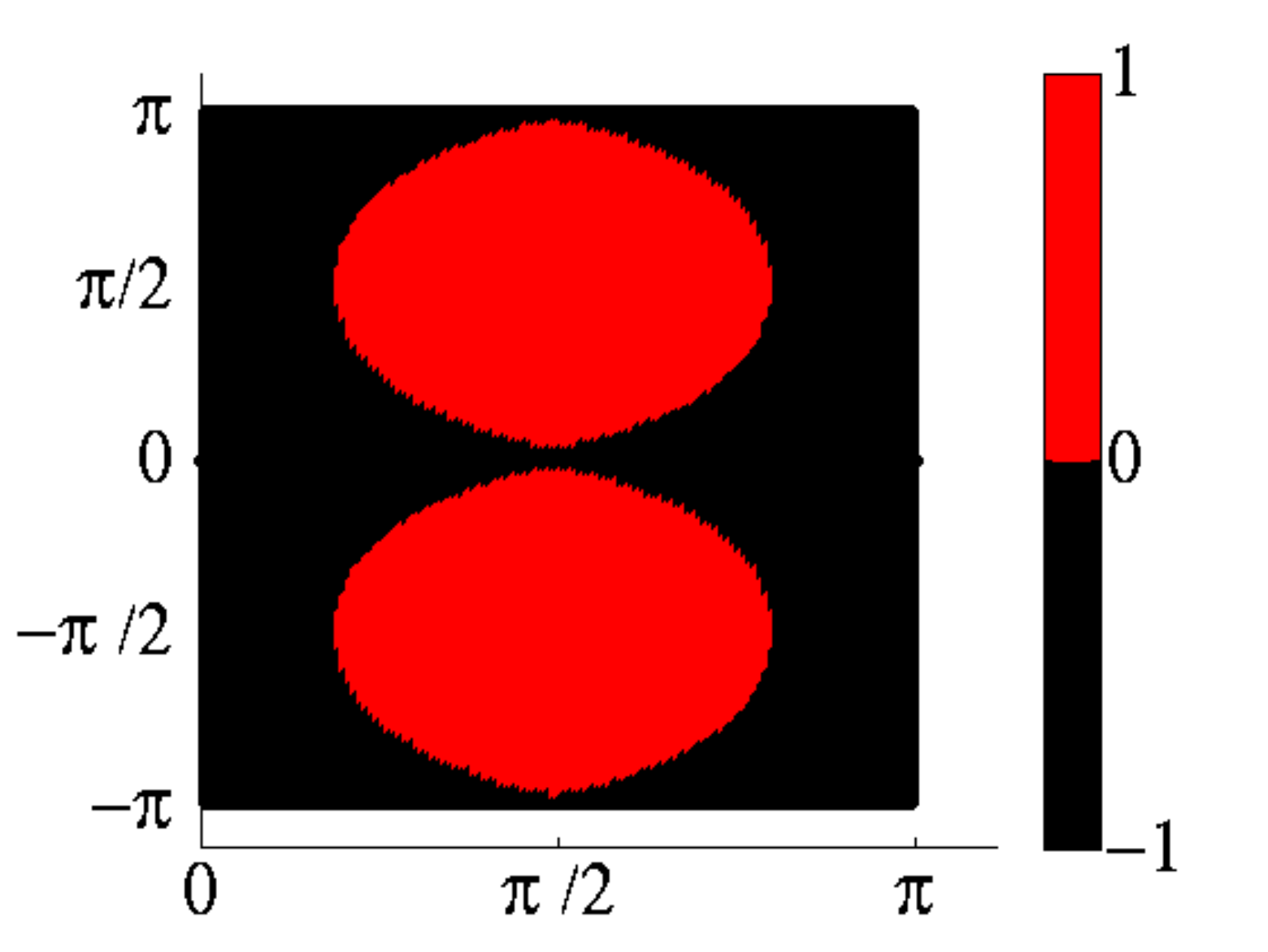}}
\put(0.0,0.2){\includegraphics[height = 5.3cm]{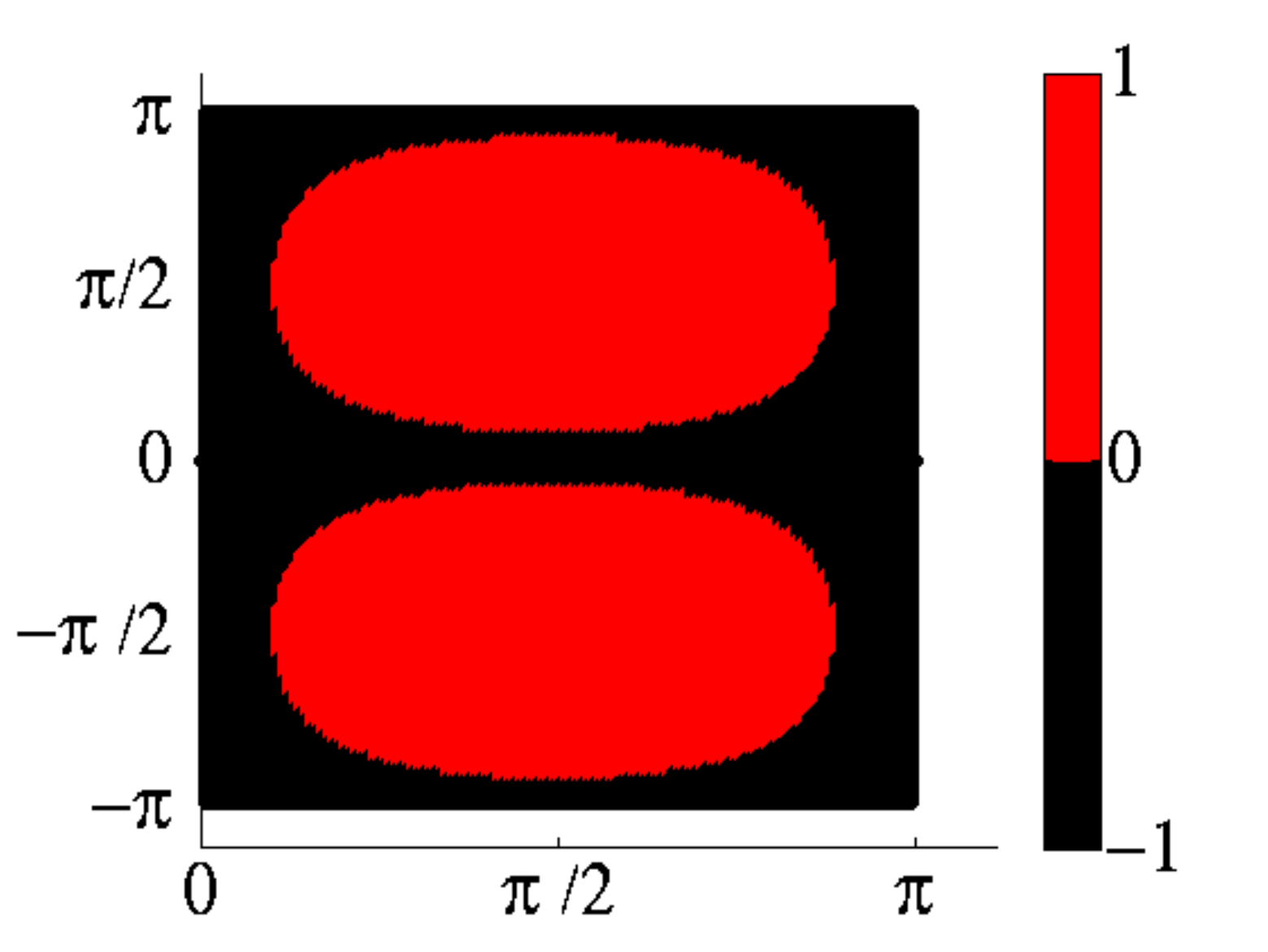}}
\put(8.0,0.2){\includegraphics[height = 5.3cm]{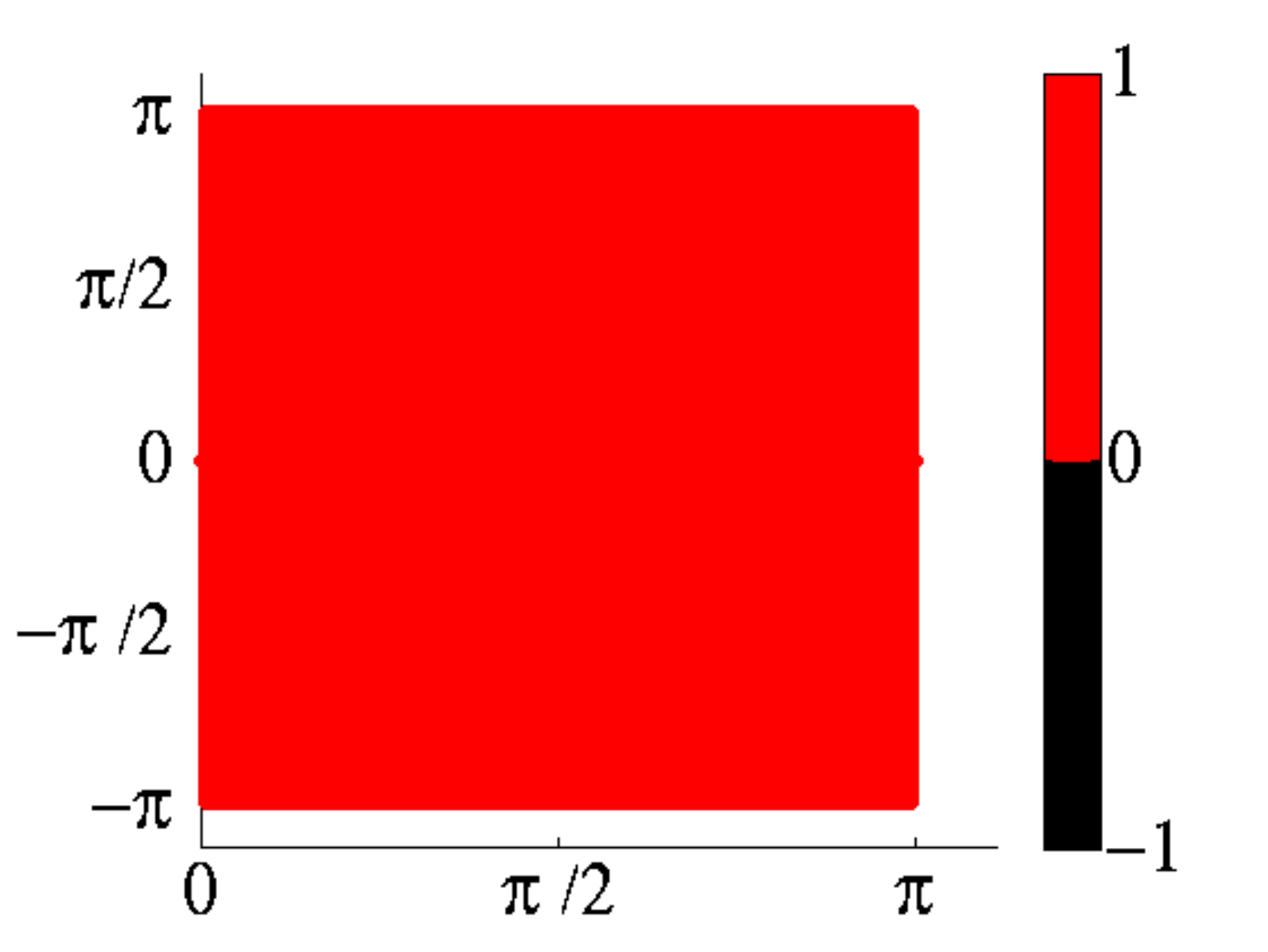}}
\put(3.02,-0.1){$\theta$}
\put(0.1,2.82){$\phi$}
\put(11.2,-0.1){$\theta$}
\put(8.1,2.82){$\phi$}
\put(3.02,5.5){$\theta$}
\put(0.1,8.42){$\phi$}
\put(11.2,5.5){$\theta$}
\put(8.1,8.42){$\phi$}
\put(3.02,11.0){$\theta$}
\put(0.1,13.95){$\phi$}
\put(11.2,11.0){$\theta$}
\put(8.1,13.95){$\phi$}
\put(3.02,16.5){$\theta$}
\put(0.1,19.55){$\phi$}
\put(11.2,16.5){$\theta$}
\put(8.1,19.55){$\phi$}
{\begin{turn}{90}
\put(1.4,-7.2){\small ${\rm sign}(\langle \bC^4,\bM\rangle-1)$}
\put(6.9,-7.2){\small ${\rm sign}(\langle \bC^3,\bM\rangle-1)$}
\put(12.4,-7.2){\small ${\rm sign}(\langle \bC^2,\bM\rangle-1)$}
\put(18.1,-7.2){\small ${\rm sign}(\langle \bC,\bM\rangle-1)$}
\put(1.4,-15.2){\small ${\rm sign}(\langle (\log\bU)^4,\bM\rangle)$}
\put(6.9,-15.2){\small ${\rm sign}(\langle (\log\bU)^3,\bM\rangle)$}
\put(12.4,-15.2){\small ${\rm sign}(\langle (\log\bU)^2,\bM\rangle)$}
\put(18.1,-15.2){\small ${\rm sign}(\langle (\log\bU),\bM\rangle)$}
\end{turn}}
\end{picture}
\setlength{\baselineskip}{11pt}
\caption{\label{fig: fiber_invariants_signum}
Plot of the sign of the invariants plotted in Fig.~\ref{fig: fiber_invariants_sphere} over the spherical coordinates.
The boundary between positive and negative values marks the transition zone of a possible case distinction.}
\end{Figure}

In the light of the previous considerations,  we may formulate the generalized strain energy functions 
\begin{align}
\psi^{\rm ti}_{\rm \mbox{\small \ding{185}}_C} &= \begin{cases} 
\dfrac{\mu_1}{2 k_1}\left\{\exp \left[k_1 (\underbrace{\langle \bC^i,\bM\rangle}_{I_4^{{\rm C}^i}}-1)^2\right]-1\right\} &\quad\mbox{if}\quad\langle \bC^i,\bM\rangle\geq 1\\
0 &\quad\mbox{if}\quad \langle \bC^i,\bM\rangle< 1\; , \label{eq: ti Ci}
\end{cases}\\
\psi^{\rm ti}_{\rm \mbox{\small \ding{186}}_H} &= \begin{cases} 
\dfrac{\mu_1}{2 k_1}\left\{\exp \left[k_1 {{\underbrace{\langle \log\bU,\bM\rangle}_{I_4^{{\rm H}^1}}}^\varepsilon\underbrace{\langle (\log\bU)^i,\bM\rangle}_{I_4^{{\rm H}^i}}}^2\right]-1\right\} &\quad\mbox{if}\quad\langle \log\bU,\bM\rangle\geq 0\\
0 &\quad\mbox{if}\quad \langle \log\bU,\bM\rangle< 0\;, \label{eq: ti Hi}
\end{cases}
\end{align}
in terms of the anisotropic invariants $I_4^{{\rm C}^i}| i =1,2,3,4$ and 
$I_4^{{\rm H}^i}| i =1,2,3,4$, defined in Eq.~(\ref{eq: I4Hi}). 
In Eq.~(\ref{eq: ti Hi}) the value of $\varepsilon$ is to be chosen positive and close to zero. 
The incorporation of ${\langle \log\bU,\bM\rangle}^\varepsilon$ in the free energy function prevents jumps in the stresses at the switch over point. 
Alternatively one may also incorporate the criterion $({\langle \bC,\bM\rangle}-1)^\varepsilon$ and formulate the case distinction based on
the quadratic fiber stretch $\langle\bC,\bM\rangle$.
As it is shown below a continuous material tangent is also ensured. 

Case distinctions, like in Eq.~(\ref{eq: ti Ci}) and Eq.~(\ref{eq: ti Hi})  may generally lead  to discontinuous functions.
In order to avoid this it is evident that the stresses and the tangent must become zero at each switch-point of the chosen criterion.
Indeed one can show that 
\begin{align}
\nonumber \pp{\psi^{\rm ti}_{\rm \mbox{\small \ding{185}}_C}}{\bC}    &= \bzero \qquad \mbox{if}  \qquad \langle \bC^i,\bM\rangle = 1\,,\\
          \pp{\psi^{\rm ti}_{\rm \mbox{\small \ding{186}}_H}}{\bC}    &= \bzero \qquad \mbox{if}  \qquad \langle \log\bU,\bM\rangle = 0\,,
\end{align}
such that no jumps in the stresses at the switchover points are possible.
Only $\psi^{\rm ti}_{\rm \mbox{\small \ding{185}}_C}$ of the introduced strain energy calsses misses continuity, since
\begin{align}
\nonumber \frac{\partial^2\psi^{\rm ti}_{\rm \mbox{\small \ding{185}}_C}}{\partial \bC\partial \bC}    &\neq \bzero \qquad \mbox{if}  \qquad \langle \bC^i,\bM\rangle = 1\,,\\
          \frac{\partial^2\psi^{\rm ti}_{\rm \mbox{\small \ding{186}}_H}}{\partial \bC\partial \bC}    &= \bzero \qquad \mbox{if}  \qquad \langle \log\bU,\bM\rangle = 0\,,
\end{align}
The material tangent  of the function $\psi^{\rm ti}_{\rm \mbox{\small \ding{185}}_C}$  at the point $\bC = \bone$ in the reference configuration with a structural tensor $\bM = {\rm diag}(0,0,1)$,  
\eb
\nonumber \IC|_{\bC=\bone}(\psi^{\rm ti}_{\rm \mbox{\small \ding{185}}_C}(I_4^{{\rm C}^i}| i =1,2,3,4)) = \begin{pmatrix}
         0&         0&             0&         0&         0&         0\\
         0&         0&             0&         0&         0&         0\\
         0&         0&           4 i^2 \mu_1&   0&         0&         0\\
         0&         0&         0&         0&         0&         0\\
         0&         0&         0&         0&         0&         0\\
         0&         0&         0&         0&         0&         0\\       
      \end{pmatrix}\,,      
\ee
is already different from $\bzero$ and therefore violates the continuity requirement.
For the class ${\psi^{\rm ti}_{\rm \mbox{\small \ding{186}}_H}}$
we find that $\IC|_{\bC=\bone}=\bzero$.

\section{\hspace{-5mm}. Parameter adjustment \label{sec: param}}

\subsection{\hspace{-5mm}. Parameter identification for soft biological tissues}

In the following the proposed transversely isotropic Hencky models are
adjusted to the  test data provided in \cite{Hol:2006:dom}. 
There, a human, abdominal aorta from a human cadaver was tested. 
The donor was female, 80 years old and suffering from congestive cardiomyopathy.
Arterial stripes were excised for two material layers, as indicated in Fig.~\ref{fig: parameter_anpassung}a). 
Therefore, we introduce the orthonormal coordinate system depending on the circumferential direction $\bN_\varphi$, the axial direction $\bN_z$ and the radial direction $\bN_r$.
For each of the layers the tissue was stretched in either circumferential (see Fig.~\ref{fig: parameter_anpassung}b)) or axial (see Fig.~\ref{fig: parameter_anpassung}c)) 
direction.

In an incompressible uniaxial tension test with two fiber families orientated in the $\bN_\varphi-\bN_z$ plane, we may 
write the tensors
\begin{align}
\nonumber
\bF =\bU &= 
\begin{pmatrix}
 \sqrt{\widehat{\lambda}_1} &0& 0 \\
 0  &\sqrt{\widehat{\lambda}_2}& 0 \\
 0  &0& \sqrt{\frac{1}{\widehat{\lambda}_1 \widehat{\lambda}_2}}
\end{pmatrix}\,,
\quad
\bC =
\begin{pmatrix}
\widehat{\lambda}_1 &0&0\\ 
 0  &\widehat{\lambda}_2& 0 \\
 0  &0& \frac{1}{\widehat{\lambda}_1 \widehat{\lambda}_2}
\end{pmatrix} 
\quad \mbox{and} \\
\nonumber
\log\bU &=
\begin{pmatrix}
\frac{1}{2}{\rm log} (\widehat{\lambda}_1)& 0&0 \\
0&\frac{1}{2}{\rm log} (\widehat{\lambda}_2)&0 \\
0&0&\frac{1}{2}{\rm log} (\frac{1}{\widehat{\lambda}_1\widehat{\lambda}_2})
\end{pmatrix}\,.
\end{align}
Because of the incompressibility we have ${\rm tr}(\log\bU)= \log(\det \bU)=0$ and the second part of ${W}_{\rm H}$ and ${W}_{\rm eH}$ becomes automatically 
zero. 
Moreover, the structural tensors are
\eb
\bM_{(1)} = 
\begin{pmatrix}
c^2 & -cs& 0 \\
-cs & s^2& 0 \\
0   & 0  & 0
\end{pmatrix}
\quad \mbox{and} \quad
\bM_{(2)} = 
\begin{pmatrix}
c^2  &cs &0 \\
cs & s^2 &0 \\
0  &  0   &0
\end{pmatrix}
\ee
with $c={\rm cos}\beta_{\rm f}$ and $s = {\rm sin}\beta_{\rm f}$
and therefore we have
\begin{align}
\nonumber I_4^{\rm H^i} &= \left(\frac{1}{2}{\rm log} (\widehat{\lambda}_1)\right)^{\rm i} (\cos \beta_{\rm f})^2 + \left(\frac{1}{2}{\rm log} (\widehat{\lambda}_2)\right)^{\rm i}(\sin \beta_{\rm f})^2\qquad \mbox{and}\\
\nonumber I_4^{\rm C^i} &= \widehat{\lambda}_1^{\rm i} (\cos \beta_{\rm f})^2 + \widehat{\lambda}_2^{\rm i}(\sin \beta_{\rm f})^2\,.
\end{align}
The angle $\beta_{\rm f}$ denotes the angle between each fiber and the local circumferential direction, while the angle between both fibers follows to $2\beta_{\rm f}$.
The second Piola-Kirchoff stresses in this case 
may be written as
\eb
\nonumber
S_{11} = 2\frac{\partial \psi}{\partial\widehat{\lambda}_1}+2\frac{p}{\widehat{\lambda}_1}\,, \quad
S_{22} = 2\frac{\partial \psi}{\partial\widehat{\lambda}_2}+2\frac{p}{\widehat{\lambda}_2}\,, \quad
S_{33} = 2\frac{\partial \psi}{\partial\frac{1}{\widehat{\lambda}_1\widehat{\lambda}_2}}+2\frac{p}{\widehat{\lambda}_1\widehat{\lambda}_2}\,.\\
\ee
The Lagrange multiplier $p$ is introduced in order to enforce the incompressibility and can directly be calculated with help of the requirement that $S_{22}$ must be equal to zero. 
The stretch $\widehat{\lambda}_{11}$ is known from the experiments and the remaining unknown $\widehat{\lambda}_{22}$ is
iterated with help of Newton's method, making use of the requirement that also $S_{33}$ must be equal to zero:
\eb
S_{33}(\widehat{\lambda}_2)  \overset{!}{=} 0 \quad \Rightarrow \widehat{\lambda}_2^{n+1} = \widehat{\lambda}_2^n - \frac{S_{33}(\widehat{\lambda}_2^n)}{{\rm Lin}S_{33}(\widehat{\lambda}_2^n)}\,, \quad\mbox{with}\quad {\rm Lin}S_{33}(\widehat{\lambda}_2^n) = \frac{\partial S_{33}(\widehat{\lambda}_2^n) }{\partial \widehat{\lambda}_2^n}\,.
\ee
The parameter fitting was performed with  help of a Sequential Quadratic Programming (SQP) algorithm for 
nonlinear numerical constrained optimization problems. The gradient needed for the optimization
procedure is calculated based on a finite difference scheme in conjunction with the above described Newton iteration.
The objective function 
\eb
f^{\rm obj}(\boldsymbol\alpha) := \sum_{{\rm e}=1}^{n_{\rm exp}}\sqrt{\frac{1}{n_{\rm mp}}\sum_{i=1}^{n_{\rm mp}}\left(\frac{\sigma^{\rm exp}_e(\widehat{\lambda}_k^i)-\sigma^{\rm sim} (\widehat{\lambda}_k^i,\boldsymbol\alpha)}{{\rm max}(\sigma^{\rm exp}_e)}\right)^2} \label{eq: fobj}
\ee
is utilized as the optimization criterion. Here $n_{\rm exp}$ and ${n_{\rm mp}}$ denote the number of experiments to be
fitted and the number of specific measuring points to be evaluated. The predefined amount of stretch associated
to each measuring point $i$ is labeled with $\widehat{\lambda}_k^i$. The simulated Cauchy stresses $\sigma^{\rm sim}$ and the error $f^{\rm obj}$ 
are dependent on the chosen
material parameter set contained in the field $\boldsymbol \alpha$.

\begin{Figure}[!htb]
\unitlength 1 cm
\begin{picture}(14,7.6)
\put(0,0.3){\includegraphics[width = 15.7cm]{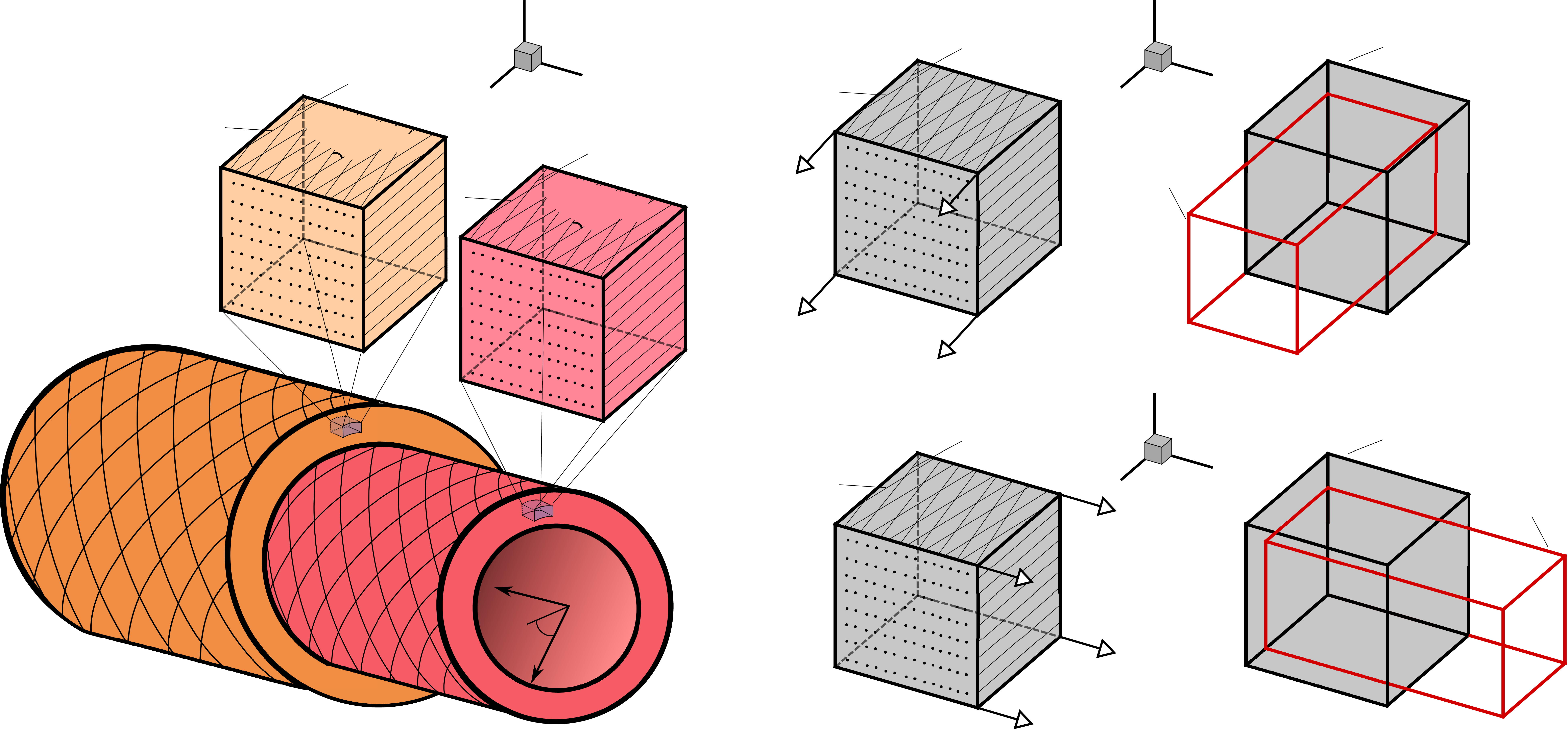}}
\put(0,0){a)}
\put(7.7,0){c)}
\put(7.7,3.4){b)}
\put(13.9,7.05){ref}
\put(13.9,3.1){ref}
\put(15.1,2.5){act.}
\put(11.4,5.8){act.}
\put(10.62,6.8){$\bN_{\varphi}$}
\put(10.62,2.85){$\bN_{\varphi}$}
\put(12.2,6.8){$\bN_{\rm z}$}
\put(12.2,2.85){$\bN_{\rm z}$}
\put(11.6,7.5){$\bN_{\rm r}$}
\put(11.6,3.53){$\bN_{\rm r}$}
\put(4.32,6.8){$\bN_{\varphi}$}
\put(5.9,6.8){$\bN_{\rm z}$}
\put(5.3,7.5){$\bN_{\rm r}$}
\put(5.3,7.5){$\bN_{\rm r}$}
\put(8.0,6.65){$\bA^{(1)}$}
\put(9.5,7.15){$\bA^{(2)}$}
\put(8.0,2.7){$\bA^{(1)}$}
\put(9.5,3.2){$\bA^{(2)}$}
\put(4.89,1.83){$\bN_{\rm z}$}
\put(5.48,0.88){$\bN_{\rm r}$}
\put(5.09,1.1){$\varphi$}
\put(1.85,6.3){$\bA^{(1)}$}
\put(3.25,6.8){$\bA^{(2)}$}
\put(4.3,5.57){$\bA^{(1)}$}
\put(5.8,6.07){$\bA^{(2)}$}
\begin{rotate}{-18.5}
\put(1.0,6.91){\small {2$\beta_{\rm f}^{\rm Med}$}}
\put(3.49,6.98){\small {2$\beta_{\rm f}^{\rm Adv}$}}
\end{rotate}
\begin{rotate}{-12}
\put(0.6,1.0){Adventitia} 
\put(4.0,1.1){Media} 
\end{rotate}
\end{picture}
\setlength{\baselineskip}{11pt}
\caption{\label{fig: parameter_anpassung}%
a) Excised tissue samples, b) uniaxial tension test in circumferential and c) in axial direction.
}
\end{Figure}

The values of the shear modulus $\mu$ for the isotropic response will be directly estimated from the experiments as the average of 
the circumferential and axial stiffness. This is possible since the initial tangent $\IC|_{\bC=\bone}$ of all adjusted strain energies is solely determined by $\mu$ and will be independent of the anisotropic response.
Then considering that
\begin{equation*}
\Delta \bS = \IC|_{\bC=\bone} :\frac{1}{2}\Delta\bC\,,
\end{equation*}
and $\IC$ being a function only of $\mu$ in the reference configuration, the value of $\mu$ can be directly computed for the first measuring point $\bC(\hat{\lambda}_k^1)$.   
This is done separately for both directions and the mean value is taken as the estimated parameter. However, $\mu$ could as well be considered as a parameter to be included in the
parameter set $\Balpha$ in Eq.~\eqref{eq: fobj}, which is to be optimized.
Due to the assumed incompressibility it is impossible to estimate the bulk modulus $\kappa$.
From the numerical point of view the bulk modulus may act like a penalty constraint to enforce quasi incompressibility in more complex
computations.

\subsection{Media}

The results of the parameter adjustment of the Media are given in Table~\ref{tab: Parameter sets Media} for 8 different models and the corresponding stress-strain 
curves are plotted in Fig.~\ref{fig: model_fit_med_1_8}. 
These models result from the combination of the two isotropic strain energy functions in Eq.~(\ref{eq: iso Hencky}) and Eq.~(\ref{eq: exponentiated Hencky}) 
and the proposed transversely isotropic functions in Eq.~(\ref{eq: ti Hi}). The exponent $\varepsilon$ for the switchover criterion is set to 0.1 for all functions.
The fiber angle $\beta_{\rm f}$ between the fiber direction and the circumferential direction was part of the optimization, while
the parameters $\kappa$ and $\hat{k}$ were excluded from the optimization, due to the above assumed quasi-incompressibility.
The isotropic shear modulus was directly estimated from the experiments and was also not optimized.

Apparently, model 1 and 5 fail to accurately fit the experimental data which is evident in the large values of the objective function  $f^{\rm obj}$. When considering the exponentiated Hencky 
energy ${W}_{\rm eH}$ instead of the classical isotropic energy ${W}_{\rm H}$ the fit quality is substantially improved for models 2 and 6. Nevertheless, one has to admit that the associated 
fiber angle becomes unsatisfactory small, which does not seem to be reasonable. 
The small fiber angle is also accompanied by the effect that the isotropic material response in this case is highly nonlinear. 
The functions which are based on invariants with even exponents generally seem to perform better, independent of the chosen isotropic strain energy function. 
In \cite{Hol:2006:dom}, a mean angle of $37.5^{\circ}$ was reported. The optimized fiber angles of model 3, 4, 7 and 8 are of this order of magnitude. When ${W}_{\rm eH}$ is used
the error generally becomes slightly smaller, since one additional parameter is available.

\begin{Table}[!htb]
\centering
\begin{tabular}{|c|c|c|c|c|c|c|c|}
\hline
 & $\psi$ in - & $\mu$ in kPa& $k$ in - & $\mu_1$ in kPa& $k_1$ in -& $\beta_{\rm f}$ in $^{\circ}$ & $f^{\rm obj}$ \\
\hline
\rule{0pt}{13pt} 1& ${W}_{\rm H}$  + $\sum_{a=1}^2\psi^{\rm ti}_{\mbox{\small \ding{185}}_{H}}(I_{4_{(a)}}^{\rm H^1})$    & 31.16 & - & 0.0001 & 948.81 & 25.36 & 0.426\\
\rule{0pt}{13pt} 2& ${W}_{\rm eH}$ + $\sum_{a=1}^2\psi^{\rm ti}_{\mbox{\small \ding{185}}_{H}}(I_{4_{(a)}}^{\rm H^1})$    & 31.16 & 10.54 & 0.50 & 107.94 & 0.73 & 0.071\\
\hline                                                                                                                              
\rule{0pt}{13pt} 3& ${W}_{\rm H}$  + $\sum_{a=1}^2\psi^{\rm ti}_{\mbox{\small \ding{185}}_{H}}(I_{4_{(a)}}^{\rm H^2})$    & 31.16 & - & 1204.86 & 1599.53 & 41.24 & 0.046 \\
\rule{0pt}{13pt} 4& ${W}_{\rm eH}$ + $\sum_{a=1}^2\psi^{\rm ti}_{\mbox{\small \ding{185}}_{H}}(I_{4_{(a)}}^{\rm H^2})$    & 31.16 & 3.38 & 726.09 & 1848.66 & 40.68 & 0.044 \\
\hline
\rule{0pt}{13pt} 5& ${W}_{\rm H}$  + $\sum_{a=1}^2\psi^{\rm ti}_{\mbox{\small \ding{185}}_{H}}(I_{4_{(a)}}^{\rm H^3})$    & 31.16 & - & 11677.63 & 3112.51 & 0.0001 & 0.386 \\
\rule{0pt}{13pt} 6& ${W}_{\rm eH}$ + $\sum_{a=1}^2\psi^{\rm ti}_{\mbox{\small \ding{185}}_{H}}(I_{4_{(a)}}^{\rm H^3})$    & 31.16 & 10.54 & 5033.61 & 17685.18 & 28.74  & 0.071\\
\hline
\rule{0pt}{13pt} 7& ${W}_{\rm H}$  + $\sum_{a=1}^2\psi^{\rm ti}_{\mbox{\small \ding{185}}_{H}}(I_{4_{(a)}}^{\rm H^4})$    & 31.16 & - & 591428.36 & 51778.23 & 38.49 & 0.100 \\
\rule{0pt}{13pt} 8& ${W}_{\rm eH}$ + $\sum_{a=1}^2\psi^{\rm ti}_{\mbox{\small \ding{185}}_{H}}(I_{4_{(a)}}^{\rm H^4})$    & 31.16 & 7.56 &  232287.68 & 174224.46 & 36.86 & 0.052 \\
\hline
\end{tabular}
\caption{Adjusted parameter sets of the Media.\label{tab: Parameter sets Media}}
\end{Table}

\begin{Figure}[!htb]
\unitlength 1 cm
\begin{picture}(14,11.3)
\put(0.0,5.9){
\begingroup
  \makeatletter
  \providecommand\color[2][]{%
    \GenericError{(gnuplot) \space\space\space\@spaces}{%
      Package color not loaded in conjunction with
      terminal option `colourtext'%
    }{See the gnuplot documentation for explanation.%
    }{Either use 'blacktext' in gnuplot or load the package
      color.sty in LaTeX.}%
    \renewcommand\color[2][]{}%
  }%
  \providecommand\includegraphics[2][]{%
    \GenericError{(gnuplot) \space\space\space\@spaces}{%
      Package graphicx or graphics not loaded%
    }{See the gnuplot documentation for explanation.%
    }{The gnuplot epslatex terminal needs graphicx.sty or graphics.sty.}%
    \renewcommand\includegraphics[2][]{}%
  }%
  \providecommand\rotatebox[2]{#2}%
  \@ifundefined{ifGPcolor}{%
    \newif\ifGPcolor
    \GPcolortrue
  }{}%
  \@ifundefined{ifGPblacktext}{%
    \newif\ifGPblacktext
    \GPblacktexttrue
  }{}%
  \let\gplgaddtomacro\g@addto@macro
  \gdef\gplbacktext{}%
  \gdef\gplfronttext{}%
  \makeatother
  \ifGPblacktext
    \def\colorrgb#1{}%
    \def\colorgray#1{}%
  \else
    \ifGPcolor
      \def\colorrgb#1{\color[rgb]{#1}}%
      \def\colorgray#1{\color[gray]{#1}}%
      \expandafter\def\csname LTw\endcsname{\color{white}}%
      \expandafter\def\csname LTb\endcsname{\color{black}}%
      \expandafter\def\csname LTa\endcsname{\color{black}}%
      \expandafter\def\csname LT0\endcsname{\color[rgb]{1,0,0}}%
      \expandafter\def\csname LT1\endcsname{\color[rgb]{0,1,0}}%
      \expandafter\def\csname LT2\endcsname{\color[rgb]{0,0,1}}%
      \expandafter\def\csname LT3\endcsname{\color[rgb]{1,0,1}}%
      \expandafter\def\csname LT4\endcsname{\color[rgb]{0,1,1}}%
      \expandafter\def\csname LT5\endcsname{\color[rgb]{1,1,0}}%
      \expandafter\def\csname LT6\endcsname{\color[rgb]{0,0,0}}%
      \expandafter\def\csname LT7\endcsname{\color[rgb]{1,0.3,0}}%
      \expandafter\def\csname LT8\endcsname{\color[rgb]{0.5,0.5,0.5}}%
    \else
      \def\colorrgb#1{\color{black}}%
      \def\colorgray#1{\color[gray]{#1}}%
      \expandafter\def\csname LTw\endcsname{\color{white}}%
      \expandafter\def\csname LTb\endcsname{\color{black}}%
      \expandafter\def\csname LTa\endcsname{\color{black}}%
      \expandafter\def\csname LT0\endcsname{\color{black}}%
      \expandafter\def\csname LT1\endcsname{\color{black}}%
      \expandafter\def\csname LT2\endcsname{\color{black}}%
      \expandafter\def\csname LT3\endcsname{\color{black}}%
      \expandafter\def\csname LT4\endcsname{\color{black}}%
      \expandafter\def\csname LT5\endcsname{\color{black}}%
      \expandafter\def\csname LT6\endcsname{\color{black}}%
      \expandafter\def\csname LT7\endcsname{\color{black}}%
      \expandafter\def\csname LT8\endcsname{\color{black}}%
    \fi
  \fi
  \setlength{\unitlength}{0.0500bp}%
  \begin{picture}(4534.40,3310.40)%
    \gplgaddtomacro\gplbacktext{%
      \csname LTb\endcsname%
      \put(748,704){\makebox(0,0)[r]{\strut{} 0}}%
      \csname LTb\endcsname%
      \put(748,1039){\makebox(0,0)[r]{\strut{} 20}}%
      \csname LTb\endcsname%
      \put(748,1373){\makebox(0,0)[r]{\strut{} 40}}%
      \csname LTb\endcsname%
      \put(748,1708){\makebox(0,0)[r]{\strut{} 60}}%
      \csname LTb\endcsname%
      \put(748,2042){\makebox(0,0)[r]{\strut{} 80}}%
      \csname LTb\endcsname%
      \put(748,2377){\makebox(0,0)[r]{\strut{} 100}}%
      \csname LTb\endcsname%
      \put(748,2711){\makebox(0,0)[r]{\strut{} 120}}%
      \csname LTb\endcsname%
      \put(748,3046){\makebox(0,0)[r]{\strut{} 140}}%
      \csname LTb\endcsname%
      \put(880,484){\makebox(0,0){\strut{} 1}}%
      \csname LTb\endcsname%
      \put(1345,484){\makebox(0,0){\strut{} 1.05}}%
      \csname LTb\endcsname%
      \put(1811,484){\makebox(0,0){\strut{} 1.1}}%
      \csname LTb\endcsname%
      \put(2276,484){\makebox(0,0){\strut{} 1.15}}%
      \csname LTb\endcsname%
      \put(2741,484){\makebox(0,0){\strut{} 1.2}}%
      \csname LTb\endcsname%
      \put(3206,484){\makebox(0,0){\strut{} 1.25}}%
      \csname LTb\endcsname%
      \put(3672,484){\makebox(0,0){\strut{} 1.3}}%
      \csname LTb\endcsname%
      \put(4137,484){\makebox(0,0){\strut{} 1.35}}%
      \put(176,1875){\rotatebox{-270}{\makebox(0,0){\strut{}\large Cauchy sress $\sigma$ in kPa}}}%
      \put(2508,154){\makebox(0,0){\strut{}\large Stretch $\sqrt{\widehat{\lambda}}$ in -}}%
    }%
    \gplgaddtomacro\gplfronttext{%
      \csname LTb\endcsname%
      \put(1936,2873){\makebox(0,0)[r]{\strut{}circumf}}%
      \csname LTb\endcsname%
      \put(1936,2653){\makebox(0,0)[r]{\strut{}axial.}}%
      \csname LTb\endcsname%
      \put(1936,2433){\makebox(0,0)[r]{\strut{}model 1}}%
      \csname LTb\endcsname%
      \put(1936,2213){\makebox(0,0)[r]{\strut{}model 2}}%
    }%
    \gplbacktext
    \put(0,0){\includegraphics{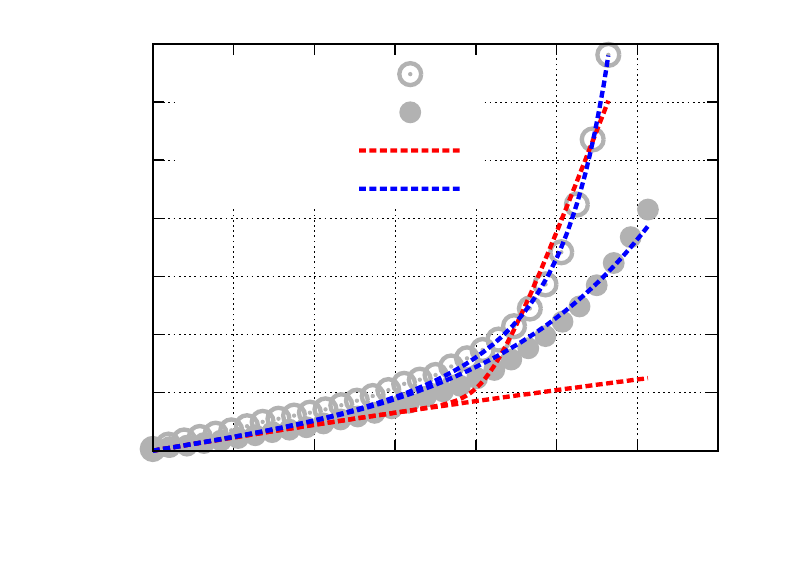}}%
    \gplfronttext
  \end{picture}%
\endgroup
}
\put(8.5,5.9){
\begingroup
  \makeatletter
  \providecommand\color[2][]{%
    \GenericError{(gnuplot) \space\space\space\@spaces}{%
      Package color not loaded in conjunction with
      terminal option `colourtext'%
    }{See the gnuplot documentation for explanation.%
    }{Either use 'blacktext' in gnuplot or load the package
      color.sty in LaTeX.}%
    \renewcommand\color[2][]{}%
  }%
  \providecommand\includegraphics[2][]{%
    \GenericError{(gnuplot) \space\space\space\@spaces}{%
      Package graphicx or graphics not loaded%
    }{See the gnuplot documentation for explanation.%
    }{The gnuplot epslatex terminal needs graphicx.sty or graphics.sty.}%
    \renewcommand\includegraphics[2][]{}%
  }%
  \providecommand\rotatebox[2]{#2}%
  \@ifundefined{ifGPcolor}{%
    \newif\ifGPcolor
    \GPcolortrue
  }{}%
  \@ifundefined{ifGPblacktext}{%
    \newif\ifGPblacktext
    \GPblacktexttrue
  }{}%
  \let\gplgaddtomacro\g@addto@macro
  \gdef\gplbacktext{}%
  \gdef\gplfronttext{}%
  \makeatother
  \ifGPblacktext
    \def\colorrgb#1{}%
    \def\colorgray#1{}%
  \else
    \ifGPcolor
      \def\colorrgb#1{\color[rgb]{#1}}%
      \def\colorgray#1{\color[gray]{#1}}%
      \expandafter\def\csname LTw\endcsname{\color{white}}%
      \expandafter\def\csname LTb\endcsname{\color{black}}%
      \expandafter\def\csname LTa\endcsname{\color{black}}%
      \expandafter\def\csname LT0\endcsname{\color[rgb]{1,0,0}}%
      \expandafter\def\csname LT1\endcsname{\color[rgb]{0,1,0}}%
      \expandafter\def\csname LT2\endcsname{\color[rgb]{0,0,1}}%
      \expandafter\def\csname LT3\endcsname{\color[rgb]{1,0,1}}%
      \expandafter\def\csname LT4\endcsname{\color[rgb]{0,1,1}}%
      \expandafter\def\csname LT5\endcsname{\color[rgb]{1,1,0}}%
      \expandafter\def\csname LT6\endcsname{\color[rgb]{0,0,0}}%
      \expandafter\def\csname LT7\endcsname{\color[rgb]{1,0.3,0}}%
      \expandafter\def\csname LT8\endcsname{\color[rgb]{0.5,0.5,0.5}}%
    \else
      \def\colorrgb#1{\color{black}}%
      \def\colorgray#1{\color[gray]{#1}}%
      \expandafter\def\csname LTw\endcsname{\color{white}}%
      \expandafter\def\csname LTb\endcsname{\color{black}}%
      \expandafter\def\csname LTa\endcsname{\color{black}}%
      \expandafter\def\csname LT0\endcsname{\color{black}}%
      \expandafter\def\csname LT1\endcsname{\color{black}}%
      \expandafter\def\csname LT2\endcsname{\color{black}}%
      \expandafter\def\csname LT3\endcsname{\color{black}}%
      \expandafter\def\csname LT4\endcsname{\color{black}}%
      \expandafter\def\csname LT5\endcsname{\color{black}}%
      \expandafter\def\csname LT6\endcsname{\color{black}}%
      \expandafter\def\csname LT7\endcsname{\color{black}}%
      \expandafter\def\csname LT8\endcsname{\color{black}}%
    \fi
  \fi
  \setlength{\unitlength}{0.0500bp}%
  \begin{picture}(4534.40,3310.40)%
    \gplgaddtomacro\gplbacktext{%
      \csname LTb\endcsname%
      \put(748,704){\makebox(0,0)[r]{\strut{} 0}}%
      \csname LTb\endcsname%
      \put(748,1039){\makebox(0,0)[r]{\strut{} 20}}%
      \csname LTb\endcsname%
      \put(748,1373){\makebox(0,0)[r]{\strut{} 40}}%
      \csname LTb\endcsname%
      \put(748,1708){\makebox(0,0)[r]{\strut{} 60}}%
      \csname LTb\endcsname%
      \put(748,2042){\makebox(0,0)[r]{\strut{} 80}}%
      \csname LTb\endcsname%
      \put(748,2377){\makebox(0,0)[r]{\strut{} 100}}%
      \csname LTb\endcsname%
      \put(748,2711){\makebox(0,0)[r]{\strut{} 120}}%
      \csname LTb\endcsname%
      \put(748,3046){\makebox(0,0)[r]{\strut{} 140}}%
      \csname LTb\endcsname%
      \put(880,484){\makebox(0,0){\strut{} 1}}%
      \csname LTb\endcsname%
      \put(1345,484){\makebox(0,0){\strut{} 1.05}}%
      \csname LTb\endcsname%
      \put(1811,484){\makebox(0,0){\strut{} 1.1}}%
      \csname LTb\endcsname%
      \put(2276,484){\makebox(0,0){\strut{} 1.15}}%
      \csname LTb\endcsname%
      \put(2741,484){\makebox(0,0){\strut{} 1.2}}%
      \csname LTb\endcsname%
      \put(3206,484){\makebox(0,0){\strut{} 1.25}}%
      \csname LTb\endcsname%
      \put(3672,484){\makebox(0,0){\strut{} 1.3}}%
      \csname LTb\endcsname%
      \put(4137,484){\makebox(0,0){\strut{} 1.35}}%
      \put(176,1875){\rotatebox{-270}{\makebox(0,0){\strut{}\large Cauchy sress $\sigma$ in kPa}}}%
      \put(2508,154){\makebox(0,0){\strut{}\large Stretch $\sqrt{\widehat{\lambda}}$ in -}}%
    }%
    \gplgaddtomacro\gplfronttext{%
      \csname LTb\endcsname%
      \put(1936,2873){\makebox(0,0)[r]{\strut{}circumf}}%
      \csname LTb\endcsname%
      \put(1936,2653){\makebox(0,0)[r]{\strut{}axial.}}%
      \csname LTb\endcsname%
      \put(1936,2433){\makebox(0,0)[r]{\strut{}model 3}}%
      \csname LTb\endcsname%
      \put(1936,2213){\makebox(0,0)[r]{\strut{}model 4}}%
    }%
    \gplbacktext
    \put(0,0){\includegraphics{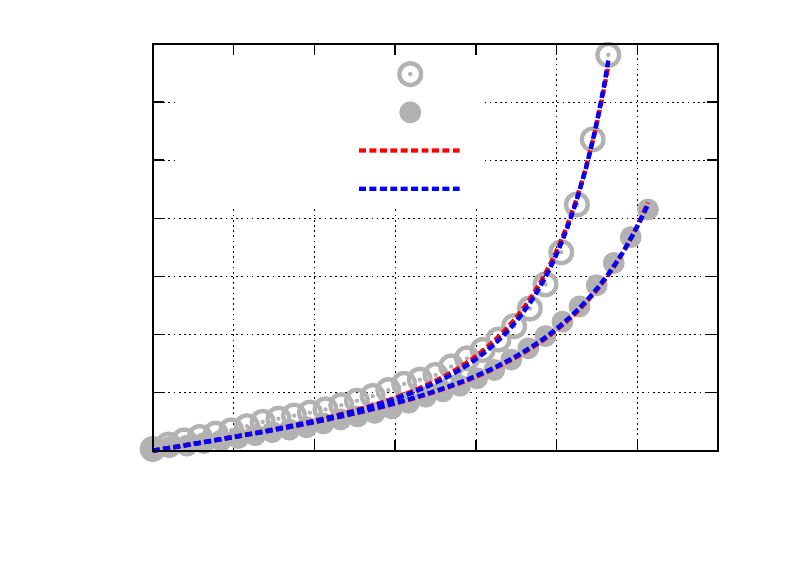}}%
    \gplfronttext
  \end{picture}%
\endgroup
}
\put(0.0,0.0){
\begingroup
  \makeatletter
  \providecommand\color[2][]{%
    \GenericError{(gnuplot) \space\space\space\@spaces}{%
      Package color not loaded in conjunction with
      terminal option `colourtext'%
    }{See the gnuplot documentation for explanation.%
    }{Either use 'blacktext' in gnuplot or load the package
      color.sty in LaTeX.}%
    \renewcommand\color[2][]{}%
  }%
  \providecommand\includegraphics[2][]{%
    \GenericError{(gnuplot) \space\space\space\@spaces}{%
      Package graphicx or graphics not loaded%
    }{See the gnuplot documentation for explanation.%
    }{The gnuplot epslatex terminal needs graphicx.sty or graphics.sty.}%
    \renewcommand\includegraphics[2][]{}%
  }%
  \providecommand\rotatebox[2]{#2}%
  \@ifundefined{ifGPcolor}{%
    \newif\ifGPcolor
    \GPcolortrue
  }{}%
  \@ifundefined{ifGPblacktext}{%
    \newif\ifGPblacktext
    \GPblacktexttrue
  }{}%
  \let\gplgaddtomacro\g@addto@macro
  \gdef\gplbacktext{}%
  \gdef\gplfronttext{}%
  \makeatother
  \ifGPblacktext
    \def\colorrgb#1{}%
    \def\colorgray#1{}%
  \else
    \ifGPcolor
      \def\colorrgb#1{\color[rgb]{#1}}%
      \def\colorgray#1{\color[gray]{#1}}%
      \expandafter\def\csname LTw\endcsname{\color{white}}%
      \expandafter\def\csname LTb\endcsname{\color{black}}%
      \expandafter\def\csname LTa\endcsname{\color{black}}%
      \expandafter\def\csname LT0\endcsname{\color[rgb]{1,0,0}}%
      \expandafter\def\csname LT1\endcsname{\color[rgb]{0,1,0}}%
      \expandafter\def\csname LT2\endcsname{\color[rgb]{0,0,1}}%
      \expandafter\def\csname LT3\endcsname{\color[rgb]{1,0,1}}%
      \expandafter\def\csname LT4\endcsname{\color[rgb]{0,1,1}}%
      \expandafter\def\csname LT5\endcsname{\color[rgb]{1,1,0}}%
      \expandafter\def\csname LT6\endcsname{\color[rgb]{0,0,0}}%
      \expandafter\def\csname LT7\endcsname{\color[rgb]{1,0.3,0}}%
      \expandafter\def\csname LT8\endcsname{\color[rgb]{0.5,0.5,0.5}}%
    \else
      \def\colorrgb#1{\color{black}}%
      \def\colorgray#1{\color[gray]{#1}}%
      \expandafter\def\csname LTw\endcsname{\color{white}}%
      \expandafter\def\csname LTb\endcsname{\color{black}}%
      \expandafter\def\csname LTa\endcsname{\color{black}}%
      \expandafter\def\csname LT0\endcsname{\color{black}}%
      \expandafter\def\csname LT1\endcsname{\color{black}}%
      \expandafter\def\csname LT2\endcsname{\color{black}}%
      \expandafter\def\csname LT3\endcsname{\color{black}}%
      \expandafter\def\csname LT4\endcsname{\color{black}}%
      \expandafter\def\csname LT5\endcsname{\color{black}}%
      \expandafter\def\csname LT6\endcsname{\color{black}}%
      \expandafter\def\csname LT7\endcsname{\color{black}}%
      \expandafter\def\csname LT8\endcsname{\color{black}}%
    \fi
  \fi
  \setlength{\unitlength}{0.0500bp}%
  \begin{picture}(4534.40,3310.40)%
    \gplgaddtomacro\gplbacktext{%
      \csname LTb\endcsname%
      \put(748,704){\makebox(0,0)[r]{\strut{} 0}}%
      \csname LTb\endcsname%
      \put(748,1039){\makebox(0,0)[r]{\strut{} 20}}%
      \csname LTb\endcsname%
      \put(748,1373){\makebox(0,0)[r]{\strut{} 40}}%
      \csname LTb\endcsname%
      \put(748,1708){\makebox(0,0)[r]{\strut{} 60}}%
      \csname LTb\endcsname%
      \put(748,2042){\makebox(0,0)[r]{\strut{} 80}}%
      \csname LTb\endcsname%
      \put(748,2377){\makebox(0,0)[r]{\strut{} 100}}%
      \csname LTb\endcsname%
      \put(748,2711){\makebox(0,0)[r]{\strut{} 120}}%
      \csname LTb\endcsname%
      \put(748,3046){\makebox(0,0)[r]{\strut{} 140}}%
      \csname LTb\endcsname%
      \put(880,484){\makebox(0,0){\strut{} 1}}%
      \csname LTb\endcsname%
      \put(1345,484){\makebox(0,0){\strut{} 1.05}}%
      \csname LTb\endcsname%
      \put(1811,484){\makebox(0,0){\strut{} 1.1}}%
      \csname LTb\endcsname%
      \put(2276,484){\makebox(0,0){\strut{} 1.15}}%
      \csname LTb\endcsname%
      \put(2741,484){\makebox(0,0){\strut{} 1.2}}%
      \csname LTb\endcsname%
      \put(3206,484){\makebox(0,0){\strut{} 1.25}}%
      \csname LTb\endcsname%
      \put(3672,484){\makebox(0,0){\strut{} 1.3}}%
      \csname LTb\endcsname%
      \put(4137,484){\makebox(0,0){\strut{} 1.35}}%
      \put(176,1875){\rotatebox{-270}{\makebox(0,0){\strut{}\large Cauchy sress $\sigma$ in kPa}}}%
      \put(2508,154){\makebox(0,0){\strut{}\large Stretch $\sqrt{\widehat{\lambda}}$ in -}}%
    }%
    \gplgaddtomacro\gplfronttext{%
      \csname LTb\endcsname%
      \put(2068,2873){\makebox(0,0)[r]{\strut{}circumf}}%
      \csname LTb\endcsname%
      \put(2068,2653){\makebox(0,0)[r]{\strut{}axial.}}%
      \csname LTb\endcsname%
      \put(2068,2433){\makebox(0,0)[r]{\strut{}model 5}}%
      \csname LTb\endcsname%
      \put(2068,2213){\makebox(0,0)[r]{\strut{}model 6}}%
    }%
    \gplbacktext
    \put(0,0){\includegraphics{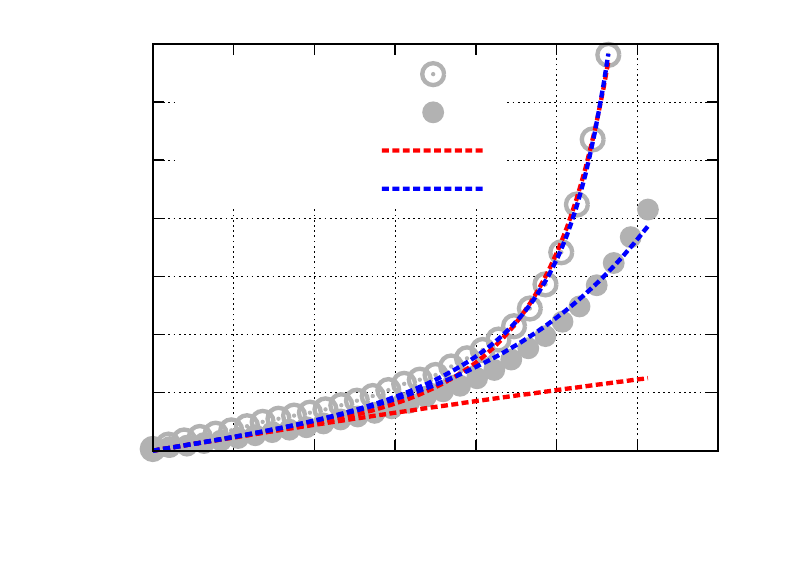}}%
    \gplfronttext
  \end{picture}%
\endgroup
}
\put(8.5,0.0){
\begingroup
  \makeatletter
  \providecommand\color[2][]{%
    \GenericError{(gnuplot) \space\space\space\@spaces}{%
      Package color not loaded in conjunction with
      terminal option `colourtext'%
    }{See the gnuplot documentation for explanation.%
    }{Either use 'blacktext' in gnuplot or load the package
      color.sty in LaTeX.}%
    \renewcommand\color[2][]{}%
  }%
  \providecommand\includegraphics[2][]{%
    \GenericError{(gnuplot) \space\space\space\@spaces}{%
      Package graphicx or graphics not loaded%
    }{See the gnuplot documentation for explanation.%
    }{The gnuplot epslatex terminal needs graphicx.sty or graphics.sty.}%
    \renewcommand\includegraphics[2][]{}%
  }%
  \providecommand\rotatebox[2]{#2}%
  \@ifundefined{ifGPcolor}{%
    \newif\ifGPcolor
    \GPcolortrue
  }{}%
  \@ifundefined{ifGPblacktext}{%
    \newif\ifGPblacktext
    \GPblacktexttrue
  }{}%
  \let\gplgaddtomacro\g@addto@macro
  \gdef\gplbacktext{}%
  \gdef\gplfronttext{}%
  \makeatother
  \ifGPblacktext
    \def\colorrgb#1{}%
    \def\colorgray#1{}%
  \else
    \ifGPcolor
      \def\colorrgb#1{\color[rgb]{#1}}%
      \def\colorgray#1{\color[gray]{#1}}%
      \expandafter\def\csname LTw\endcsname{\color{white}}%
      \expandafter\def\csname LTb\endcsname{\color{black}}%
      \expandafter\def\csname LTa\endcsname{\color{black}}%
      \expandafter\def\csname LT0\endcsname{\color[rgb]{1,0,0}}%
      \expandafter\def\csname LT1\endcsname{\color[rgb]{0,1,0}}%
      \expandafter\def\csname LT2\endcsname{\color[rgb]{0,0,1}}%
      \expandafter\def\csname LT3\endcsname{\color[rgb]{1,0,1}}%
      \expandafter\def\csname LT4\endcsname{\color[rgb]{0,1,1}}%
      \expandafter\def\csname LT5\endcsname{\color[rgb]{1,1,0}}%
      \expandafter\def\csname LT6\endcsname{\color[rgb]{0,0,0}}%
      \expandafter\def\csname LT7\endcsname{\color[rgb]{1,0.3,0}}%
      \expandafter\def\csname LT8\endcsname{\color[rgb]{0.5,0.5,0.5}}%
    \else
      \def\colorrgb#1{\color{black}}%
      \def\colorgray#1{\color[gray]{#1}}%
      \expandafter\def\csname LTw\endcsname{\color{white}}%
      \expandafter\def\csname LTb\endcsname{\color{black}}%
      \expandafter\def\csname LTa\endcsname{\color{black}}%
      \expandafter\def\csname LT0\endcsname{\color{black}}%
      \expandafter\def\csname LT1\endcsname{\color{black}}%
      \expandafter\def\csname LT2\endcsname{\color{black}}%
      \expandafter\def\csname LT3\endcsname{\color{black}}%
      \expandafter\def\csname LT4\endcsname{\color{black}}%
      \expandafter\def\csname LT5\endcsname{\color{black}}%
      \expandafter\def\csname LT6\endcsname{\color{black}}%
      \expandafter\def\csname LT7\endcsname{\color{black}}%
      \expandafter\def\csname LT8\endcsname{\color{black}}%
    \fi
  \fi
  \setlength{\unitlength}{0.0500bp}%
  \begin{picture}(4534.40,3310.40)%
    \gplgaddtomacro\gplbacktext{%
      \csname LTb\endcsname%
      \put(748,704){\makebox(0,0)[r]{\strut{} 0}}%
      \csname LTb\endcsname%
      \put(748,1039){\makebox(0,0)[r]{\strut{} 20}}%
      \csname LTb\endcsname%
      \put(748,1373){\makebox(0,0)[r]{\strut{} 40}}%
      \csname LTb\endcsname%
      \put(748,1708){\makebox(0,0)[r]{\strut{} 60}}%
      \csname LTb\endcsname%
      \put(748,2042){\makebox(0,0)[r]{\strut{} 80}}%
      \csname LTb\endcsname%
      \put(748,2377){\makebox(0,0)[r]{\strut{} 100}}%
      \csname LTb\endcsname%
      \put(748,2711){\makebox(0,0)[r]{\strut{} 120}}%
      \csname LTb\endcsname%
      \put(748,3046){\makebox(0,0)[r]{\strut{} 140}}%
      \csname LTb\endcsname%
      \put(880,484){\makebox(0,0){\strut{} 1}}%
      \csname LTb\endcsname%
      \put(1345,484){\makebox(0,0){\strut{} 1.05}}%
      \csname LTb\endcsname%
      \put(1811,484){\makebox(0,0){\strut{} 1.1}}%
      \csname LTb\endcsname%
      \put(2276,484){\makebox(0,0){\strut{} 1.15}}%
      \csname LTb\endcsname%
      \put(2741,484){\makebox(0,0){\strut{} 1.2}}%
      \csname LTb\endcsname%
      \put(3206,484){\makebox(0,0){\strut{} 1.25}}%
      \csname LTb\endcsname%
      \put(3672,484){\makebox(0,0){\strut{} 1.3}}%
      \csname LTb\endcsname%
      \put(4137,484){\makebox(0,0){\strut{} 1.35}}%
      \put(176,1875){\rotatebox{-270}{\makebox(0,0){\strut{}\large Cauchy sress $\sigma$ in kPa}}}%
      \put(2508,154){\makebox(0,0){\strut{}\large Stretch $\sqrt{\widehat{\lambda}}$ in -}}%
    }%
    \gplgaddtomacro\gplfronttext{%
      \csname LTb\endcsname%
      \put(2068,2873){\makebox(0,0)[r]{\strut{}circumf}}%
      \csname LTb\endcsname%
      \put(2068,2653){\makebox(0,0)[r]{\strut{}axial.}}%
      \csname LTb\endcsname%
      \put(2068,2433){\makebox(0,0)[r]{\strut{}model 7}}%
      \csname LTb\endcsname%
      \put(2068,2213){\makebox(0,0)[r]{\strut{}model 8}}%
    }%
    \gplbacktext
    \put(0,0){\includegraphics{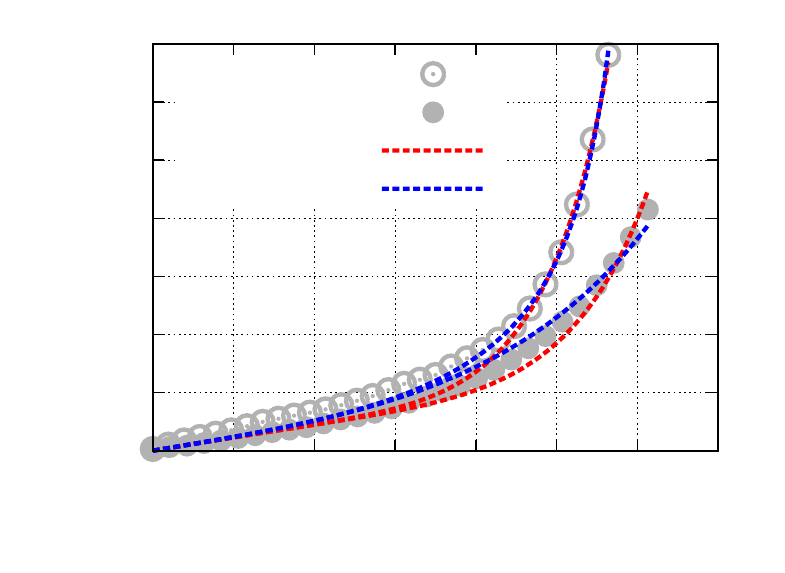}}%
    \gplfronttext
  \end{picture}%
\endgroup
}
\end{picture}
\setlength{\baselineskip}{11pt}
\caption{Adjusted stress-strain curves for the media for model 1-8. \label{fig: model_fit_med_1_8}}
\end{Figure}

\subsection{Adventitia}

The case distinction in the isotropic  response for the parameter fitting is neglected for the Adventitia. Only four models are tested in the following.
The optimized parameters are shown in Table~\ref{tab: Parameter sets Adventitia} and the corresponding stress-strain curves are given in Fig.~\ref{fig: model_fit_adv}. Please note that in this case 
the constraint $\beta_f\leq75^{\circ}$ was added to the optimization procedure, which is in the range of experimentally obtained mean angles, see \cite{Hol:2006:dom}. 
Again, only model 4 and 8 are able to reproduce the experimental curves.

\begin{Table}[!htb]
\centering
\begin{tabular}{|c|c|c|c|c|c|c|c|}
\hline
 & $\psi$ in - & $\mu$ in kPa& $k$ in - & $\mu_1$ in kPa& $k_1$ in -& $\beta_{\rm f}$ in $^{\circ}$ & $f^{\rm obj}$ \\
\hline
\rule{0pt}{13pt} 2& ${W}_{\rm eH}$ + $\sum_{a=1}^2\psi^{\rm ti}_{\mbox{\small \ding{185}}_{H}}(I_{4_{(a)}}^{\rm H^1})$    & 15.90 & 3.50 & 7.4e-06 & 246.90 & 75.00 & 0.127\\
\hline                                                                                                                              
\rule{0pt}{13pt} 4& ${W}_{\rm eH}$ + $\sum_{a=1}^2\psi^{\rm ti}_{\mbox{\small \ding{185}}_{H}}(I_{4_{(a)}}^{\rm H^2})$    & 15.90 & 1.0e-08 & 0.05 & 3707.99 & 50.30 & 0.045\\   
\hline
\rule{0pt}{13pt} 6& ${W}_{\rm eH}$ + $\sum_{a=1}^2\psi^{\rm ti}_{\mbox{\small \ding{185}}_{H}}(I_{4_{(a)}}^{\rm H^3})$    & 15.90 & 3.58 & 6.00 & 17045.18 & 63.28 & 0.128\\
\hline
\rule{0pt}{13pt} 8& ${W}_{\rm eH}$ + $\sum_{a=1}^2\psi^{\rm ti}_{\mbox{\small \ding{185}}_{H}}(I_{4_{(a)}}^{\rm H^4})$    & 15.90 & 1.0e-08 & 3973.68 & 56653.51 & 54.74 &  0.071\\
\hline
\end{tabular}
\caption{Adjusted parameter sets of the Adventitia.\label{tab: Parameter sets Adventitia}}
\end{Table}

\begin{Figure}[!htb]
\unitlength 1 cm
\begin{picture}(14,5.8)
\put(0.0,0.0){
\begingroup
  \makeatletter
  \providecommand\color[2][]{%
    \GenericError{(gnuplot) \space\space\space\@spaces}{%
      Package color not loaded in conjunction with
      terminal option `colourtext'%
    }{See the gnuplot documentation for explanation.%
    }{Either use 'blacktext' in gnuplot or load the package
      color.sty in LaTeX.}%
    \renewcommand\color[2][]{}%
  }%
  \providecommand\includegraphics[2][]{%
    \GenericError{(gnuplot) \space\space\space\@spaces}{%
      Package graphicx or graphics not loaded%
    }{See the gnuplot documentation for explanation.%
    }{The gnuplot epslatex terminal needs graphicx.sty or graphics.sty.}%
    \renewcommand\includegraphics[2][]{}%
  }%
  \providecommand\rotatebox[2]{#2}%
  \@ifundefined{ifGPcolor}{%
    \newif\ifGPcolor
    \GPcolortrue
  }{}%
  \@ifundefined{ifGPblacktext}{%
    \newif\ifGPblacktext
    \GPblacktexttrue
  }{}%
  \let\gplgaddtomacro\g@addto@macro
  \gdef\gplbacktext{}%
  \gdef\gplfronttext{}%
  \makeatother
  \ifGPblacktext
    \def\colorrgb#1{}%
    \def\colorgray#1{}%
  \else
    \ifGPcolor
      \def\colorrgb#1{\color[rgb]{#1}}%
      \def\colorgray#1{\color[gray]{#1}}%
      \expandafter\def\csname LTw\endcsname{\color{white}}%
      \expandafter\def\csname LTb\endcsname{\color{black}}%
      \expandafter\def\csname LTa\endcsname{\color{black}}%
      \expandafter\def\csname LT0\endcsname{\color[rgb]{1,0,0}}%
      \expandafter\def\csname LT1\endcsname{\color[rgb]{0,1,0}}%
      \expandafter\def\csname LT2\endcsname{\color[rgb]{0,0,1}}%
      \expandafter\def\csname LT3\endcsname{\color[rgb]{1,0,1}}%
      \expandafter\def\csname LT4\endcsname{\color[rgb]{0,1,1}}%
      \expandafter\def\csname LT5\endcsname{\color[rgb]{1,1,0}}%
      \expandafter\def\csname LT6\endcsname{\color[rgb]{0,0,0}}%
      \expandafter\def\csname LT7\endcsname{\color[rgb]{1,0.3,0}}%
      \expandafter\def\csname LT8\endcsname{\color[rgb]{0.5,0.5,0.5}}%
    \else
      \def\colorrgb#1{\color{black}}%
      \def\colorgray#1{\color[gray]{#1}}%
      \expandafter\def\csname LTw\endcsname{\color{white}}%
      \expandafter\def\csname LTb\endcsname{\color{black}}%
      \expandafter\def\csname LTa\endcsname{\color{black}}%
      \expandafter\def\csname LT0\endcsname{\color{black}}%
      \expandafter\def\csname LT1\endcsname{\color{black}}%
      \expandafter\def\csname LT2\endcsname{\color{black}}%
      \expandafter\def\csname LT3\endcsname{\color{black}}%
      \expandafter\def\csname LT4\endcsname{\color{black}}%
      \expandafter\def\csname LT5\endcsname{\color{black}}%
      \expandafter\def\csname LT6\endcsname{\color{black}}%
      \expandafter\def\csname LT7\endcsname{\color{black}}%
      \expandafter\def\csname LT8\endcsname{\color{black}}%
    \fi
  \fi
  \setlength{\unitlength}{0.0500bp}%
  \begin{picture}(4534.40,3310.40)%
    \gplgaddtomacro\gplbacktext{%
      \csname LTb\endcsname%
      \put(748,704){\makebox(0,0)[r]{\strut{} 0}}%
      \csname LTb\endcsname%
      \put(748,1172){\makebox(0,0)[r]{\strut{} 20}}%
      \csname LTb\endcsname%
      \put(748,1641){\makebox(0,0)[r]{\strut{} 40}}%
      \csname LTb\endcsname%
      \put(748,2109){\makebox(0,0)[r]{\strut{} 60}}%
      \csname LTb\endcsname%
      \put(748,2578){\makebox(0,0)[r]{\strut{} 80}}%
      \csname LTb\endcsname%
      \put(748,3046){\makebox(0,0)[r]{\strut{} 100}}%
      \csname LTb\endcsname%
      \put(880,484){\makebox(0,0){\strut{} 1}}%
      \csname LTb\endcsname%
      \put(1531,484){\makebox(0,0){\strut{} 1.1}}%
      \csname LTb\endcsname%
      \put(2183,484){\makebox(0,0){\strut{} 1.2}}%
      \csname LTb\endcsname%
      \put(2834,484){\makebox(0,0){\strut{} 1.3}}%
      \csname LTb\endcsname%
      \put(3486,484){\makebox(0,0){\strut{} 1.4}}%
      \csname LTb\endcsname%
      \put(4137,484){\makebox(0,0){\strut{} 1.5}}%
      \put(176,1875){\rotatebox{-270}{\makebox(0,0){\strut{}\large Cauchy sress $\sigma$ in kPa}}}%
      \put(2508,154){\makebox(0,0){\strut{}\large Stretch $\sqrt{\widehat{\lambda}}$ in -}}%
    }%
    \gplgaddtomacro\gplfronttext{%
      \csname LTb\endcsname%
      \put(1936,2873){\makebox(0,0)[r]{\strut{}circumf}}%
      \csname LTb\endcsname%
      \put(1936,2653){\makebox(0,0)[r]{\strut{}axial.}}%
      \csname LTb\endcsname%
      \put(1936,2433){\makebox(0,0)[r]{\strut{}model 2}}%
      \csname LTb\endcsname%
      \put(1936,2213){\makebox(0,0)[r]{\strut{}model 4}}%
    }%
    \gplbacktext
    \put(0,0){\includegraphics{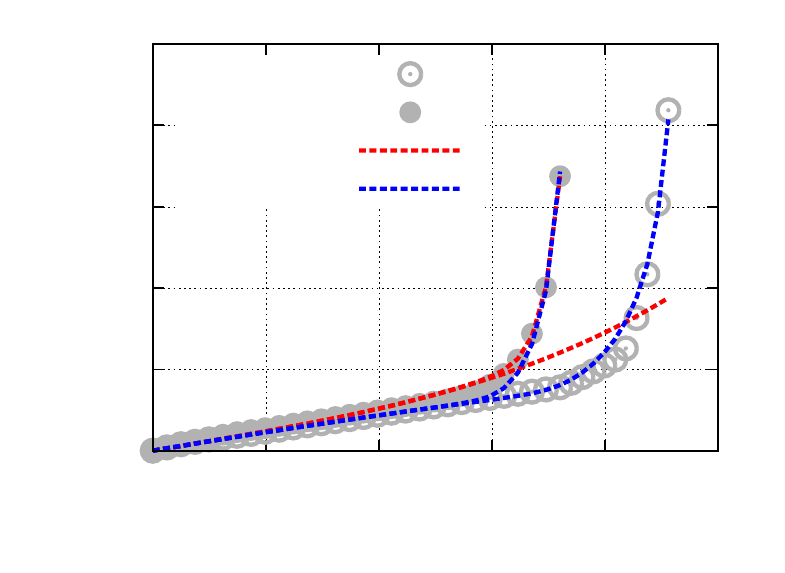}}%
    \gplfronttext
  \end{picture}%
\endgroup
}
\put(8.5,0.0){
\begingroup
  \makeatletter
  \providecommand\color[2][]{%
    \GenericError{(gnuplot) \space\space\space\@spaces}{%
      Package color not loaded in conjunction with
      terminal option `colourtext'%
    }{See the gnuplot documentation for explanation.%
    }{Either use 'blacktext' in gnuplot or load the package
      color.sty in LaTeX.}%
    \renewcommand\color[2][]{}%
  }%
  \providecommand\includegraphics[2][]{%
    \GenericError{(gnuplot) \space\space\space\@spaces}{%
      Package graphicx or graphics not loaded%
    }{See the gnuplot documentation for explanation.%
    }{The gnuplot epslatex terminal needs graphicx.sty or graphics.sty.}%
    \renewcommand\includegraphics[2][]{}%
  }%
  \providecommand\rotatebox[2]{#2}%
  \@ifundefined{ifGPcolor}{%
    \newif\ifGPcolor
    \GPcolortrue
  }{}%
  \@ifundefined{ifGPblacktext}{%
    \newif\ifGPblacktext
    \GPblacktexttrue
  }{}%
  \let\gplgaddtomacro\g@addto@macro
  \gdef\gplbacktext{}%
  \gdef\gplfronttext{}%
  \makeatother
  \ifGPblacktext
    \def\colorrgb#1{}%
    \def\colorgray#1{}%
  \else
    \ifGPcolor
      \def\colorrgb#1{\color[rgb]{#1}}%
      \def\colorgray#1{\color[gray]{#1}}%
      \expandafter\def\csname LTw\endcsname{\color{white}}%
      \expandafter\def\csname LTb\endcsname{\color{black}}%
      \expandafter\def\csname LTa\endcsname{\color{black}}%
      \expandafter\def\csname LT0\endcsname{\color[rgb]{1,0,0}}%
      \expandafter\def\csname LT1\endcsname{\color[rgb]{0,1,0}}%
      \expandafter\def\csname LT2\endcsname{\color[rgb]{0,0,1}}%
      \expandafter\def\csname LT3\endcsname{\color[rgb]{1,0,1}}%
      \expandafter\def\csname LT4\endcsname{\color[rgb]{0,1,1}}%
      \expandafter\def\csname LT5\endcsname{\color[rgb]{1,1,0}}%
      \expandafter\def\csname LT6\endcsname{\color[rgb]{0,0,0}}%
      \expandafter\def\csname LT7\endcsname{\color[rgb]{1,0.3,0}}%
      \expandafter\def\csname LT8\endcsname{\color[rgb]{0.5,0.5,0.5}}%
    \else
      \def\colorrgb#1{\color{black}}%
      \def\colorgray#1{\color[gray]{#1}}%
      \expandafter\def\csname LTw\endcsname{\color{white}}%
      \expandafter\def\csname LTb\endcsname{\color{black}}%
      \expandafter\def\csname LTa\endcsname{\color{black}}%
      \expandafter\def\csname LT0\endcsname{\color{black}}%
      \expandafter\def\csname LT1\endcsname{\color{black}}%
      \expandafter\def\csname LT2\endcsname{\color{black}}%
      \expandafter\def\csname LT3\endcsname{\color{black}}%
      \expandafter\def\csname LT4\endcsname{\color{black}}%
      \expandafter\def\csname LT5\endcsname{\color{black}}%
      \expandafter\def\csname LT6\endcsname{\color{black}}%
      \expandafter\def\csname LT7\endcsname{\color{black}}%
      \expandafter\def\csname LT8\endcsname{\color{black}}%
    \fi
  \fi
  \setlength{\unitlength}{0.0500bp}%
  \begin{picture}(4534.40,3310.40)%
    \gplgaddtomacro\gplbacktext{%
      \csname LTb\endcsname%
      \put(748,704){\makebox(0,0)[r]{\strut{} 0}}%
      \csname LTb\endcsname%
      \put(748,1172){\makebox(0,0)[r]{\strut{} 20}}%
      \csname LTb\endcsname%
      \put(748,1641){\makebox(0,0)[r]{\strut{} 40}}%
      \csname LTb\endcsname%
      \put(748,2109){\makebox(0,0)[r]{\strut{} 60}}%
      \csname LTb\endcsname%
      \put(748,2578){\makebox(0,0)[r]{\strut{} 80}}%
      \csname LTb\endcsname%
      \put(748,3046){\makebox(0,0)[r]{\strut{} 100}}%
      \csname LTb\endcsname%
      \put(880,484){\makebox(0,0){\strut{} 1}}%
      \csname LTb\endcsname%
      \put(1531,484){\makebox(0,0){\strut{} 1.1}}%
      \csname LTb\endcsname%
      \put(2183,484){\makebox(0,0){\strut{} 1.2}}%
      \csname LTb\endcsname%
      \put(2834,484){\makebox(0,0){\strut{} 1.3}}%
      \csname LTb\endcsname%
      \put(3486,484){\makebox(0,0){\strut{} 1.4}}%
      \csname LTb\endcsname%
      \put(4137,484){\makebox(0,0){\strut{} 1.5}}%
      \put(176,1875){\rotatebox{-270}{\makebox(0,0){\strut{}\large Cauchy sress $\sigma$ in kPa}}}%
      \put(2508,154){\makebox(0,0){\strut{}\large Stretch $\sqrt{\widehat{\lambda}}$ in -}}%
    }%
    \gplgaddtomacro\gplfronttext{%
      \csname LTb\endcsname%
      \put(1936,2873){\makebox(0,0)[r]{\strut{}circumf}}%
      \csname LTb\endcsname%
      \put(1936,2653){\makebox(0,0)[r]{\strut{}axial.}}%
      \csname LTb\endcsname%
      \put(1936,2433){\makebox(0,0)[r]{\strut{}model 6}}%
      \csname LTb\endcsname%
      \put(1936,2213){\makebox(0,0)[r]{\strut{}model 8}}%
    }%
    \gplbacktext
    \put(0,0){\includegraphics{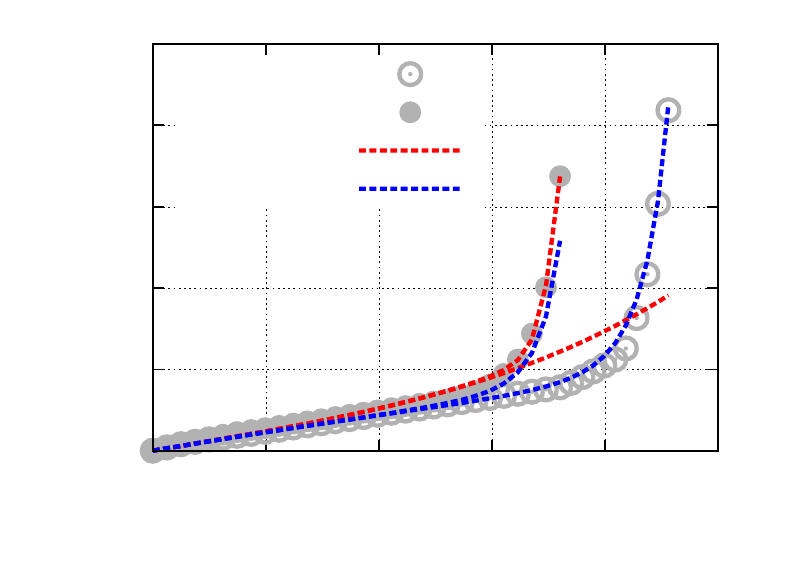}}%
    \gplfronttext
  \end{picture}%
\endgroup
}
\end{picture}
\setlength{\baselineskip}{11pt}
\caption{Evolution of the stress ratio induced by different anisotropic measures under biaxial-tension-compression. \label{fig: model_fit_adv}}
\end{Figure}

\newpage
\subsection{Artery \label{sec: artery}}

In this section the calibrated transversely isotropic model is applied to three-dimensional simulations of 
coronary patient-specific arterial walls in order to prove the robustness of the novel material formulation. 
Since the parameter fitting was based on an diseased abdominal artery, the results may merely be discussed in a qualitative context and conclusion related to clinical interpretation are highly limited.
An arterial geometry is reconstructed based on two sequenced 
two-dimensional \textit{virtual histology (VH) intravascular ultrasound (IVUS)} images.
For a detailed description of the three-dimensional reconstruction the reader is referred to \cite{BalBoeBraErbKlaRheSch:2011:pso}.

The considered artery consists of two layers, see Fig.~\ref{fig: Artery_geometry}. The outer layer is the Adventitia, the inner layer
the Media. This artery was loaded in a finite element simulation with an inner pressure of 16~kPa. An augmented Lagrange strategy was applied
to enforce quasi-incompressibility with an allowed tolerance of $1\%$ in change of volume.
For additional information the reader is referred to \cite{Hes:1969:mag}, \cite{Pow:1969:amf}, \cite{GloLet:1984:fea}, \cite{GloLet:1988:alm} and
\cite{GloLet:1989:ala}. Tetrahedron finite elements with ten nodes and quadratic shape functions are used. 

In a first simulation model 4, based on  $I_4^{\rm H^2}$, was used with the estimated parameters from the adjustment, see Table~\ref{tab: Parameter sets Media} and Table~\ref{tab: Parameter sets Adventitia}. 
These results are to be compared with a second simulation, where the strain energy function and parameters 
according to \cite{SchBri:2014:ans} were used. There a Mooney-Rivlin model was used for the isotropic part and the function from \cite{HolGasOgd:2000:anc}, given in
Eq.~(\ref{eq: calssical HGO}), for the superimposed transversely isotropic parts. Both parts  refer to the strain measure $\bC$. 
The parameters were adjusted to the same 
experimental data. The deformed configurations for different stresses are shown in Fig.~\ref{fig: Artery_defo}.
While the general stress distributions are comparable the increase in the volume of the lumen is significantly
larger for the Hencky model.

The same effect for different strain energy functions, which were adjusted to the same data and then used for numerical
simulations of arterial segments was noticed in \cite{BraKlaRheSch:2008:mac}.

\begin{Figure}[!htb]
\unitlength 1 cm
\begin{picture}(14,4.0)
\put(0.8,0.0){\includegraphics[width = 5.5cm]{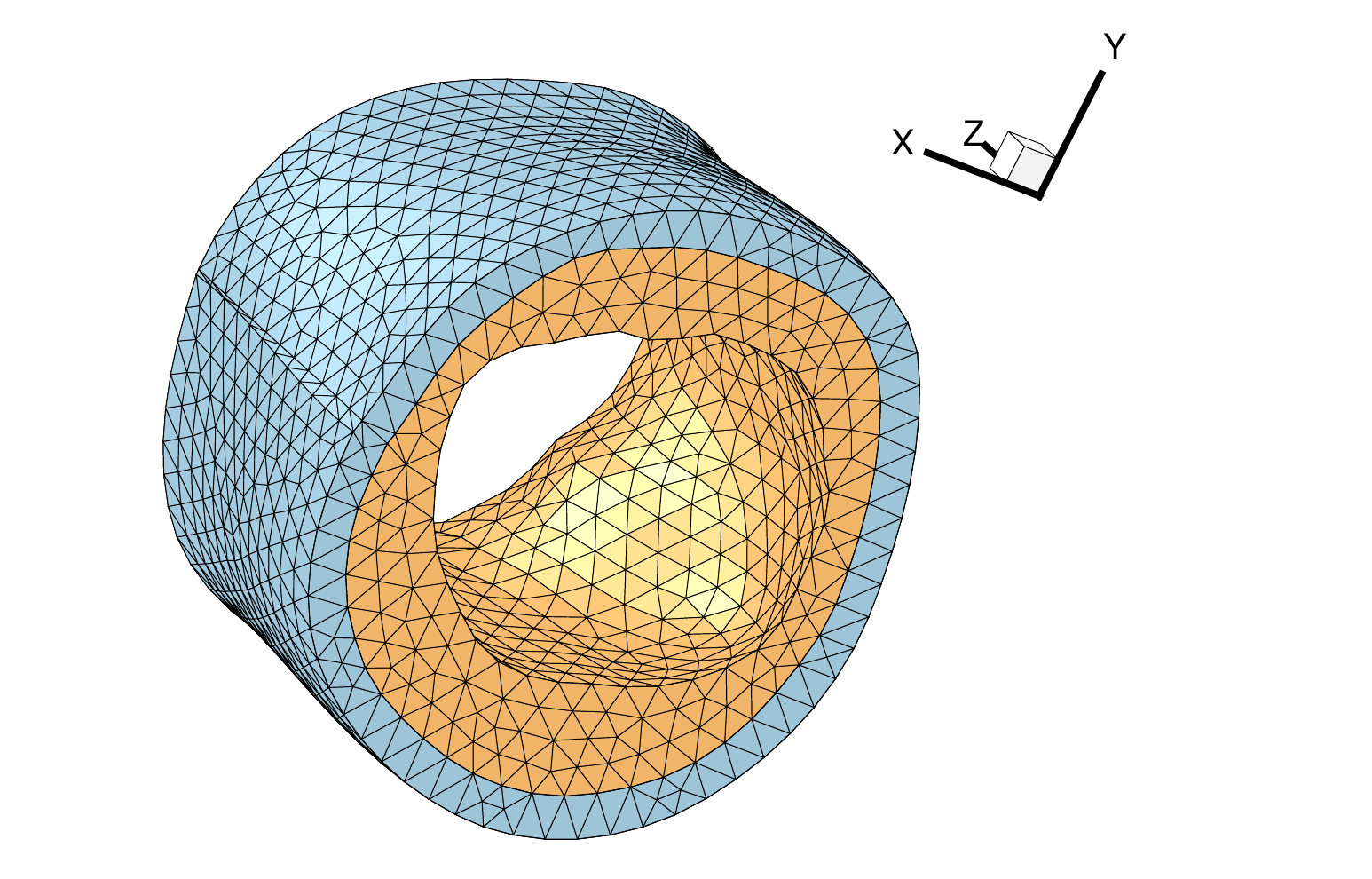}}
\put(5.3,0.0){\includegraphics[width = 5.5cm]{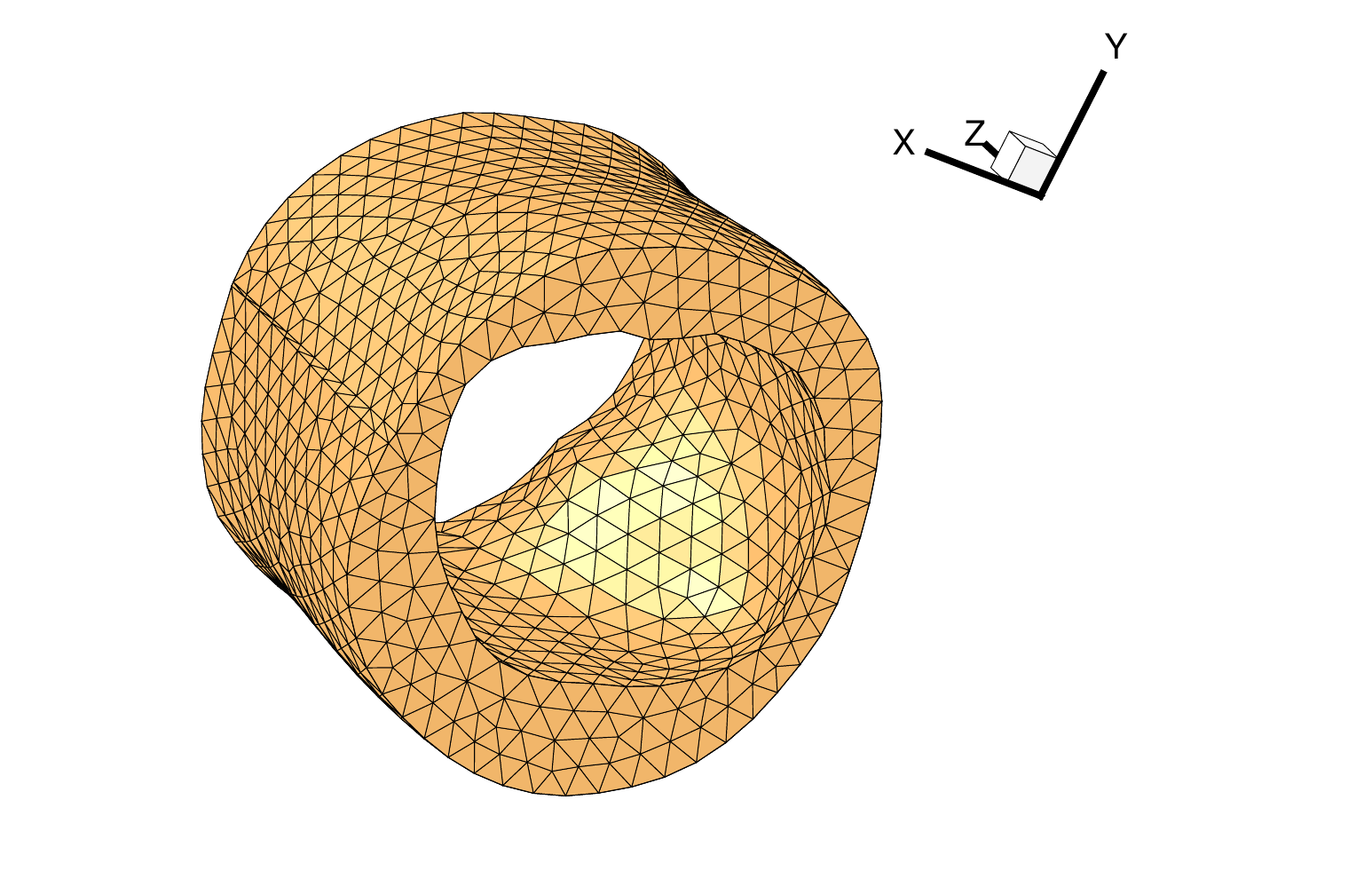}}
\put(9.8,0.0){\includegraphics[width = 5.5cm]{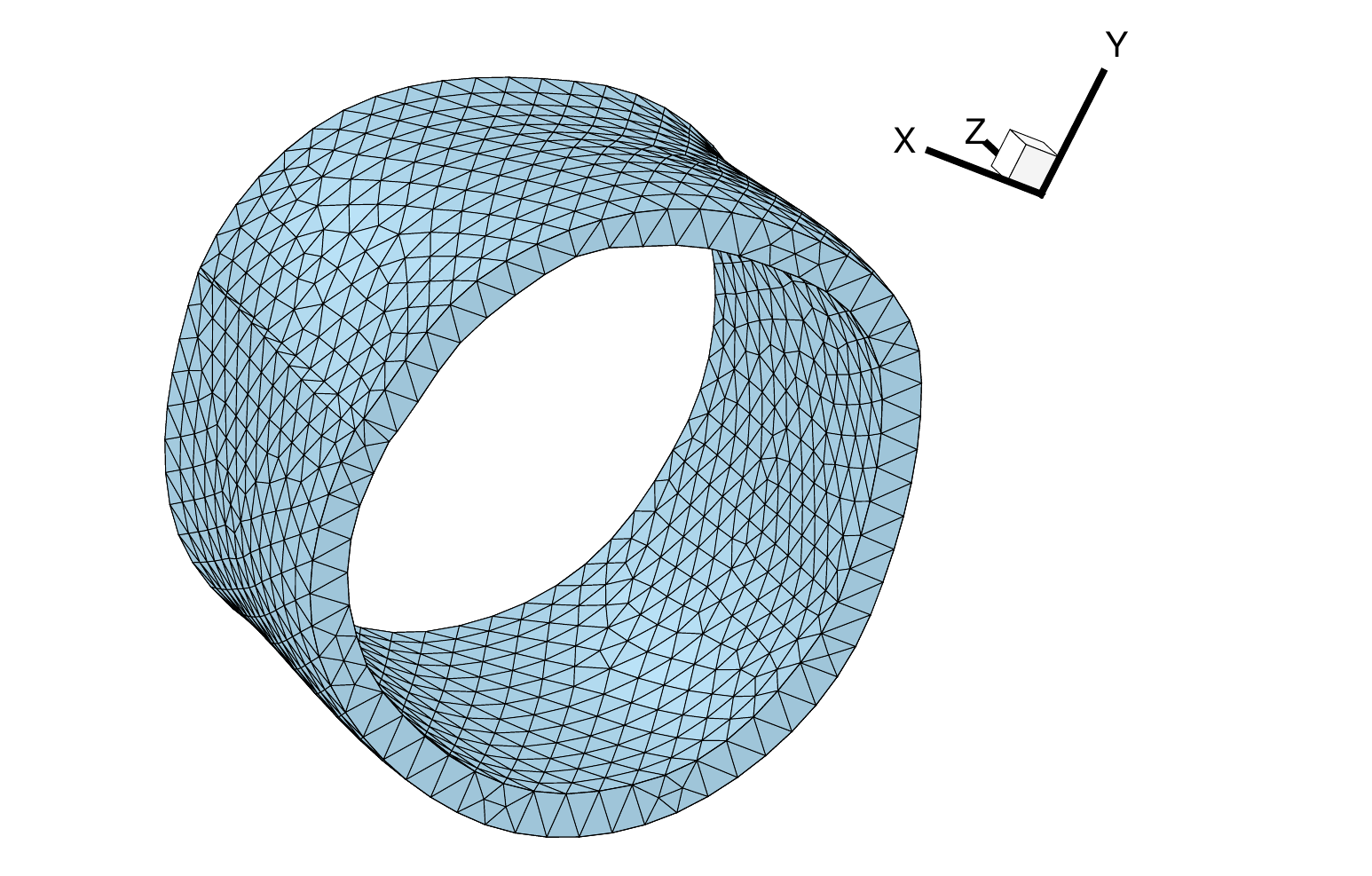}}
\end{picture}
\setlength{\baselineskip}{11pt}
\caption{\label{fig: Artery_geometry}%
Reconstructed artery consisting of Adventitia (outer layer) and Media (inner layer).
}
\end{Figure}

\begin{Figure}[!htb]
\unitlength 1 cm
\begin{picture}(14,8.0)
\put(-0.2,4.0){\includegraphics[width = 5.5cm]{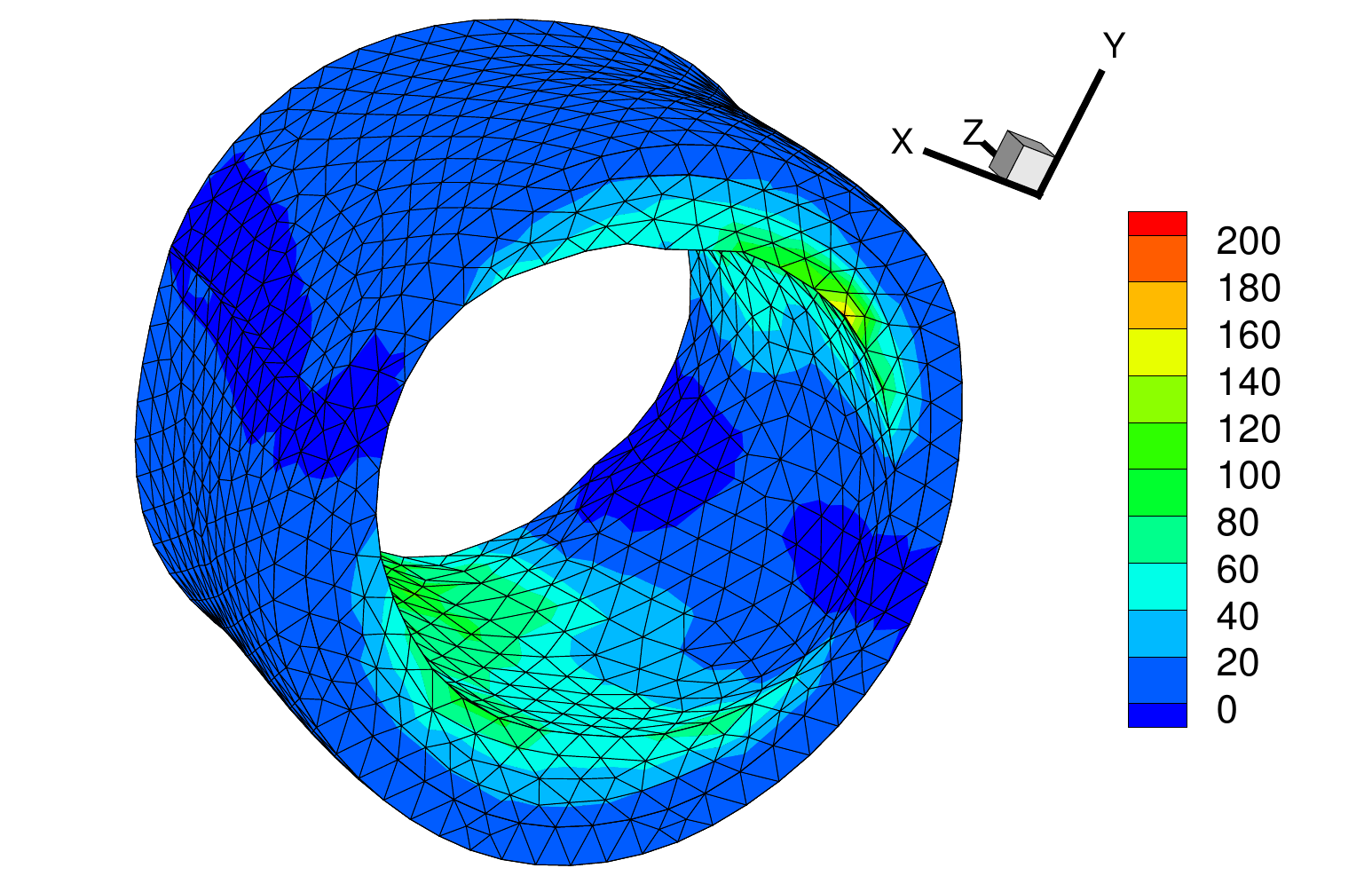}}
\put(5.3,4.0){\includegraphics[width = 5.5cm]{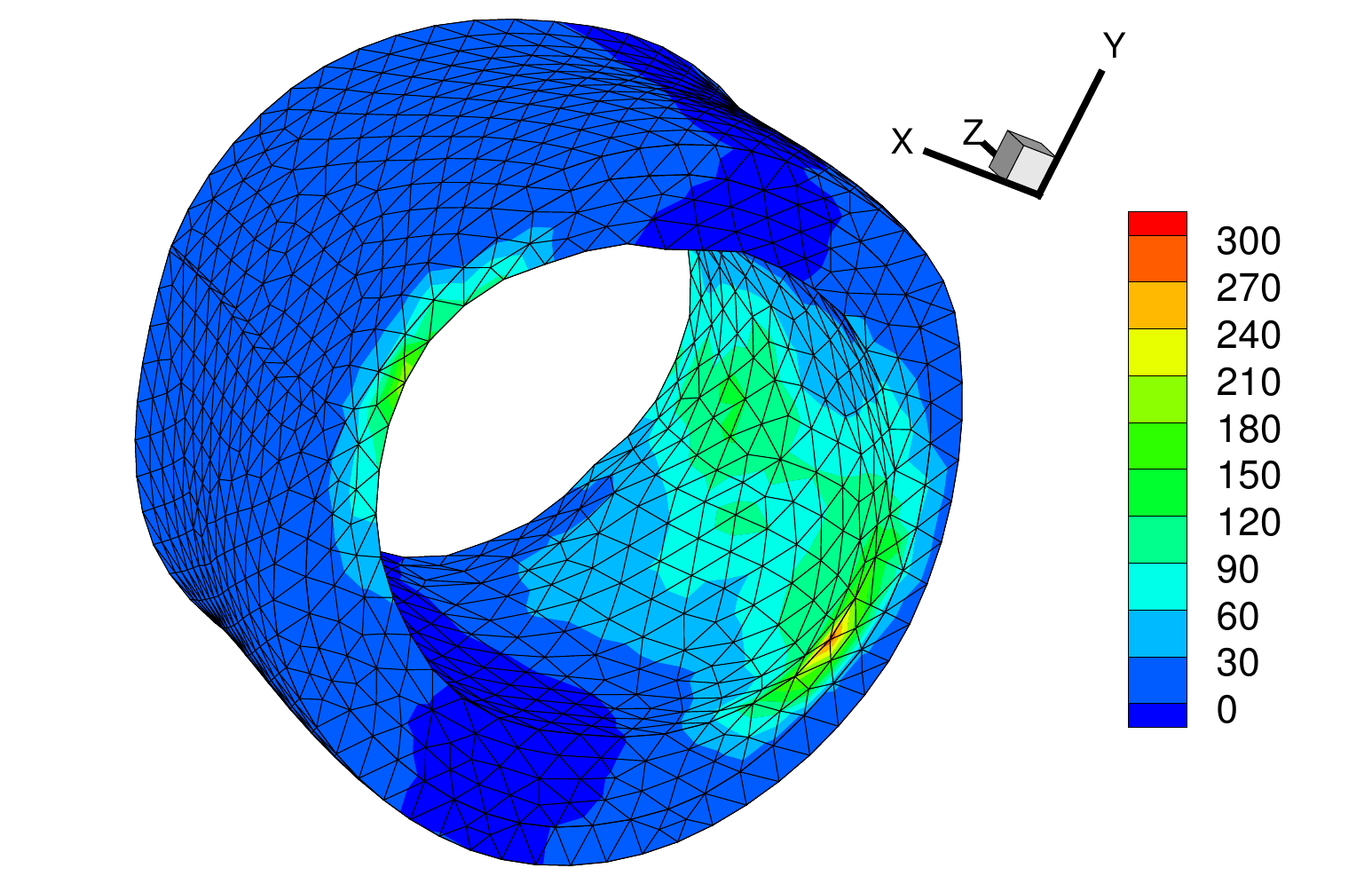}}
\put(10.8,4.0){\includegraphics[width = 5.5cm]{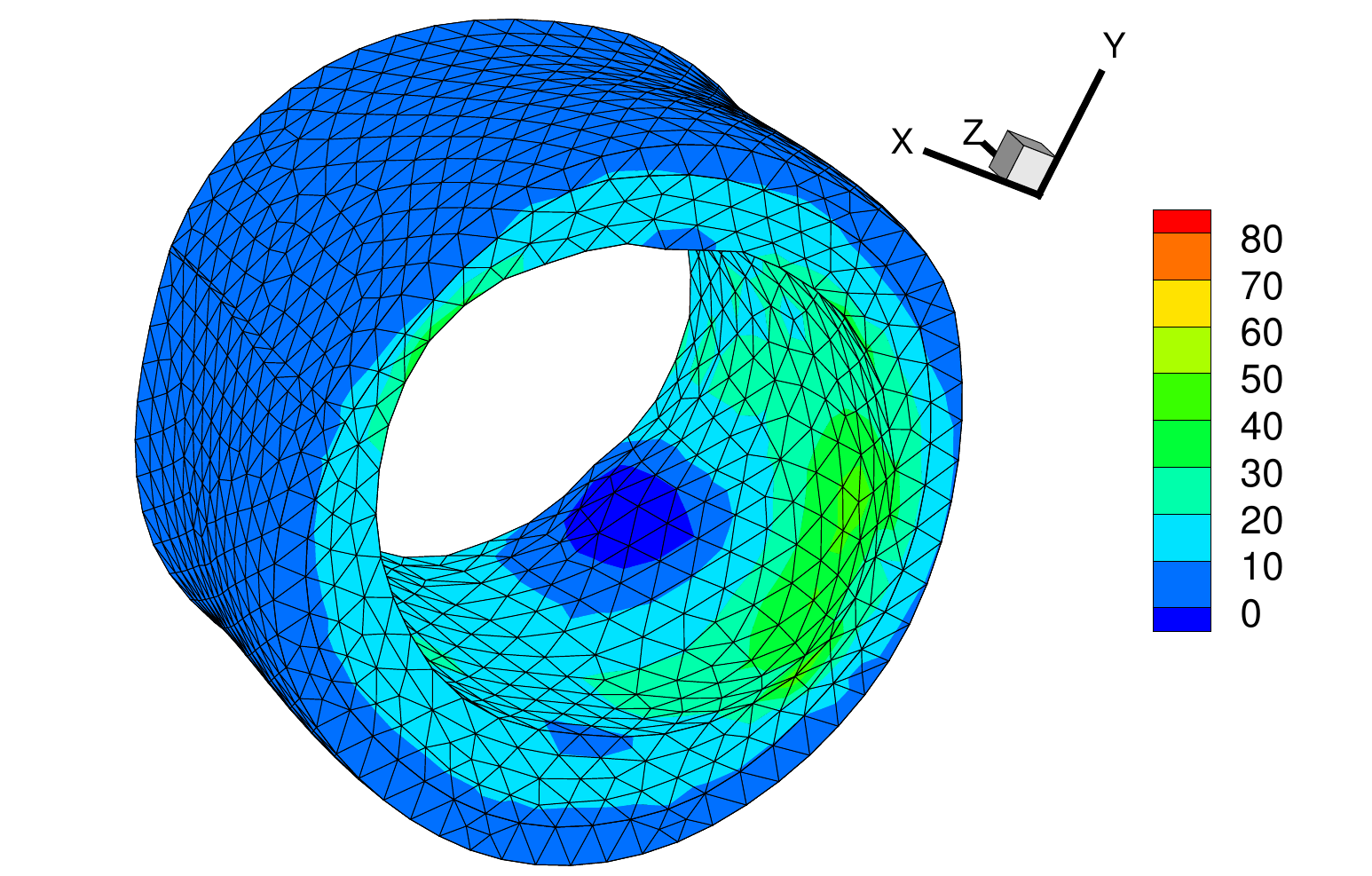}}
\put(-0.2,0.0){\includegraphics[width = 5.5cm]{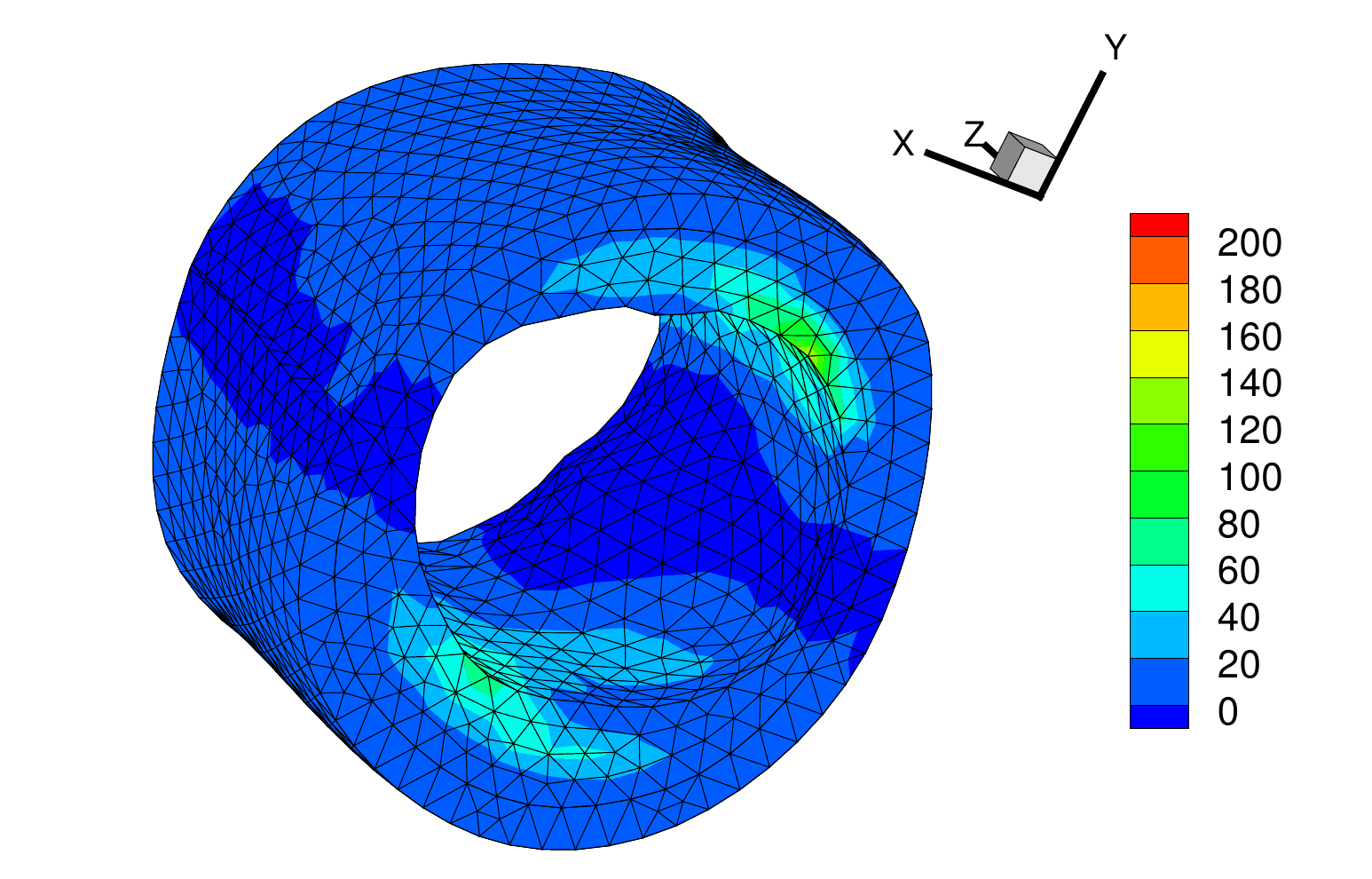}}
\put(5.3,0.0){\includegraphics[width = 5.5cm]{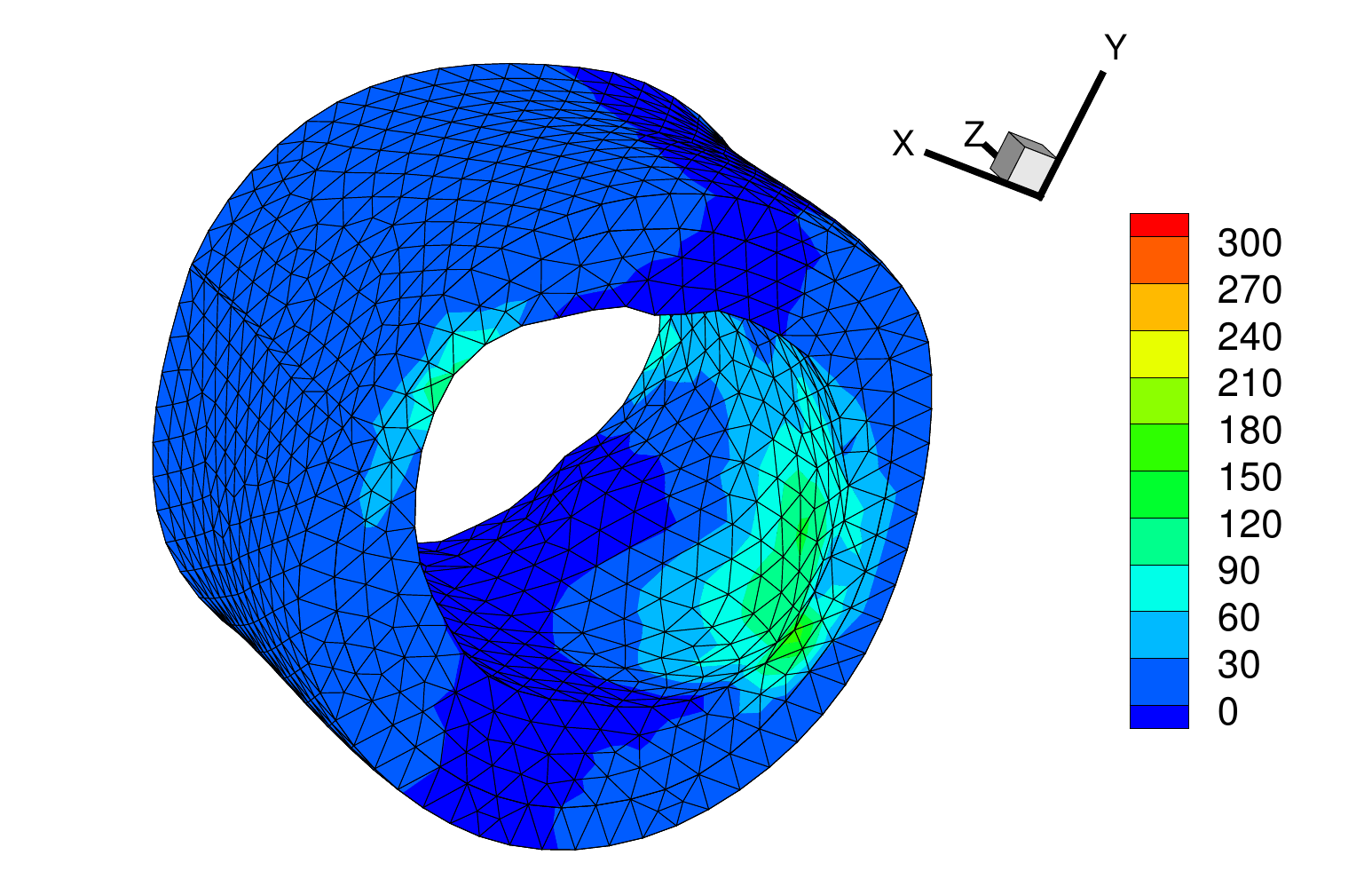}}
\put(10.8,0.0){\includegraphics[width = 5.5cm]{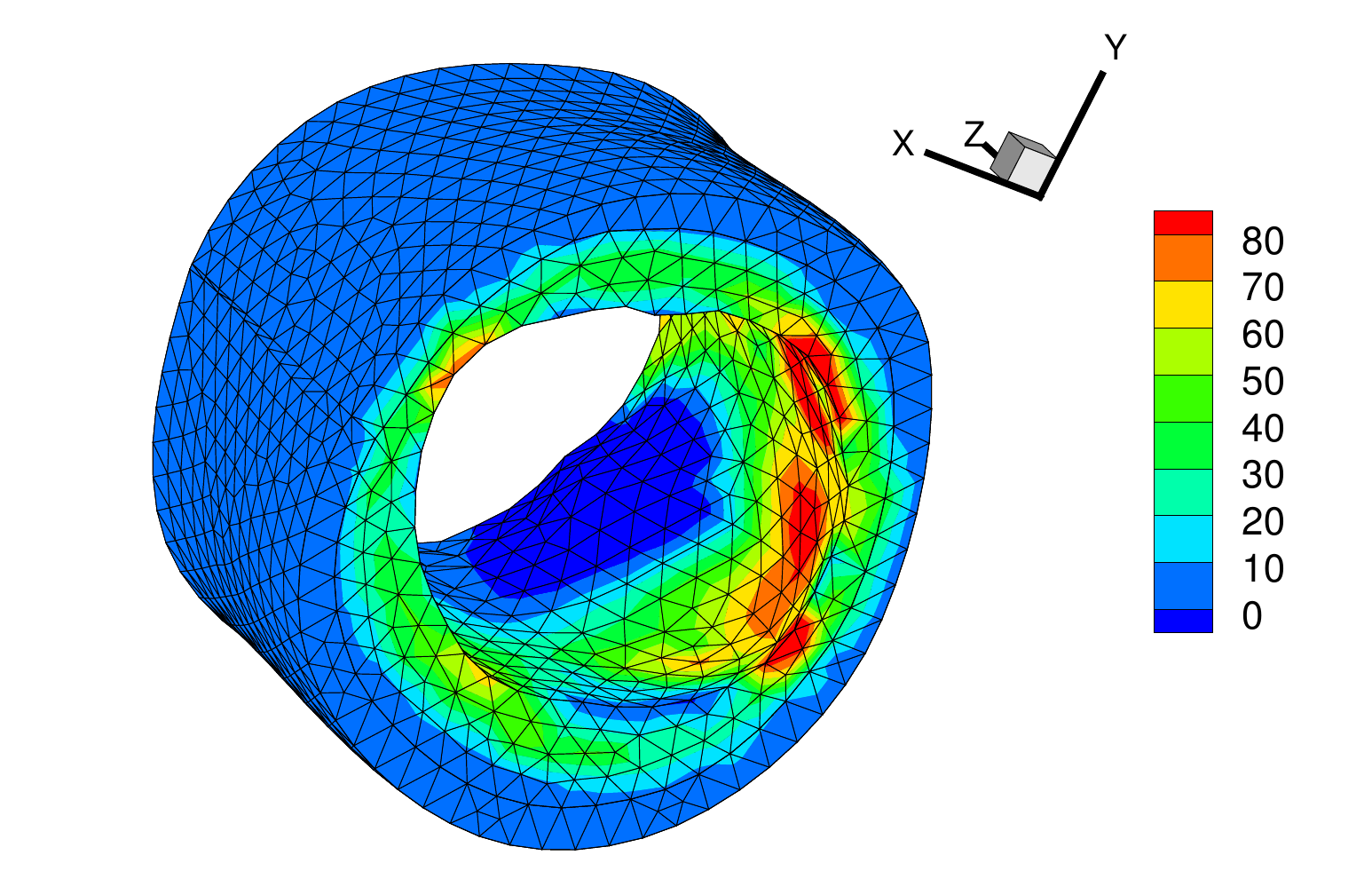}}
\put(4.3,6.8){$\sigma_{11}$}
\put(4.3,2.8){$\sigma_{11}$}
\put(9.8,6.8){$\sigma_{22}$}
\put(9.8,2.8){$\sigma_{22}$}
\put(15.4,6.8){$\sigma_{33}$}
\put(15.4,2.8){$\sigma_{33}$}
\put(14.7,0){in kPa}
\put(0,4){a)}
\put(0,0){b)}
\end{picture}
\setlength{\baselineskip}{11pt}
\caption{\label{fig: Artery_defo}%
Deformed configuration and stress distributions in an arterial segment, a) simulated with model 4, based on $I_4^{\rm H^2}$ and b) the material 
model and parameters used in \cite{SchBri:2014:ans} based on $I_4^{\rm C^1}$  for an internal pressure $p$ of 16~kPa.
}
\end{Figure}

\section{Open problems}

While the ellipticity domain of the isotropic exponentiated Hencky energy is reasonably well understood (it is an extremely  large cone in stretch space) the same is not clear for the anisotropic logarithmic energy terms.
The problem which has to be faced is due to the analytical difficulties in treating derivatives of the right Hencky strain tensor; we need to calculate for  example 
$\partial^2_{\bF} (\exp\langle \log \bU, \bM\rangle^2).(\Bxi\otimes \Beta,\Bxi\otimes \Beta)$
which is manageable along the lines of the presented algorithmic treatment in this paper; but it remains to identify a suitable ellipticity region. The experience with the isotropic exponentiated Hencky energy suggests, however, that the ellipticity domain largely contains the physical range of arteries,
i.e. principal stretches in a suitable $\lambda_k\in[1/2,2]$.
Therefore, our model proposal  is elliptic in the physiological range of arteries and this is all that must be required on mathematical grounds. \\

Since our extension of the exponentiated Hencky energy towards anisotropy is done on an ad hoc basis, it remains to study whether the differential-geometric program presented in \cite{NefEidMar:2016:gol} may be extended to the anisotropic case. Of course, major technical difficulties have to be solved. 
The benefit, however, cannot be overemphasized: There would result finite strain anisotropic energies having a clear physical meaning. We will pursue this issue in the future.

{\bf Acknowledgments} 
The first two authors gratefully acknowledge support by the Deutsche Forschungsgemeinschaft in the Priority Program 1748
under the project ``Novel finite elements for anisotropic media at finite strain'' (SCHR 570/23-1), section \ref{sec: intro}-\ref{sec: academic}. 
Further, the first two authors would like to acknowledge support by the Deutsche Forschungsgemeinschaft within the framework of the project 
``Domain-decomposition-based fluid structure interaction algorithms for highly nonlinear and anisotropic elastic arterial wall models in 3D`` (SCHR 570/15-2)
under the D-A-CH agreement, section \ref{sec: bio}-\ref{sec: param}.

\bibliography{biomechanik_07}
\begin{appendix}
\sect{Appendix}
\ssect{Notes on the Hencky tensor}
The Hencky strain tensor is defined through
\begin{equation}
\log \bU = \frac{1}{2} \log \bC
\end{equation}
and 
\begin{equation}
{\rm tr} (\log \bU) = \log ({\rm det} \bU)\,.
\end{equation}

The symmetric right Cauchy-Green tensor $\bC$ in spectral decomposition is given by
\eb
\bC = \sum_{k=1}^3 \widehat{\lambda}_k \bN^k\otimes\bN^k = \sum_{k=1}^3 \widehat{\lambda}_k \bP^k\,,\quad \mbox{with}\quad \bP^k = \bN^k\otimes\bN^k 
\ee
and for the Hencky strain we obtain
\eb
\log\bU = \sum_{k=1}^3 \frac{1}{2}\log(\widehat{\lambda}_k) \bN^k\otimes\bN^k = \sum_{k=1}^3 \frac{1}{2}\log(\widehat{\lambda}_k) \bP_k\,,\quad \mbox{with}\quad \bP_k = \bN^k\otimes\bN^k \,.
\ee
The first derivative of $\log \bU$ with respect to  $\bC$ can be computed  as
\begin{align}
\nonumber  \pp{\log\bU}{\bC}=& \sum_{k=1}^3 \bP_k\otimes\pp{\frac{1}{2}\log\widehat{\lambda}_k}{\bC} + \frac{1}{2}\log\widehat{\lambda}_k\pp{\bP_k}{\bC} \\
\nonumber                   =& \sum_{k=1}^3 \bP_k\otimes\pp{\frac{1}{2}\log\widehat{\lambda}_k}{\widehat{\lambda}_k}\,\pp{\widehat{\lambda}_k}{\bC} + \frac{1}{2}\log\widehat{\lambda}_k\pp{\bP_k}{\bC}\\
                            =& \sum_{k=1}^3 \frac{1}{2}\widehat{\lambda}_k^{-1}\bP_k\otimes\pp{\widehat{\lambda}_k}{\bC} + \frac{1}{2}\log\widehat{\lambda}_k\pp{\bP_k}{\bC} 
\end{align}

Considering that 
\eb
\pp{\widehat{\lambda}_k}{\bC} = \bP_k\qquad \mbox{and} \qquad \frac{\partial \bP_k }{\partial \bC} = \sum_{j=1, j\neq k}^3 \frac{\bP_k\boxtimes\bP_j^T + \bP_j\boxtimes\bP_k^T}{\widehat{\lambda}_{k}-\widehat{\lambda}_{j}}\,,
\ee
see for instance \cite{Jog:2006:dot},  
we find that
\eb
\frac{\partial \log \bU}{\partial \bC} = \sum_{k=1}^3  \frac{1}{2}\widehat{\lambda}_{k}^{-1} \bP_k \otimes \bP_k  + 
 \sum_{k=1}^3 \sum^3_{\substack{j=1\\ k\neq j}}  \frac{\frac{1}{2}(\log\widehat{\lambda}_k)-\frac{1}{2}(\log\widehat{\lambda}_j)}{\widehat{\lambda}_{k}-\widehat{\lambda}_{j}}(\bP_k \boxtimes \bP_j + \bP_j\boxtimes\bP_k)\,.
\ee
The second derivative for the linearization is given by
\begin{align}
 \nonumber \frac{\partial^2 \log \bU}{\partial \bC \partial \bC} =& \sum_{k=1}^3  \bP_k\otimes\bP_k\otimes\pp{\frac{1}{2}\widehat{\lambda}_{k}^{-1}}{\bC} + \frac{1}{2}\widehat{\lambda}_{k}^{-1}\left[\left(\frac{\partial \bP_k}{\partial \bC}\otimes\bP_k\right)^{\substack{35\\T}\substack{46\\T}} + \bP_k\otimes\frac{\partial\bP_k}{\partial \bC}\right]\\
 \nonumber                                          &+ \frac{1}{2}({\rm log}\widehat{\lambda}_k) \frac{\partial^2 \bP_k}{\partial \bC \partial \bC}  + \pp{\bP_k}{\bC}\otimes\pp{\frac{1}{2}\widehat{\lambda}_k^{-1}}{\bC}\\
 \nonumber                                         =& \sum_{k=1}^3 -\frac{1}{2}\widehat{\lambda}_{k}^{-2} \bP_k\otimes\bP_k\otimes\bP_k + \frac{1}{2}\widehat{\lambda}_k^{-1}\left[\left(\frac{\partial \bP_k}{\partial \bC}\otimes\bP_k\right)^{\substack{35\\T}\substack{46\\T}} + \bP_k\otimes\frac{\partial\bP_k}{\partial \bC}\right]\\
                                                    &+ \frac{1}{2}({\rm log}\widehat{\lambda}_k) \frac{\partial^2 \bP_k}{\partial \bC \partial \bC}  + \frac{1}{2} \widehat{\lambda}_k^{-1} \pp{\bP_k}{\bC}\otimes\bP_k\,,
\end{align}
where
\begin{align}
 \nonumber \frac{\partial^2 \bP_k}{\partial \bC \partial \bC} =& \sum^3_{\substack{j=1\\ k\neq j}} \frac{1}{\widehat{\lambda}_{k}-\widehat{\lambda}_{j}}\left[ \left(\bP_j^T \otimes \left(\pp{\bP_k}{\bC}\right)^{\substack{12\\T}}\right)^{\substack{23\\T}} + \left(\bP_k\otimes\pp{\bP_j}{\bC}\right)^{\substack{23\\T}} \right.\\
 \nonumber                                                    &+ \left. \left(\bP_k^T \otimes \left(\pp{\bP_j}{\bC}\right)^{\substack{12\\T}}\right)^{\substack{23\\T}} + \left(\bP_j\otimes\pp{\bP_k}{\bC}\right)^{\substack{23\\T}} \right] \\
                                                              &+ \frac{1}{\left(\widehat{\lambda}_k -\widehat{\lambda}_j\right)^2} \left(\bP_k\boxtimes\bP_j + \bP_j \boxtimes \bP_k\right)\otimes\left(\bP_j-\bP_k\right)\,.
\end{align}
The exponent and the logarithm of an arbitrary symmetric tensor may also be be expressed with help of a Taylor expansion of the form 
\begin{align}
{\rm exp}(\bullet) &= \bone + \sum_{m =1}^\infty  \frac{1}{m!}(\bullet)^m\qquad \mbox{and} \\
{\rm log}(\bullet) &=\sum_{k=1}^\infty \frac{-1^{k-1}}{k}[(\bullet)-\bone]^k\,,
\end{align}
where the latter is convergent in a neighborhood of $\bone$.

\subsection{\hspace{-5mm}. Conjugate stress tensors}

The following considerations are adapted from \cite{Ogd:1997:nle}.

The constitutive equation for the stresses are derived form the (isothermal) entropy inequality 
\eb
\dot{\psi} - \langle\bP,\dot{\bF}\rangle\geq0\,,
\ee
From the latter we deduce the constitutive relation $\bP = \partial_{\bF}\psi$.
Let the generalized Lagrangean  strain measures 
\eb
\bE^{(m)} = \begin{cases}
          \frac{1}{2}\left(\bU^{m}-\bone\right) \quad & m\neq 0 \\
          \frac{1}{2}{\rm log}\,\bU\quad &m=0
          \end{cases}
\ee
and Eulerian strain measures
\eb
\bK^{(m)} = \begin{cases}
          \frac{1}{m}\left(\bV^{m}-\bone\right) \quad & m\neq 0 \\
          {\rm log}\bV\quad &m=0
          \end{cases}
\ee
be given, we aim to find the corresponding constitutive equations. 
The so called stress power may be written as
\eb
\langle\bP,\dot{\bF}\rangle= \langle\Btau,\bD\rangle = \langle\bS,\dot{\bE}\rangle=\langle\partial_{\bE^{(m)}}\psi,\dot{\bE}^{(m)}\rangle=\langle\partial_{\bK^{(m)}}\psi,\dot{\bK}^{(m)}\rangle \,,  \label{eq: stress power}
\ee
where $\bD = \frac{1}{2}(\bL+\bL^T)$ and $\bL = \grad\dot{\bx}$.
Considering that $\dot{\bE}=\frac{1}{2}\left(\dot{\bF}^T\bF+\bF^T\dot{\bF}\right)$, we obtain the relations 
\eb
\bP = \bF\bS = \Btau\bF^{-T}\,.
\ee
The pairs in Eq.~(\ref{eq: stress power}) are said to be work conjugate.
By making use of the fact that $\bR^T\dot{\bR}= -\dot{\bR}^T\bR$, we may rewrite  $\dot{\bE}= \frac{1}{2}\left(\bU\dot{\bU}+\dot{\bU}\bU\right)$ and we are able to reformulate
\eb
\Big\langle\bS,\frac{1}{2}\left(\bU\dot{\bU}+\dot{\bU}\bU\right)\Big\rangle = \Big\langle\underbrace{\frac{1}{2}\left(\bS\bU+\bU\bS\right)}_{\bT_{\rm Biot}},\dot{\bU}\Big\rangle\,,
\ee
such that we directly obtain the Biot stress $\bT_{\rm Biot}=\partial_{\bU}\psi^\#(\bU)$, work conjugate to $\bU$ from the entropy inequality. 
With $\bU = \bR^T\bV\bR$ we are able to relate $\bE^{(m)}$ and $\bK^{(m)}$ and the corresponding time derivatives as follows:
\eb
\bE^{(m)} = \bR^T\bK^{(m)}\bR \quad \mbox{and} \quad \dot{\bE}^{(m)} = \bR^T\dot{\bK}^{(m)}\bR + \bE^{(m)}\bR^T\dot{\bR} -\bR^T\dot{\bR}\bE^{(m)}\,.
\ee
Regarding the generalized  stress-power it follows
\eb
\langle\partial_{\bE^{(m)}}\psi,\dot{\bE}^{(m)}\rangle = \langle\bR(\partial_{\bE^{(m)}}\psi)\bR^T,\dot{\bK}^{(m)}\rangle+\langle\left[(\partial_{\bE^{(m)}}\psi)\bE^{(m)}-\bE^{(m)}(\partial_{\bE^{(m)}}\psi)\right],\bR^T\dot{\bR}\rangle 
\ee
\textbf{Only if} $\bE^{(m)}\partial_{\bE^{(m)}}\psi = \partial_{\bE^{(m)}}\psi \bE^{(m)}$, i.e. $\partial_{\bE^{(m)}}\psi$ \textbf{is coaxial with} $\bE^{(m)}$, it immediately follows that
the constitutive law results in
\eb
\pp{\psi}{\bK^{(m)}} = \bR(\partial_{\bE^{(m)}}\psi)\bR^T\label{eq: K_E_coax}
\ee
and the stress power is expressed through $\langle\dot{\bK}^{(m)}, \bR(\partial_{\bE^{(m)}}\psi)\bR^T\rangle$,
which is identical to Eq.~(\ref{eq: stress power}). \\
The case that $\partial_{\bE^{(m)}}\psi$ is coaxial with $\bE^{(m)}$ implies that also  $\partial_{\bE^{(m)}}\psi$ and $\bU$ are coaxial. 
Under this assumption one may show that 
\eb
\partial_{\bE^{(m)}}\psi = \bT_{\rm Biot} \bU^{-(m-1)} = \bU^{-(m-1)} \bT_{\rm Biot}\,,
\ee
and it follows that 
\eb
\partial_{\bE^{(m)}}\psi = \bT_{\rm Biot}\bU^{-m+1} \quad \mbox{and} \quad \partial_{\bE^{(0)}}\psi = \partial_{\log\bU}\psi = \bT_{\rm Biot} \bU = \bR^T\Btau\bR\,,
\ee
\textbf{for isotropic materials}. Inserting the latter result in Eq.~(\ref{eq: K_E_coax})
we obtain the relation 
\eb
\Btau = \pp{\psi({\rm log}\bV)}{{\rm log}\bV}\,.
\ee
Regarding the conjugate stress to $\log\bU$ the reader is also referred to \cite{Hil:1970:cif,Hil:1978:aoi}  and \cite{Hog:1987:tsc}.

\end{appendix}
\end{document}